\documentclass[12pt,preprint]{aastex}
\usepackage[backref,breaklinks,colorlinks,citecolor=blue]{hyperref}
\usepackage[all]{hypcap}
\usepackage{epsfig}
\usepackage{xcolor}

\def\apj{ApJ}
\def\aj{AJ}
\def\apjl{ApJ}
\def\apjs{ApJ}

\def\pasp{PASP}
\def\araa{Ann. Rev. Astr. Ap}
\def\mnras{MNRAS}

\def\aap{A\&A}

\def\araa{ARA\&A}
\def\pasa{PASA}

\begin{document}


\title{Compact Bright Radio-loud AGNs -- III.
A Large VLBA Survey at 43~GHz}

\author{X.-P. Cheng\altaffilmark{1,2,3}, T. An\altaffilmark{1,4}, S. Frey\altaffilmark{5,6}, X.-Y. Hong\altaffilmark{1,2,4}, X. He\altaffilmark{1}, K.~I. Kellermann\altaffilmark{7}, M.~L. Lister\altaffilmark{8}, B.-Q. Lao\altaffilmark{1}, X.-F. Li\altaffilmark{1,2}, P. Mohan\altaffilmark{1}, J. Yang\altaffilmark{1,9}, X.-C. Wu\altaffilmark{1}, Z.-L. Zhang\altaffilmark{1}, Y.-K. Zhang\altaffilmark{1,2}, W. Zhao\altaffilmark{1,4}}

\altaffiltext{1}{Shanghai Astronomical Observatory, Chinese Academy of Sciences, Shanghai 200030, China; e-mail: antao@shao.ac.cn}
\altaffiltext{2}{University of Chinese Academy of Sciences, 19A Yuquanlu, Beijing 100049, China}
\altaffiltext{3}{Korea Astronomy and Space Science Institute, 776 Daedeok-daero, Yuseong-gu, Daejeon 34055, Korea}
\altaffiltext{4}{Key Laboratory of Radio Astronomy, Chinese Academy of Sciences, 210008 Nanjing, China}
\altaffiltext{5}{Konkoly Observatory, Research Centre for Astronomy and Earth Sciences, Konkoly Thege Mikl\'os \'ut 15-17, H-1121 Budapest, Hungary}
\altaffiltext{6}{Institute of Physics, E\"otv\"os Lor\'and University, P\'azm\'any P\'eter s\'et\'any 1/A, H-1117 Budapest, Hungary}
\altaffiltext{7}{National Radio Astronomy Observatory, 520 Edgemont Rd., Charlottesville, VA 22903, USA}\author{}
\altaffiltext{8}{Department of Physics, Purdue University, 525 Northwestern Avenue, West Lafayette, IN 47907, U.S.A.}
\altaffiltext{9}{Department of Earth and Space Sciences, Chalmers University of Technology, Onsala Space Observatory, SE-43992 Onsala, Sweden}

\begin{abstract}
We present the observational results from the 43-GHz Very Long Baseline Array (VLBA) observations of 124 compact radio-loud active galactic nuclei (AGNs) that were conducted between 2014 November and 2016 May.
The typical dimensions of the restoring beam in each image are about 0.5~mas $\times$ 0.2~mas.
The highest resolution of 0.2 mas corresponds to a physical size of 0.02 pc for the lowest redshift source in the sample.
The 43-GHz very long baseline interferometry (VLBI) images of 97 AGNs are presented for the first time.
We study the source compactness on
mas 
and sub-mas scales, and suggest
that 95 sources in our sample
are suitable for future space VLBI observations.
By analyzing our data supplemented with other VLBA AGN surveys from literature, we find that the core brightness temperature increases with increasing frequency below a break frequency $\sim$7~GHz, and decreases between $\sim$7--240~GHz but
increases 
again above~240 GHz in the rest frame of the sources.
This
indicates 
that the synchrotron opacity changes from optically thick to thin.
We also find a strong statistical correlation between radio and $\gamma$-ray flux densities.
Our correlation is tighter than those in literature derived from
lower-frequency 
VLBI data, suggesting that the $\gamma$-ray emission is produced more co-spatially with the 43-GHz VLBA core emission.
This correlation can also be extrapolated to the un-beamed AGN population,
implying that a universal $\gamma$-ray production mechanism
might be 
at work for all types of AGNs.
\end{abstract}


\section{Introduction}
\label{intro}
Active galactic nuclei (AGNs) host the most powerful natural particle accelerators, producing also high-energy cosmic rays and neutrino emission.
Compact jets emanating from around the central supermassive black holes (SMBHs) in radio-active AGNs are prominent sources in the widest range of electromagnetic wavebands, from radio wavelengths to $\gamma$-rays.
AGN jets are in the forefront of modern multi-messenger astrophysical research
\citep[see][for a recent review]{2019ARA&A..57..467B}. Surveys of AGNs supply a wealth of information for advancing our understanding of jet physics. Of particular importance are high-resolution radio interferometric surveys that zoom directly into the parsec (pc) and sub-pc scale regions of AGN jets which are closely related to the extreme astrophysical phenomena.

Recently, the source of high-energy cosmic neutrinos ($\sim$300~TeV) detected by IceCube on 2017 September 22 was identified as a distant $\gamma$-ray blazar, TXS~0506+056, which is an intermediate synchrotron-peaked BL Lac object at a redshift of $z=0.34$ \citep{2018arXiv181107439H,2018Sci...361.1378Ia,2018Sci...361..147Ib}.  This first confirmed that blazars are sources of high-energy astrophysical-origin neutrinos, opening a new window of studying the Universe using the unobstructed messenger neutrino \citep{2018Sci...361..147Ib}. It is widely speculated that high-energy cosmic rays are accelerated in the magnetic fields in the innermost jets of blazars. The generated cosmic rays interact with nearby gas, photons or other cosmic rays producing charged mesons that decay into high-energy neutrinos, $\gamma$-rays and other particles \citep{1960ARNPS..10....1R}.

Very long baseline interferometry (VLBI) is an elegant observing technique which provides the highest angular resolution. Blazars are among the most powerful objects in the Universe. Images from high-frequency high-resolution VLBI surveys of blazars are essential to test jet models \citep[e.g.,][]{1995PNAS...9211439M}, and to investigate the innermost region of compact jets where the acceleration and collimation of the relativistic plasma flow takes place \citep{1969ApJ...155L..71K,2004ApJ...613..794G,2016AJ....152...12L} and where the high-energy neutrinos and $\gamma$-rays are produced \citep{2019arXiv191201743R}.

The Large Area Telescope on-board the {\it Fermi} $\gamma$-ray space telescope ({\it Fermi}-LAT) has already detected 2683 AGNs, listed in the fourth {\it Fermi}-LAT catalog \citep{2019F}. Most of the sources are classified as blazars (comprising BL Lacs and optically violently variable quasars), which show prominent emission over a broad range of electromagnetic radiation, from radio to $\gamma$-ray energies. The spatial localization of the region where the $\gamma$-rays are emitted, $\gamma$-ray radiation mechanism in blazars, and correlation between radio and $\gamma$-rays are key questions to understand the blazar activity at multiple bands. Synchrotron radiation is responsible for the bump at radio to X-ray frequencies in the $\log \nu F_{\nu}$ versus $\log \nu$ spectral energy distribution (SED). Another dominant mechanism responsible for the bump from X-ray to TeV $\gamma$-ray regions is inverse-Compton (IC) radiation. Two possible scenarios are attributed to this: synchrotron self-Compton (SSC) radiation which results from IC scattering of synchrotron radiation by the same relativistic electrons that produced the synchrotron radiation, and external inverse-Compton (EC) radiation where the photons available for IC scattering in the inner jet are seeded from external sources such as the broad-line region \citep{1994ApJ...421..153S} and the accretion disk \citep{1993ApJ...416..458D}. One method to verify the process of SSC radiation is using the correlation between the mm-wavelength radio luminosity of the core and its $\gamma$-ray luminosity.

Some VLBI observations of blazars at high frequencies \citep[e.g.,][]{2008AJ....136..159L,2010ApJS..189....1A,2010ApJ...723.1150P} have revealed complex inner jet morphology and kinematics at sub-milliarcsecond (sub-mas) scales. New statistical studies show that the jets are accelerated in the sub-pc regions from the central engine \citep{2016ApJ...826..135L}.
However, non-ballistic motion of the jet is found within a few parsec, e.g., NRAO~150 \citep[][]{2007A&A...476L..17A} from observations with the Global mm-VLBI Array (GMVA) and the Very Long Baseline Array (VLBA) at 86 and 43 GHz, respectively.
Thus, high-resolution VLBI observations and large complete surveys at high frequencies are important to study the jet structures and kinematics at sub-pc scales.

Ground-based VLBI observations show typical apparent core brightness temperature ($T_{\rm b}$) in the range of $10^{11}- 10^{12}$~K \citep[e.g., ][]{1996AJ....111.2174M}. However, the Japanese VLBI Space Observatory Programme \citep[VSOP,][]{1998Sci...281.1825H} detected some AGNs with core brightness temperature higher than $10^{12}$~K \citep{2001ApJ...549L..55T,2008ApJS..175..314D}. The highest brightness temperature that has been measured with VSOP is $5.8 \times 10^{13}$~K for AO\,0235+164 at 5~GHz \citep{2000PASJ...52..975F}. The maximum brightness temperature is determined by the longest baseline length, regardless of the observing frequency \citep{2005AJ....130.2473K}. The Russia-led {\it RadioAstron} mission \citep{2013ARep...57..153K} detected core brightness temperatures even higher than $10^{13}$~K \citep[e.g.,][]{2016ApJ...820L...9K,2018MNRAS.474.3523P, 2018MNRAS.475.4994K}, about two orders of magnitude above the equipartition \citep{1994ApJ...426...51R} and inverse Compton limits \citep{1969ApJ...155L..71K}.
The interpretation of extremely high brightness temperatures is a challenge to AGN physics. Moreover, \citet{2008AJ....136..159L} showed that the core brightness temperature will be small at lower frequencies due to opacity effect between 2~GHz and 15~GHz. However, their 86-GHz core brightness temperatures are significantly lower than those measured at 15~GHz. Therefore 43-GHz VLBI observations, straddling 15 and 86 GHz, are crucial to explore the maximum core brightness temperature and its corresponding frequency and to study core brightness temperature distribution along the jet which can be used to test models of the inner jet \citep{1995PNAS...9211439M}.

Despite more than 30 years of research of radio-loud AGNs at high frequencies with VLBI, only 163 sources were observed and imaged at 43~GHz \citep{2001ApJ...562..208L,2010AJ....139.1713C,2017ApJ...846...98J,2018ApJS..234...17C} and 263 AGNs have been successfully imaged at 86~GHz \citep{2008AJ....136..159L,2018ApJS..234...17C,2019A&A...622A..92N}.
Most of these observations were carried out in snapshot mode, and high-quality images are scarce. Even fewer sources have multi-epoch imaging observations.
In addition, the resolution is still not enough to explore the innermost jet emission region where the jet is formed and accelerated.
Future space VLBI missions would leap forward in the direction of improving both the resolution and the imaging capability.
A systematic survey of a large AGN sample is necessary to make  progress of the future space VLBI proposals.

In this paper, we present the results of 43-GHz VLBA imaging of 124 AGNs and a statistical study of their compactness, core brightness temperatures, and the correlation between the radio and $\gamma$-ray emissions. The observations and the data reduction process are described in Sect.~\ref{observ}.
We present the main results and comment on selected individual sources in Sect.~\ref{results}.
Section~\ref{discussion} contains the discussion of the properties of our sample, and a summary appears in Sect.~\ref{summary}.
Throughout this paper, the standard $\Lambda$CDM cosmological model with H$_{0}$ = 73~km\,s$^{-1}$\,Mpc$^{-1}$, $\Omega_{\rm M} = 0.27$, and $\Omega_{\Lambda} = 0.73$ is adopted.

\section{VLBA Observations and Data Reduction}
\label{observ}

\subsection{Sample Selection}

As mentioned in Section \ref{intro}, a large sample of compact and bright AGNs is important for the successful detection and imaging to support the scientific operation of ground-based mm-wavelength and space VLBI.
To enlarge and complete the existing database, we selected and observed 134 AGNs. The selection criteria are described in details in Paper~I \citep{2014Ap&SS.352..825A} and Paper~II \citep{2018ApJS..234...17C}. Ten bright AGNs were selected from our sample and observed at 43 and 86 GHz with long integration time and good $(u,v)$ coverage, in order to further study the inner jet structures \citep{2018ApJS..234...17C}. The remaining 124 sources consist of 97 quasars, 15 BL Lac objects, 6 radio galaxies, and 6 objects with no optical identification. We carried out an imaging survey at 43 GHz using the VLBA.
Combined with the previous 43-GHz VLBI observations \citep{1992vob..conf..205K,2001ApJ...562..208L,2002ApJ...577...85M,2010AJ....139.1695L}, we obtain a large sample containing 260 AGNs to study brightness temperatures of AGN cores and to explore the jet acceleration mechanism.
Table \ref{tab:sample} lists general information of the sample: IAU source name (B1950.0), commonly used other source name, observation session, right ascension (RA), declination (Dec), redshift, optical classification, an indication of comments on the source in Sect.~\ref{comments}, and $\gamma$-ray photon flux (where available).
The photon flux of the $\gamma$-ray bright AGNs are provided by the {\it Fermi}-LAT in the 100~MeV -- 300~GeV energy range \citep{2015ApJS..218...23A}.

\subsection{Observations}
To ensure the success of the observations, we split the sources into two sub-samples, based on the source flux densities measured at lower radio frequencies.
The first sub-sample consists of 40 bright sources for which we acquired good images, while the second sub-sample includes 84 relatively weaker sources which were observed less thoroughly.
These sub-samples were observed in the corresponding sessions B and C, as listed in Column~3 of Table~\ref{tab:sample}.
Millimeter-wavelength VLBI observations are easily affected by tropospheric fluctuations, so the observing periods were carefully chosen to have exceptionally favorable weather conditions. The observations were spread over a period of almost one year from 2015 June to 2016 April as shown in Column~3 of Table~\ref{tab:log} and in Column~2 of Table~\ref{tab:image}.
Each of the 40 sources in session B was observed for four scans of 7~min duration. Each source in session C was observed for two 7-min scans separated by long time spans in order to obtain relatively uniform $(u, v)$ coverage.
Table \ref{tab:log} summarizes the observation setups.
All ten VLBA antennas were used for the 43-GHz observations.
At the beginning and the end of each observation, calibrator scans were scheduled on some of these bright quasars: 3C\,84, 3C\,273, 3C\,279, 3C\,345, 3C\,454.3, 4C\,39.25, 1749+096, or BL Lac.

\begin{figure}
 \centering
 \begin{tabular}{cc}
 \includegraphics[width=0.5\textwidth]{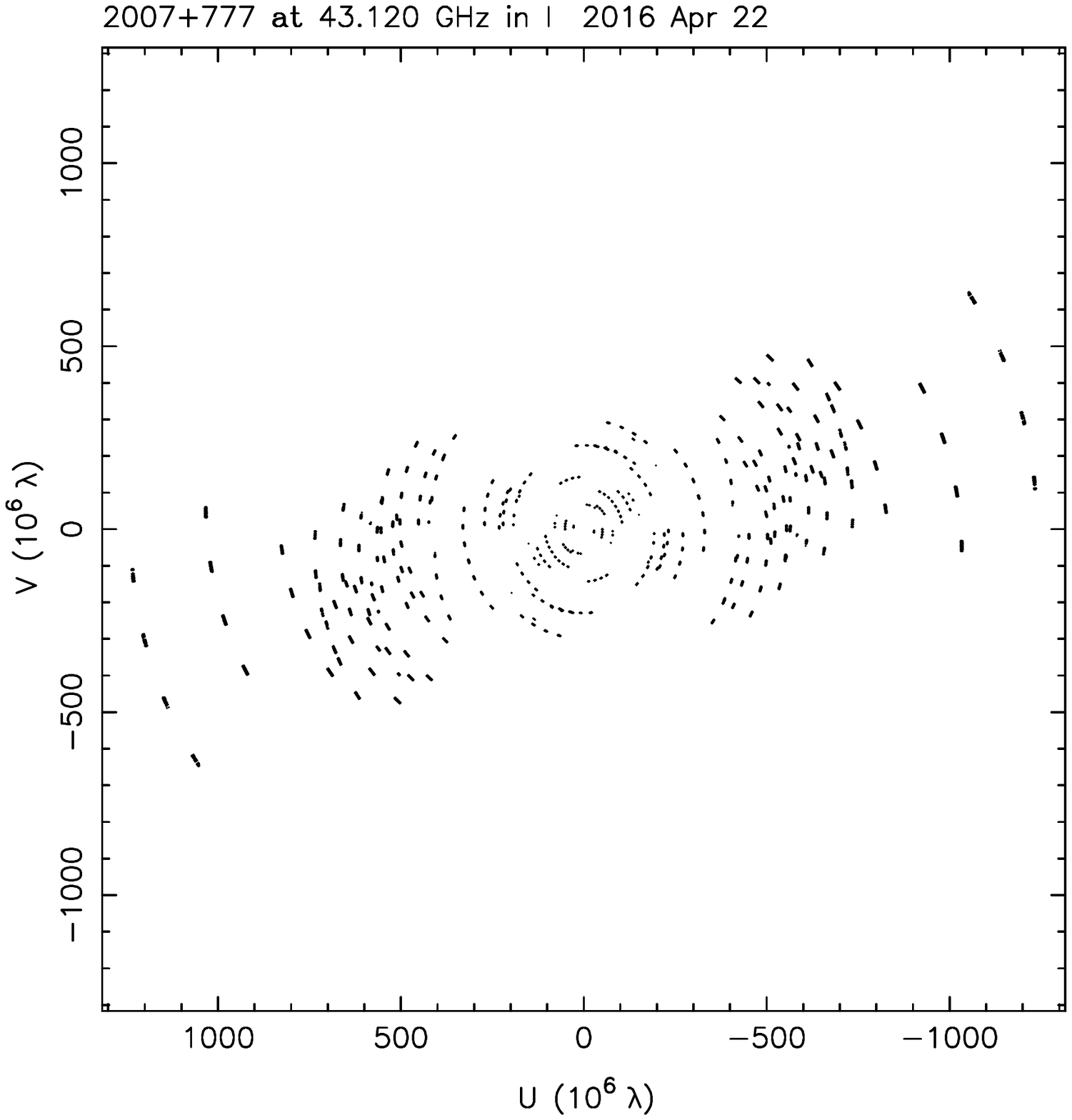} & \includegraphics[width=0.5\textwidth]{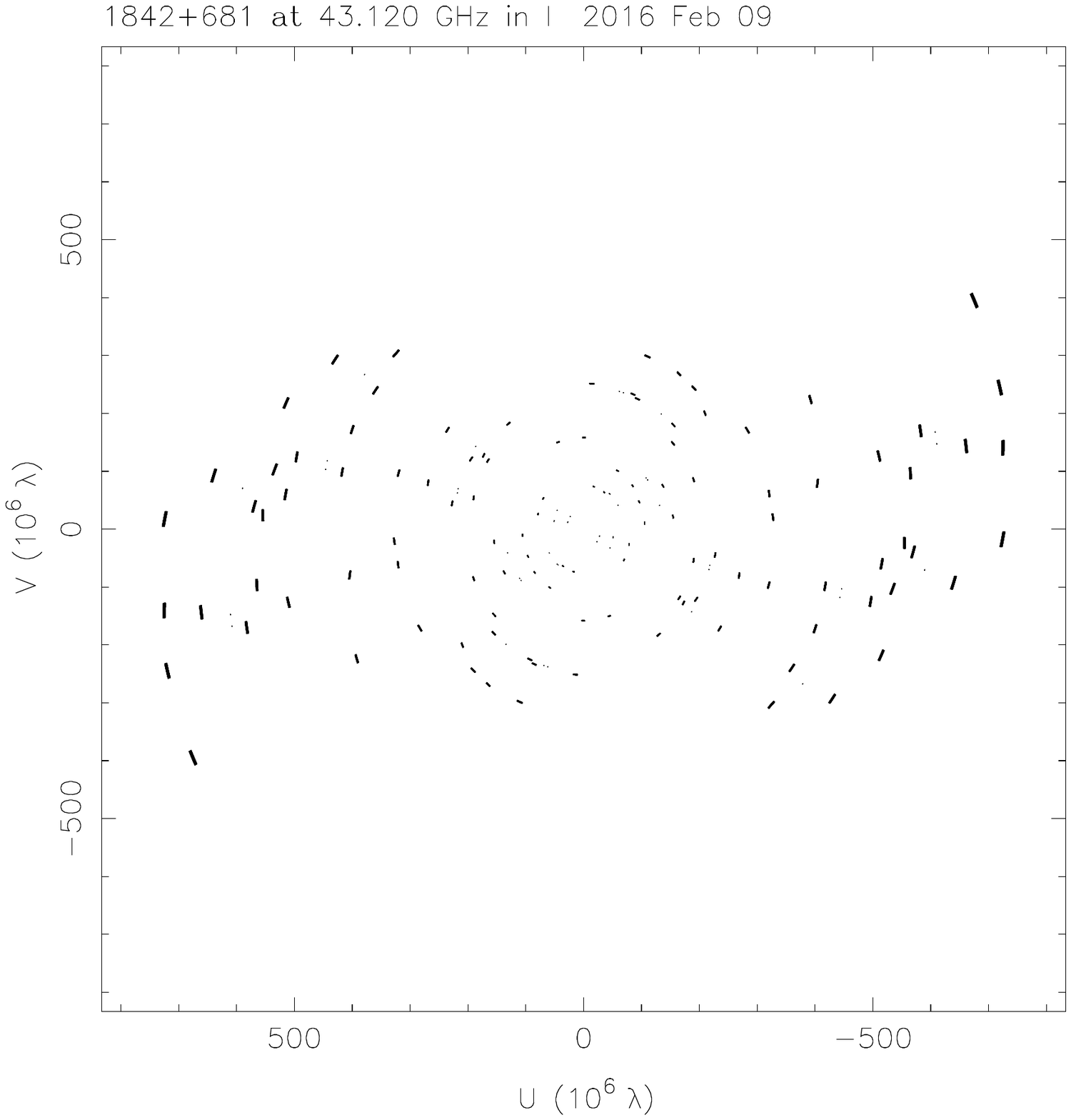}\\
    \end{tabular}
  \caption{Example $(u,v)$ coverages at 43~GHz for a source in the first sample (left panel) and in the second sample (right panel). The labels on the vertical and horizontal axes are in units of million times the observing wavelength ($\lambda$).}
  \label{uv}
  \end{figure}

The data were recorded with 2-bit sampling at an aggregate data rate of 2~Gbps, using 8 intermediate frequency (IF) channels of 32~MHz bandwidth each.
The observations were made in both left and right circular polarizations.
One scan on fringe finder sources every 1.5 h was used to check the recording, pointing, and calibration.
Figure~\ref{uv} shows a typical $(u,v)$ coverage for sources observed in the first (session B) and second (session C) samples in the left and right panels, respectively.
The raw VLBA data were correlated using the DiFX software correlator \citep{2007PASP..119..318D,2011PASP..123..275D} at the Array Operations Center in Socorro (New Mexico, U.S.), with 2~s averaging time, 128 frequency channels per IF, and uniform weighting.
The correlated data were transferred to computer clusters in the China SKA Regional Centre prototype \citep{2019NatAs...3.1030A} where the calibration and imaging analysis were carried out.

\subsection{Data Reduction}

We processed the data in the standard way with the U.S. National Radio Astronomy Observatory (NRAO) Astronomical Imaging Processing Software ({\sc AIPS}) package \citep{2003ASSL..285..109G}.
First we loaded the data to AIPS by the task {\sc fitld}, and flagged the bad data points (typically due to inclement weather) before proceeding further.
We selected Pie Town (PT) as the reference antenna for most of the data.
Fort Davis (FD) was used as the alternative reference telescope. In the amplitude calibration, we first removed the sampler bias with the task {\sc accor}, then calibrated the correlator output with {\sc apcal} and fitted for the ionospheric opacity correction using weather information, antenna system temperature and gain curve tables.
Then we fringe-fitted a short (generally 1 min) scan of calibrator source data to determine the phase and single-band delay offsets and applied the solutions to the calibration table.
After removing instrumental phase offsets from each IF, we performed a global fringe-fit using the task {\sc fring} by combining all IFs to determine the frequency- and time-dependent phase corrections for each antenna and removed them from the data.
To avoid false detections, fringe-fitting was done with a minimum signal-to-noise ratio (SNR) of 5.
After fringe-fitting, the solutions for the sources were applied to their own data by linear phase connection using rates to resolve phase ambiguities.
In the final step, we used the task {\sc bpass} to calibrate the bandpass shapes by fitting a short data scan on a calibrator, and applied the solutions to all data.
Then we made single-source calibrated data sets with {\sc splat}/{\sc split} and used the task {\sc fittp} to export the calibrated visibility data files to the external work space in the clusters. These single source data files were then imported in the {\sc Difmap} program \citep{1997ASPC..125...77S} to carry out self-calibration and imaging. A few runs of phase-only self-calibration were made to eliminate the residual phase errors. Amplitude self-calibration was performed with the amplitude normalization when the {\sc clean} models reached an integrated flux density close to the correlated amplitude on the shortest baselines. The determined overall telescope gain correction factors were found to be small, typically within 10\%, in agreement with other similar 43- and 86-GHz VLBA surveys \citep{2017ApJ...846...98J,2019A&A...622A..92N}.

\subsection{Model Fitting}

We used a number of circular Gaussian components to model the brightness distribution structure of each source in the visibility domain in {\sc Difmap}. Typically 1 to 6 components were used to represent the detected features of the source structure. The radio core is identified as the bright and compact component at the apparent jet base. The fitted parameters are cataloged in Table~\ref{model}. The sizes of these compact jet components are usually smaller than the synthesized beam. The minimum resolvable size of a component in a VLBI observation is given by \citet{2005astro.ph..3225L}:
\begin{equation}
  d_{\rm lim}=\frac{\rm 2^{\rm 1+\rm \beta/2}}{\rm \pi}[\rm \pi \rm B_{\rm maj}B_{\rm min} \ln2 \ln\frac{(\rm S/N)}{(\rm S/N)-1}]^{\rm 1/2},
\end{equation}
where $B_{\rm maj}$ and $B_{\rm min}$ are the major and minor axes of the restoring beam, respectively (full width at half-maximum, FWHM), S/N is the signal-to-noise ratio, and $\beta$ is the weighting function, which is 0 for natural weighting or 2 for uniform weighting.
We took $d_{\rm lim}$ as the upper limit for the components with the fitted size $d < d_{\rm lim}$.

\section{Results}
\label{results}

\subsection{VLBA Images}

\begin{figure}
  \centering
  \begin{tabular}{lcr}
  \includegraphics[width=0.3\textwidth]{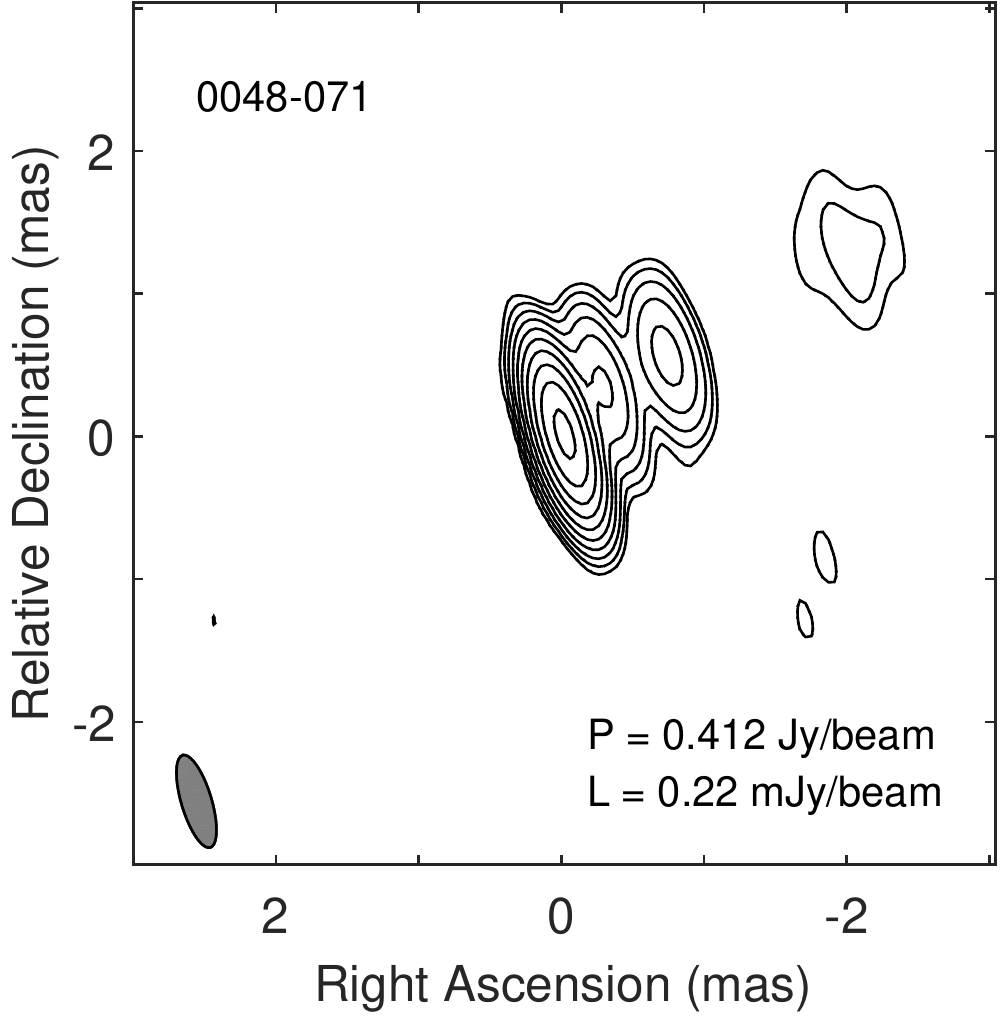} & \includegraphics[width=0.3\textwidth]{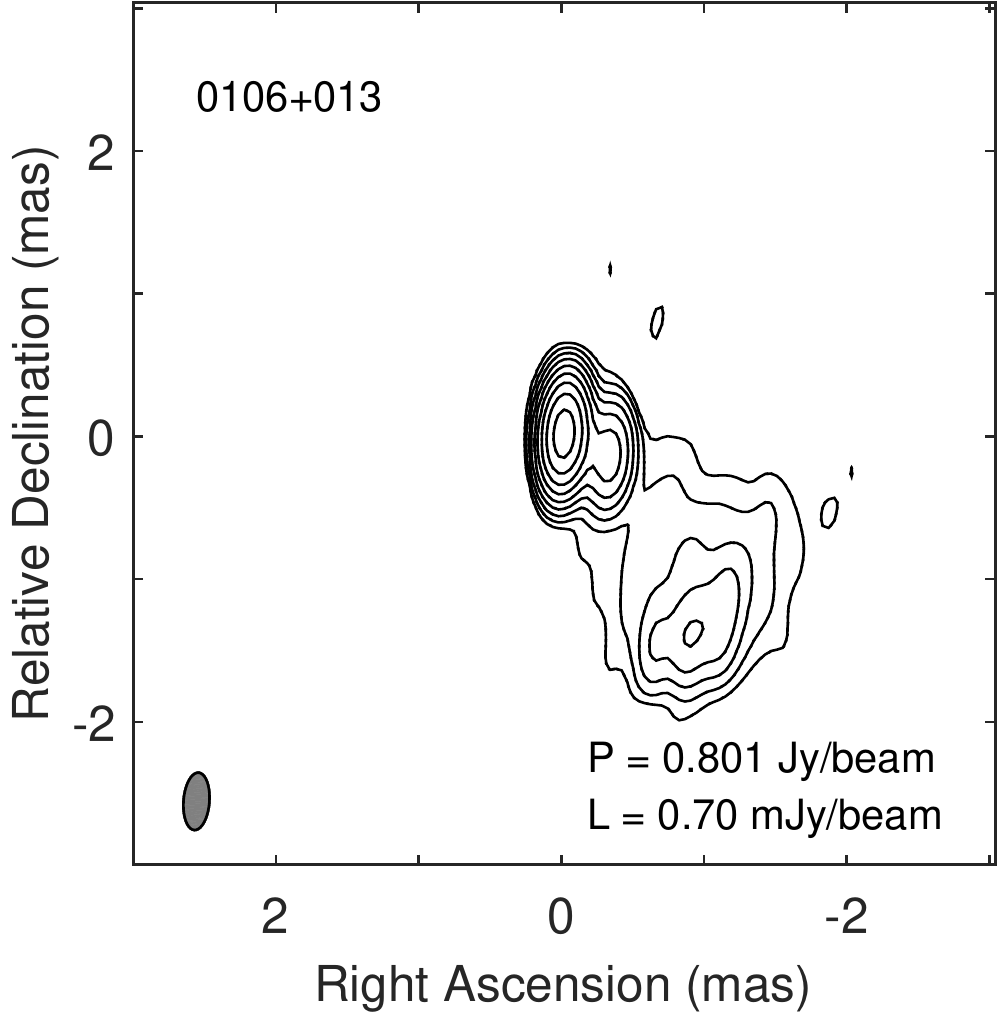}  & \includegraphics[width=0.3\textwidth]{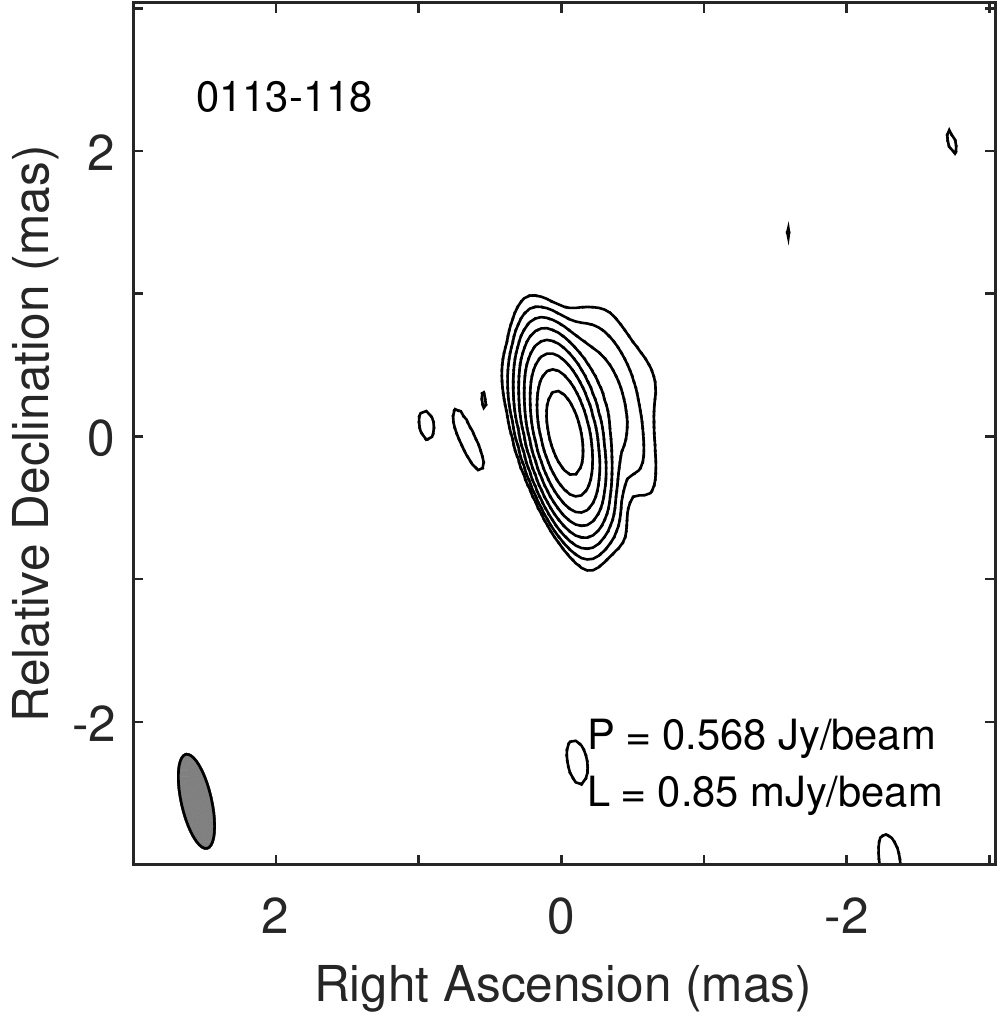}\\ \includegraphics[width=0.3\textwidth]{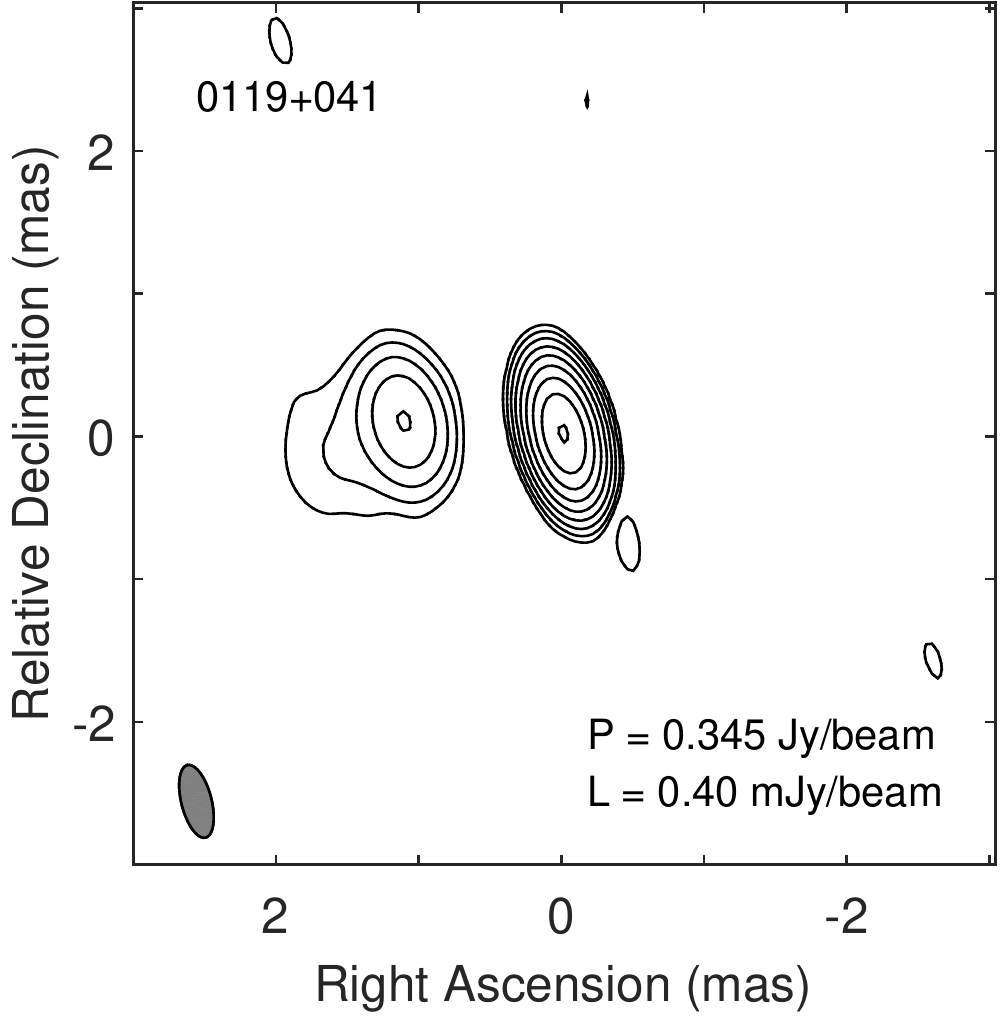} & \includegraphics[width=0.3\textwidth]{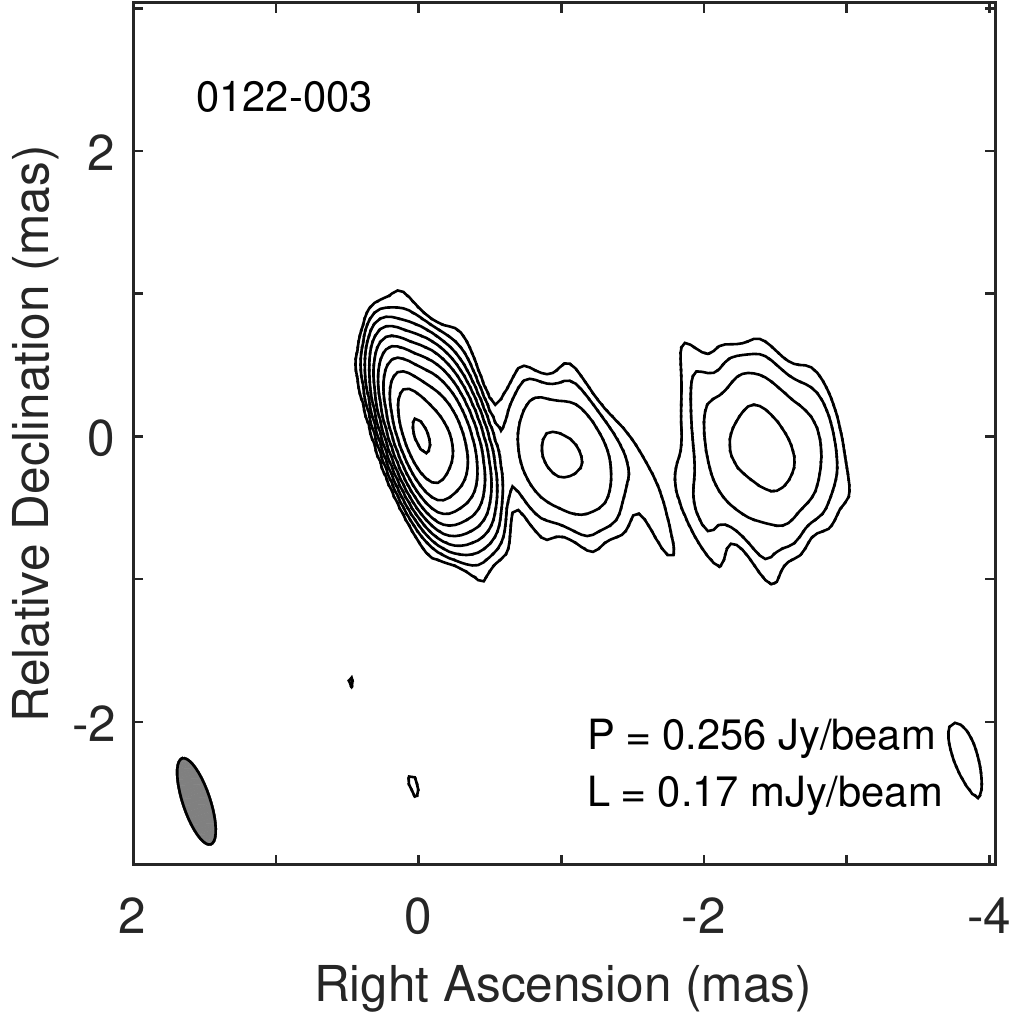}  & \includegraphics[width=0.3\textwidth]{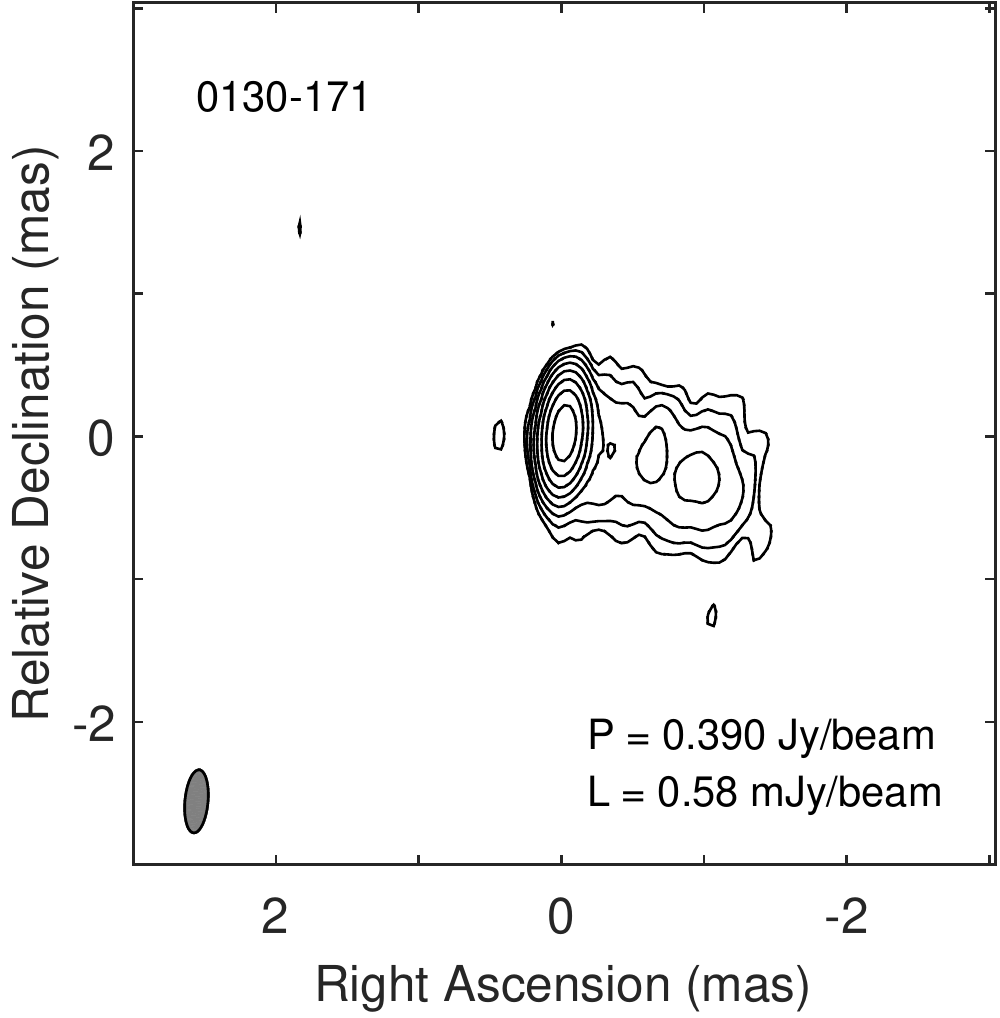}\\
  \includegraphics[width=0.3\textwidth]{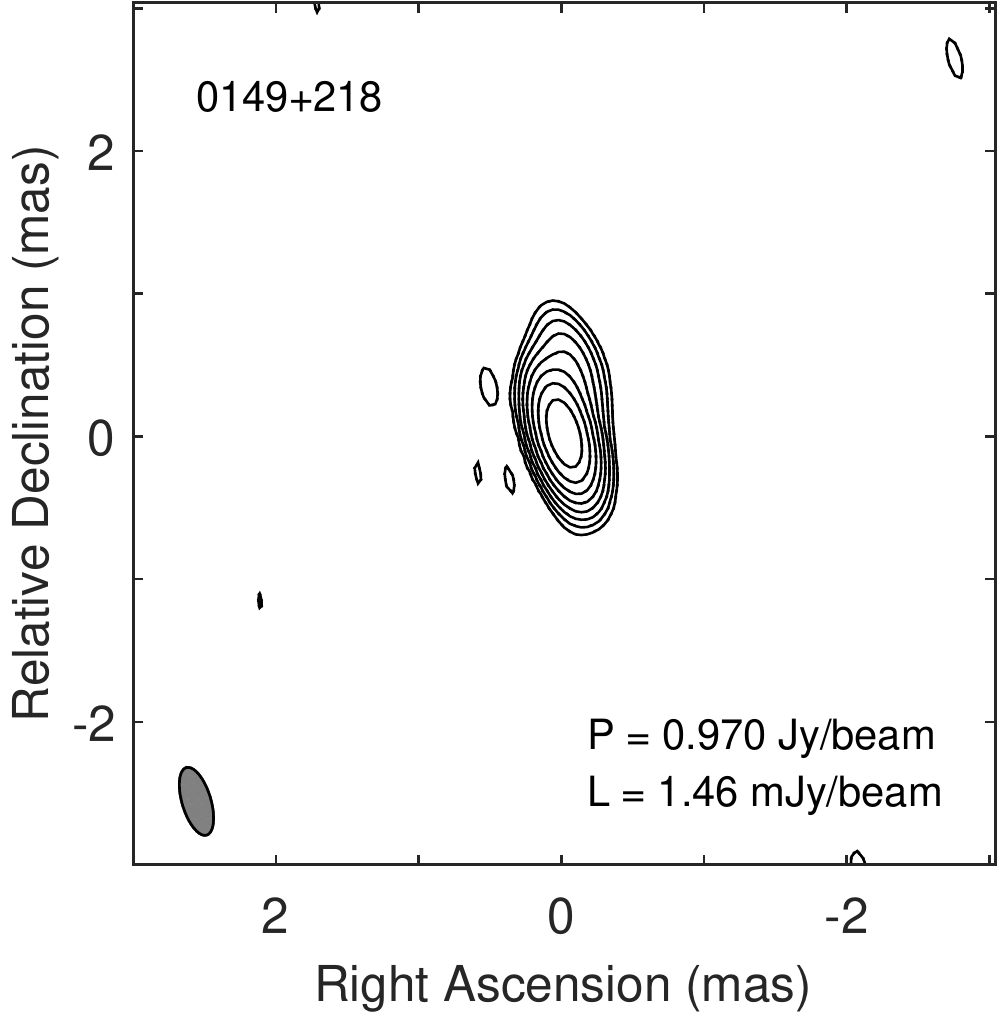} & \includegraphics[width=0.3\textwidth]{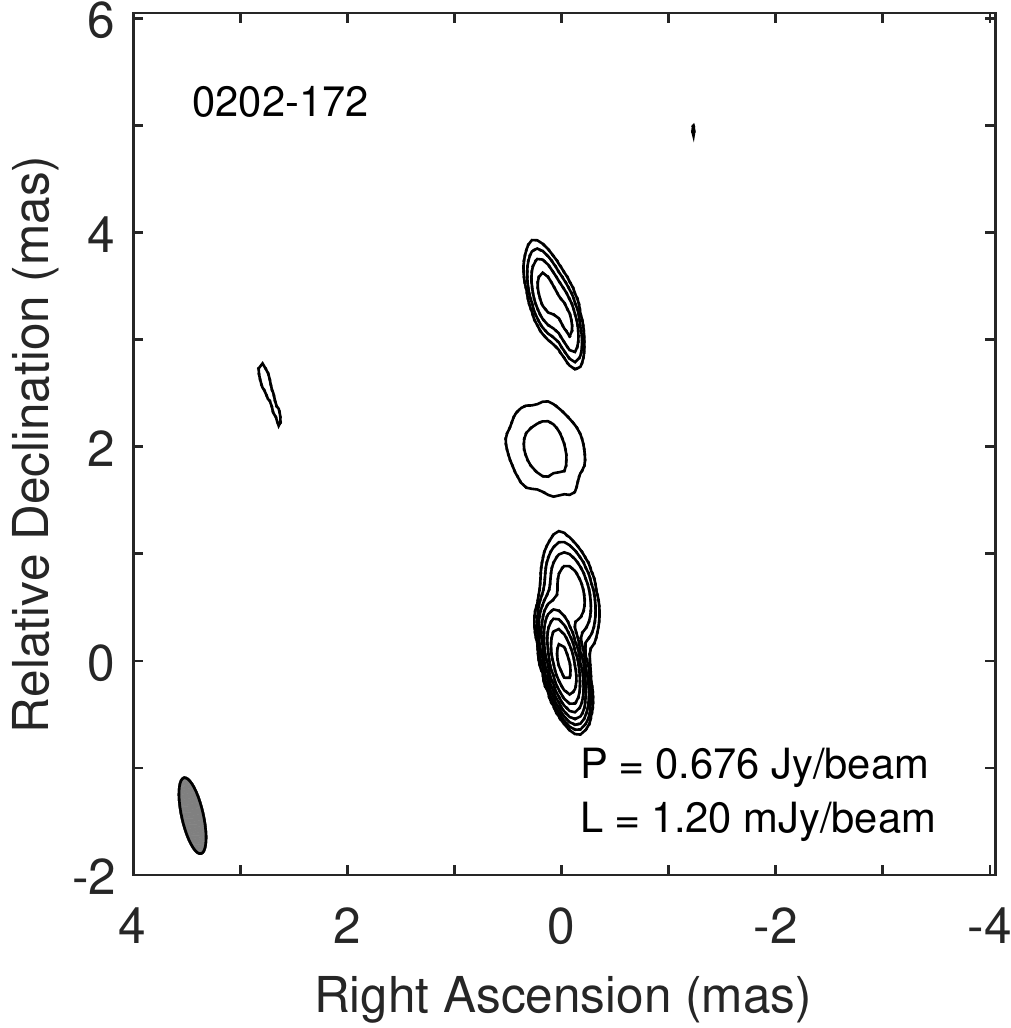}  & \includegraphics[width=0.3\textwidth]{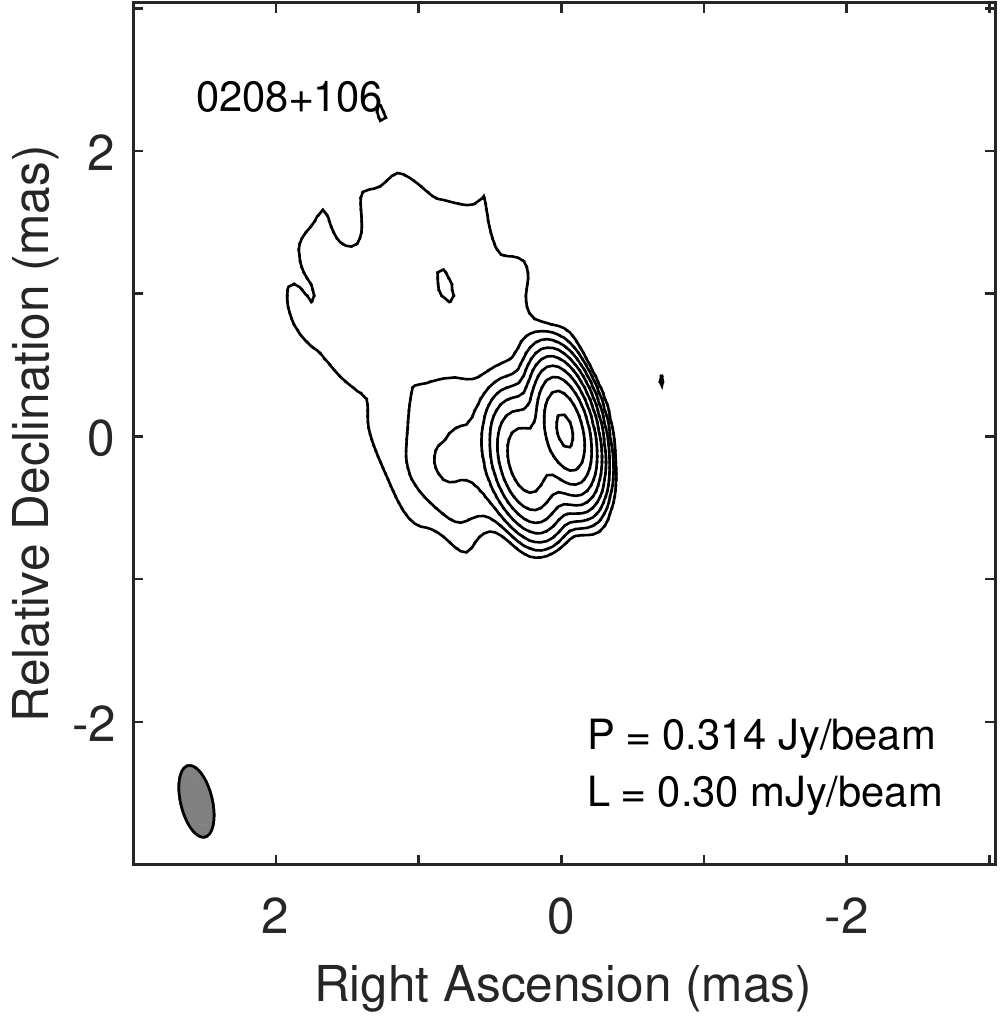}\\ \includegraphics[width=0.3\textwidth]{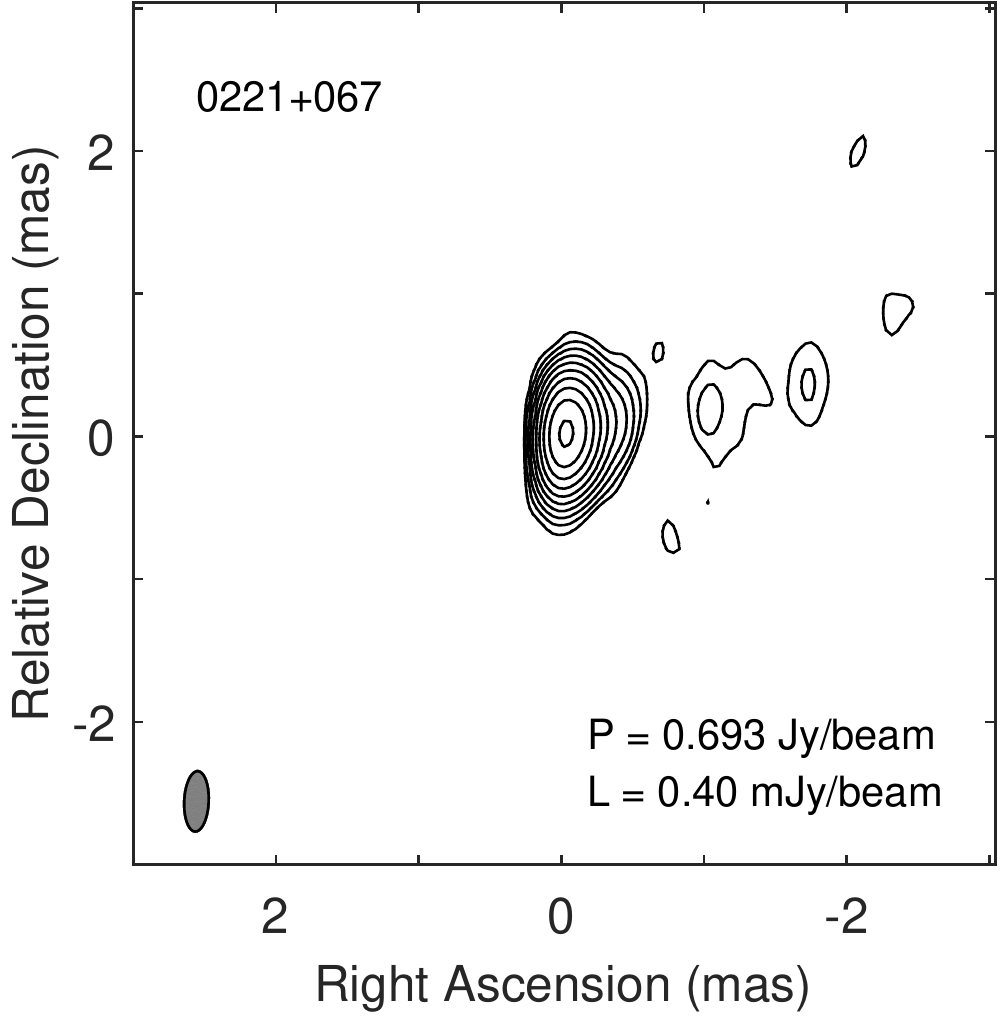} & \includegraphics[width=0.3\textwidth]{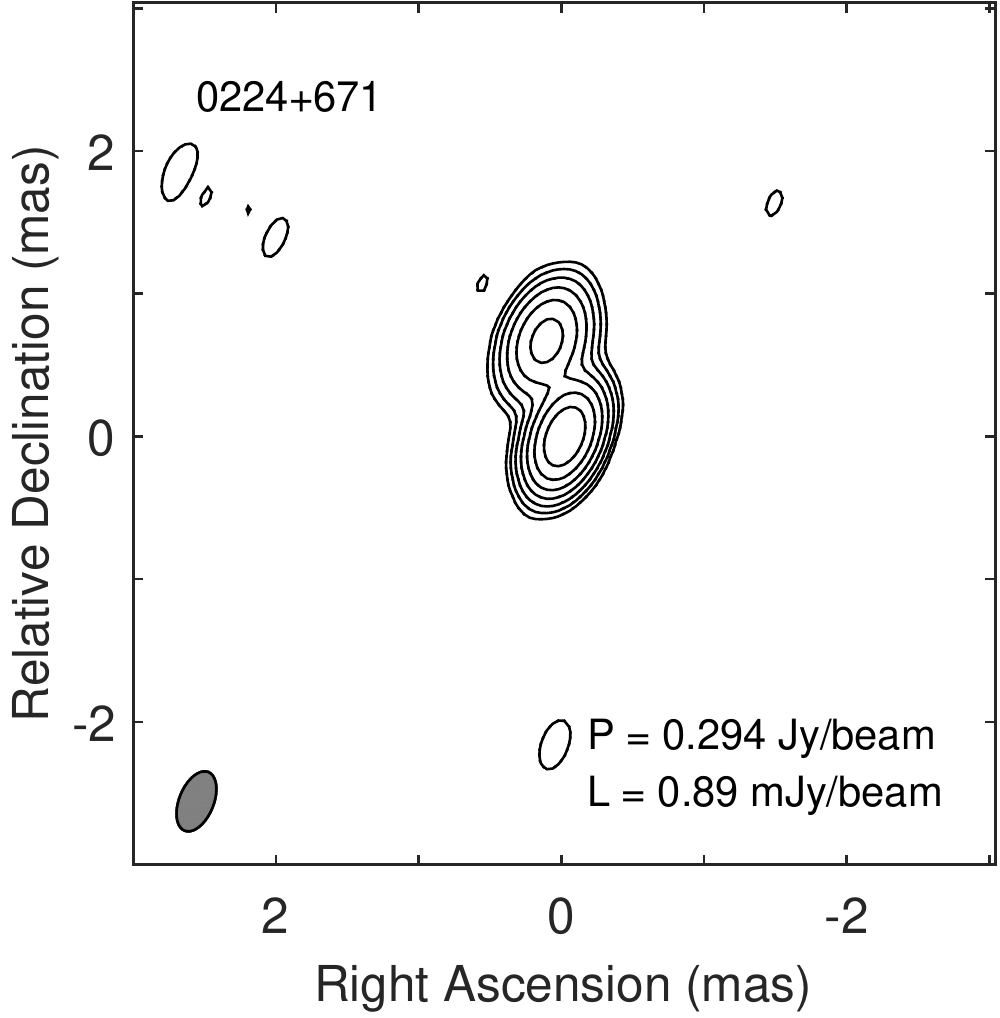}  & \includegraphics[width=0.3\textwidth]{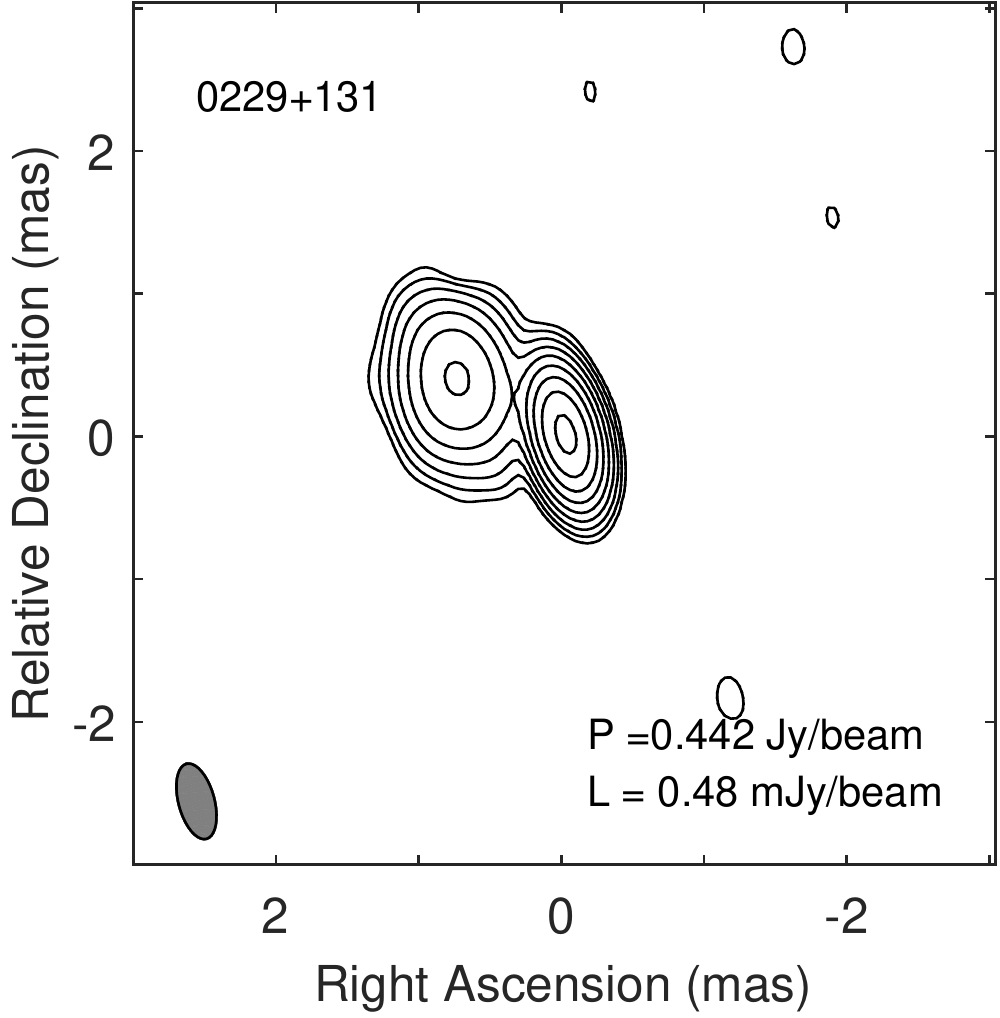}\\
    \end{tabular}
  \caption{Naturally weighted total intensity VLBA images of 124 sources at 43~GHz. The legends in the image: P -- peak intensity (in units of $\rm Jy \, beam^{-1}$); L -- rms noise (in units of $\rm mJy \, beam^{-1}$). The lowest contours are at $\pm 3$ times rms noise and further positive contours are drawn at increasing steps of 2. The image parameters are given in Table~\ref{tab:image}. The gray ellipse in the lower left corner of each image is the restoring beam.}
  \label{image-1}
  \end{figure}

Figure~\ref{image-1} shows contour plots of the final naturally weighted total intensity images for all the 124 sources. For 97 sources, the survey provides the first-ever VLBI image made at 43~GHz, extending the existing database by about 60\%. Most of the sources shown here have secondary features or have complex structures. For these objects, we give brief comments on the characteristics of the radio structure, comparing our observations with other lower-frequency VLBI observations from the literature as discussed in the next subsection. The typical image size is 3~mas $\times$ 3~mas and the dimensions of the restoring beam in each image are 0.5~mas $\times$ 0.2~mas. The elliptical Gaussian restoring beam size  is indicated in the bottom left corner of the maps in Fig.~\ref{image-1}. The peak intensity and the rms noise level are given in the bottom right corner of the maps. The lowest contour represents 3 times the off-source image noise level, and the contours are drawn at $-1$, 1, 2, 4, ..., $2^{\rm n}$ times the lowest level. Table~\ref{tab:image} provides the parameters relevant to the images: source name, observation date, beam size, integrated flux density, peak intensity, off-source rms noise in the {\sc clean} image, correlated flux density on the shortest baselines, the length of the shortest baseline, correlated flux density on the longest baselines, and the length of the longest baseline.

Among the most important parameters derived from high-frequency VLBI surveys are the source compactness and core brightness temperature.
Blazars often shows a one-sided core--jet structure, and the core is usually the most compact and bright component at one end of the jet.
Eight other sources in our sample are identified as compact symmetric objects \citep[CSOs,][]{2012ApJ...760...77A} or gigahertz-peaked spectrum \citep[GPS,][]{1998A&AS..131..303S} sources:  0738+313, 0742+103, 0743$-$006, 1435+638, 2021+614, 2126$-$158, 2134+004, and 2209+236. All these sources, except 2021+614, are GPS sources, which show a sharp low-frequency spectral cutoff near 1~GHz.
The exceptional source 2021+614 is a well-known CSO \citep{2003PASA...20...69P}, but we are unable to identify the core component from our image. Comments on selected sources, including the designation of the core we used for calculating the core brightness temperature and the reliability of faint features in the images are given in Sec.~\ref{comments}. Table \ref{model} lists the model-fitting parameters: source name, observation epoch, component identifier, model flux density, peak intensity, angular separation from the core, size of the components, position angle of the components.

\subsection{Parsec-scale Morphology}

The sources in our sample can be divided into four basic classes based on their morphology: simple core, one-sided core--jet, compact symmetric object, and complex structure.
Eight sources (0847$-$120, 1049+215, 1257+519, 1329$-$049, 1417+385, 1657$-$261, 2325+093, 2342$-$161) have a single core component.
One hundred and twelve sources show a core and a one-sided jet structure. However, 8 sources (0113$-$118, 0221+067, 0708+506, 0736+017, 0805$-$077, 1124$-$186, 1219+044, 2227$-$088) have faint extended radio emission and for 5 sources the cores (0605$-$085, 1036+054, 1045$-$188, 2021+317, 2223+210) are not the brightest but the most compact among the fitted components.
Three sources (0241+622, 0743$-$006, 2021+614) show compact symmetric jet structure. We detect a component exceeding 7$\sigma$ noise level in the counter-jet direction in 0241+622. 0743$-$006 is GPS source that shows a triple structure. 2021+614 is a well-known CSO, showing two-sided structure.
One source (0354+559) shows a complex structure, as shown in Sect. \ref{comments}.

Given our selection criteria, all sources have VLBI images at lower frequencies, at least at 2.3 and 8.4~GHz. Almost all sources also have radio images in literature showing their kpc-scale structure based on Very Large Array (VLA) observations. By comparing the radio structures from pc to kpc scales, we find that there are 6 sources showing completely oppositely-directed jet structures on mas and arcsec scales, and 23 sources display a relatively large change in the apparent jet direction.

\subsection{Comments on Selected Individual Sources}
\label{comments}

0048$-$071 (OB $-$082): Our 43-GHz image shows a single-sided jet to the northwest, in an opposite direction of the kpc-scale lobe (Kharb et al., in prep.)\footnote{http://www.physics.purdue.edu/astro/MOJAVE/sourcepages/0048-071.shtml}.

0106$+$013 (4C $+$01.02): Our image shows that the innermost jet structure ($<1$~mas) is along the east--west direction (in a position angle $\sim -110\degr$), then it bends towards the southwest ($\sim -140\degr$) at about 2~mas from the core. This is in agreement with the low-frequency VLBI images \citep{2009AJ....138.1874L}.

0113$-$118: This source is very compact at 43~GHz, we only detect a core and a faint extension within 1~mas. However, it was not detected on space--ground baselines in the 5-GHz VSOP AGN Survey \citep{2008ApJS..175..314D}.

0119$+$041 (OC $+$033): The brightness temperature of the core is $1.6 \times 10^{10}$~K at 43~GHz, the derived Doppler factor is below unity which is consistent with slow jet motion found in the literature \citep{2009AJ....138.1874L,2012ApJ...758...84P}. Neither the {\it Compton} Gamma Ray Observatory Energetic Gamma Ray Experiment Telescope (EGRET) nor {\it Fermi}/LAT has detected this source. The 15-GHz light curve from the Owens Valley Radio Observatory (OVRO) 40-m radio telescope shows that the flux density does not have significant variation from 2008 to mid-2017 \citep{2011ApJS..194...29R}. The VLBI structure (the core and the eastern jet component) does not show noticeable change \citep{2009AJ....138.1874L,2012ApJ...758...84P}. The overall radio spectrum peaks at about 7~GHz. All these pieces of evidence imply that this is a GPS source, rather than a blazar.

0122$-$003 (UM 321): Our image shows a compact core and three jet components along a straight line to the west. It is in good agreement with the 5- and 15-GHz VLBA images \citep{2000ApJS..131...95F,2013AJ....146..120L}.

0130$-$171 (OC $-$150): The compact core is 0.08~mas in size. We detected a series of jet components extending to the southwest.

0149$+$218: This source shows a compact structure at 43~GHz. A relatively weak component extending to the north is detected.

0202$-$172: The jet shows a centrally symmetric S-shaped morphology within 5~mas. An extended feature is located at $\sim$2 mas from the core, sitting in the gap between the inner jet and the outer 3 mas jet.

0208$+$106 : The jet points to southeast up to 1~mas from the core, then it bends towards northeast seen in the 15-GHz VLBA image obtained in the Monitoring Of Jets in Active galactic nuclei with VLBA Experiments (MOJAVE) survey \citep{2009AJ....138.1874L}, with a 70$\degr$ change in position angle. The outer ($>1$~mas) bent jet is diffuse at 43~GHz.

0221$+$067 (4C $+$06.11): The source shows a  compact core--jet structure within 1 mas. Our high dynamic range image detected two weaker jet components to the west.

0224$+$671 (4C $+$67.05): The core is 0.13 mas in size. A jet component to the north is detected and has the similar flux density with the core at 43 GHz.

0229$+$131 (4C $+$13.14): The source displays considerable emission on both sides of the core at arcsec scales \citep{1995AJ....109.1555P}. Our image shows a bright core and an inner jet pointing to the northeast which is consistent with the high-resolution 5-GHz VSOP image \citep{2008ApJS..175..314D}.

0239$+$108 (OD $+$166): It shows a compact core--jet structure. According to the total flux density light curve from the 15-GHz OVRO monitoring program \citep{2011ApJS..194...29R}, our 43-GHz observation was made in a fading phase of the source.

0241$+$622 : This is a low-redshift ($z = 0.045$) Seyfert 1.2 galaxy \citep{2006A&A...455..773V}. The eastern jet component is consistent with the 15-GHz VLBA image \citep{2009AJ....138.1874L}. However, there is another component appearing on the opposite (western) side of the core with an intensity in excess of 7$\sigma$ image noise, indicating a reliable detection. A detailed study of this component would require future high-resolution and high-sensitivity VLBI observations.

0306$+$102 (OE $+$110): The source is very compact in lower-frequency VLBI images. A faint radio emission is detected in the northeast.

0309$+$411 (NRAO 128): This is a strongly core-dominated broad-line radio galaxy showing core and double lobes morphology \citep{1989A&A...226L..13D,1996MNRAS.281..425M}.  Our 43~GHz VLBA image reveals a prominent core and a straight jet extending to 1~mas, aligning with the brighter and advancing kpc-scale jet.

0354$+$559: The source shows a complex jet in low-frequency VLBI images \citep{2003AJ....126.2562F}. In our 43-GHz image, the jet points to the northwest and then bends to the southwest within 1~mas. The source has rich structure even at mas and probably sub-mas scale, calling for a detailed study.

0400$+$258 (CTD 026): The core size is 0.21 mas. The jet extends to the southeast up to about 2~mas; this is also seen by \citet{2008ApJS..175..314D}. Further out, the jet becomes diffuse and bends to northeast, as seen only at larger scales with lower-frequency VLBI imaging \citep[e.g.,][]{2000ApJS..131...95F,2009AJ....138.1874L}.

0403$-$132 (OF $-$105): Our 43-GHz VLBA image shows a very compact core and faint jet emission extending to the southeast. \citet{2007ApJS..171..376C} detected an unresolved core and radio emission to the southwest. The VLBI data at 2.3 and 15~GHz exhibit a bright core and jet emission extending to the southern direction \citep{2000ApJS..131...95F,2009AJ....138.1874L}.

0405$-$123 (OF $-$109): This source is the second Seyfert 1.2 galaxy in our sample \citep{2006A&A...455..773V}. The VLA image shows two hot spots in the north--south direction, and only the one in the northern lobe was detected in X-rays and optical \citep{2004ApJ...608..698S}. The lower-frequency VLBI images show a core--jet structure extending $\sim$30 mas to the north. Our image shows a resolved structure within 2~mas, with the south component corresponding to the core.

0507$+$179: \citet{2000A&AS..143..181D} identified it as a BL Lac object. The core brightness temperature is the maximum in our sample.

0529$+$075 (OG 050): Our image shows the inner jet extending to the northwest within 3~mas. However, the kpc-scale structure is pointing to the opposite direction \citep{2007ApJS..171..376C}.

0605$-$085 (OC $-$010): The most important concern for this source is the core identification, considering that our observation took place during a flare \citep{2011ApJS..194...29R}. Two components along the east--west direction have equal brightness. Referring to the 15-GHz VLBA image \citep{2009AJ....138.1874L}, we assume that the most compact and upstream component, i.e., the western component, is the core, even if it is not the brightest one. Further high-frequency VLBI observations are needed to confirm this.

0648$-$165: The source shows a large jet bending from the northwest as indicated by our image to the west-southwest which can be better seen in lower-frequency VLBI images \citep{2009AJ....138.1874L,2012A&A...544A..34P}

0657$+$172: The jet direction of the source at pc scale in our image (west-northwest) is different from that seen in low-frequency (2.3- and 8.6-GHz) VLBI images \citep{2012A&A...544A..34P}. Based also on radio spectrum data\footnote{NASA/IPAC Extragalactic Database, http://ned.ipac.caltech.edu/}, we suggest this source is a GPS or high-frequency peaker (HFP) candidate.

0723$-$008: Our image shows a jet pointing to the northeast at $\sim 35\degr$ position angle, in a good agreement with the 15-GHz VLBA image \citep{2009AJ....138.1874L}.

0736$+$017 (OI $+$061): We only detect a compact core and a new weak jet component to the northeast at a distance of 0.41 mas.

0738$+$313 (OI $+$363): This is a GPS quasar \citep{1998A&AS..131..303S}. Our 43-GHz VLBA image shows a core--jet structure, consistently with the 15 GHz-image \citep{2001A&A...377..377S}. Although the position of the radio core in this AGN is uncertain, we assumed the most compact feature at the base of the jet as the core. A more extended component appears at 3.5 mas south of the core.

0742$+$103 (OI $+$171): This is a high-redshift ($z=2.624$) GPS quasar \citep{1998A&AS..131..303S} which shows a large jet bending. We detect a bright component and two diffuse inner jet components in the northwest within 3~mas at 43~GHz. Further out, the jet bends to the northeast in the 15-GHz VLBA image \citep{2009AJ....138.1874L}. The 1.4-GHz\footnote{Here and elsewhere, when we refer to 1.4-GHz VLBA images, we used the unpublished data observed in the project BG196 (PI: D. Gabuzda); calibrated data were downloaded from the Astrogeo database (http://astrogeo.org/).} and 2.3-GHz \citep{2012A&A...544A..34P} VLBA images show the jet bending from northeast to southeast. Although the position of the radio core in this GPS source is uncertain, we assumed the most compact feature at the base of the jet as the core. The coherent jet bending from northwest (inner $\sim$2 mas) to northeast (inner 4 mas), to east (12 mas), then to southeast ($\sim$200 mas) suggests a 180$\degr$ curved trajectory.

0743$-$006 (OI $-$072): This is a GPS quasar \citep{1998A&AS..131..303S}. Although the core position is uncertain, we assumed the most compact feature at the base of the jet as the core.

0838$+$133 (3C\,207): This is a powerful Fanaroff--Riley II (FR II) radio galaxy \citep{1983MNRAS.204..151L}. The 1.4- and 8.4-GHz VLA images show a fairly symmetric triple structure \citep{1994A&AS..105...91B}. Our 43-GHz VLBA image shows a bright core and a one-sided extended jet towards the east. This is in good agreement with what was found previously at 1.4 and 15~GHz \citep{2009AJ....138.1874L}.

0859$-$140 (OJ $-$199): The source shows a compact core and two lobes aligned in the north--south direction at 408~MHz \citep{1996A&A...308..415B}. Our image shows a faint and smoothly curved jet to the south-southeast within 2~mas, which is in good agreement with the previous 15-GHz VLBA image \citep{1998AJ....115.1295K}.

0906$+$015 (4C $+$01.24): The VLA image at 1.6~GHz shows a compact core and a bright component $12\arcsec$ east of the core on kpc scale \citep{1993MNRAS.264..298M}. The 2.3-, 8.6-, and 15-GHz VLBI images show a jet toward the northeast from 5 to 30~mas \citep{2000ApJS..128...17F,2009AJ....138.1874L}. There are 3 jet components detected towards the northeast within 2~mas in our 43-GHz image.

0945$+$408 (4C $+$40.24): The large-scale structure of this source is resolved into a very compact core with a one-sided jet extending over $4\arcsec$ ($\sim 18$~kpc) in the northeast direction using the VLA at 5~GHz and the Multi-Element Radio-Linked Interferometer Network (MERLIN) at 408 and 1666~MHz \citep{1995A&AS..110..213R}. The inner jet structure extends to southeast in our image, which appears to have a 90$\degr$ misalignment with respect to the large-scale structure.

1036$+$054: The 1.4-GHz VLA image shows a bright core and extended jet emission structure in the northeast--southwest direction until $\sim 18\arcsec$ \citep{2010ApJ...710..764K}. The 1.4-GHz VLBA image detected a one-sided jet pointing to the northeast over $\sim$150~mas. The position of the 15-GHz core \citep{2009AJ....138.1874L} coincides with the southernmost component in our image. The brightening jet component might be associated with the major outburst in late 2014 seen in the OVRO 40m light curve \citep{2011ApJS..194...29R}. Taking into account of the jet proper motion of 0.22 mas yr$^{-1}$ \citep{2019ApJ...874...43L} and the time interval between the outburst and our VLBA observation (2015 October 10), the flare-generated shock should have moved about 0.4~mas downstream, roughly consistently with the bright jet component seen in our 43-GHz image.

1045$-$188 (OL $-$176): The 1.4-GHz VLA image shows a bright core and extended jet emission structure in the northwest--southeast direction until $\sim 14\arcsec$ \citep{2010ApJ...710..764K}. The 1.4-GHz VLBA image shows a one-sided jet pointing to the northeast over $\sim$55~mas. The typical beam size at 15~GHz in the MOJAVE survey is 1.5~mas $\times$ 0.5~mas \citep{2009AJ....138.1874L}. In our new 43-GHz image, we detected two components within the area of the 15-GHz beam. We assumed the most compact and northernmost component as the core. Our J1 component corresponds to the bright 15-GHz core. More high-frequency VLBI observations are needed to clarify this.

1124$-$186 (OM $-$148): \citet{2015AJ....150...58F} only detect a compact core at 2.3 and 8.6~GHz. We detect a core and an extended faint emission feature to the south, in agreement with the 15-GHz image \citep{2009AJ....138.1874L}.

1150$+$497 (4C $+$49.22): The 1.5-GHz VLA image shows a complex triple source with a halo in the north--south direction \citep{1981AJ.....86.1010U}. The source is not detected on space--Earth baselines at 5~GHz with the VSOP \citep{2008ApJS..175..314D}. In our image, we detect a bright core and a series of  jet components to the southwest.

1219$+$044 (4C $+$04.42): The source shows a compact core and extended jet emission aligned in the north--south direction until $\sim 5\arcsec$ \citep{2010ApJ...710..764K}. The 15-GHz VLBA image only detected a one-sided jet pointing to the south until 7~mas \citep{2009AJ....138.1874L}. Our new image shows a bright core and a faint emission to the south.

1228$+$126 (M87): This is a well-known low-luminosity FR I radio galaxy \citep{1983MNRAS.204..151L}. The large-scale image of M87 observed with the VLA at 90~cm wavelength suggests that the outward flow from the nucleus extends well beyond the 2~kpc radio jet \citep{2000ApJ...543..611O}. The 15-GHz VLBA image displays an unresolved core and complex jet structure with an extent of 22~mas. Our 43-GHz VLBA image suggests a limb-brightening morphology with two ridge lines extending to the northwest and west directions. The jet opening angle, estimated from the northern and southern bright jet knots, is about 45\degr{} in projection, as is approximately consistent with the value reported in \citet{2016ApJ...817..131H} and \citet{2019Galax...7...86Z}.

1324$+$224: The deep VLA image at 1.4 GHz only detected a compact core \citep{2007ApJS..171..376C}. The 1.4-GHz VLBA image shows the jet pointing to the northwest up to 10~mas, then bending towards northeast until 90~mas from the core. In the 15-GHz VLBA image, the source shows a very compact core and a weak emission feature approximately 3~mas to the southwest \citep{2009AJ....138.1874L}. However, our new 43-GHz image indicates an inner jet towards the northwest on an angular scale of 2~mas, at a position angle consistent with the 1.4-GHz VLBA image.

1435$+$638 (VIPS 0792): The 1.4-GHz VLA image presents a faint lobe separated from the core by $15.4\arcsec$ in the southwest. It was not detected at 5~GHz \citep{1995A&AS..110..213R}. Previous studies presumed the radio core to lie at the northernmost end of the jet, based on the 5-GHz and 15-GHz VLBA maps \citep{2007ApJ...658..203H,2016AJ....152...12L}. We also associated the northeastern component with the radio core in our image.

1504$-$166 (OR $-$107): The 1.4-GHz VLA map exhibits only radio core emission \citep{2010ApJ...710..764K}. The 1.4-GHz VLBA image shows the jet pointing to the west up to $\sim 150$~mas. The 8- and 15-GHz VLBI images show a compact core and extended structure to the south and southeast \citep{2009AJ....138.1874L,2012A&A...544A..34P}. Our new 43-GHz image shows the inner jet pointing to the south which suggests a jet bending.

1514$+$004 (4C $+$00.56): The source is a nearby radio galaxy $(z = 0.052)$. The 1.4-GHz NRAO VLA Sky Survey (NVSS) image shows a symmetric triple source extending in the northwest--southeast direction \citep{1998AJ....115.1693C}. In our image, we detect a core and bright jet component pointing to the northwest, in good agreement with the 15-GHz VLBA image \citep{2009AJ....138.1874L}.

1548$+$056 (4C $+$05.64): \citet{2010ApJ...710..764K} detected a bright core and a relatively faint extended radio emission to the north with the VLA at 1.5~GHz. On mas scales, the source is dominated by a compact core with a jet extending to the north, as was seen previously at 1.4~GHz. In our image, we see a complex and curved jet extending to north and then bending to the northeast at 2~mas from the core.

1637$+$826 (NGC 6251): This is a well-studied FR I radio galaxy which shows both a bright core and large extended asymmetric jet emission \citep{1984ApJS...54..291P}. Lower-frequency VLBI observations only detected the jet extending to the southwest direction \citep{2004AJ....127.3587F,2009AJ....138.1874L}. Our 43-GHz image shows the jet aligned well with the kpc-scale jet.

1716$+$686: The 4.5-GHz VLA observation shows a diffuse halo of 10\arcsec\ extension surrounding the core \citep{1996ApJS..107...37T}. \citet{2009AJ....138.1874L} presented a jet extending to the northwest up to $\sim 10$~mas, in a position angle in agreement with our image within 2~mas.

1926$+$611: The 1.5-GHz VLA image only detected a core \citep{1998AJ....115.1693C}, but the 1.7-GHz VLBI image shows a bright core and a jet structure extended to the south \citep{1995ApJS...98....1P}. However, our image exhibits two jet components to the southeast, in agreement with the 15-GHz VLBA image \citep{2009AJ....138.1874L}. This indicates a jet bending from the south to the southeast, with nearly 70$\degr$ change in the position angle.

2007$+$777: The image made with the VLA at 1.5~GHz shows two-sided radio emission in the east--west direction \citep{1986AJ.....92....1A}. The eastern component is a prominent hot spot \citep{1993MNRAS.264..298M}. Our image shows the jet extending to the west, corresponding to the western side of the jet in the VLA image.

2021$+$317 (4C $+$31.56): The 1.4- and 15-GHz VLBA images display the jet extending towards the south \citep{2009AJ....138.1874L}, consistently with our new 43-GHz VLBA image. However, we also detect a new component off-axis from the persistent jet, on the northwestern side. We believe this northwestern component is the core, and therefore the jet undergoes a sudden bending within 1~mas. Future observations can confirm the properties of this intriguing object.

2021$+$614 (OW $+$637): The source was classified as a CSO \citep{2000A&A...360..887T}. No radio emission was detected on scales larger than 0.2\arcsec{} with the VLA at 1.4~GHz \citep{2010ApJ...710..764K}. \citet{2009AJ....138.1874L} identified the core feature located in the end of the southwestern component. We also use the same component as the core for our image, even though the J3 component at $\sim$2.5 mas is more compact than the core.

2029$+$121 (OW $+$149): The image made with the VLA in A-array at 1.4~GHz shows a one-sided, edge-brightened morphology pointing to the northwest \citep{2001AJ....122..565R}. The jet in our image extends to the southwest in $-130\degr$ position angle, in agreement with the 2.3- and 8.6-GHz VLBA images \citep{2015AJ....150...58F}.

2126$-$158 (OX $-$146): This is the highest-redshift AGN in our sample ($z = 3.268$), a GPS quasar \citep{1997A&A...325..943S}. The peak emission component in our image has a flat spectrum between 5 and 15~GHz and is also identified as the core by \citet{2001A&A...377..377S}. Our image shows a bright core and a faint jet extending to the southwest.

2134$+$004 (OX $+$057): \citet{2005A&A...435..839T} identified this source as a GPS quasar. \citet{2006A&A...450..959O} found that the core of the source is located in the easternmost component. The core is also the brightest component in the 43-GHz radio structure.

2141$+$175 (OX $+$169): Large-scale VLA observations only detected a core at 1.4~GHz \citet{1998AJ....115.1693C}. In the 2.3- and 8.6-GHz VLBI images, the jet extends to the north, out to a distance of 25~mas ($\sim 85$~pc) \citep{2005AJ....129.1163P}. We detect the jet initially pointing to the west and then changing its position angle to the northwest, indicating that it has a large ($\sim 90\degr$) bending that starts at $\sim 0.2$~mas.

2155$-$152 (OX $-$192): The image made with the VLA at 1.4~GHz shows a triple structure with a size of $6\arcsec$ surrounding a central compact component in the north--south direction \citep{2007ApJS..171..376C}. Our image shows a jet towards the southwest up to 2~mas, in good agreement with the 5- and 15-GHz VLBA images \citep{2000ApJS..131...95F,2009AJ....138.1874L}.

2209$+$236: \citet{2000A&A...363..887D} identified the source as a HFP. The 5-GHz VLBA image \citep{2000ApJS..131...95F} shows no indication of extended emission. \citet{2016AJ....152...12L} found a jet component to the northeast and determined a maximum apparent proper motion 1.35$c$. Our image shows the inner jet bending to the northeast.

2223$+$210 (DA 580): The jet structure extends to southwest in the 2.3-, 8.6-, and 15-GHz VLBI images \citep{2002ApJS..141...13B,2009AJ....138.1874L}. We detect two compact components inside the low-frequency VLBI core region. Although the eastern one is not the brightest component, it appears in the upstream direction. Therefore we identify this as the radio core.

2227$-$088: The source shows an integrated flux density of 2.7~Jy in the core at 43~GHz, and only 21~mJy in the extended emission. \citet{2003PASJ...55..351T} found this source highly variable at 4.8~GHz with Australia Telescope Compact Array (ATCA). The many VLBI images made at lower frequencies\footnote{see http://astrogeo.org/} indicate a faint, wiggling jet towards the north.

2234$+$282 (CTD 135): \citet{2016AJ....152...12L} identified the core in this BL Lac object with the northernmost jet feature in their 15-GHz VLBA images. Earlier \citet{2016AN....337...65A} claimed this source to be a rare $\gamma$-ray emitting CSO candidate with a two-sided jet, based on a comparison of VLBA maps at 8.4 and 15~GHz. In our new 43-GHz image, the southwestern component is the brightest and most compact, suggesting that this is the true core, instead of the northeastern feature \citep{2016AJ....152...12L}. Also, our new observations are at odds with the CSO interpretation \citep{2016AN....337...65A}.

2243$-$123: The jet points to the northeast in the previous 5- and 15-GHz VLBA images \citep{2000ApJS..131...95F,2009AJ....138.1874L}. In our image, the innermost section of the jet points closer to the north, within 4~mas from the core.

2318$+$049 (OZ $+$031): VLA images show a barely resolved structure along $-40\degr$ position angle \citep{1998PASP..110..111H}. In our VLBA image, the source shows a compact core--jet structure extending to the northwest in about the same direction. This is consistent with lower-frequency VLBI images \citep[e.g.,][]{2000ApJS..131...95F, 2009AJ....138.1874L}. Note that the 5-GHz VSOP Survey image also shows emission within 2~mas extended in the same direction, albeit with no clear indication of the core \citep{2004ApJS..155...33S}.

\section{Discussion}
\label{discussion}

\subsection{Source Compactness}

\begin{figure}
  \centering
  \begin{tabular}{c}
  \includegraphics[width=1.0\textwidth]{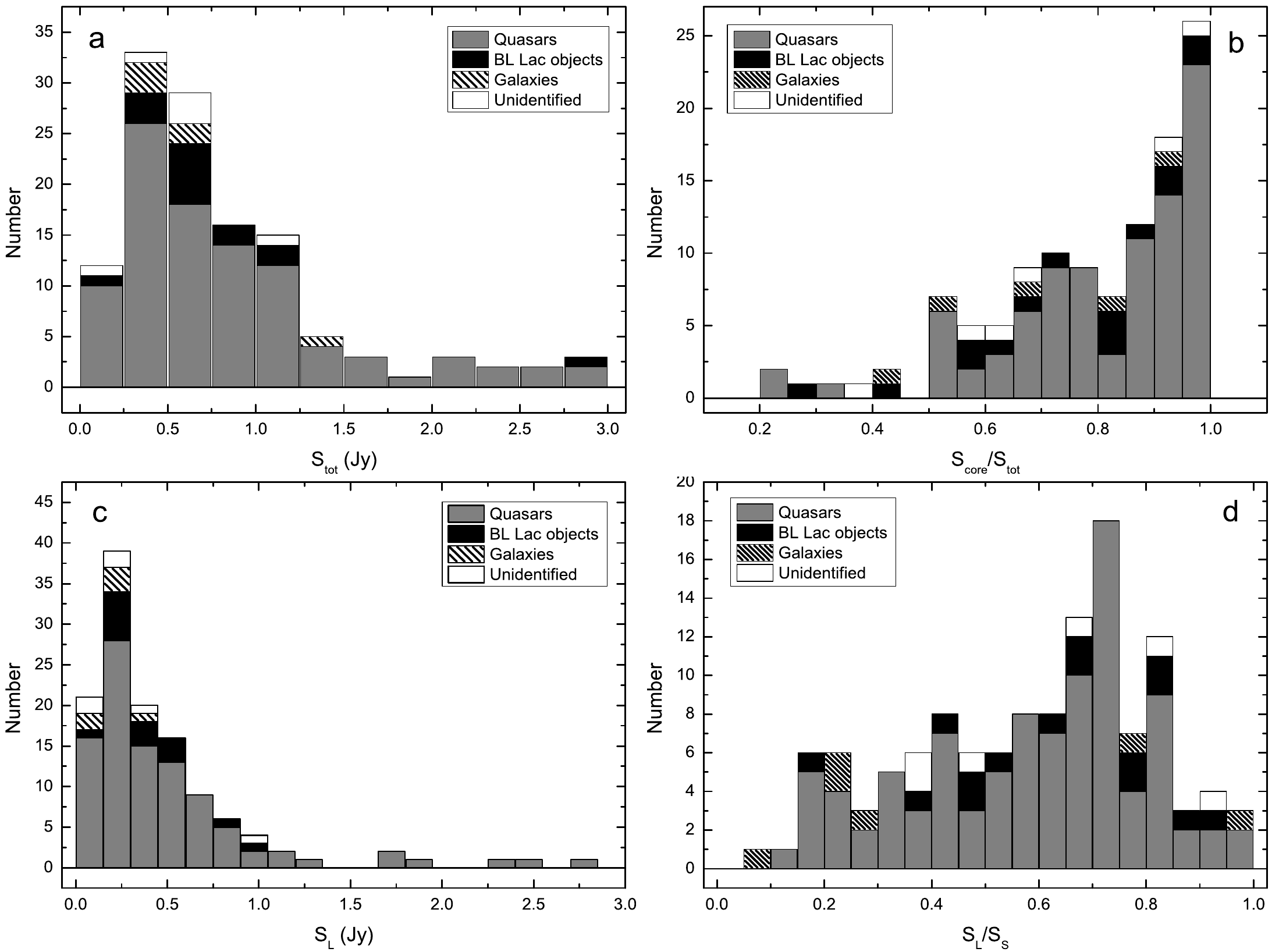} \\
    \end{tabular}
  \caption{Histograms of the total flux density, $S_{\rm tot}$ (panel a), the compactness parameter on mas scales $S_{\rm core}/S_{\rm tot}$ (panel b), the correlated flux density on the longest baselines  $S_{\rm L}$ (panel c), and the compactness parameter on sub-mas scales $S_{\rm L}/S_{\rm S}$ (panel d).}
  \label{compactness}
  \end{figure}

Only sources with cores that are compact enough can be successfully detected with the high angular resolution provided by high radio frequency VLBI observations. The source compactness is often
expressed in two ways:  the ratio of the VLBI core flux density to the total
flux density, or alternatively, the ratio of the correlated flux density on the longest baselines
to that on the shortest baselines \citep{2008AJ....136..159L,2005AJ....130.2473K,2018ApJS..234...17C}.
Figure~\ref{compactness} shows the distributions of the total flux density $S_{\rm tot}$,
the ratio $S_{\rm core}/S_{\rm tot}$, the correlated flux density on longest baselines
derived from our data $S_{\rm L}$, and the ratio $S_{\rm L}/S_{\rm S}$, using the values listed in Table~\ref{tab:image}.
$S_{\rm core}$ is obtained from fitting the brightest core component with a circular
Gaussian model in {\sc Difmap} (see Table~\ref{model}). $S_{\rm tot}$ is estimated by integrating
the flux density contained in the emission region in the VLBI image. $S_{\rm L}$ and $S_{\rm S}$ are
obtained from the correlated flux density on the longest and shortest baselines. The total flux
density of these sources $S_{\rm tot}$ (Fig.~\ref{compactness}a) ranges from 0.10 to 2.91~Jy. The correlated
flux density on the longest baselines $S_{\rm L}$ (Fig.~\ref{compactness}c) ranges from 0.02
to 2.38~Jy. The median values of the flux densities $S_{\rm tot}$ and $S_{\rm L}$ in our sample are
0.63~Jy and 0.26~Jy, respectively.

Figure~\ref{compactness}b gives the distribution of $S_{\rm core}/S_{\rm tot}$. Since most cores are even unresolved at 43 GHz with 0.5 mas $\times$ 0.2 mas resolution, the ratio $S_{\rm core}/S_{\rm tot}$ denotes the source compactness on mas scales, which varies in the range of 0.21--1.0 with a mean value of
0.85. This is also a measure of the core dominance in the VLBI image. For 112 sources (accounting for 90\% of the sample), the
compactness parameter $S_{\rm core}/S_{\rm tot}$ is larger than 0.5, indicating that a substantial fraction of the 43-GHz VLBI emission is from the core.
There are 96 sources
with a core flux density exceeding 0.30~Jy (Fig.~\ref{compactness}a). Figure~\ref{compactness}d gives the distribution
of the source compactness  $S_{\rm L}/S_{\rm S}$, ranging from 0.08
to 0.99 with a mean value of 0.65.
The visibility amplitude on the longest baseline is associated with the most compact and unresolved part of the core, therefore it is used as an indicator of the compactness on sub-mas scales.
A total of 86 sources have
$S_{\rm L}/S_{\rm S} > 0.5$.
In general, the ratios $S_{\rm core}/S_{\rm tot}$ are higher than
$S_{\rm L}/S_{\rm S}$, consistent with the observational results that some cores are further resolved at higher sub-mas resolutions.

The source compactness parameters are useful for planning mm-wavelength ground-based and/or space VLBI observations. For the proposed Chinese SMVA project \citep{2014AcAau.102..217H}, the expected highest angular resolution will be 20~$\mu$as at 43~GHz,
and the baseline sensitivity is $\sim$17~mJy (1$\sigma$, assuming 512~MHz bandwidth,
60~s integration) when a space-borne 10-m radio telescope works in tandem with a 25-m VLBA telescope.
To explore the innermost jets using space VLBI at even higher resolution, the extremely compact AGNs would be the preferred targets.
However, some of the objects in this sample do not contain a bright, compact component and thus are not suitable for imaging with the future space VLBI network.
Based on the flux densities and model-fitting results in Table~\ref{tab:image}, 95 sources have
$S_{\rm L} > 0.17$ Jy (i.e. about 10 times the baseline sensitivity of the SMVA), $S_{\rm core}/S_{\rm tot} > 0.5$ and $S_{\rm tot} > 0.30$~Jy.
These sources remain compact at the longest available ground-based baselines and have sufficiently high flux densities. Therefore, combined with the previous 10 sources in paper II \citep{2018ApJS..234...17C}, there are 105 sources suitable for future space VLBI missions \citep{2017ARep...61..310K,2020An-ASR}.

\subsection{Core Brightness Temperature}
\label{brightnesstemp}

We estimated the brightness temperature of the core components, and listed the values in Column~9 of Table~\ref{model}. We used the results of our brightness distribution model fitting, and the following formula: 
\begin{equation}
  T_{\rm b} = 1.22\times10^{\rm 12}\frac{S}{ \nu^{2}d^{2}}(1+z),
\end{equation}
where $S$ is the flux density of the core in Jy, $z$ is the redshift, $\nu$ is the observing frequency in GHz, and $d$ is the fitted Gaussian size (FWHM) of the component in mas.

The median value of the core brightness temperatures (Table~\ref{model}) is $7.92 \times 10^{10}$~K, with a maximum of $2.54 \times 10^{12}$~K. The maximum value is higher than the equipartition $T_{\rm b}$ limit \citep{1994ApJ...426...51R,1994ApJ...429L..57B} but is approximately equal to the maximum brightness temperature set by the inverse Compton limit \citep{1969ApJ...155L..71K}.
Apparent brightness temperatures exceeding the equipartition value are probably due to Doppler boosting of the relativistically beamed jet but could also be due to an intrinsic excess of particle energy over magnetic energy \citep[e.g.,][]{2000ARep...44..719K,2002PASA...19...77K}.

\begin{figure}
  \centering
  \begin{tabular}{cc}
  \includegraphics[width=0.48\textwidth]{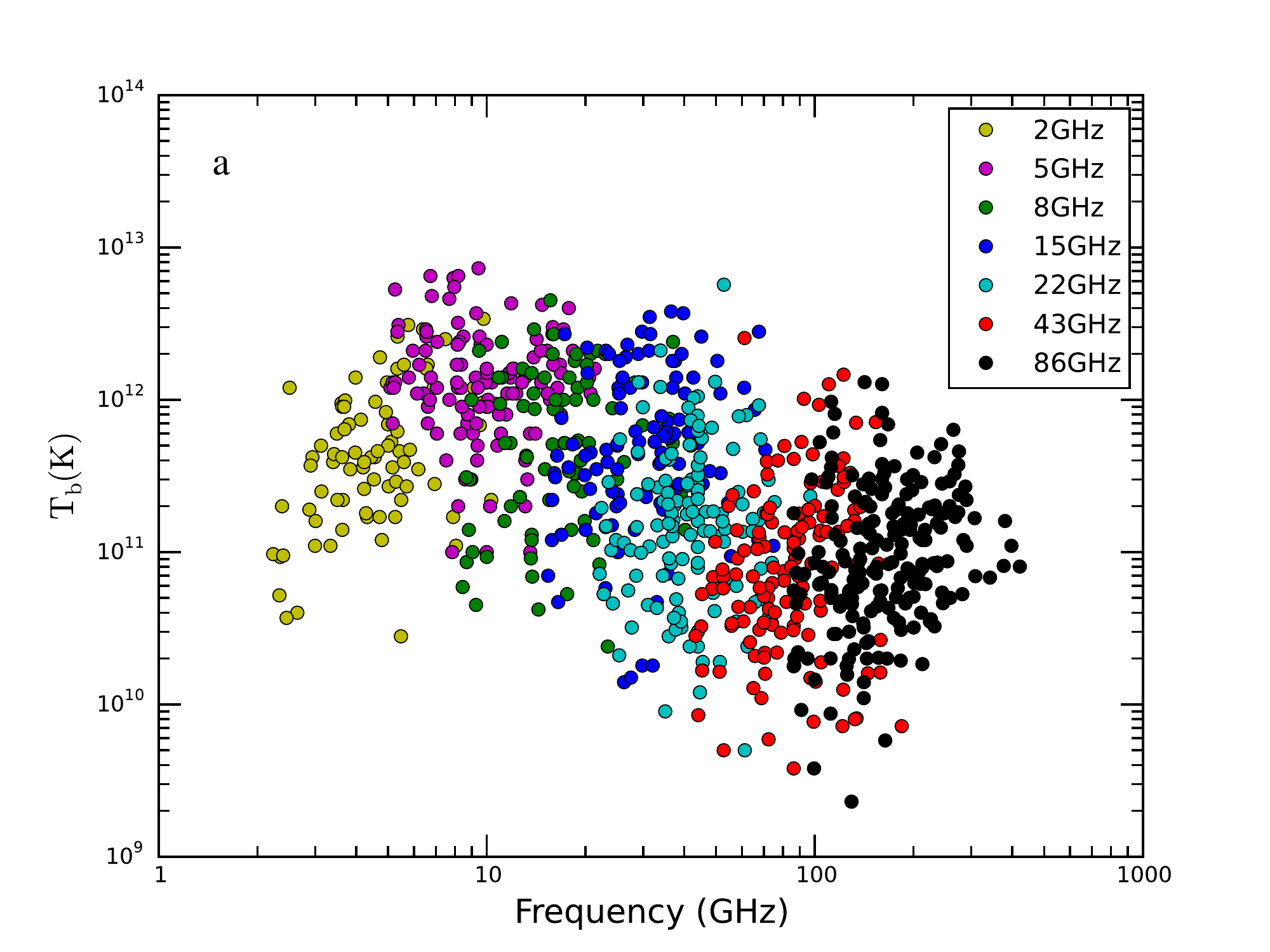} & \includegraphics[width=0.45\textwidth]{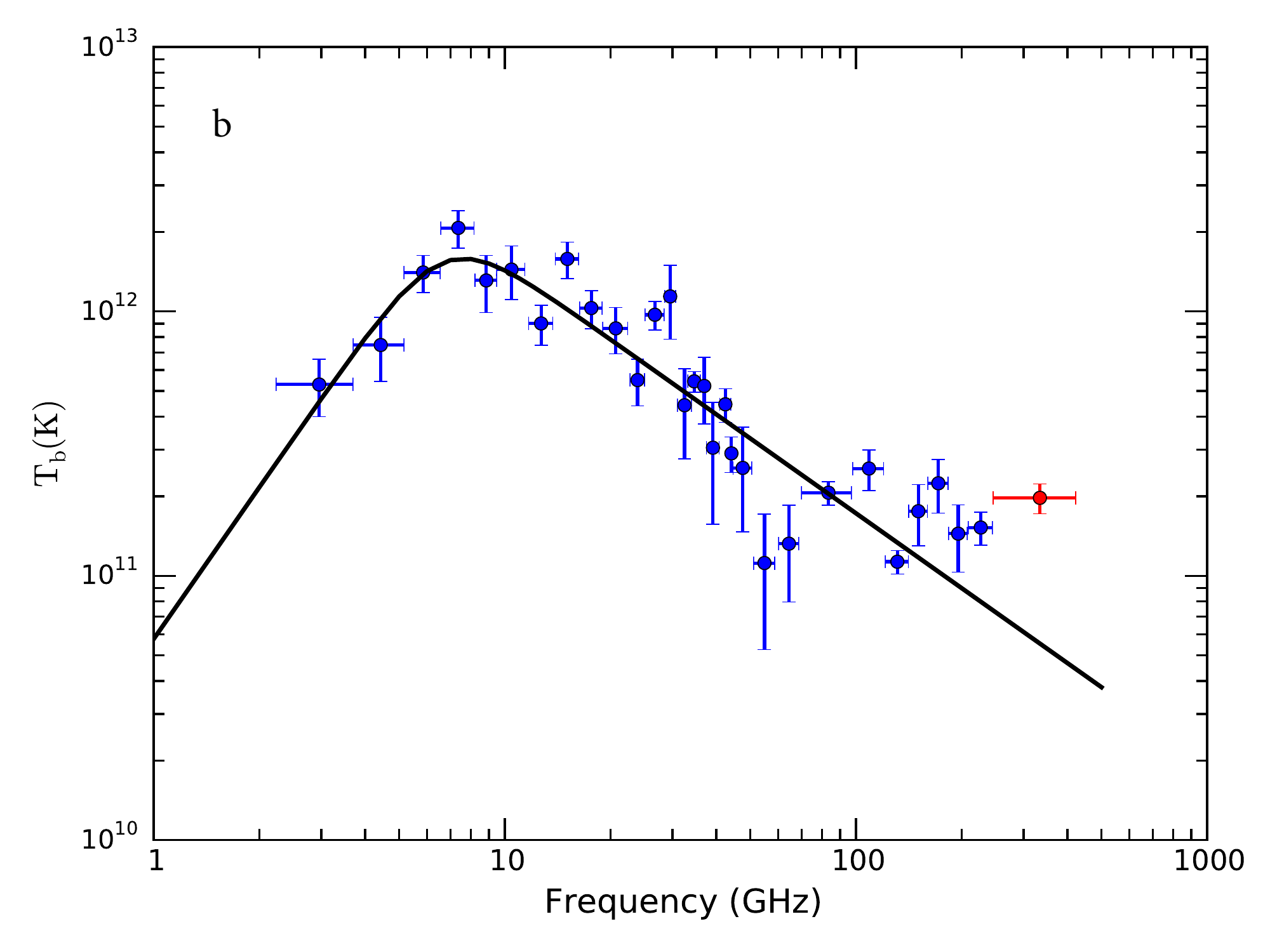}\\
    \end{tabular}
  \caption{Distribution of core brightness temperature as a function of frequency (left panel) and the best fit of the relationship (right panel). Coloured symbols represent core brightness temperatures measured at various observed frequencies:
  yellow \citep[2~GHz,][]{2012A&A...544A..34P},
  purple \citep[5~GHz,][]{2004ApJ...616..110H},
  green  \citep[8~GHz,][]{2012A&A...544A..34P},
  blue  \citep[15~GHz,][]{2005AJ....130.2473K},
  cyan  \citep[22~GHz,][]{1996AJ....111.2174M},
  red \citep[43~GHz,][and this paper]{2018ApJS..234...17C},
  black \citep[86~GHz,][]{2008AJ....136..159L,2018ApJS..234...17C,2019A&A...622A..92N}.
  The frequencies have been converted into the source rest frame.
  The fitting in the right panel is represented with a broken power-law function as described in Sect.~\ref{brightnesstemp}.}
  \label{Tb}
  \end{figure}

Figure~\ref{Tb}a shows the observed brightness temperatures ($T_{\rm b}$) as a function of frequency in the rest frame of source, $\nu$ = $\nu_{\rm obs} (1 + z)$. We adopted VLBI core brightness temperatures measured at observing frequencies $\nu_{\rm obs}$ at 2, 5, 8, 15, and 22~GHz from the literature \citep{1996AJ....111.2174M,2004ApJ...616..110H,2005AJ....130.2473K,2012A&A...544A..34P}, at 43~GHz (present data and Paper~II), and at 86~GHz \citep{2008AJ....136..159L,2018ApJS..234...17C,2019A&A...622A..92N}, to study the relation between brightness temperature and frequency.
From Fig.~\ref{Tb}a, it is obvious that the core brightness temperatures at 43 and 86~GHz are much lower than those at 2, 5, 8, 15, and 22~GHz.
The $T_{\rm b}$ values at lower frequencies, although with large scatter, seem to change only slightly with increasing frequency, until about 20~GHz. After that, a remarkable drop of $T_{\rm b}$ is seen above $\sim 30$~GHz.
In order to quantify the variation of the core brightness temperature, we divided all 889 data points into 30 frequency bins, each containing 28--30 data points.
Figure~\ref{Tb}b shows the data averaged within the bins with the error bars representing the scatter. The horizontal error bars indicate the frequency range within the bin, and the vertical error bars show the standard deviation of the mean values of core brightness temperature. The $T_{\rm b}$ distribution can be fitted with a smoothly broken power-law function between 2--240~GHz (eq. \ref{eqn:powerlaw}), while the $T_{\rm b}$ values between 240--420~GHz (red-colored data points) seem to deviate from this relation and thus are excluded from the fitting:
 \begin{equation}
   \label{eqn:powerlaw}
   T_{\rm b}(\nu) = T_{\rm b,j}\left[\left(\frac{\nu}{\nu_{\rm j}}\right)^{\alpha_1 n}
           + \left(\frac{\nu}{\nu_{\rm j}}\right)^{\alpha_2 n}\right] ^{-1/n},
   \end{equation}
where $T_{\rm b,j}$ is the observed brightness temperature at the break frequency $\nu_{\rm j}$, $-\alpha_1$ and $-\alpha_2$ are the slopes at higher and lower frequencies than the break frequency, respectively,
and $n$ is a numerical factor controlling the sharpness of the break.
The frequencies (horizontal axis) have been corrected to the source rest frame by multiplying the observing frequencies by ($1+z$). We choose the numerical factor $n = 1.97 $ and use the nonlinear least-squares algorithm to fit the function. Figure~\ref{Tb}b shows the best fit with the following parameters: $T_{\rm b,j}=(218.14\pm5.27)\times10^{10}$~K, $\nu_{\rm j}=6.78\pm0.43$~GHz, $\alpha_1=1.91\pm0.03$, and $\alpha_2=-0.94\pm0.06$.

From the best fitting parameters, we find that the core brightness temperature is increasing below the break frequency $\sim$7 GHz, and decreasing between 7--240~GHz. $T_{\rm b,j}$ is $\sim 2\times10^{12}$~K at $\sim$7 GHz, which is slightly higher than the inverse Compton catastrophe limit \citep{1969ApJ...155L..71K} and significantly higher than the equipartition value \citep{1994ApJ...426...51R,1994ApJ...429L..57B}, consistent with Doppler boosting in AGN jets. The inferred broken power-law  distribution can be explained in terms of opacity affecting the emission mechanism at different resolutions. Below 7 GHz, due to relatively lower resolutions, the core emission is mixed with jet emission, resulting in lower  $T_{\rm b}$ at lower frequencies. From 2 to 7 GHz, with the resolution increasing, this dilution effect gradually becomes weaker, thus the $T_{\rm b}$ shows an increasing trend. The decreasing trend of $T_{\rm b}$ starting from 7 GHz to higher frequencies reflects the synchrotron opacity changing from optically thick to thin. A resulting lower brightness temperature has been found by opacity core-shift observations in many sources \citep{2008A&A...483..759K,2009MNRAS.400...26O,2011A&A...532A..38S}.
Previous compactness studies suggest that most, if not all, AGN cores are resolved into sub-components at higher frequency and with higher resolution. However, the highest-frequency data (red-colored data point) in our Fig.~\ref{Tb}b show a clear deviation from the fitted trend (black line) and display a flattening. The inferred brightness temperatures are substantially higher than the values expected from the fit. We caution that this may be because of the sample selection bias caused by the flux density limited surveys. These high-frequency ($>240$~GHz in rest frame) data correspond to high-redshift ($z \gtrsim 3$) AGNs. Among those, the survey samples include the few brightest ones only, and the majority of the (much weaker) high-redshift population is missed. Future higher-sensitivity VLBI surveys at and above 86~GHz containing weaker high-redshift AGNs would be important for identifying the distribution trend at the very high frequency end of Fig.~\ref{Tb}b.

\subsection{Correlation Between Radio and $\gamma$-ray Emission}
\label{corelation}

We cross-matched our sample with the third {\it Fermi}-LAT AGN catalog \citep{2015ApJ...810...14A} and found 73 sources (59\% of our sample) detected in $\gamma$-rays. Combining with our previously observed ten sources in Paper~II, we obtain a total of 79 sources, including 61 QSOs, 14 BL Lac objects, 3 radio galaxies (M87, NGC\,6251, 3C\,371), and 1 object (0648$-$165) with no optical identification. In our sample, BL Lac objects have a higher $\gamma$-ray detection rate (88\%) than quasars (58\%).

\begin{figure}
  \centering
  \begin{tabular}{cc}
  \includegraphics[width=0.45\textwidth]{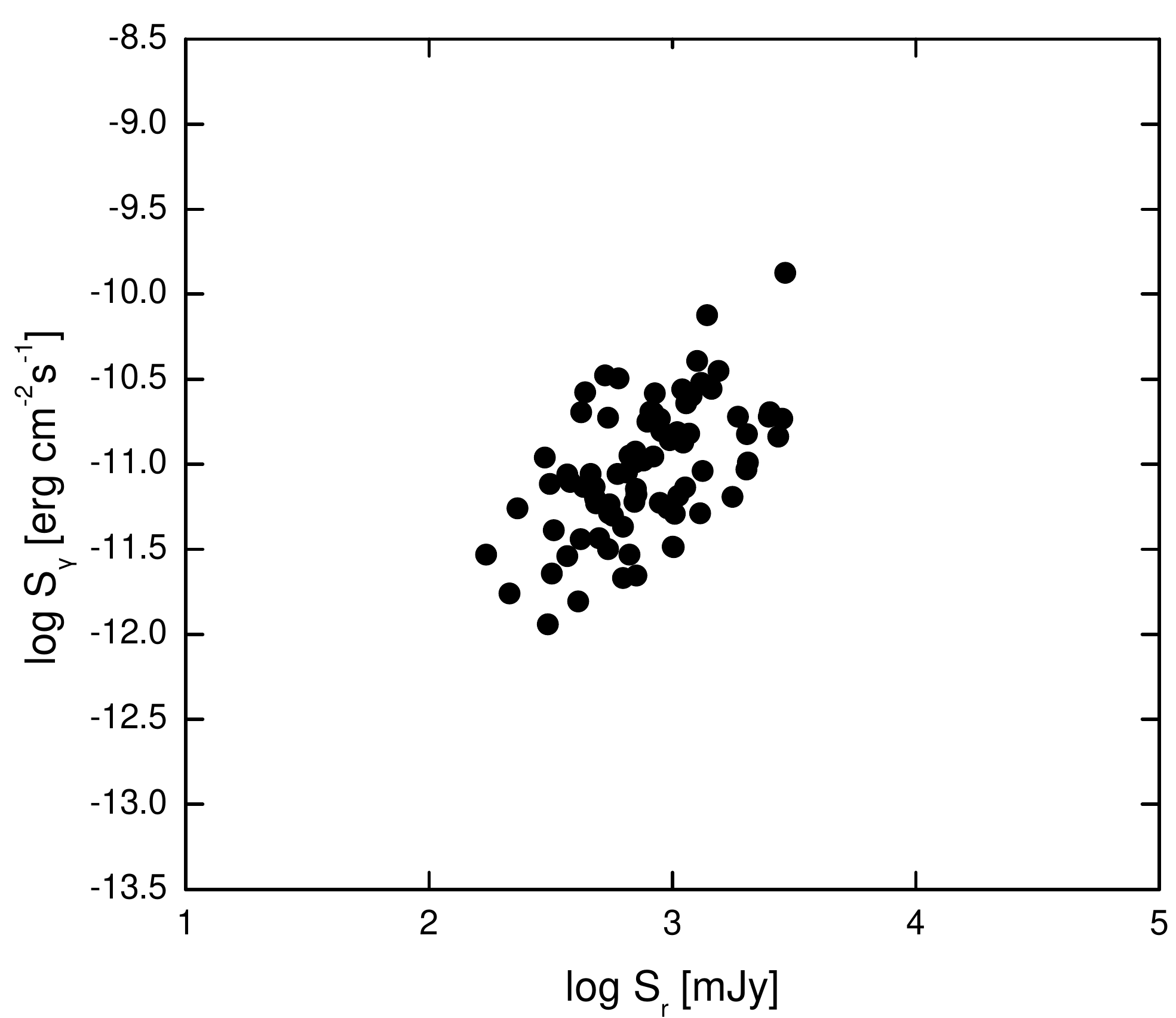}
    \end{tabular}
  \caption{Flux density correlations between radio 43 GHz and $\gamma$-ray bands. The correlation coefficient is 0.550.   The error bars are omitted for the clarity of the plot. }
  \label{correlation}
  \end{figure}

Figure~\ref{correlation} presents the flux density correlation between $\gamma$-rays and radio for all the 79 sources. We used the Pearson correlation test to reveal the flux density  correlation between the radio and $\gamma$-rays. That gives a correlation with $\tau = 0.550$ and $P<0.001$, where $\tau = 0$ means no correlation and $\tau = 1$ means strong correlation, and $P$ gives the probability of no correlation.
We should note that the three non-beamed AGNs (radio galaxies: M87, NGC\,6251, and 3C\,371) in our sample also follow the same $\gamma$-ray/radio flux density correlation, indicating that probably a universal $\gamma$-ray production mechanism, regardless of their difference in the central engine and jet power, is at work for diverse types of AGNs.

According to \citet{2010MNRAS.407..791G}, there is a strong correlation between the radio flux density at 20~GHz and the $\gamma$-ray flux above 100 MeV. \citet{2011A&A...535A..69N} even found a more significant correlation between both the flux densities and luminosities in $\gamma$-rays and 37-GHz single-dish radio data.
Recently \citet{2016ApJS..226...20F} found a correlation between $\gamma$-ray and 1.4-GHz monochromatic radio luminosities. \citet{2011ApJ...741...30A} found that there is a correlation between $\gamma$-ray and 8.4-GHz (VLA and ATCA) / 15-GHz (OVRO single dish) radio luminosities. In the above studies, the authors used single-dish or low-resolution interferometer measurements of the radio flux densities. In contrast, our analysis is based on high-resolution high-frequency VLBI data; the correlation derived from our study is consistent with the previous works cited above but shows higher correlation coefficient.
All these studies support the notion that the $\gamma$-ray emission zone is close to or same with the compact VLBI core.

In conclusion, our results give support to the strong correlation between the $\gamma$-ray and radio emission in AGNs, suggesting that the $\gamma$-ray emission zone is in the parsec-scale jet revealed by the 43 and 86 GHz VLBI images. The correlation is valid for diverse types of AGN over six orders of magnitude in radio luminosity and eight orders of magnitude in $\gamma$-rays, indicating that the general AGN population have a common $\gamma$-ray production mechanism, regardless of their different central engine and jet power.

\section{Summary}
\label{summary}

We observed a sample of 124 bright and compact radio sources with the VLBA at 43~GHz between 2014 November and 2016 May. We achieved a highest angular resolution of $\sim$0.2~mas and a typical image noise level of 0.5~mJy\,beam$^{-1}$. In our sample, 8 sources remain unresolved and are only detected with a compact core, 112 sources show one-sided jet structure, 3 sources have compact symmetric structures, and one source (0354+559) shows complex structure.
We present the 43-GHz contour images of all the 124 sources and give comments on selected individual sources to highlight their properties in the context with other information from literature. The majority of these radio-loud AGNs have not been previously imaged with VLBI at this frequency.

One of the main motivations of our project was to identify suitable target sources for the future mm-wavelength space VLBI program. From the distribution of source compactness on mas scales, $S_{\rm core}/S_{\rm tot}$, and sub-mas scales, $S_{\rm L}/S_{\rm S}$, 95 of the 124 sources have $S_{\rm L} > 0.17$ Jy (about 10 times the baseline sensitivity of the SMVA), $S_{\rm core}/S_{\rm tot} > 0.5$ and $S_{\rm tot} > 0.30$~Jy, and therefore there are 105 sources to be considered as targets for the SMVA.

For the 124 sources, we calculated the core brightness temperatures. Their median value is $7.92 \times 10^{10}$~K, with a maximum of $2.54 \times 10^{12}$~K at 43~GHz. This is somewhat higher than the inverse Compton catastrophe $T_{\rm b}$ limit and the equipartition limit, which can be explained by Doppler boosting of the relativistically beamed jet. We investigated the core brightness temperatures obtained from our project and other high-resolution VLBI observations from the literature at 2, 5, 8, 15, 22, 43, and 86 GHz. We find that the core brightness temperature is increasing below the break frequency $\sim 7$~GHz, and decreasing above 7 GHz. The break core brightness temperature value is $2 \times 10^{12}$~K at $\sim 7$~GHz. The change of brightness temperature with (source rest frame) observing frequency is related to the resolution and synchrotron opacity changes with frequency.

We used 79 sources to test the correlation between radio and $\gamma$-ray flux densities and found tighter correlation compared to previous works. Our result supports the scenario that the location of the $\gamma$-ray emission is close to the 43 GHz VLBI core.
Moreover our result also indicate that the radio--$\gamma$-ray correlation is universal and can be applicable to different types of AGNs.

\section*{Acknowledgments}
This work was supported by the SKA pre-research funding granted by the National Key R\&D Programme of China (2018YFA0404602, 2018YFA0404603), the Chinese Academy of Sciences (CAS, 114231KYSB20170003), and the Hungarian National Research, Development and Innovation Office (grant 2018-2.1.14-T\'ET-CN-2018-00001). X.-P. Cheng was supported by Korea Research Fellowship Program through the National Research Foundation of Korea (NRF) funded by the Ministry of Science and ICT (2019H1D3A1A01102564). The VLBA observations were sponsored by Shanghai
Astronomical Observatory through an MoU with the NRAO (Project code: BA111). This research has
made use of data from the MOJAVE database that is maintained by the MOJAVE
team \citep{2009AJ....138.1874L}. The MOJAVE program is supported under
NASA-{\it Fermi} grants NNX15AU76G and NNX12A087G. The Very Long Baseline Array is
a facility of the National Science Foundation operated under cooperative
agreement by Associated Universities, Inc. This research has made use of
data from the OVRO 40-m monitoring program \citep{2011ApJS..194...29R},
which is supported in part by NASA grants NNX08AW31G, NNX11A043G and
NNX14AQ89G and NSF grants AST-0808050 and AST-1109911. This work has made
use of NASA Astrophysics Data System Abstract Service, and the NASA/IPAC
Extragalactic Database (NED) which is operated by the Jet Propulsion
Laboratory, California Institute of Technology, under contract with the
National Aeronautics and Space Administration.

\vspace{5mm}



\label{lastpage}


\begin{deluxetable}{ccccccccc}
\tablecolumns{9}
\tabletypesize{\footnotesize}
\tablewidth{0pt}
\tablecaption{
General Source Properties
\label{tab:sample}}
\tablehead{
\colhead{IAU}  & \colhead {Alias}   & \colhead {Sess.}  & \colhead {R.A.}     & \colhead {Dec.}     & \colhead{$z$} & \colhead{Opt.}    & \colhead{Comm.}    & \colhead{$S_{\rm Fermi}$}  \\
\colhead{Name} & \colhead {}        & \colhead {}      & \colhead {(J2000)}  & \colhead {(J2000)}  & \colhead{}      & \colhead{Class}   & \colhead{}             & \colhead{($\rm 10^{-10}  ph\,cm^{-2}\,s^{-1}$)}   \\
\colhead{(1)}  & \colhead{(2)}      & \colhead{(3)}           & \colhead{(4)}       & \colhead{(5)}       & \colhead{(6)}      & \colhead{(7)}     & \colhead{(8)}          & \colhead{(9)}  }
\startdata
0048$-$071     & OB $-$082 &   C    & 00 51 08.20982 & $-$06 50 02.2291 & 1.975 &  QSO   & Y & 12  \\
0106$+$013     & 4C +01.02 &   B    & 01 08 38.77110 & $+$01 35 00.3713 & 2.099 &  QSO   & Y & 64  \\
0113$-$118     &           &   C    & 01 16 12.52203 & $-$11 36 15.4348 & 0.671 &  QSO   & Y & 15  \\
0119$+$041     & OC +033   &   C    & 01 21 56.86170 & $+$04 22 24.7342 & 0.637 &  QSO   & Y & ... \\
0122$-$003     & UM 321    &   C    & 01 25 28.84383 & $-$00 05 55.9322 & 1.077 &  QSO   & Y & ... \\
0130$-$171     & OC $-$150 &   B    & 01 32 43.48746 & $-$16 54 48.5218 & 1.020 &  QSO   & Y & 21  \\
0149$+$218     &           &   C    & 01 52 18.05904 & $+$22 07 07.6997 & 1.320 &  QSO   & Y & ... \\
0202$-$172     &           &   C    & 02 04 57.67434 & $-$17 01 19.8405 & 1.739 &  QSO   & Y & 7   \\
\enddata
\tablecomments{
(1) Source name (B1950.0);
(2) Other commonly used source name;
(3) Observation session;
(4) Right ascension (J2000);
(5) Declination (J2000);
(6) Redshift;
(7) Optical classification, where QSO = quasar, BL Lac = BL Lac object, G = active galaxy, and U = unidentified;
(8) Source comments available in Sect.~\ref{comments};
(9) $\gamma$-ray flux
from \citet{2015ApJS..218...23A}.
}
\end{deluxetable}

\begin{deluxetable}{ccccccccc}
\tablecolumns{7}
\tabletypesize{\footnotesize}
\tablewidth{0pt}
\tablecaption{
Log of VLBA Observations
\label{tab:log}}
\tablehead{
\colhead{Session}  &
\colhead{Observation Code}   &
\colhead {Observation Date}   &
\colhead {$\rm \nu$}   &
\colhead {$\rm \tau_{int}$}    &
\colhead {No. of sources}     &
\colhead {Reference}  \\
\colhead{}         &
\colhead{}         &
\colhead{yyyy/mm/dd}          &
\colhead{GHz}          &
\colhead{min}                  &
\colhead{}                    &
\colhead{}     \\
\colhead{(1)}      & \colhead{(2)}                 & \colhead{(3)}          & \colhead{(4)}                  & \colhead{(5)}                 & \colhead{(6)} & \colhead{(7)}    }
\startdata

A                  &   BA111A -- BA111F            &  2014/11/21 -- 2016/05/06     &  43/86                 &  80                         &  10                                       & \citet{2018ApJS..234...17C} \\
B                  &   BA111G -- BA111I            &  2015/06/19 -- 2016/04/22     &   43                   &  28                         &  40                                       & This paper                  \\
C                  &   BA111J -- BA111M            &  2015/10/10 -- 2016/02/09     &   43                   &  14                         &  84                                       & This paper                  \\
\enddata
\tablecomments{
(1) Observing session;
(2) VLBA observation code;
(3) Range of observation dates;
(4) Observing frequency;
(5) Integration time for each individual source;
(6) Number of sources observed and imaged;
(7) References.
}
\end{deluxetable}

\begin{deluxetable}{cccccccccccc}
\tablecolumns{12}
\tabletypesize{\footnotesize}
\tablewidth{0pt}
\tablecaption{
Image Parameters
\label{tab:image}}
\tablehead{
\colhead{IAU}  & \colhead {Observation}  & \colhead{$B_{\rm maj}$} & \colhead{$B_{\rm min}$}  & \colhead{$B_{\rm P.A.}$}  & \colhead {$S_{\rm tot}$} & \colhead {$S_{\rm peak}$}  & \colhead{$\sigma$} &
\colhead{$S_{\rm S}$}  & \colhead{$B_{\rm S}$}  & \colhead{$S_{\rm L}$}  & \colhead{$B_{\rm  L }$} \\
\colhead{Name} & \colhead {yyyy-mm-dd}       & \colhead{(mas)}  & \colhead{(mas)}  & \colhead{($\degr$)} & \colhead {(Jy)}  & \colhead {(Jy beam$^{-1}$)} & \colhead{(mJy beam$^{-1}$)} & \colhead{(Jy)} & \colhead{(M$\lambda$)} &  \colhead{(Jy)} &  \colhead{(M$\lambda$)} \\
\colhead{(1)}  & \colhead{(2)} & \colhead{(3)} & \colhead{(4)} & \colhead{(5)} & \colhead{(6)}  & \colhead{(7)} & \colhead{(8)} & \colhead{(9)}    & \colhead{(10)}  &  \colhead{(11)}  &  \colhead{(12)}
}
\startdata
0048$-$071  & 2015-11-30 & 0.67 & 0.21 &   16.7  &  0.480    & 0.412  & 0.22  & 0.48  &  27  &  0.41  &  270  \\
0106$+$013  & 2015-06-19 & 0.44 & 0.18 &$-$4.4   &  1.202    & 0.801  & 0.70  & 1.09  &  24  &  0.75  &  406  \\
0113$-$118  & 2015-11-30 & 0.67 & 0.21 &   12.7  &  0.698    & 0.568  & 0.85  & 0.70  &  27  &  0.55  &  260  \\
0119$+$041  & 2015-11-30 & 0.52 & 0.21 &   14.2  &  0.546    & 0.345  & 0.40  & 0.54  &  30  &  0.33  &  296  \\
0122$-$003  & 2015-11-30 & 0.63 & 0.19 &   18.4  &  0.425    & 0.256  & 0.17  & 0.40  &  27  &  0.30  &  286  \\
0130$-$171  & 2015-06-19 & 0.44 & 0.16 &$-$5.1   &  0.597    & 0.390  & 0.58  & 0.59  &  20  &  0.23  &  438  \\
0149$+$218  & 2015-11-30 & 0.49 & 0.20 &   16.8  &  1.160    & 0.970  & 1.46  & 1.14  &  33  &  0.81  &  336  \\
0202$-$172  & 2015-11-30 & 0.72 & 0.20 &   12.6  &  1.015    & 0.676  & 1.20  & 0.82  &  31  &  0.65  &  244  \\
\enddata

\tablecomments{
(1) Source name (B1950.0);
(2) Observing date;
(3) Major axis of the restoring beam (FWHM);
(4) Minor axis of the restoring beam (FWHM);
(5) Position angle of the major axis, measured from north through east;
(6) Integrated flux density;
(7) Peak specific intensity;
(8) Off-source rms noise in the clean image;
(9) Correlated flux density on the shortest baseline;
(10) Shortest baseline length in million wavelengths;
(11) Correlated flux density on the longest baseline;
(12) Longest baseline length in million wavelengths.
}
\end{deluxetable}

\begin{deluxetable}{cccccccccc}
\tablecolumns{9}
\tabletypesize{\footnotesize}
\tablewidth{0pt}
\tablecaption{
Model Fit Parameters of Sources
\label{model}}
\tablehead{
\colhead{IAU}  & \colhead {Session}  & \colhead {Comp.} & \colhead {$S_{\rm tot}$} &  \colhead {$S_{\rm peak}$} & \colhead{R} & \colhead{P.A.}    & \colhead{$d$}   & \colhead{$T_{b}$}    \\
\colhead{Name} & \colhead {}       & \colhead {}      & \colhead {(Jy)}  &    \colhead {$\rm Jy\, beam^{-1}$}   & \colhead{(mas)}  & \colhead{($\degr$)}     & \colhead{(mas)} & \colhead{($10^{10}$K)}  \\
\colhead{(1)}  & \colhead{(2)}     & \colhead{(3)}    & \colhead{(4)}        & \colhead{(5)}    & \colhead{(6)}           & \colhead{(7)}   & \colhead{(8)}   & \colhead{(9)}   }
\startdata
0048$-$071...    &  C  &  C   &  0.424  &  0.412  &    ...     &     ...    &  0.05  &  33.1        \\
                 &     &  J3  &  0.028  &  0.024  &    0.43    &  $-$39.97  &  0.13  &              \\
                 &     &  J2  &  0.021  &  0.014  &    0.90    &  $-$52.83  &  0.21  &              \\
                 &     &  J1  &  0.007  &  0.002  &    2.36    &  $-$57.45  &  0.53  &              \\
0106$+$013...    &  B  &  C   &  0.869  &  0.801  &    ...     &     ...    &  0.05  &  70.7        \\
                 &     &  J2  &  0.118  &  0.103  &    0.35    &  $-$116.37 &  0.10  &              \\
                 &     &  J1  &  0.215  &  0.036  &    1.36    &  $-$136.52 &  1.04  &              \\
0113$-$118...    &  C  &  C   &  0.657  &  0.569  &    ...     &     ...    &  0.12  &  5.0         \\
\enddata
\tablecomments{
(1) Source name (B1950.0);
(2) Observing session;
(3) Component label;
(4) Integrated flux density of the fitted Gaussian model component;
(5) Peak intensity;
(5) Distance from the core;
(8) Position angle with respect to the core, measured from north through east;
(9) Core brightness temperature.
}
\end{deluxetable}

\begin{figure}
  \centering
  \begin{tabular}{lcr}
  \includegraphics[width=0.3\textwidth]{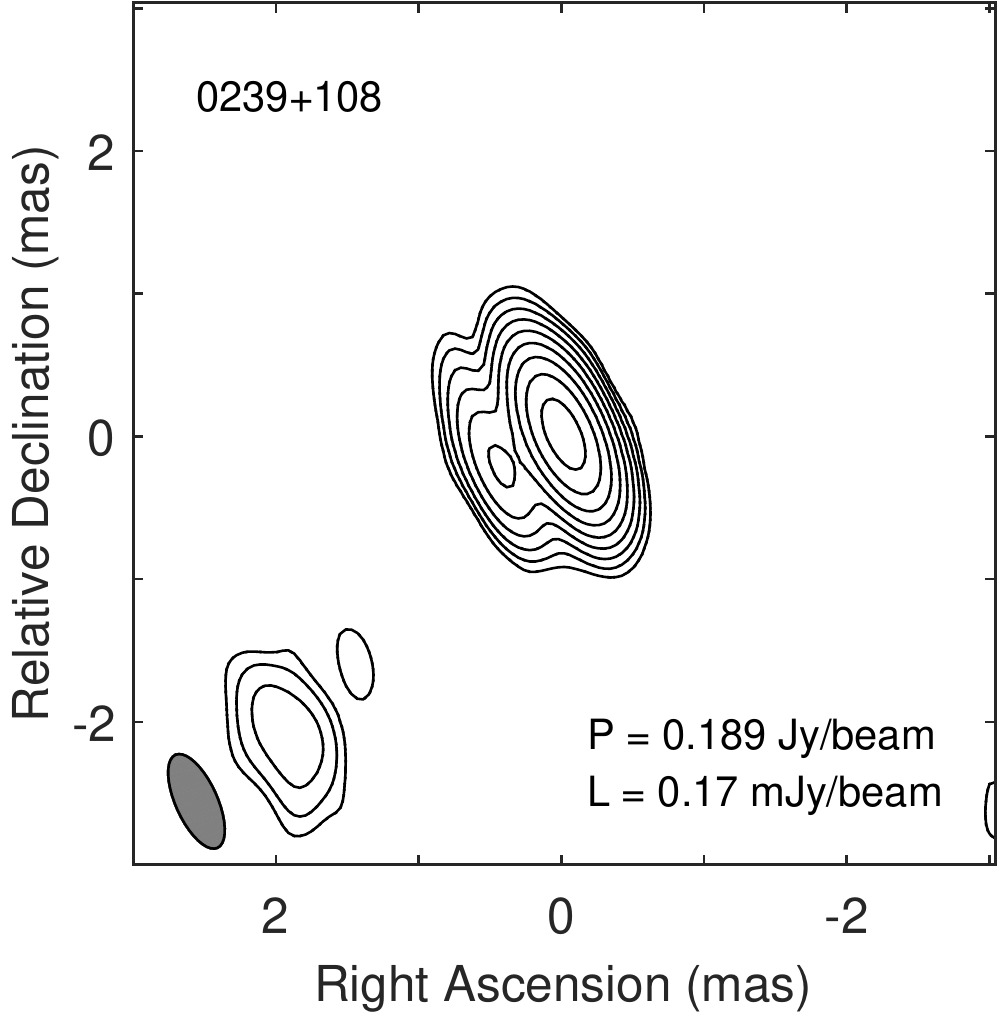} & \includegraphics[width=0.3\textwidth]{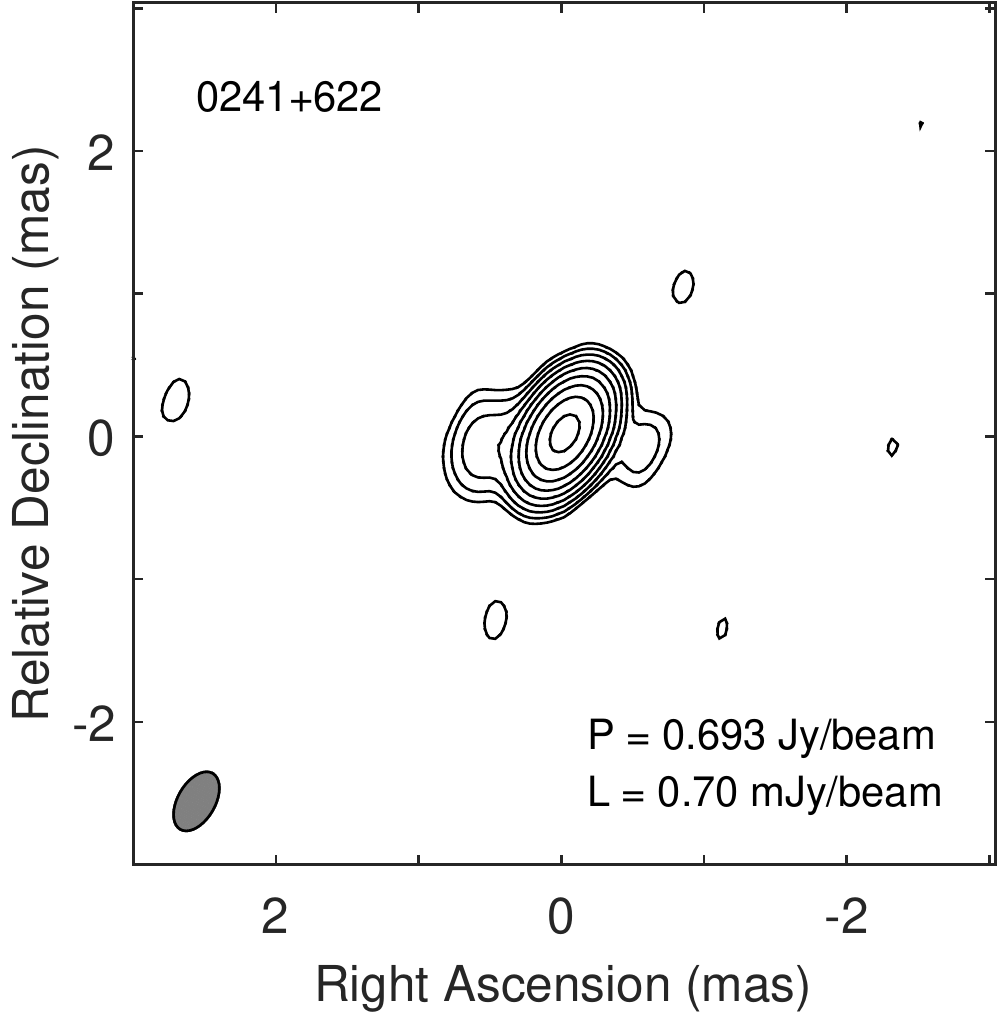}  & \includegraphics[width=0.3\textwidth]{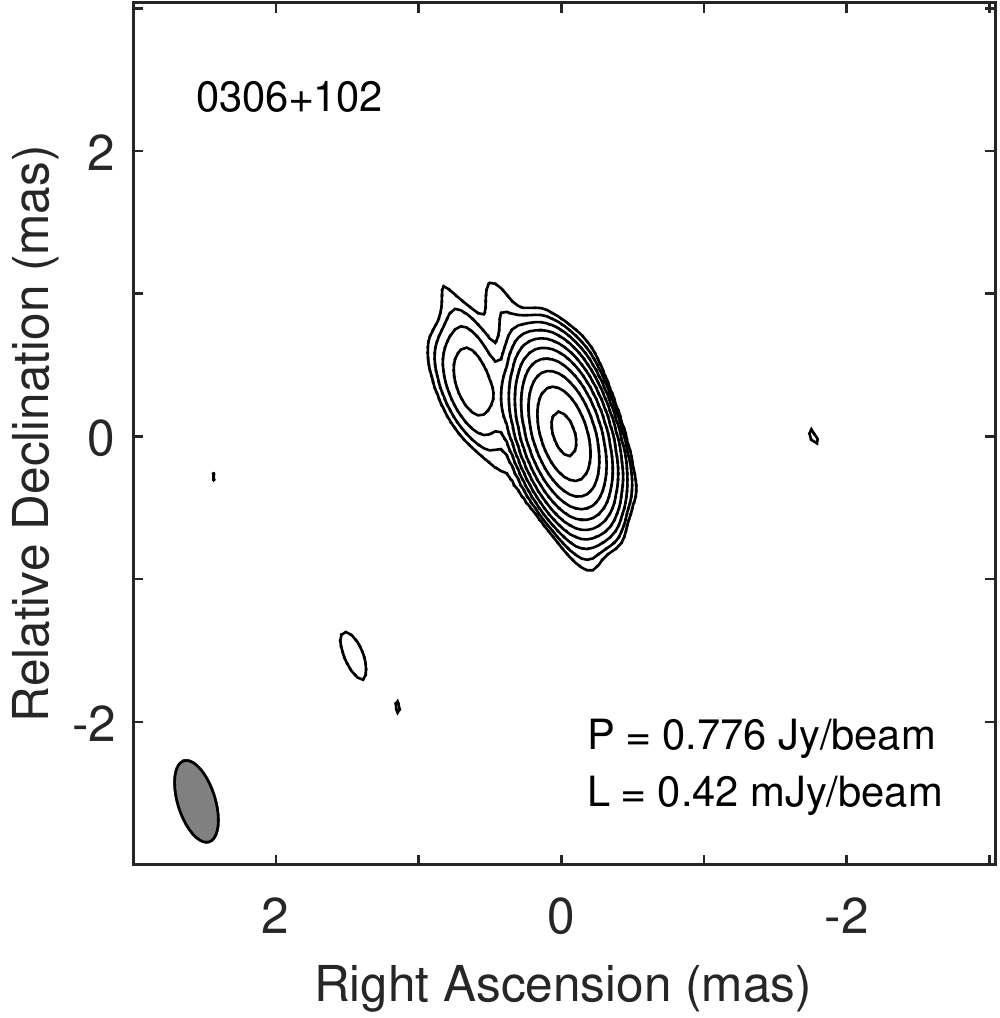}\\ \includegraphics[width=0.3\textwidth]{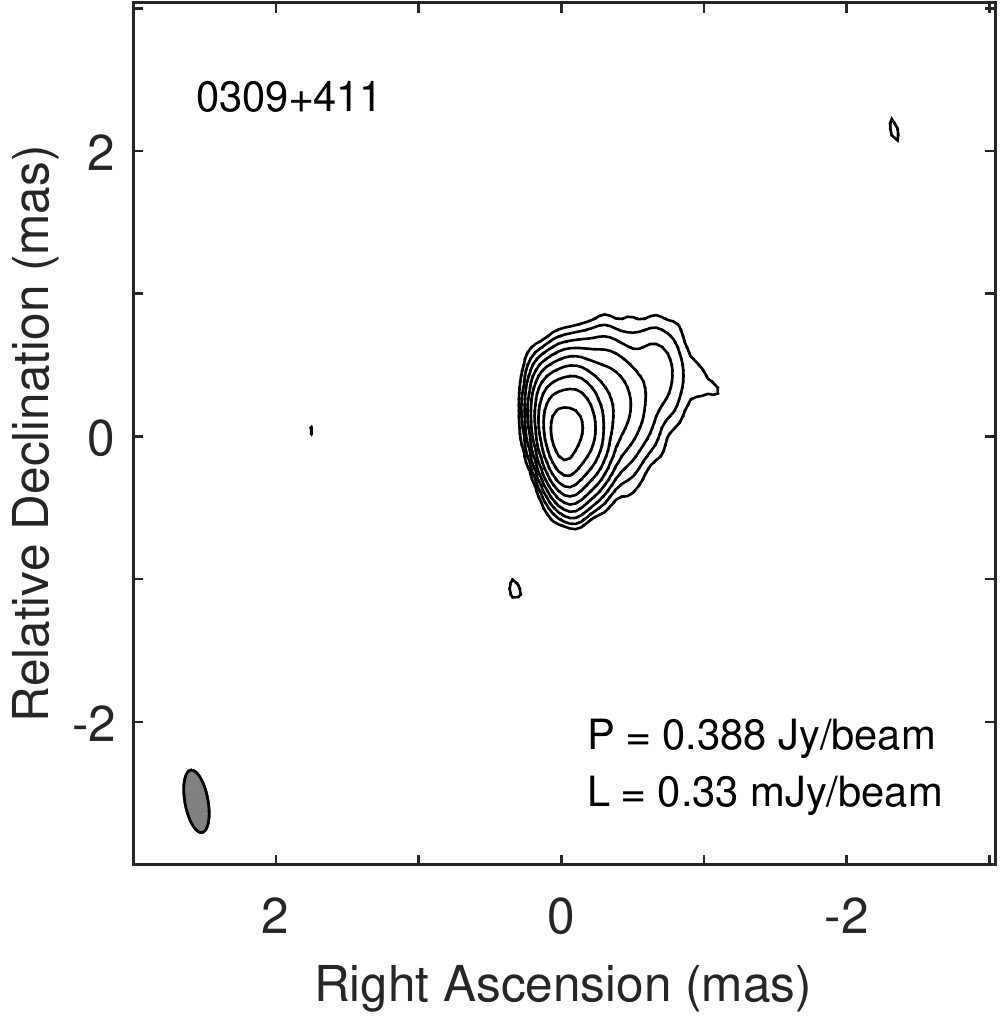} & \includegraphics[width=0.3\textwidth]{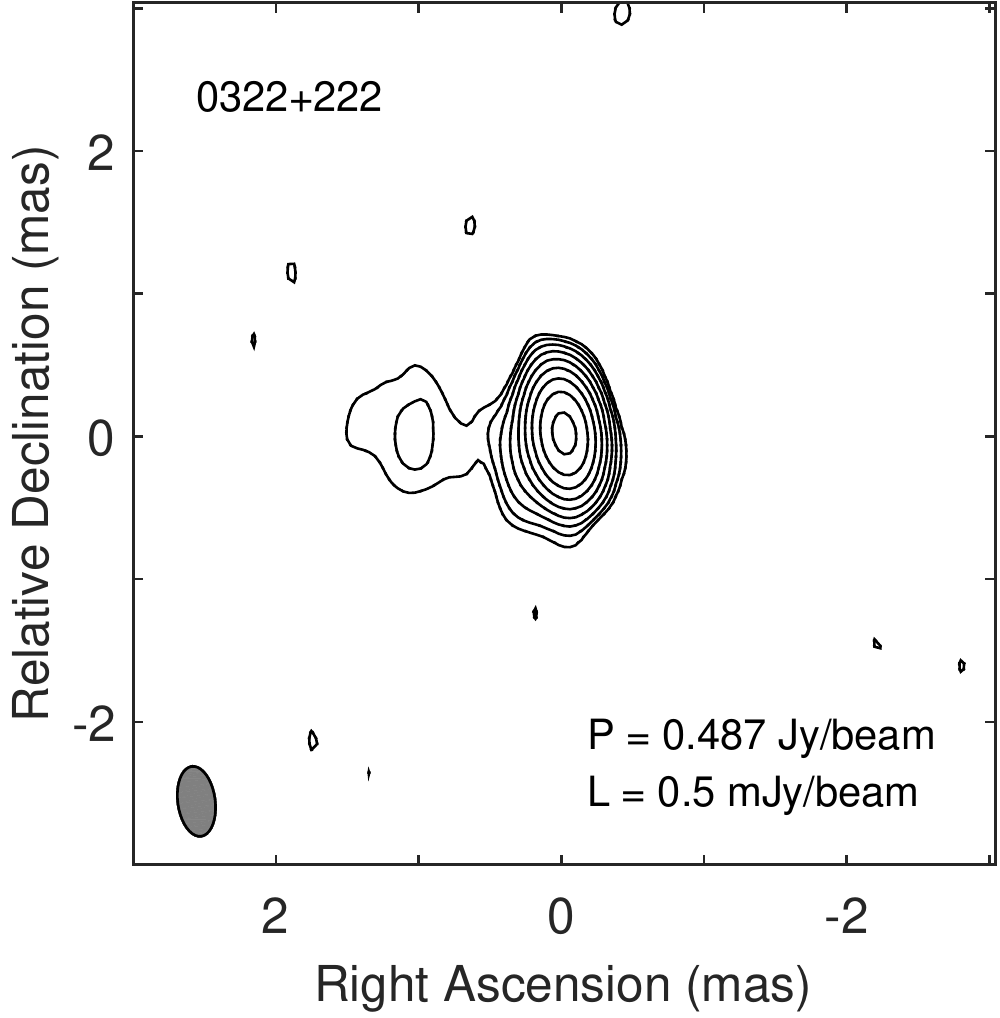}  & \includegraphics[width=0.3\textwidth]{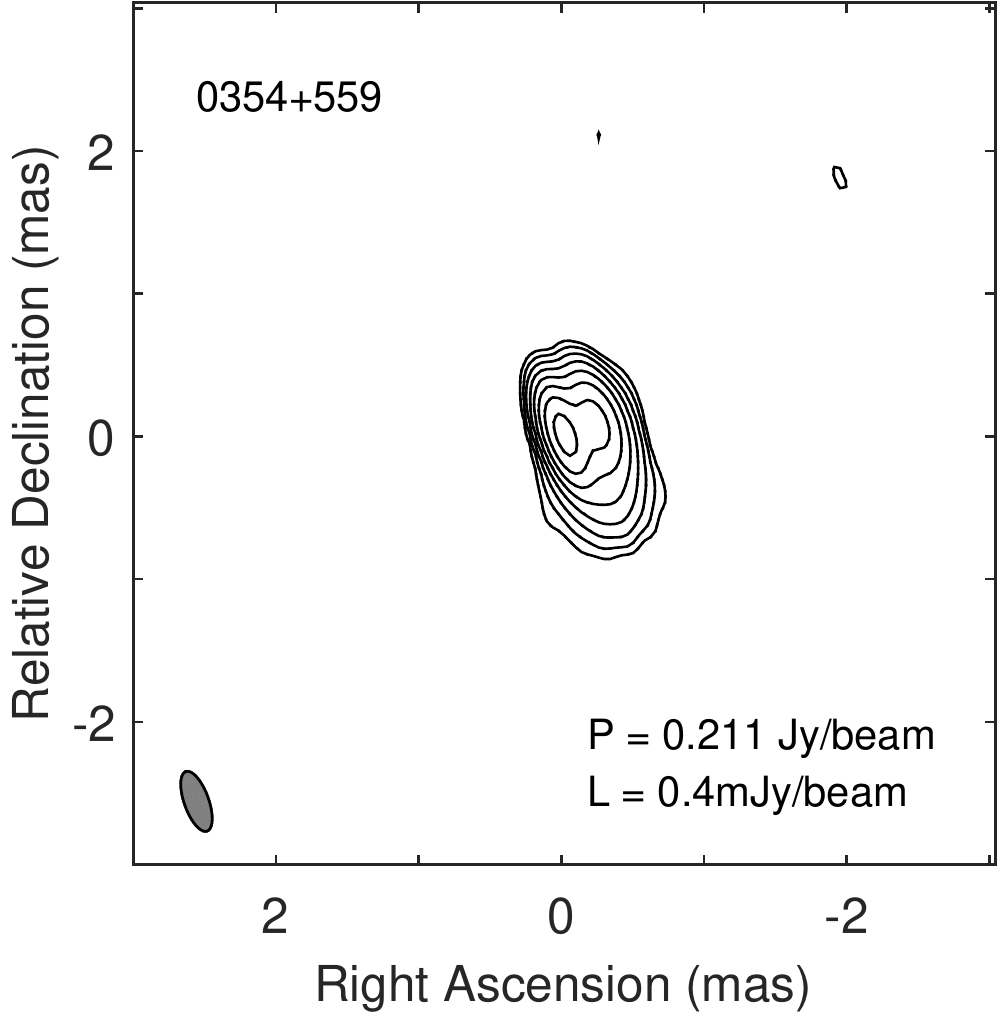}\\
  \includegraphics[width=0.3\textwidth]{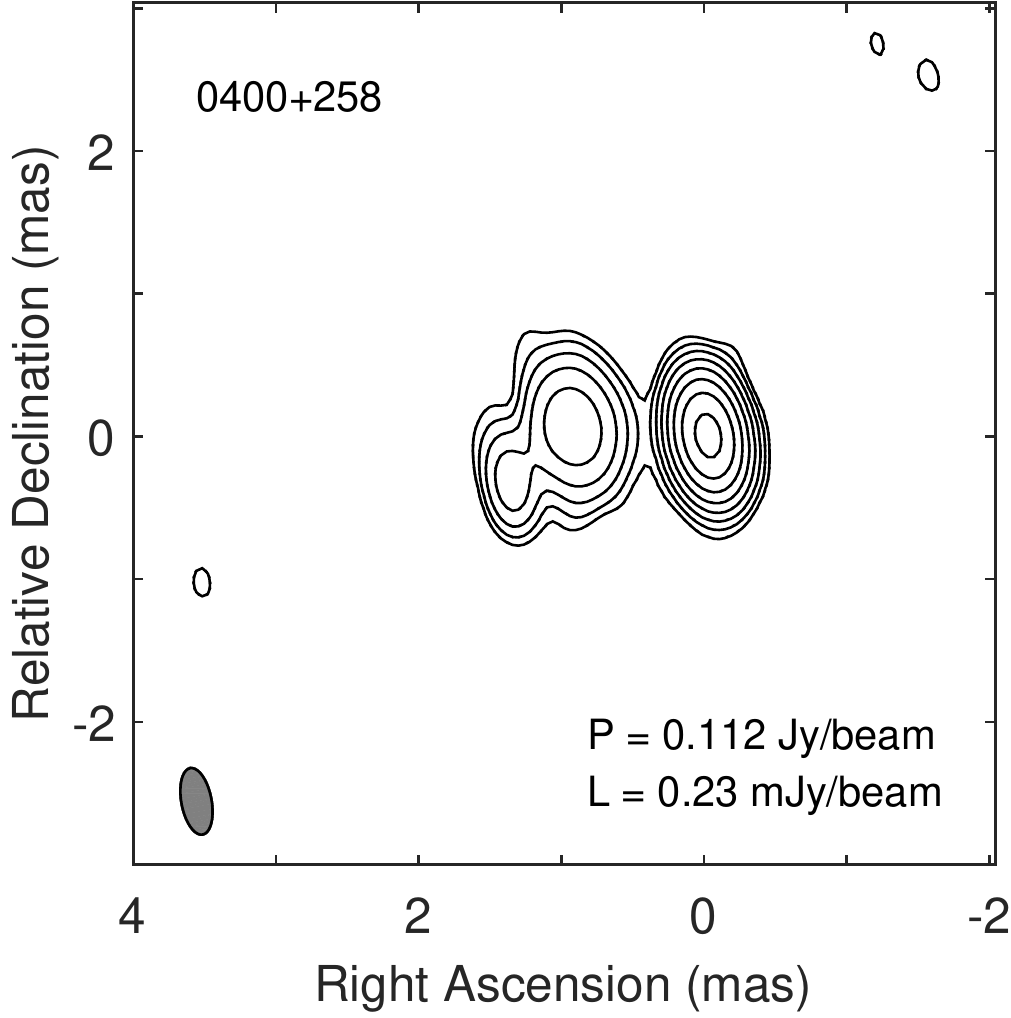} & \includegraphics[width=0.3\textwidth]{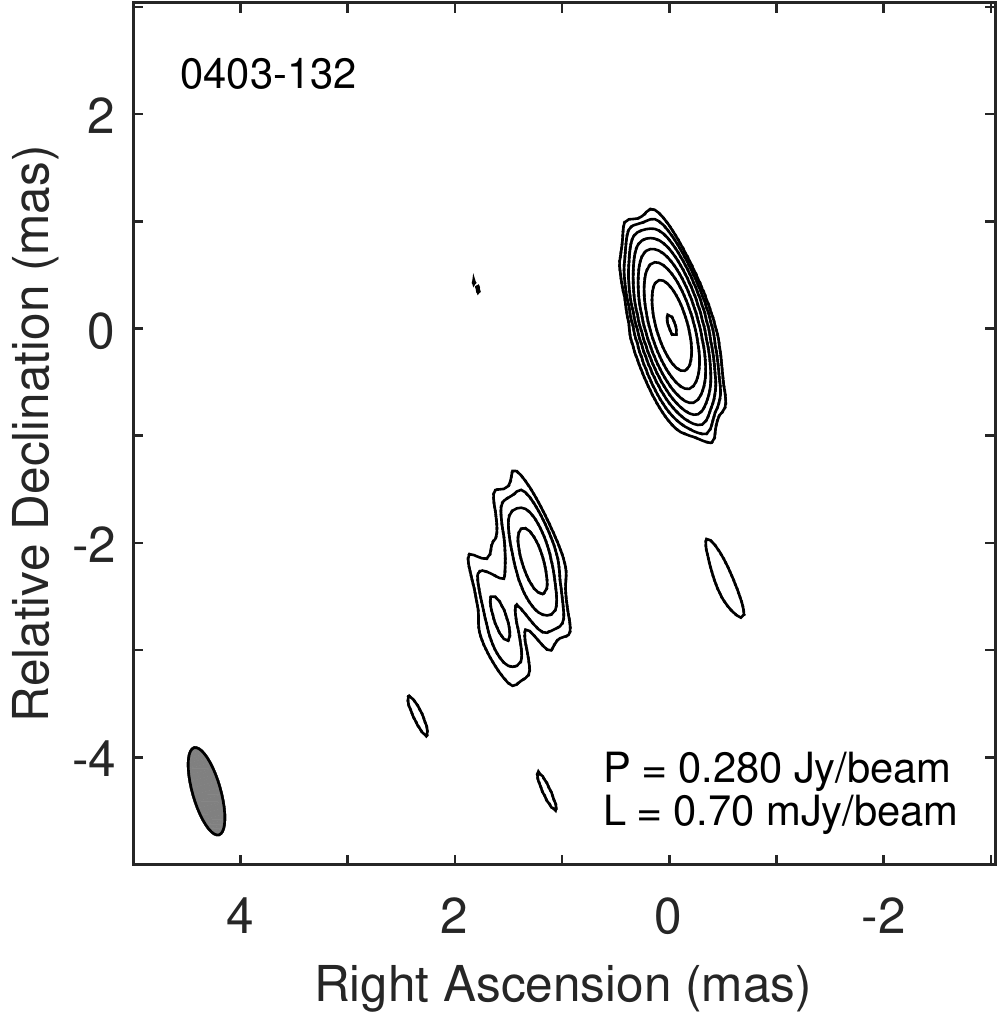}  & \includegraphics[width=0.3\textwidth]{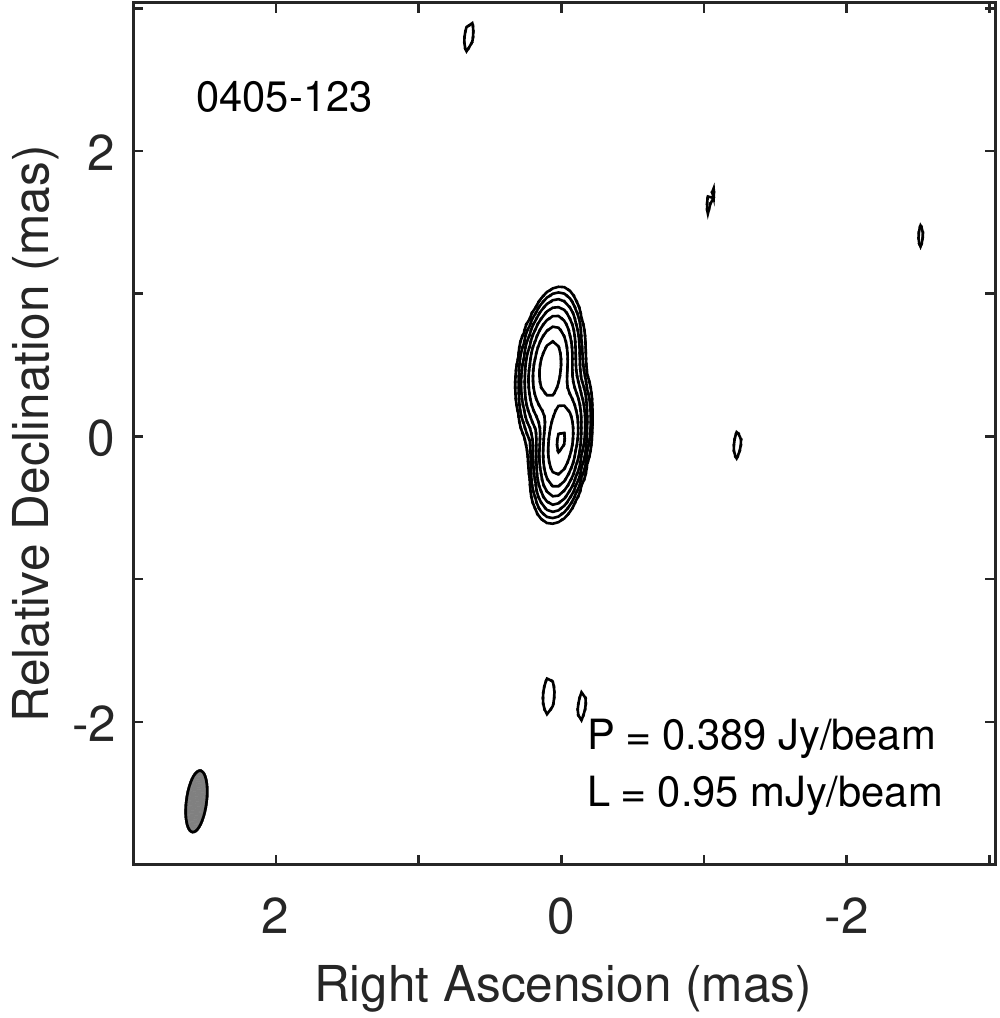}\\ \includegraphics[width=0.3\textwidth]{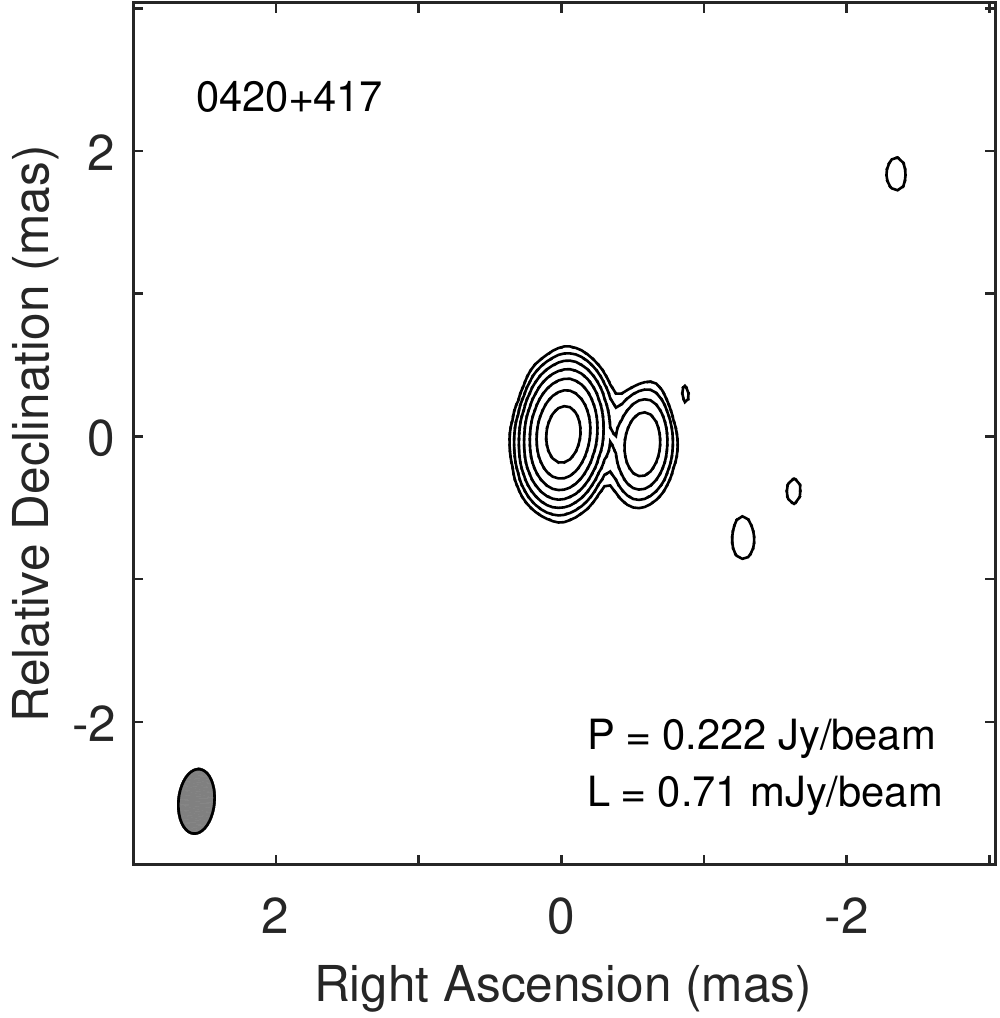} & \includegraphics[width=0.3\textwidth]{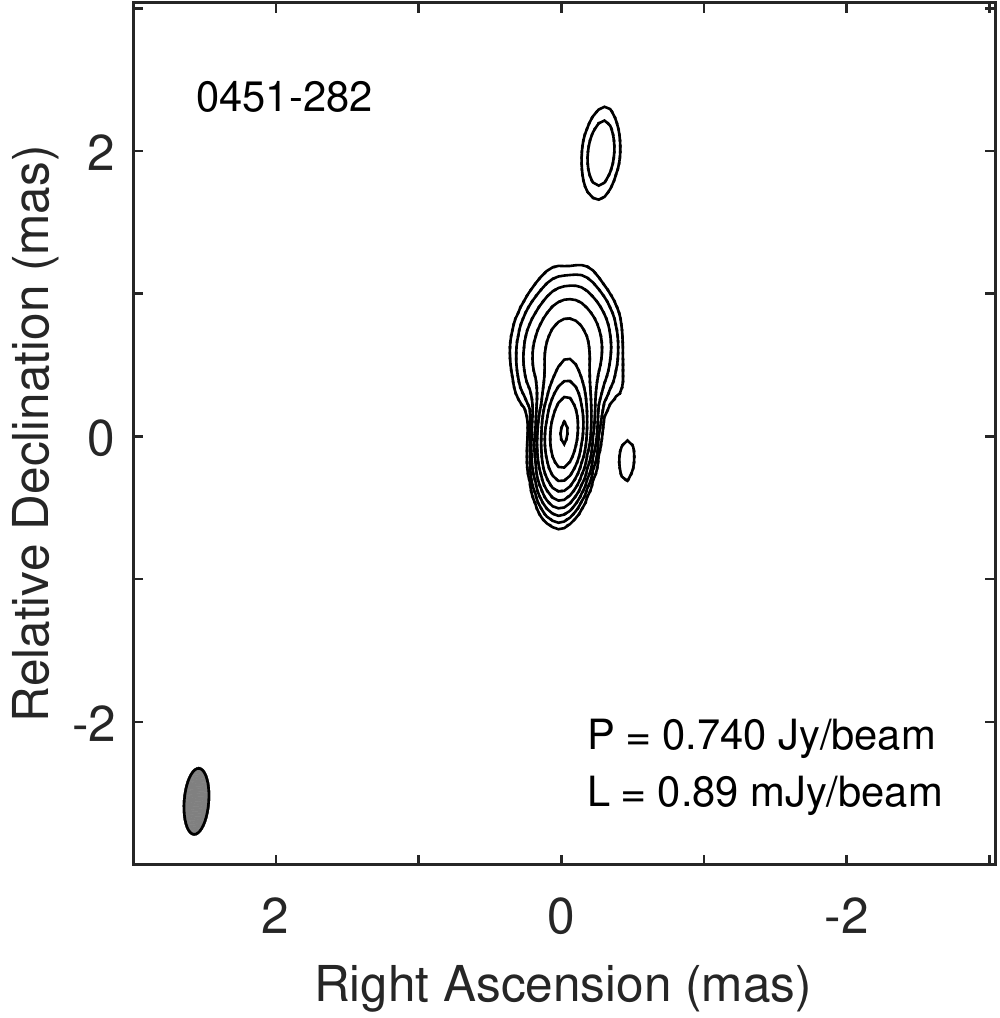}  & \includegraphics[width=0.3\textwidth]{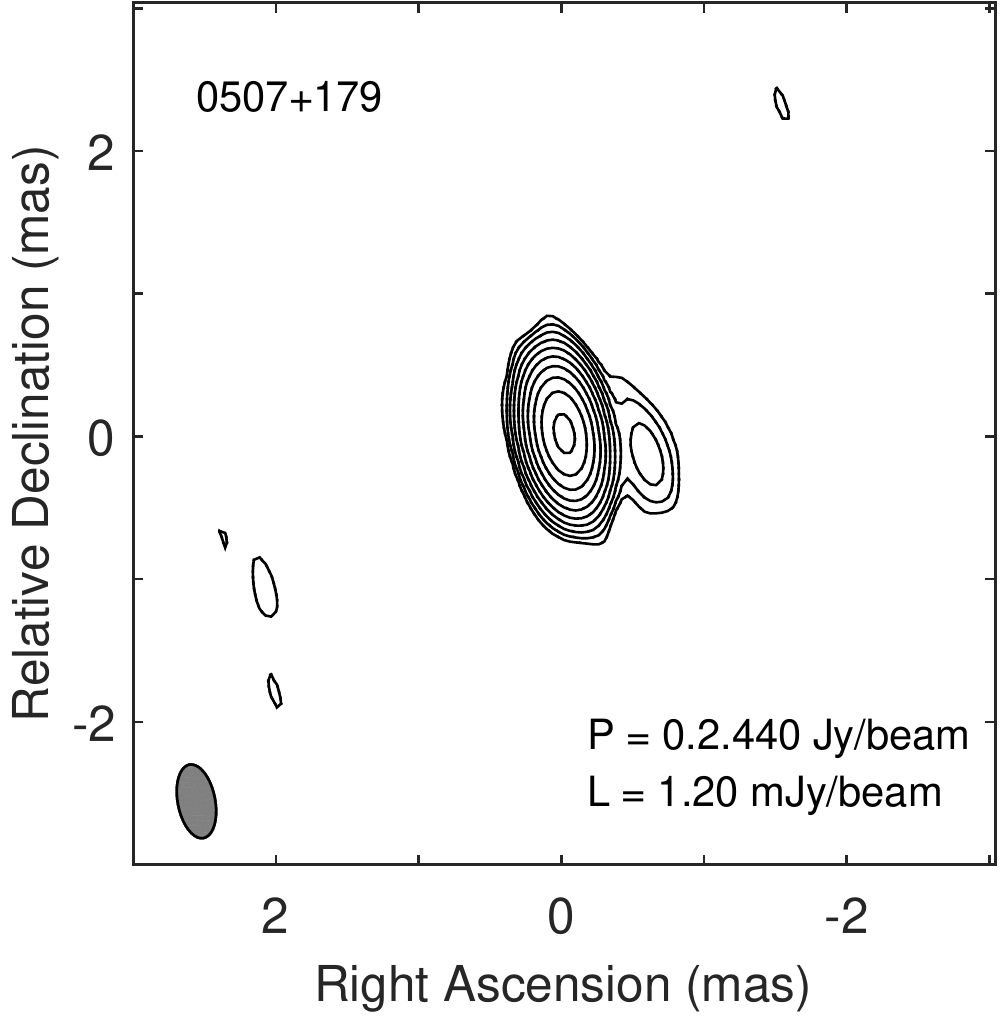}\\
    \end{tabular}
  \caption{VLBA images --- Continued}
  \end{figure}

 \begin{figure}
  \centering
  \begin{tabular}{lcr}
  \includegraphics[width=0.3\textwidth]{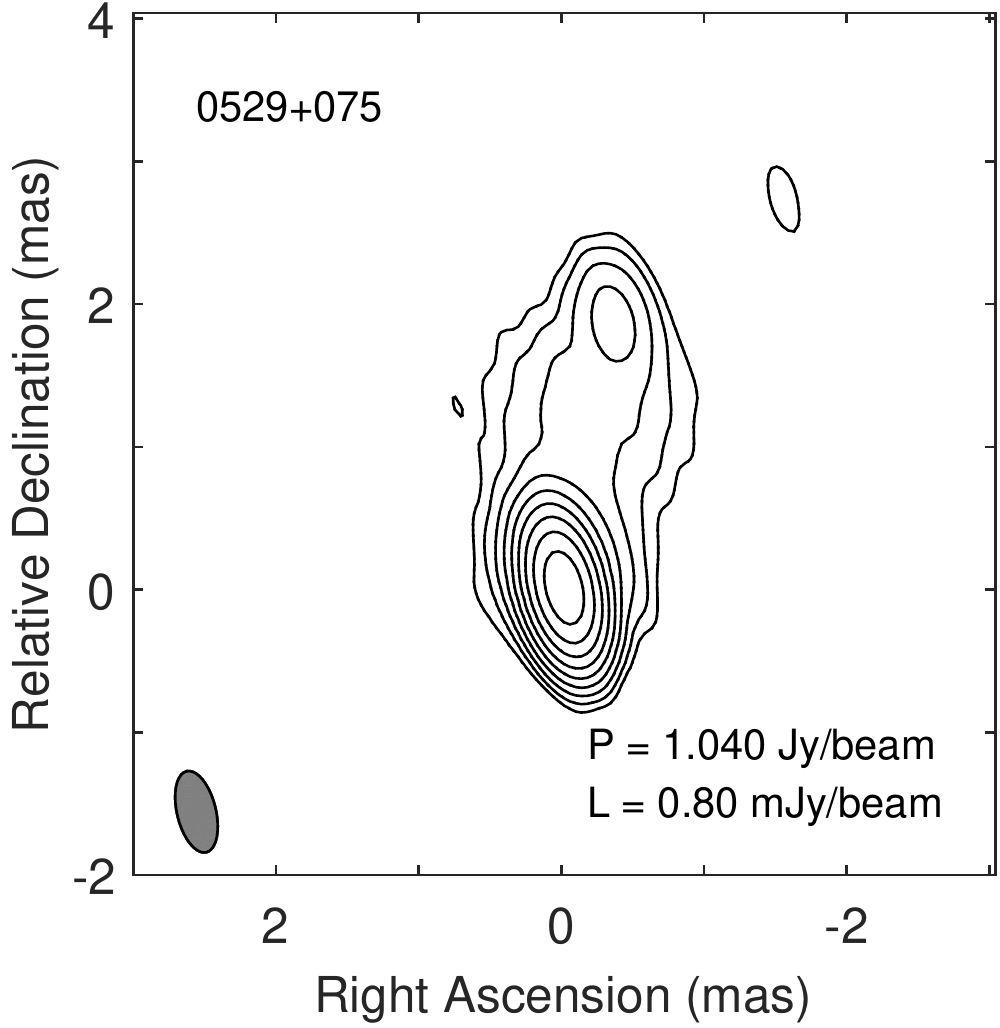} & \includegraphics[width=0.3\textwidth]{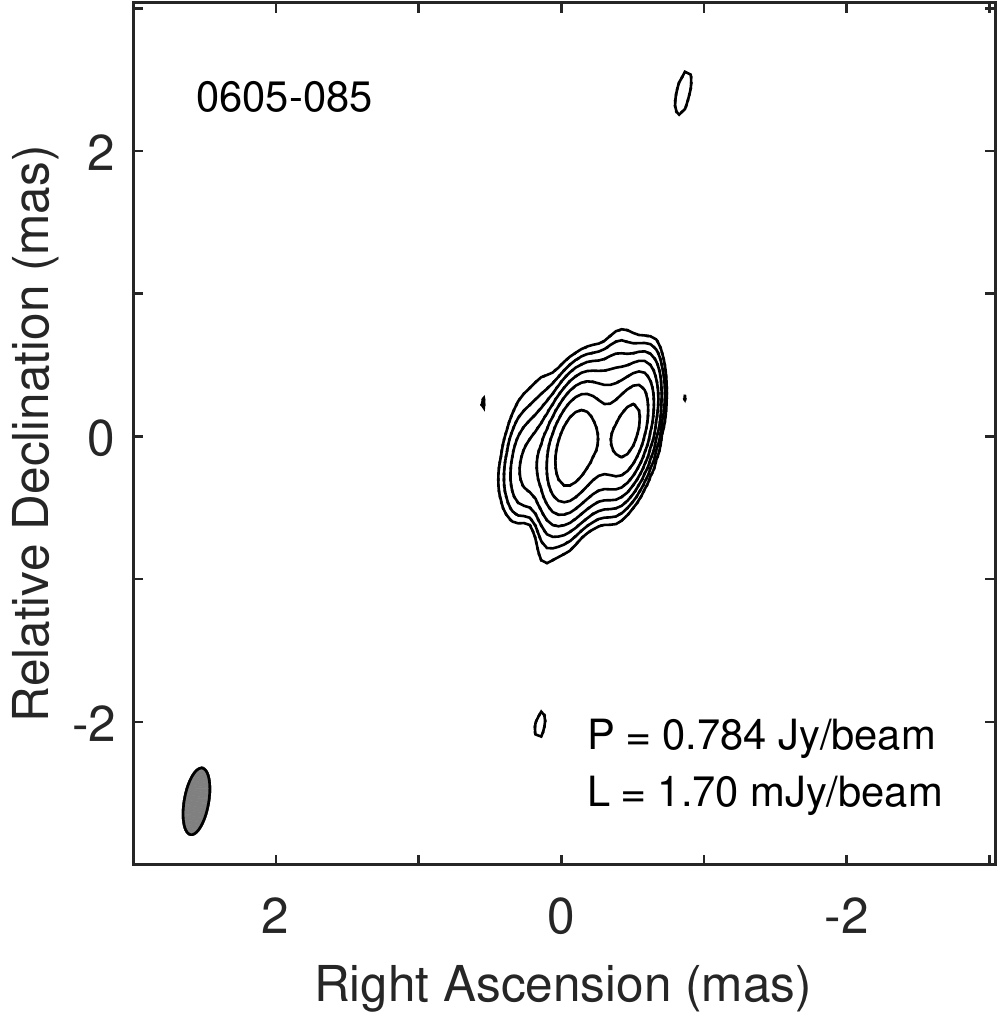}  & \includegraphics[width=0.3\textwidth]{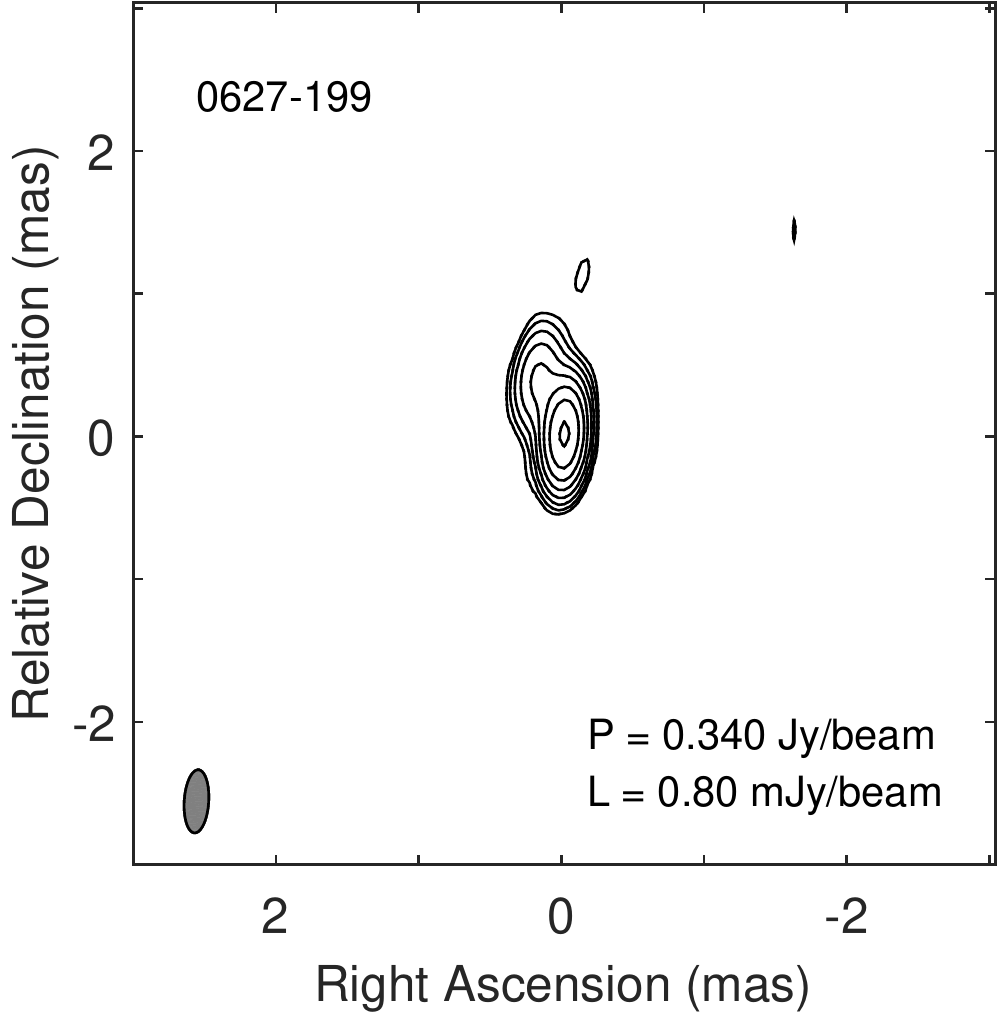}\\ \includegraphics[width=0.3\textwidth]{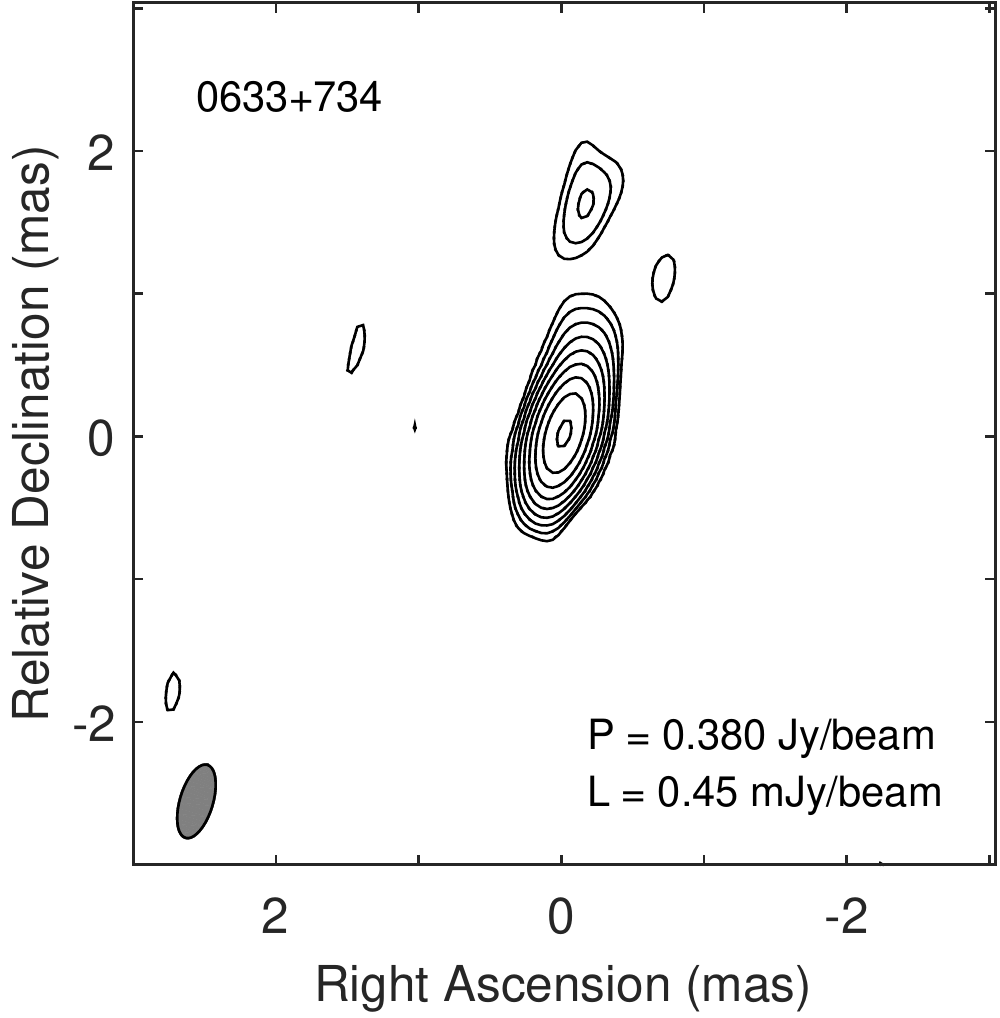} & \includegraphics[width=0.3\textwidth]{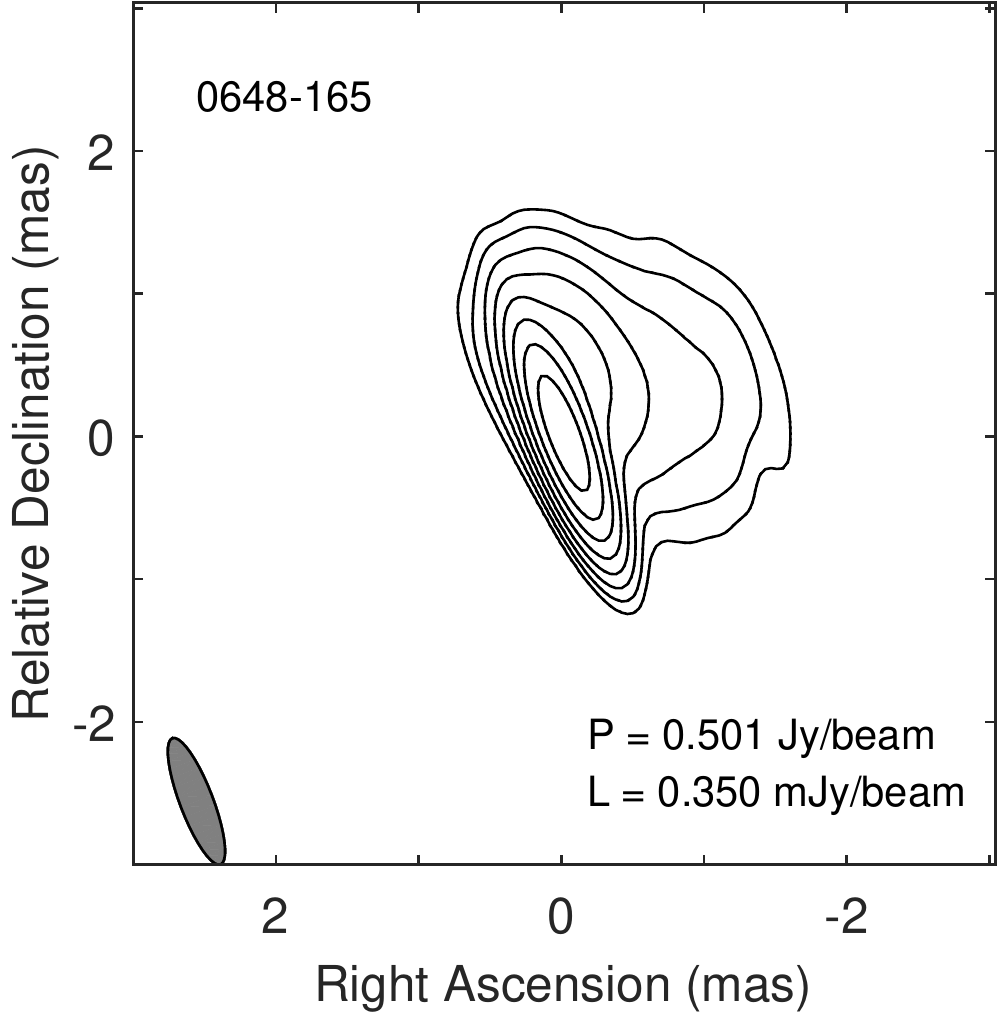}  & \includegraphics[width=0.3\textwidth]{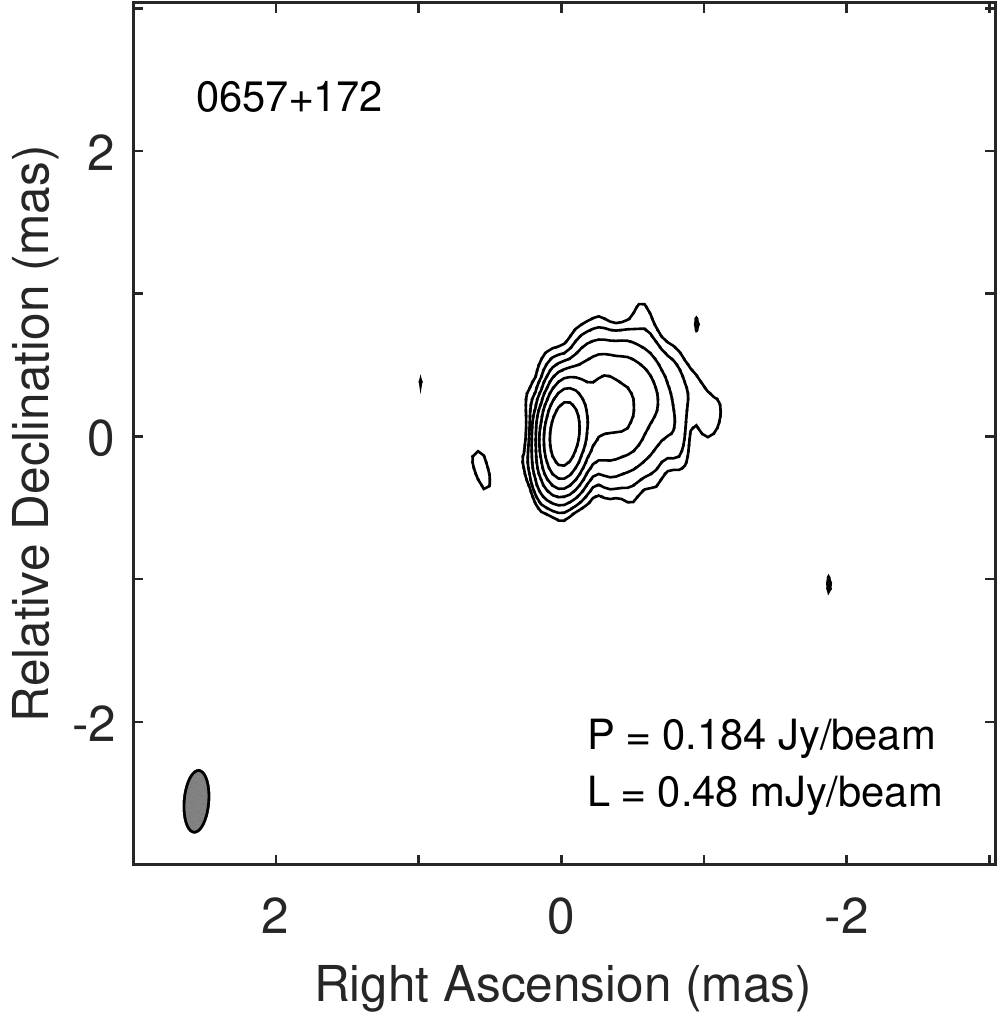}\\
  \includegraphics[width=0.3\textwidth]{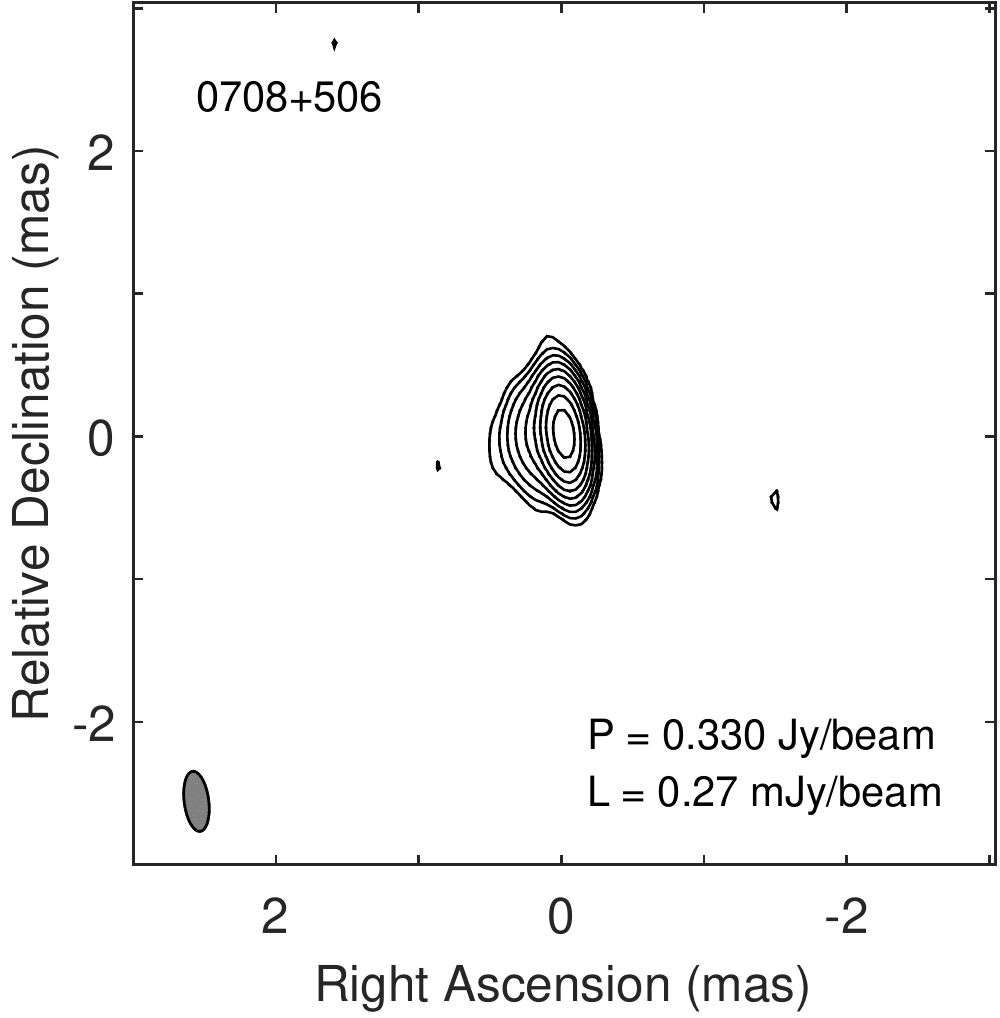} & \includegraphics[width=0.3\textwidth]{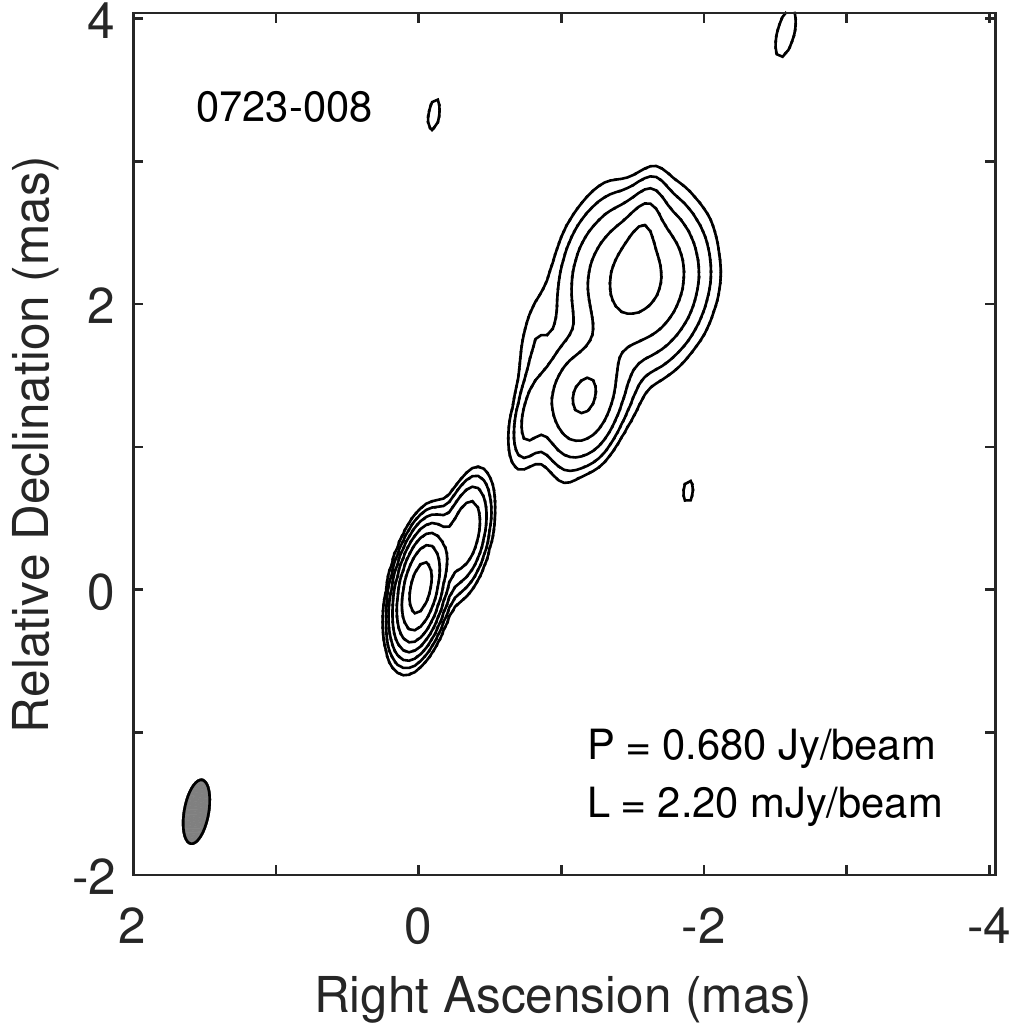}  & \includegraphics[width=0.3\textwidth]{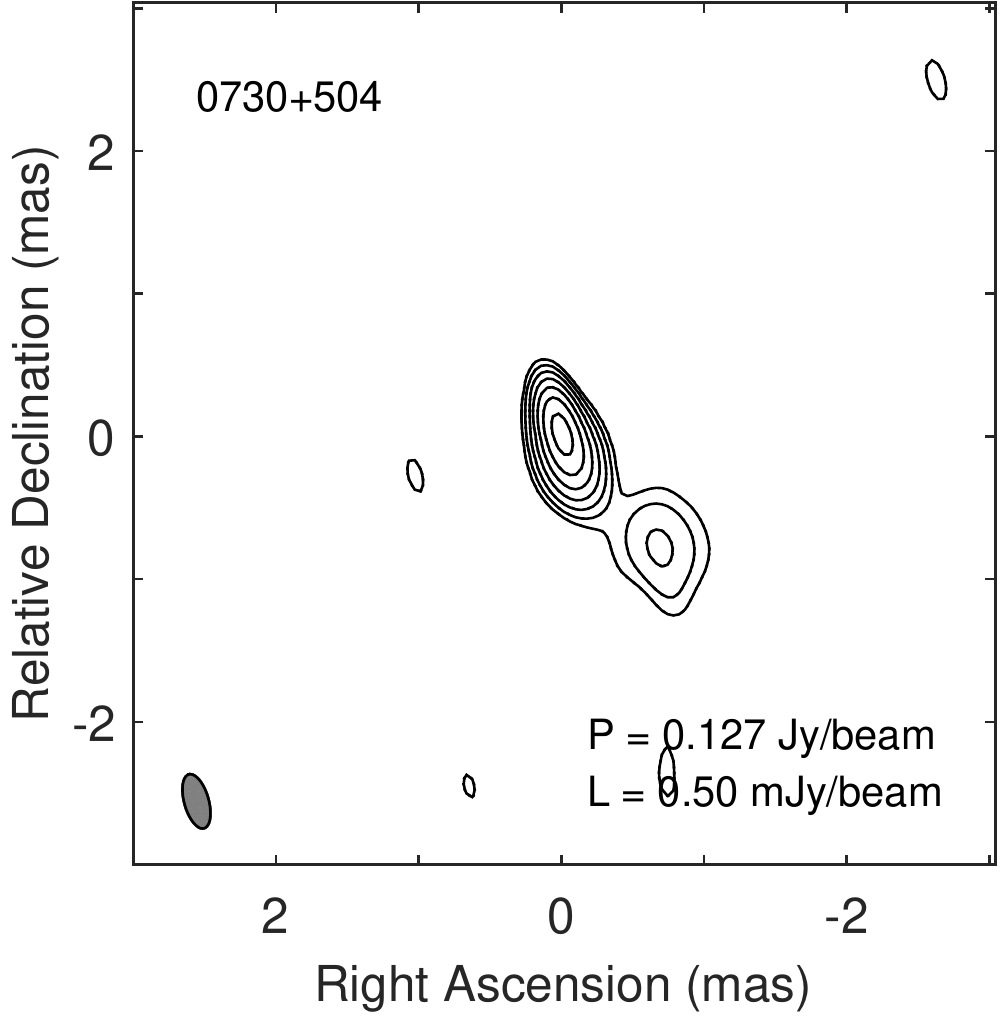}\\ \includegraphics[width=0.3\textwidth]{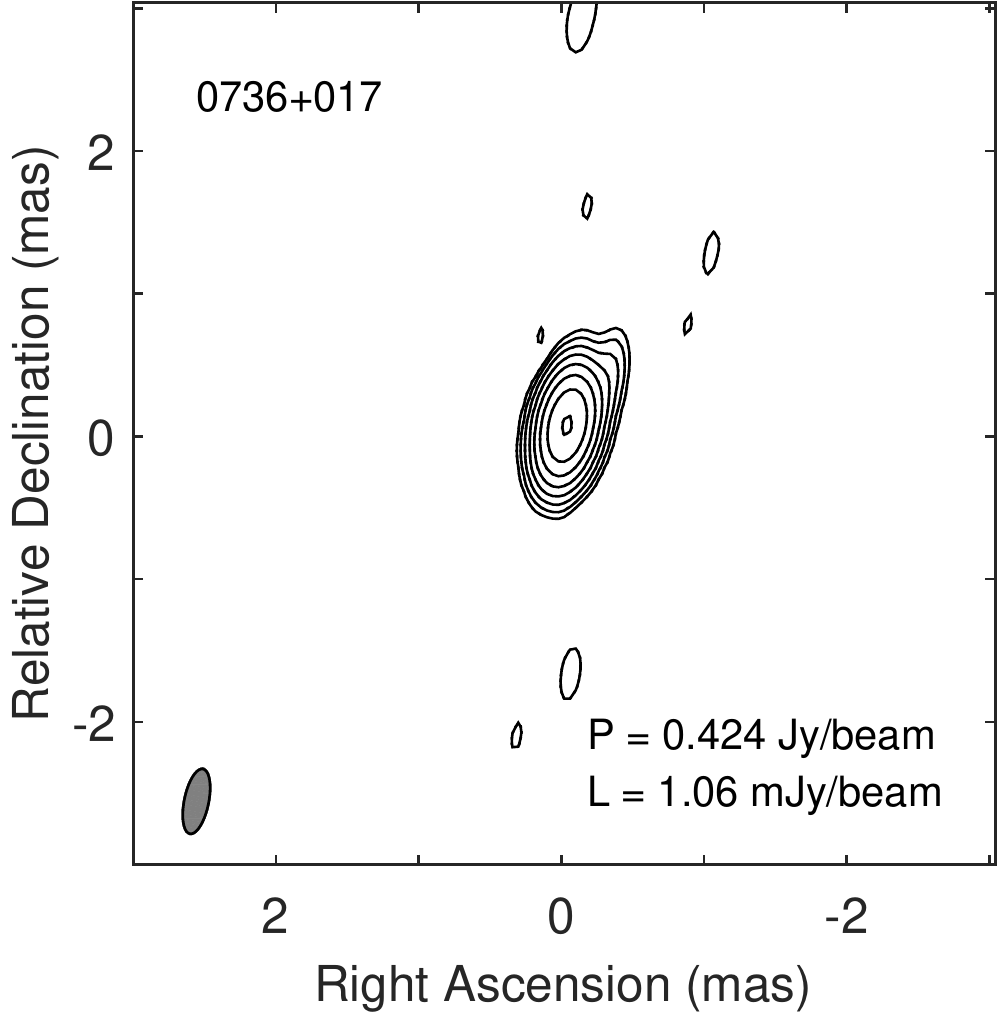} & \includegraphics[width=0.3\textwidth]{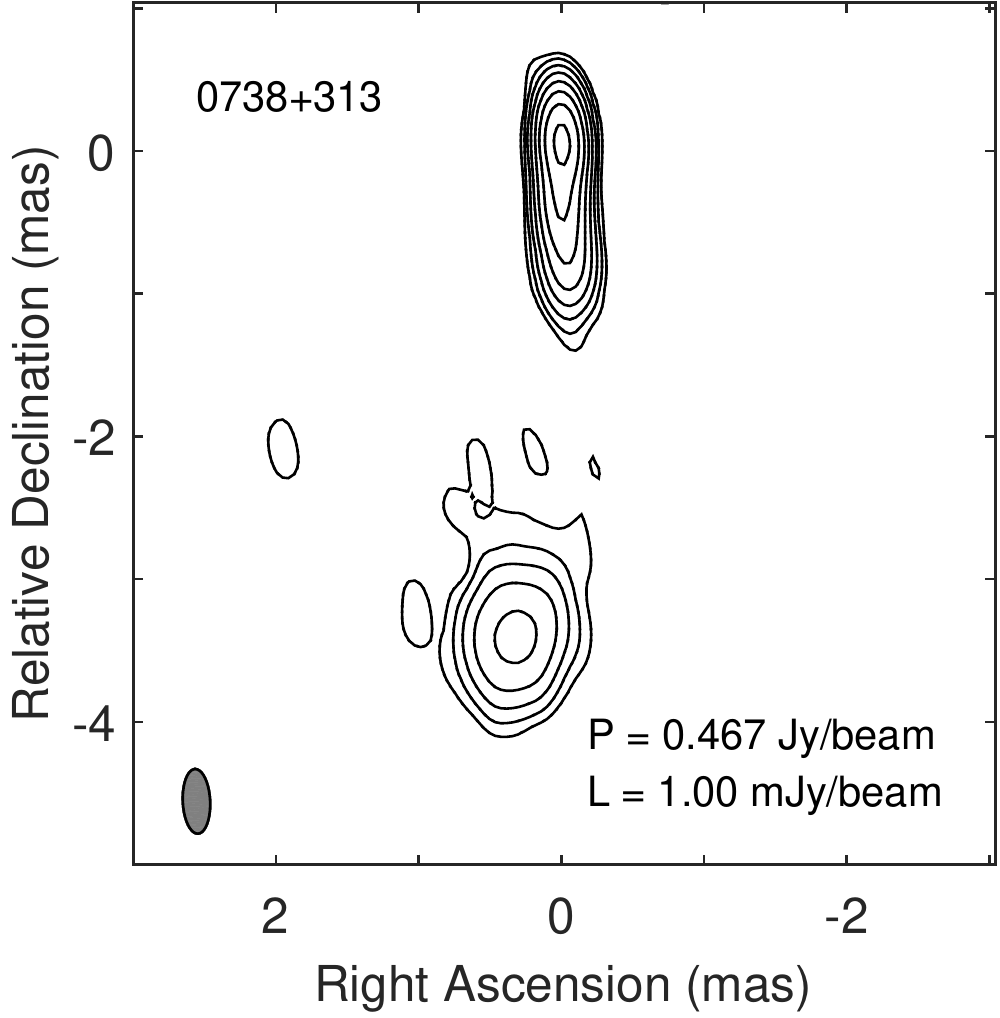}  & \includegraphics[width=0.3\textwidth]{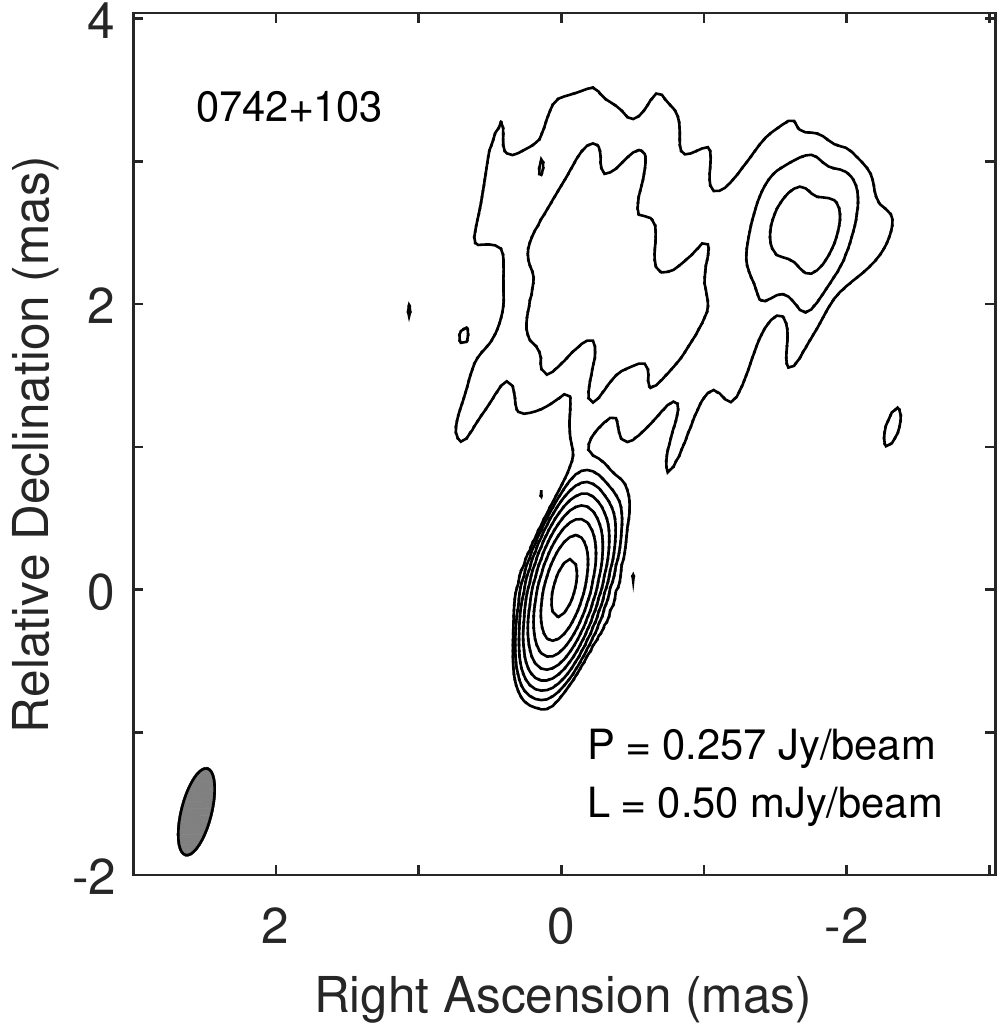}\\
    \end{tabular}
   \caption{VLBA images --- Continued}
  \end{figure}

\begin{figure}
  \centering
  \begin{tabular}{lcr}
  \includegraphics[width=0.3\textwidth]{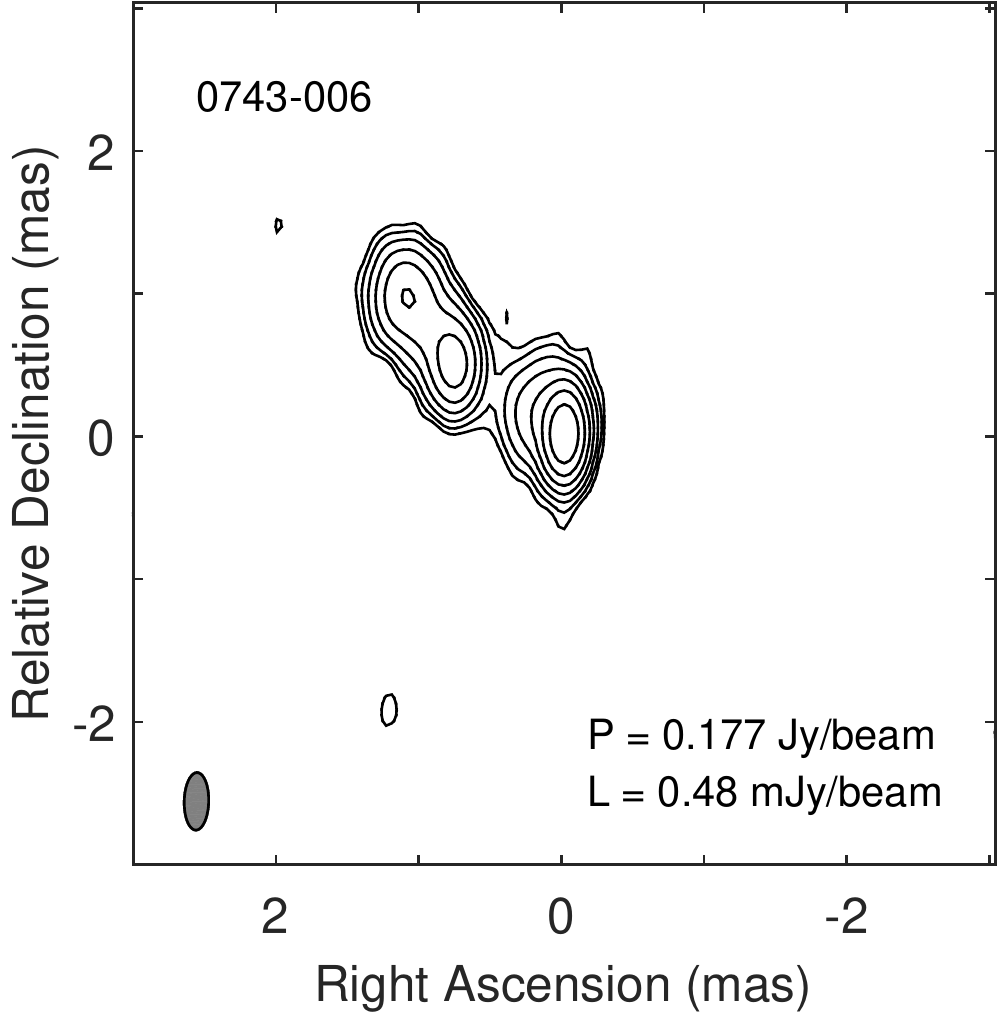} & \includegraphics[width=0.3\textwidth]{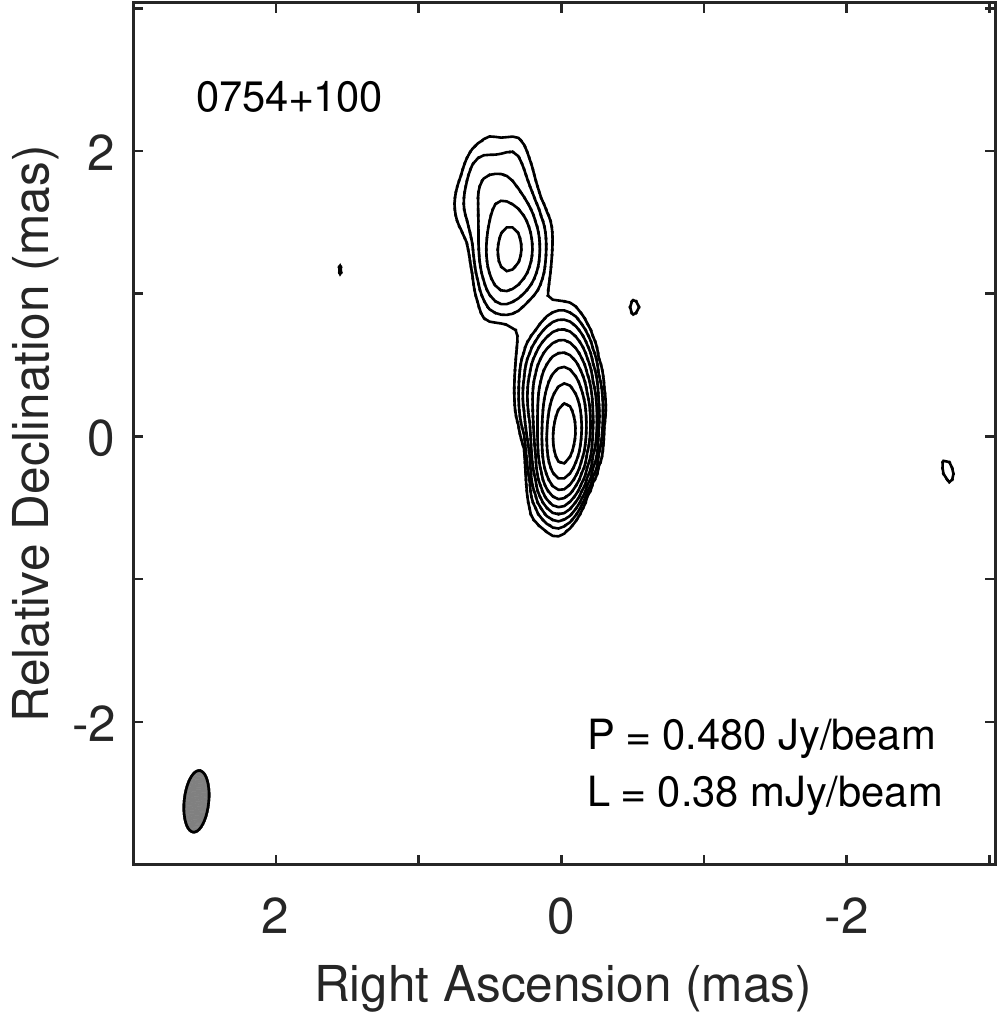}  & \includegraphics[width=0.3\textwidth]{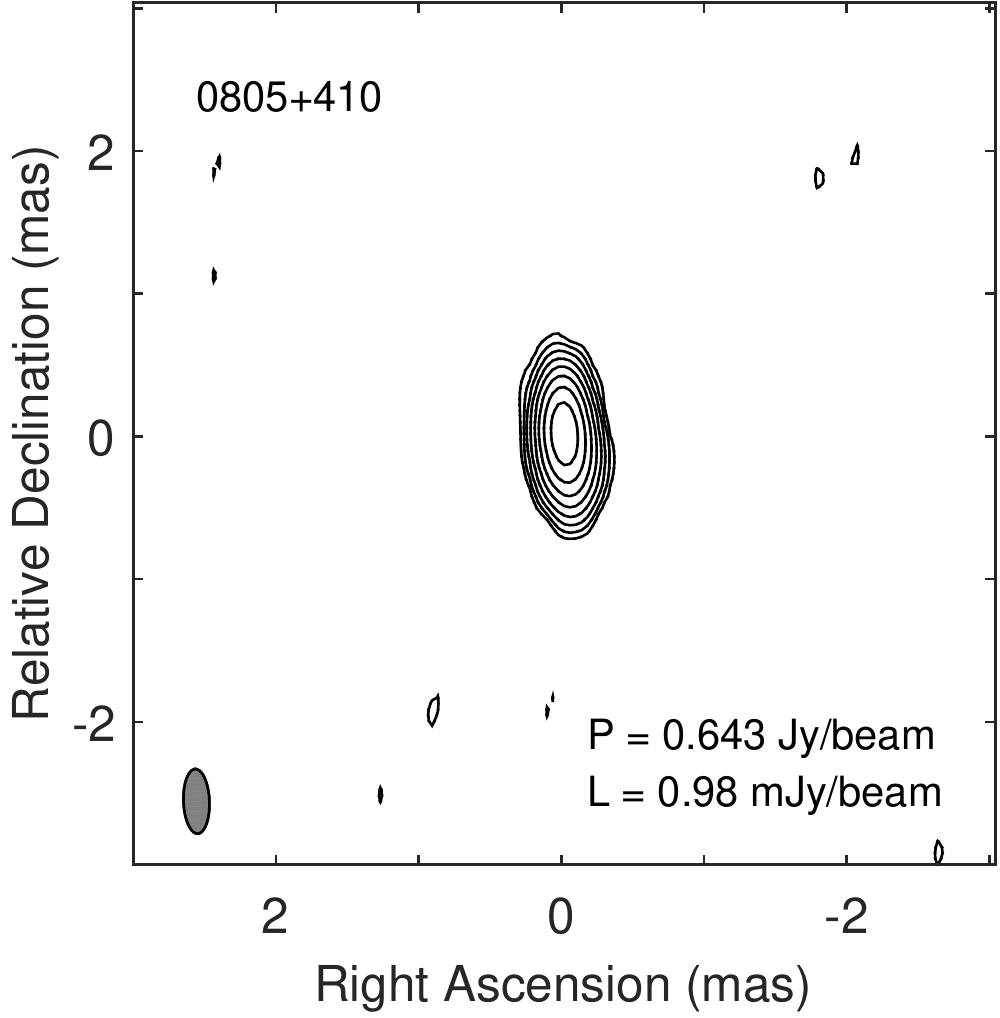}\\ \includegraphics[width=0.3\textwidth]{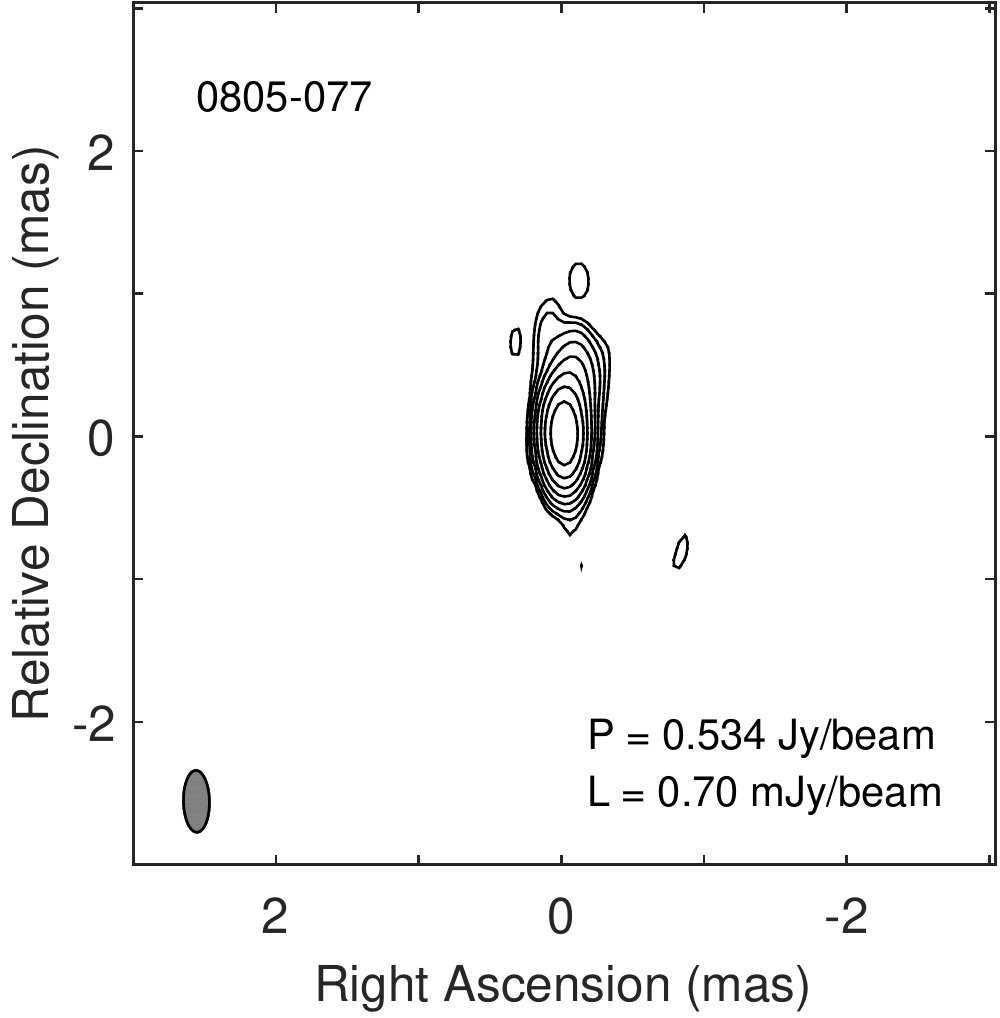} & \includegraphics[width=0.3\textwidth]{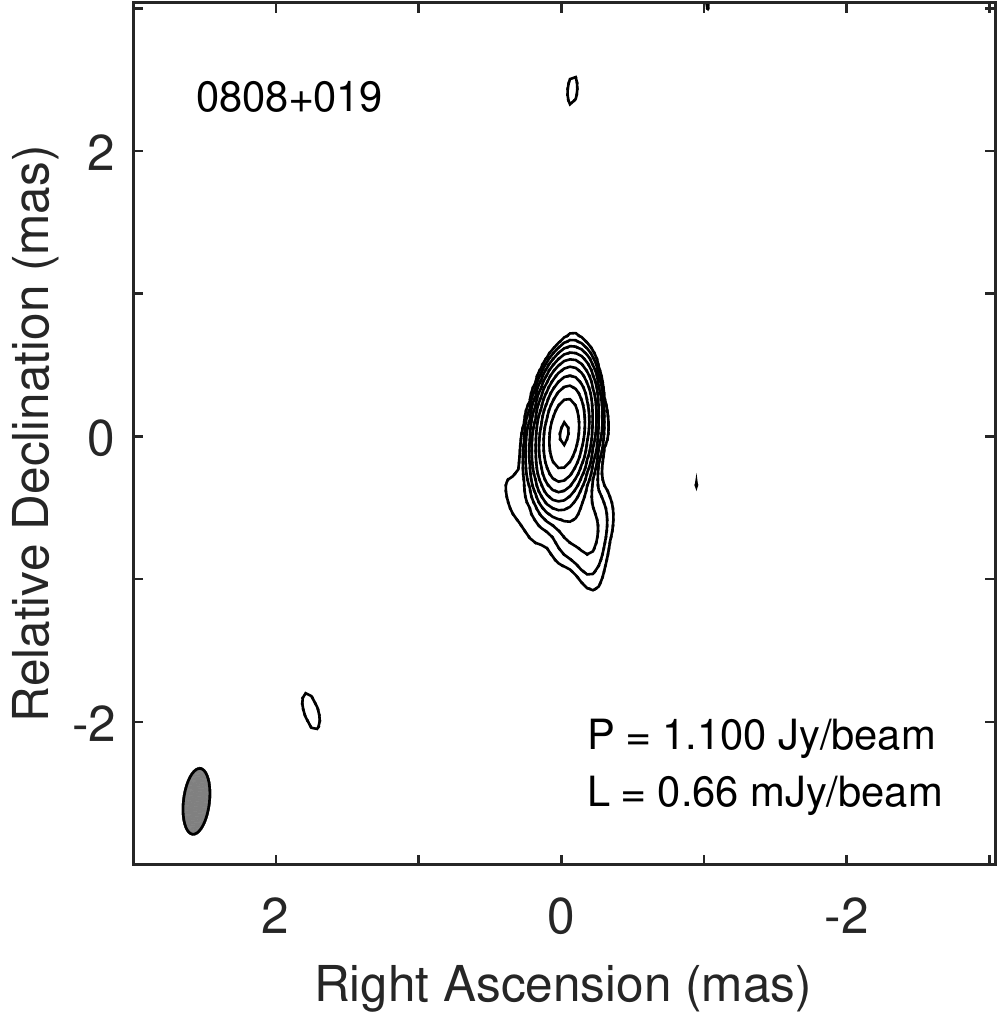}  & \includegraphics[width=0.3\textwidth]{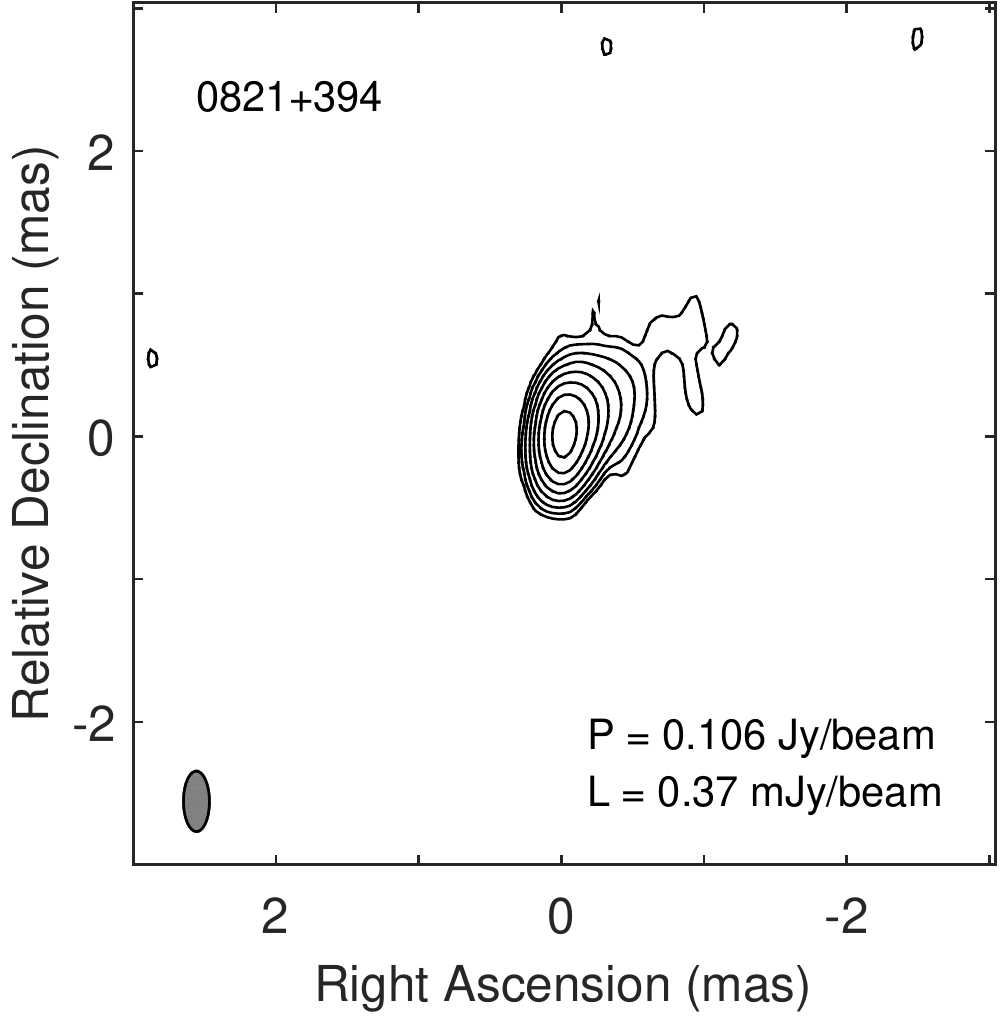}\\
  \includegraphics[width=0.3\textwidth]{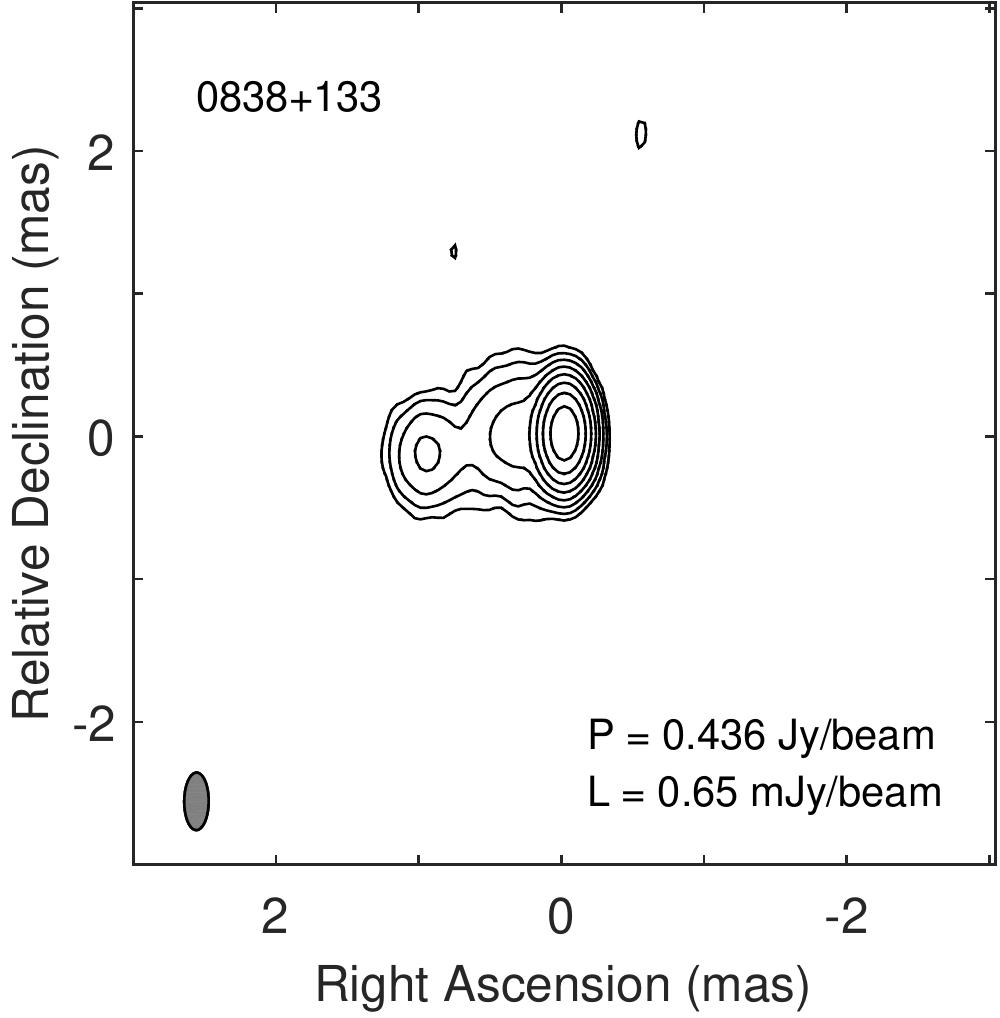} & \includegraphics[width=0.3\textwidth]{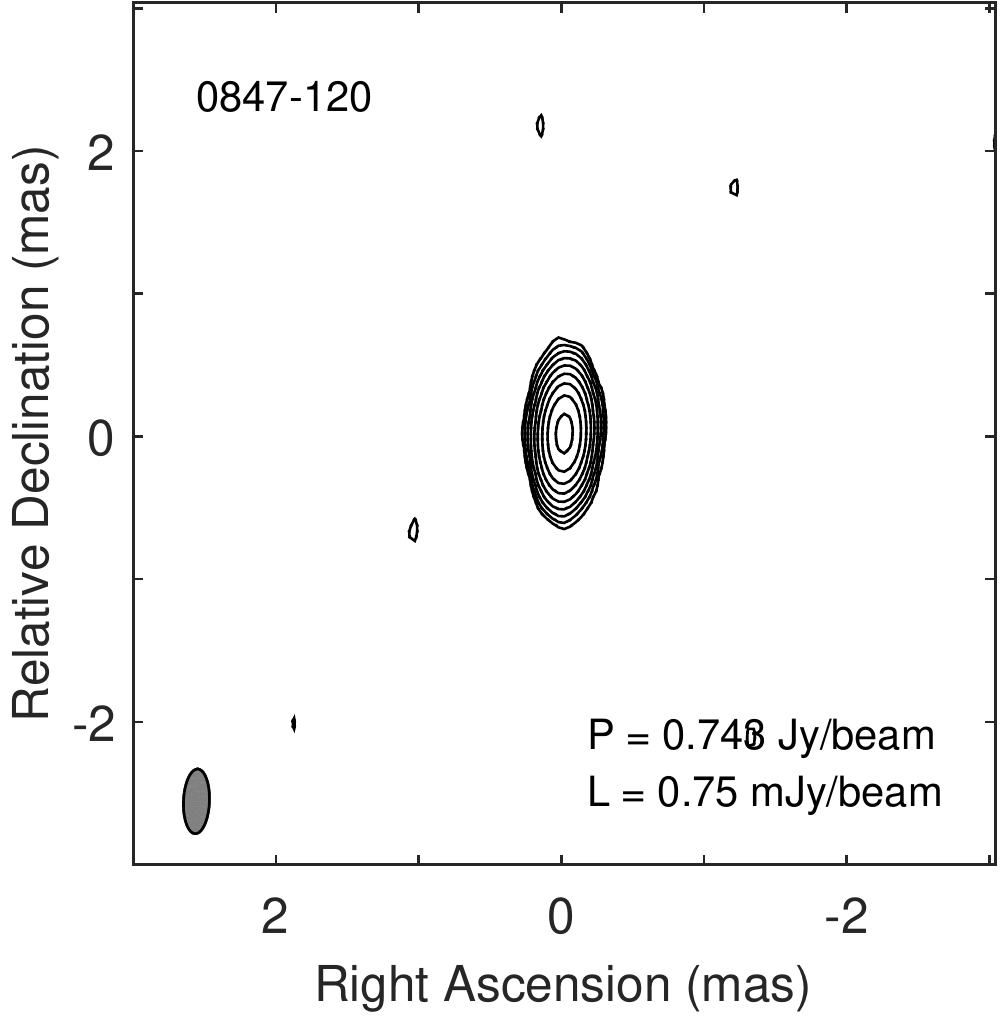}  & \includegraphics[width=0.3\textwidth]{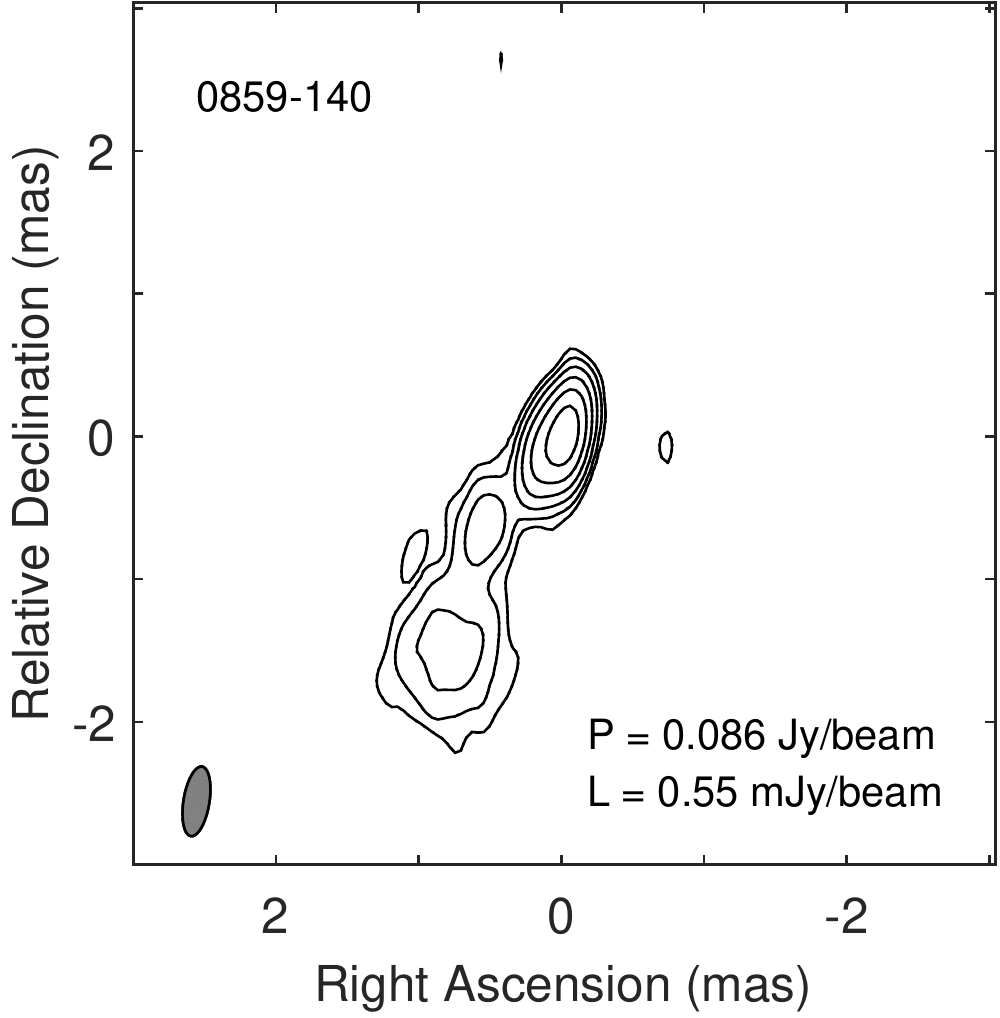}\\ \includegraphics[width=0.3\textwidth]{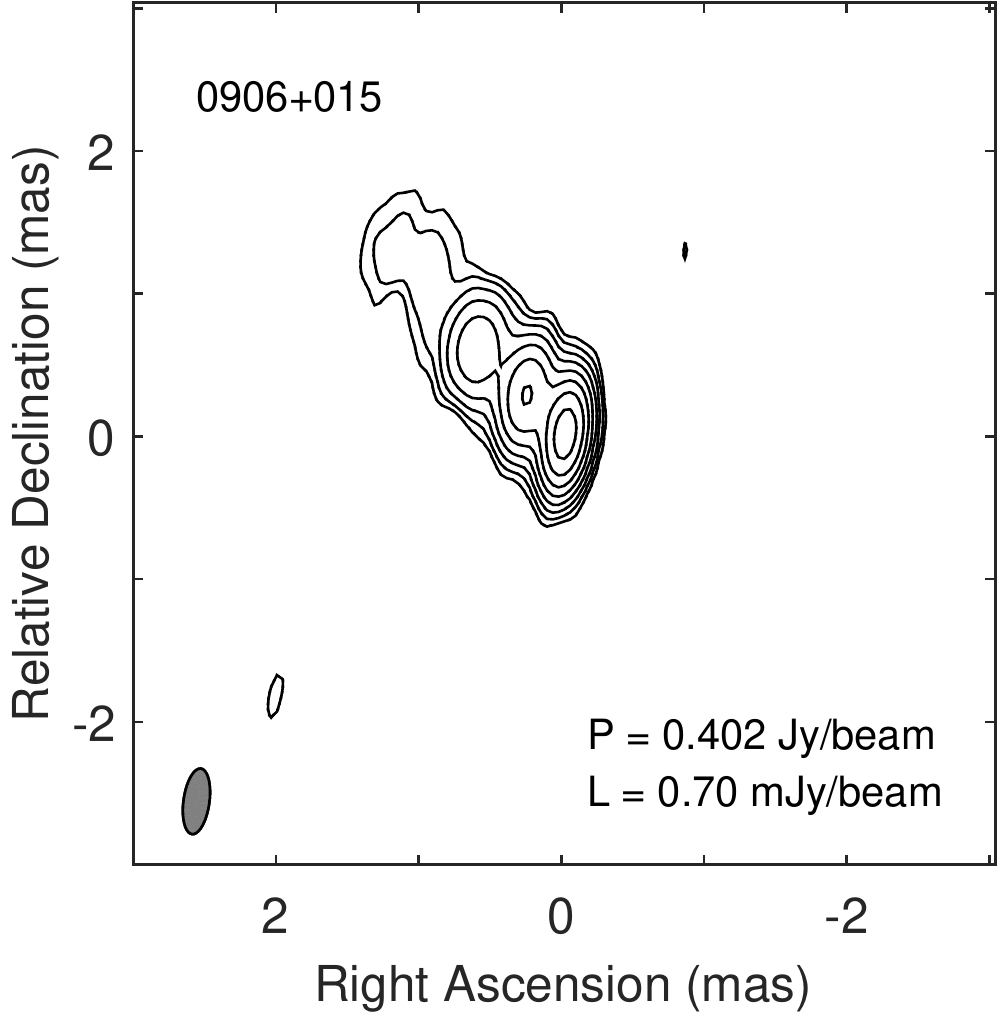} & \includegraphics[width=0.3\textwidth]{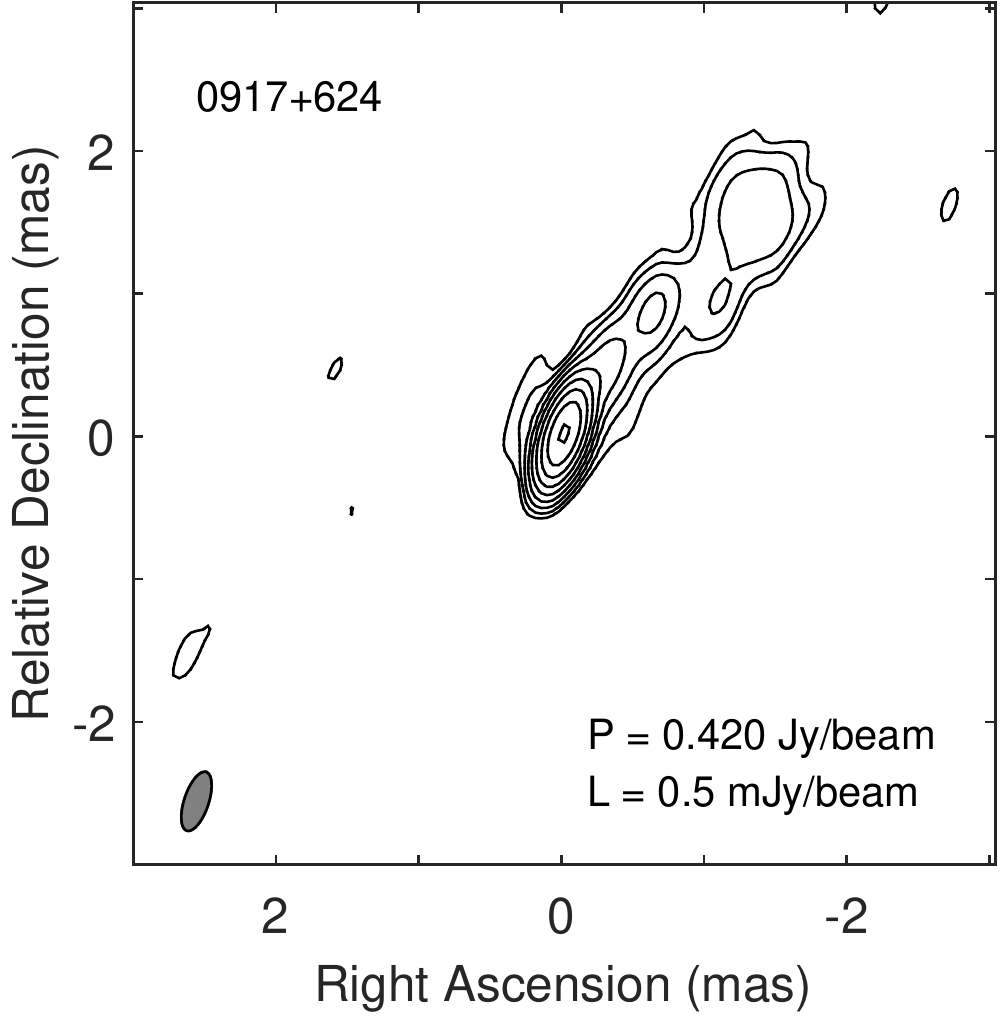}  & \includegraphics[width=0.3\textwidth]{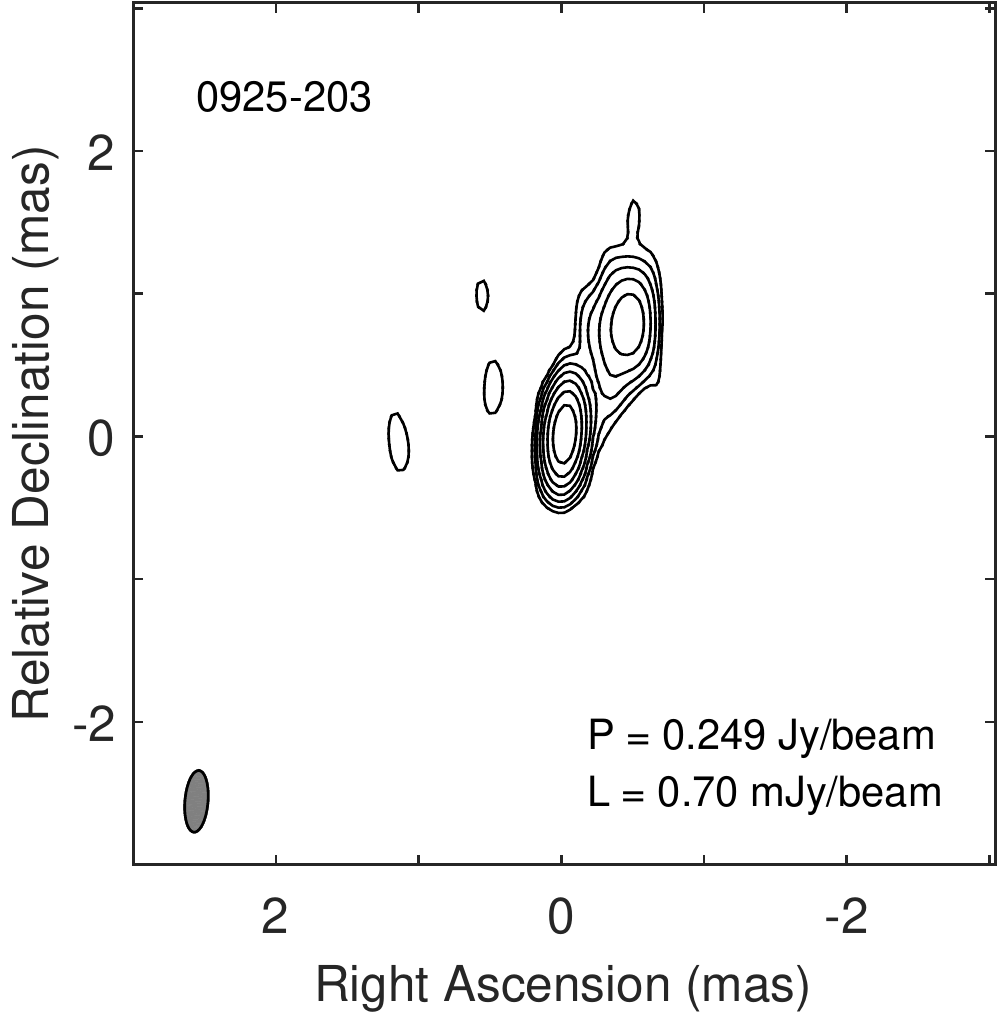}\\
    \end{tabular}
   \caption{VLBA images --- Continued}
  \end{figure}

\begin{figure}
  \centering
  \begin{tabular}{lcr}
  \includegraphics[width=0.3\textwidth]{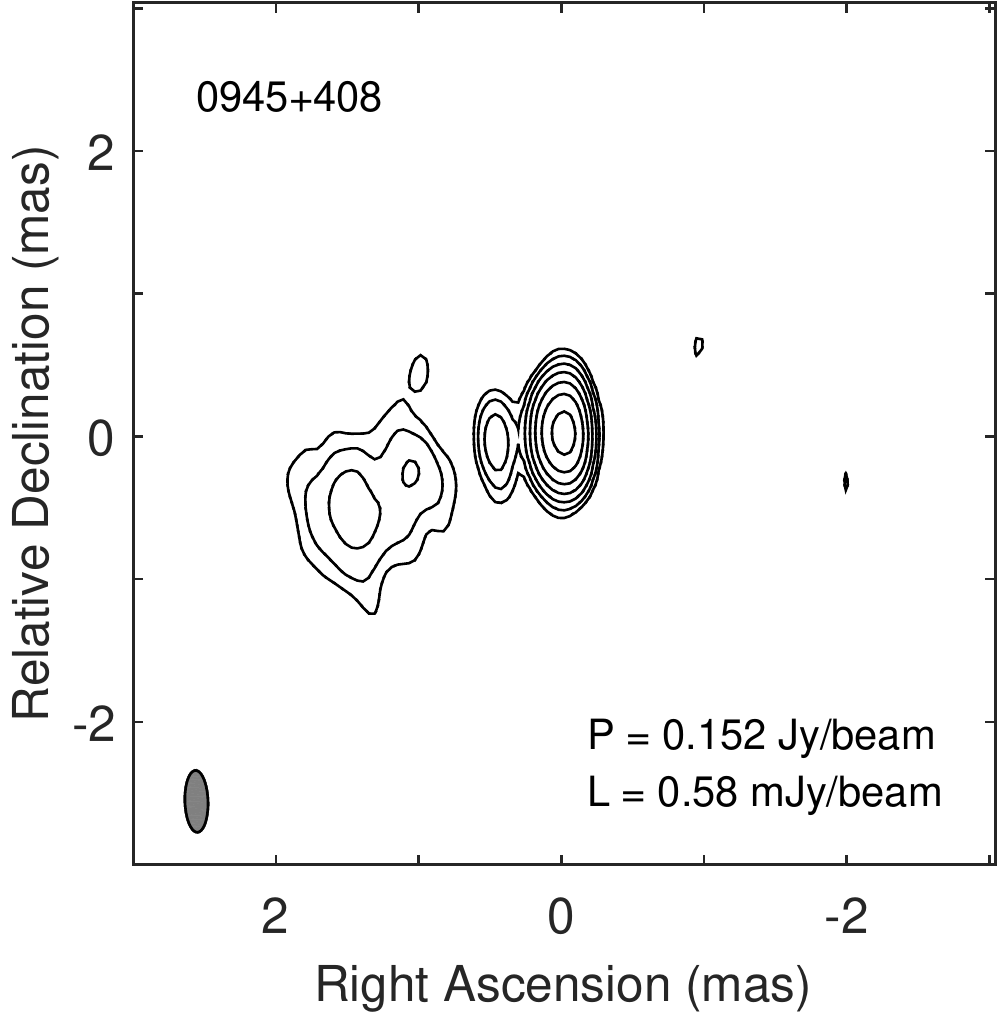} & \includegraphics[width=0.3\textwidth]{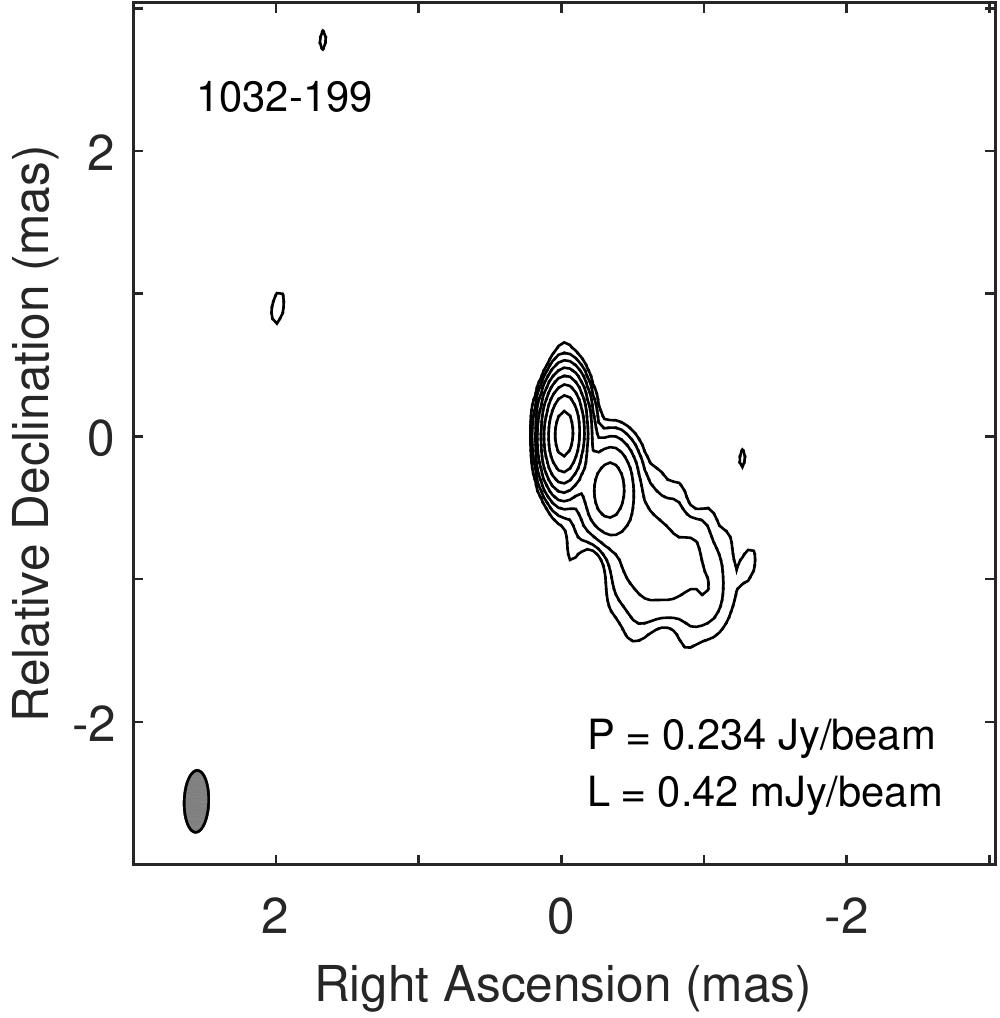}  & \includegraphics[width=0.3\textwidth]{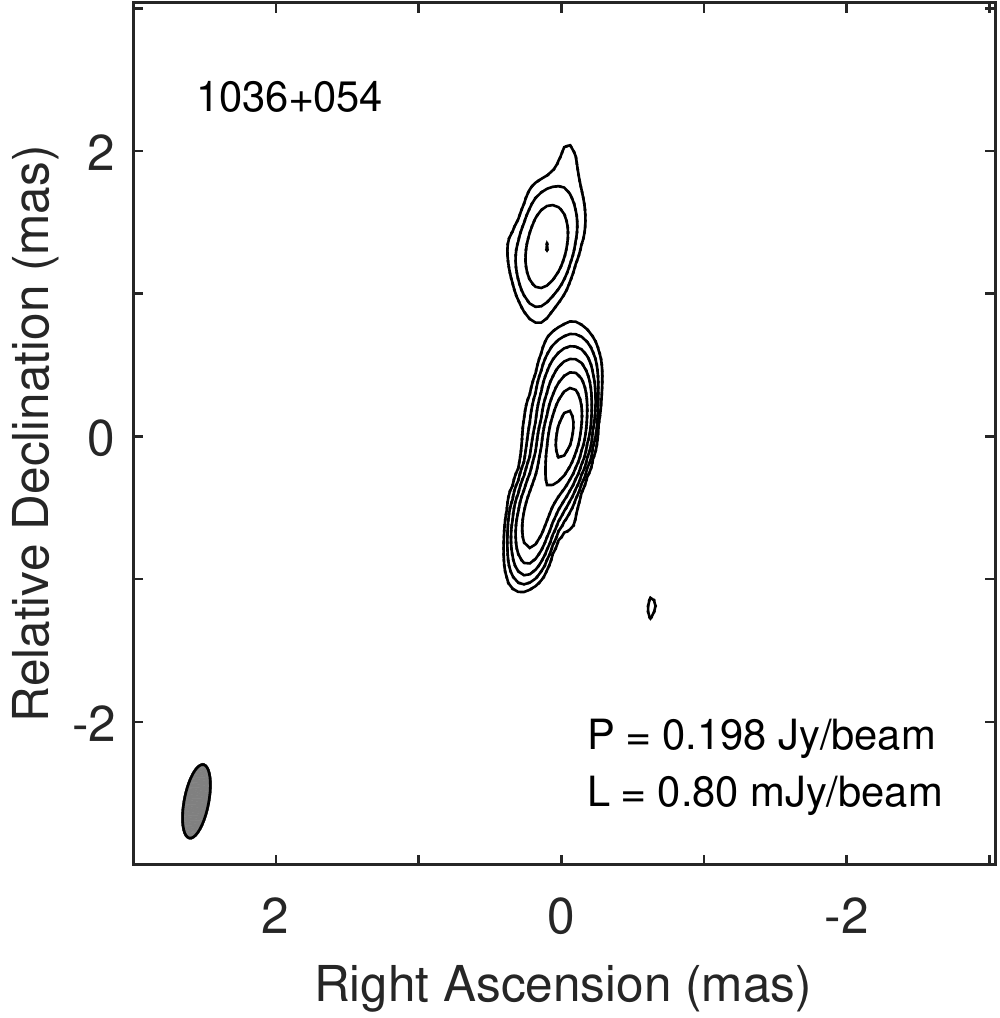}\\
  \includegraphics[width=0.3\textwidth]{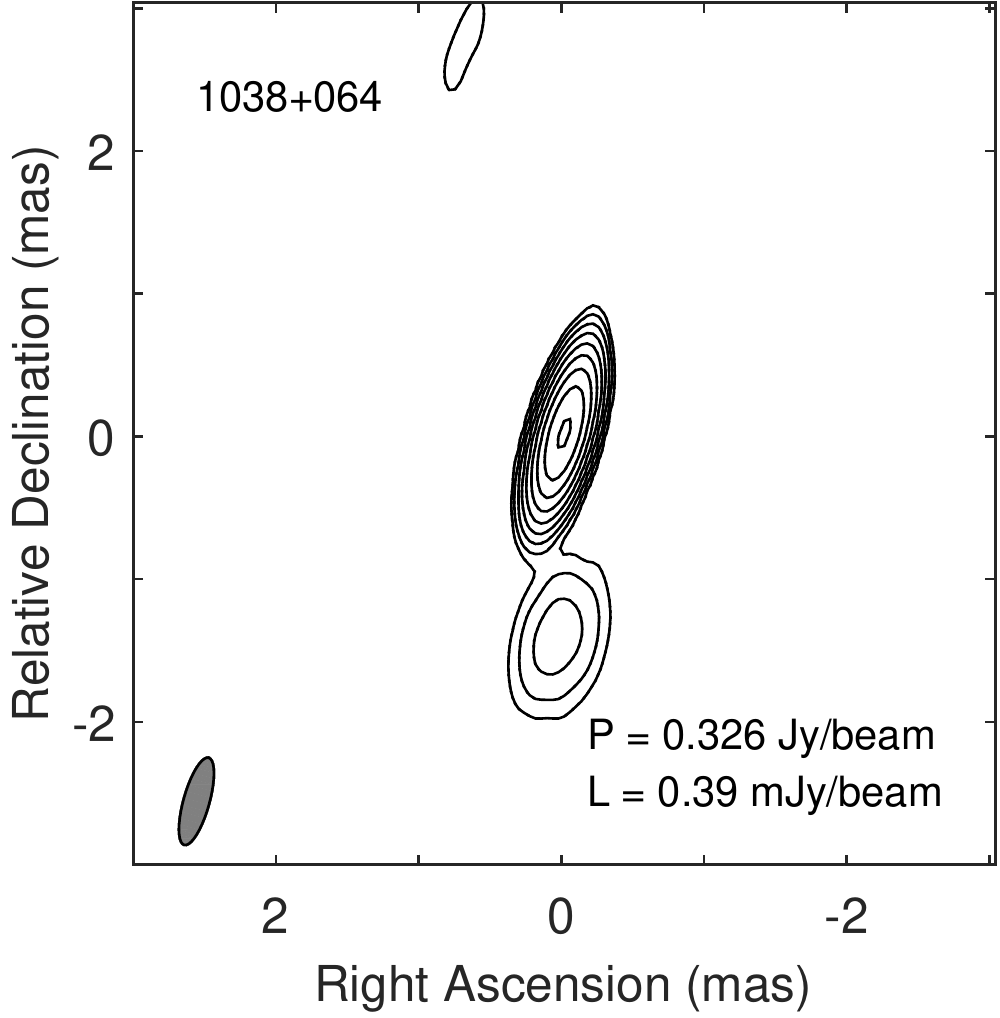} & \includegraphics[width=0.3\textwidth]{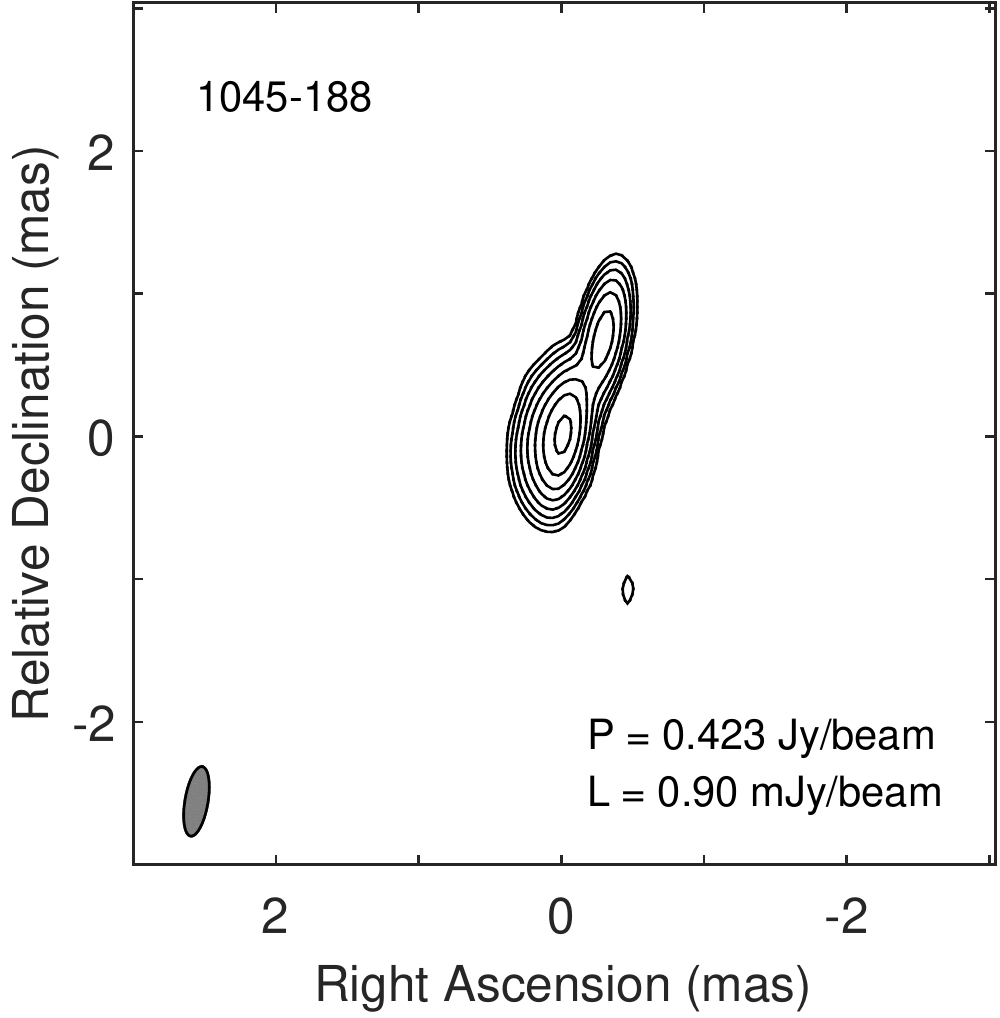}  & \includegraphics[width=0.3\textwidth]{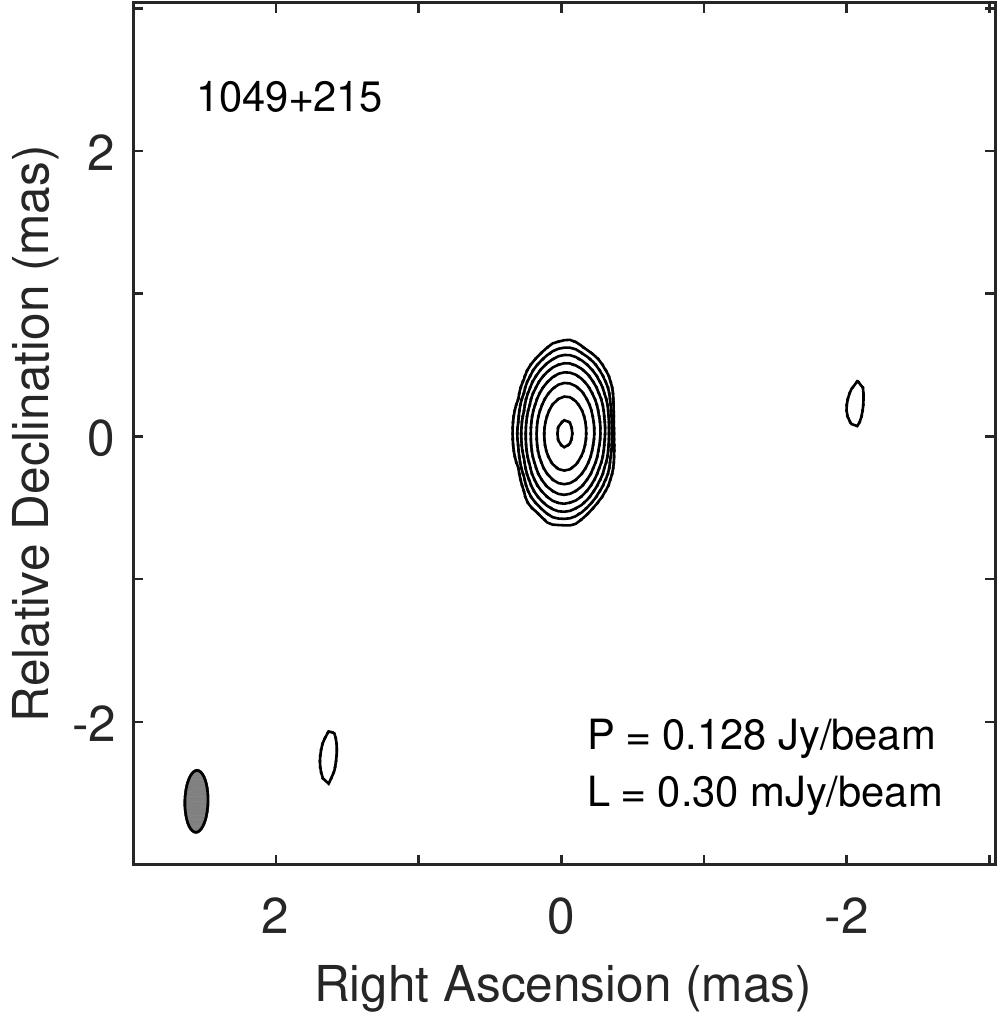}\\ \includegraphics[width=0.3\textwidth]{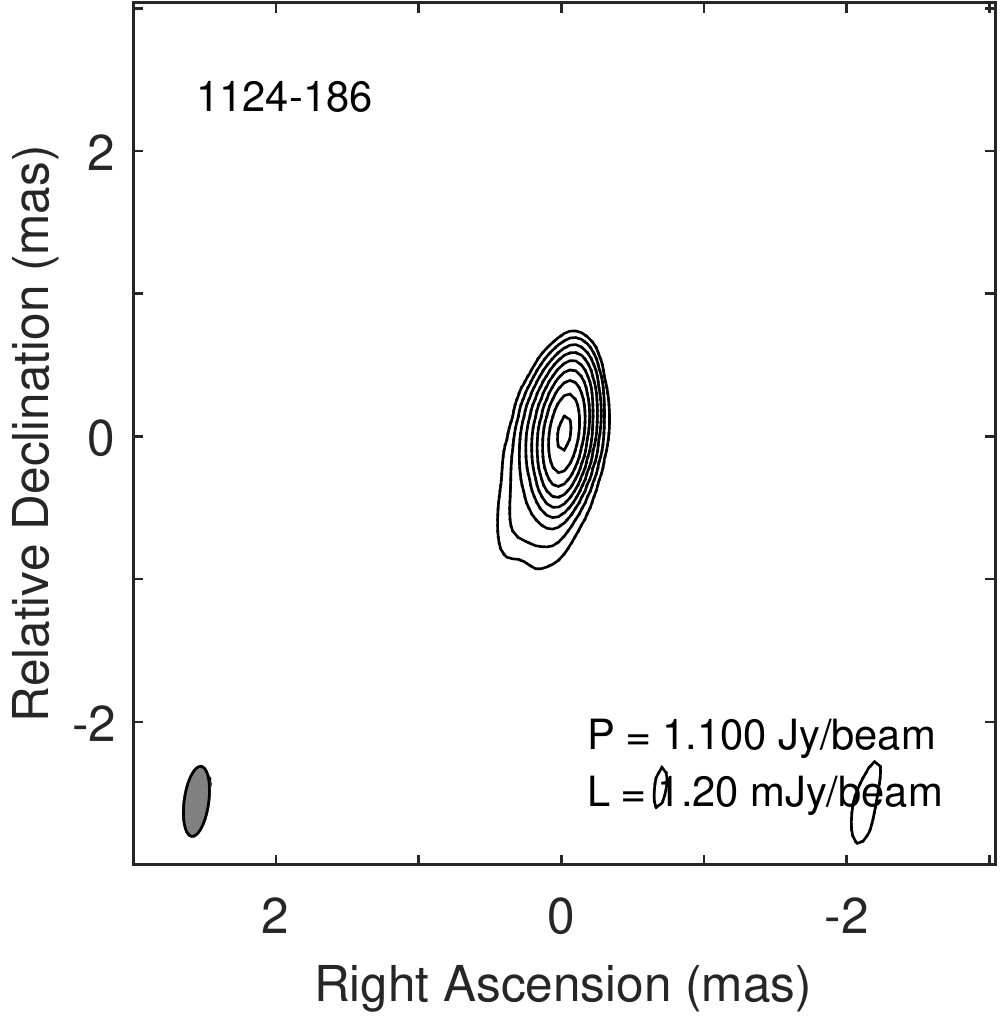} & \includegraphics[width=0.3\textwidth]{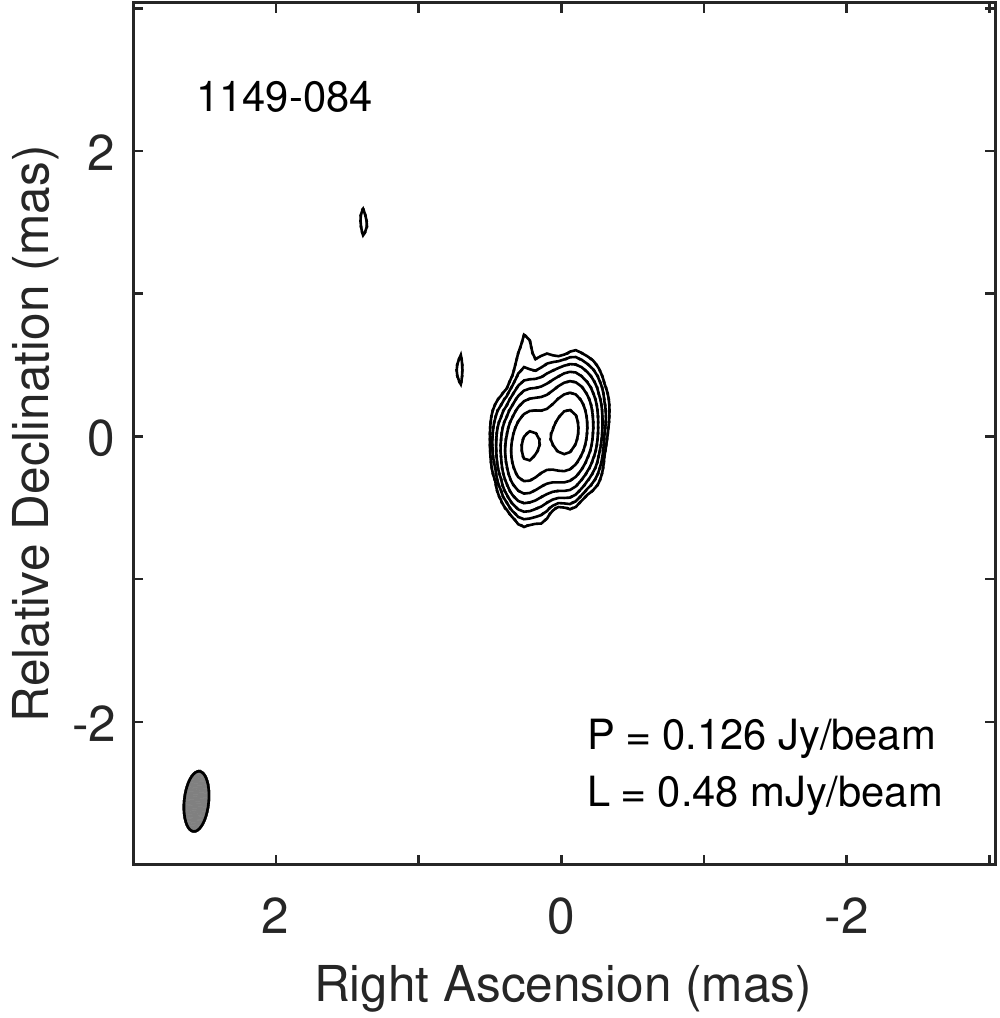}  & \includegraphics[width=0.3\textwidth]{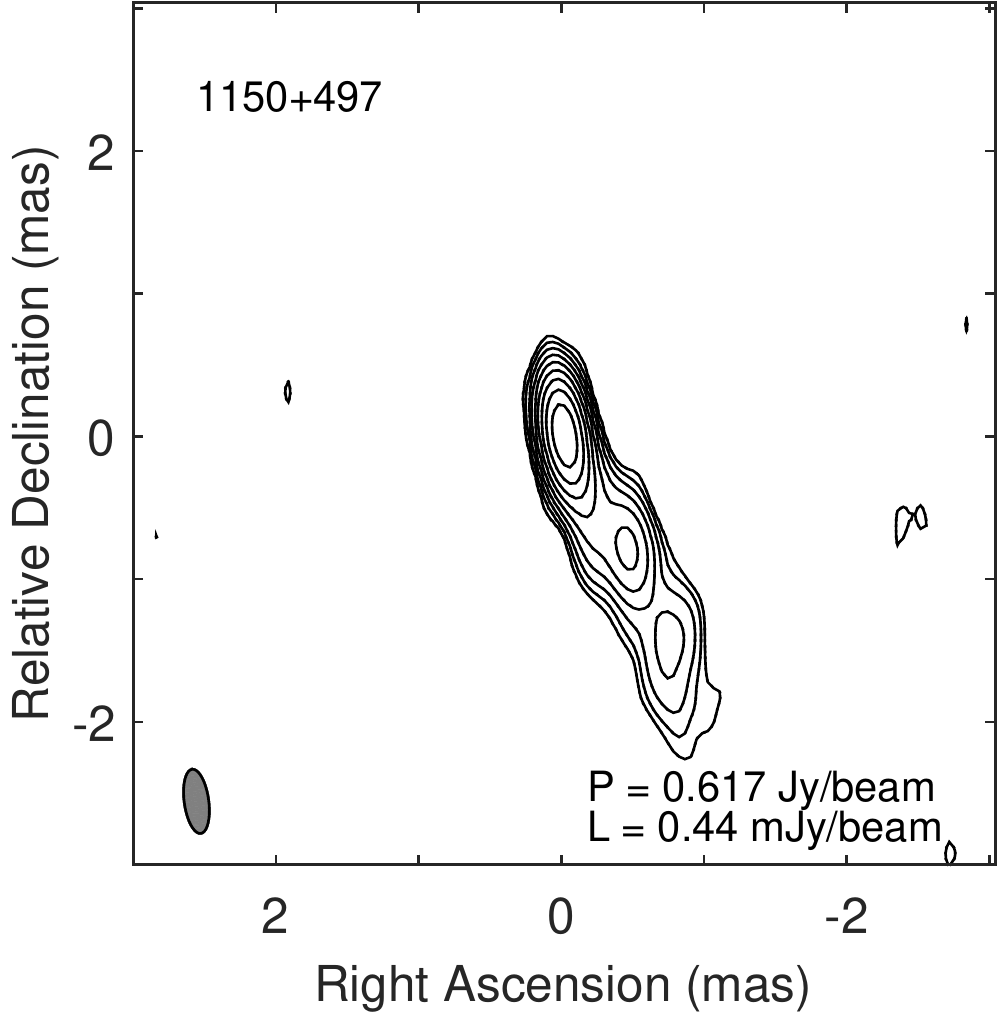}\\
  \includegraphics[width=0.3\textwidth]{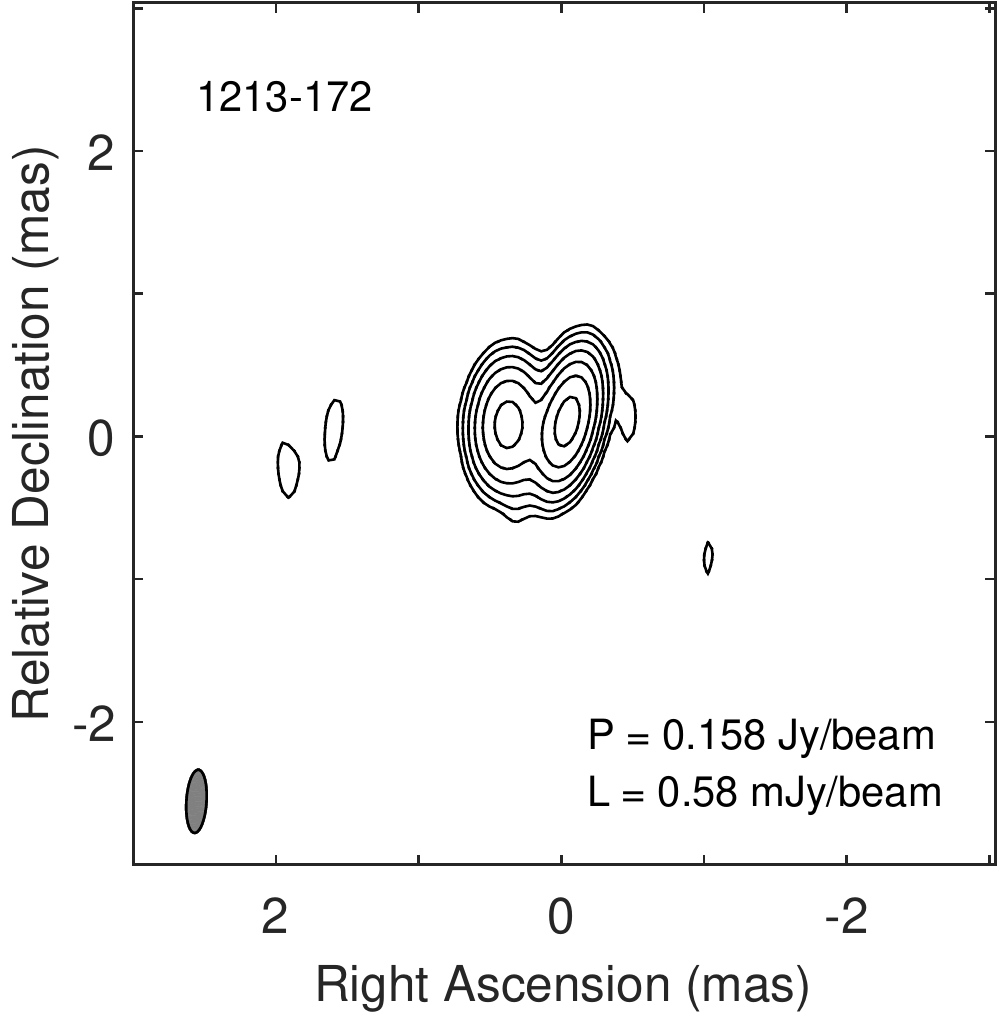} & \includegraphics[width=0.3\textwidth]{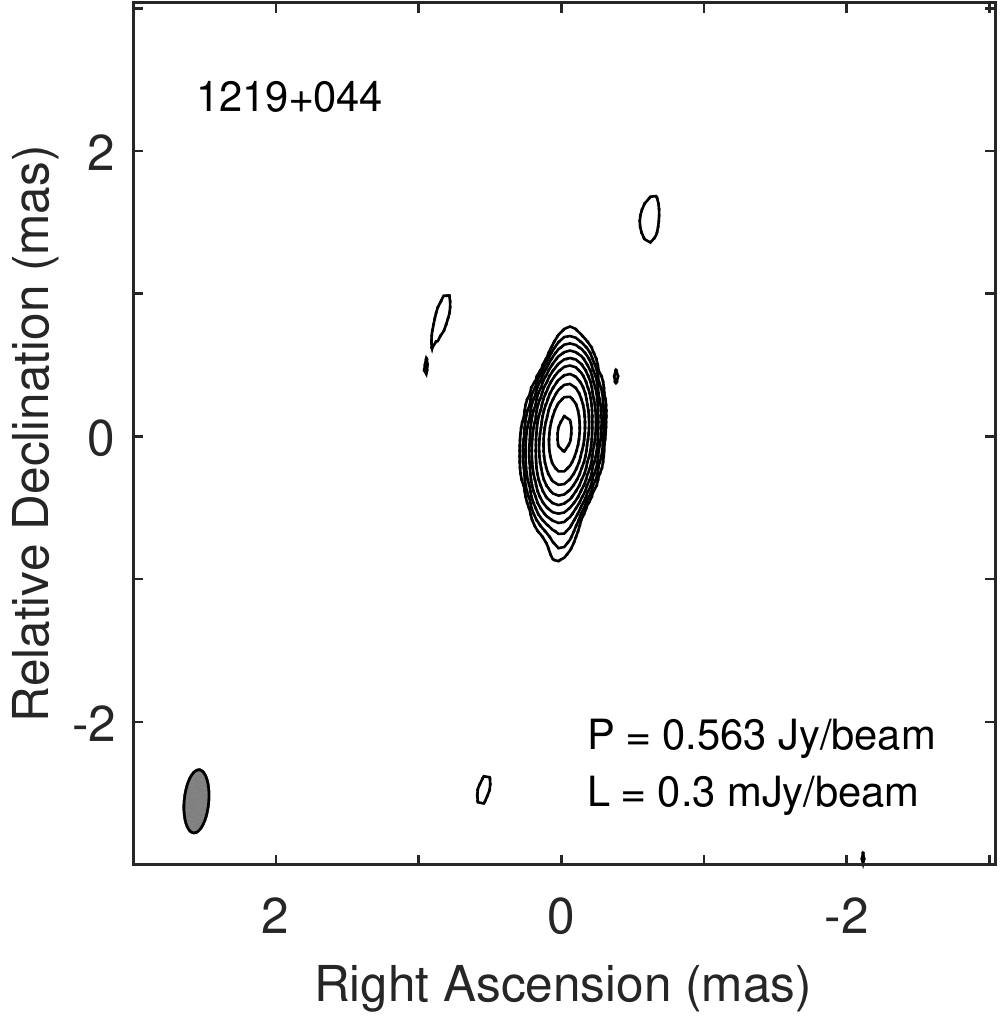}  & \includegraphics[width=0.3\textwidth]{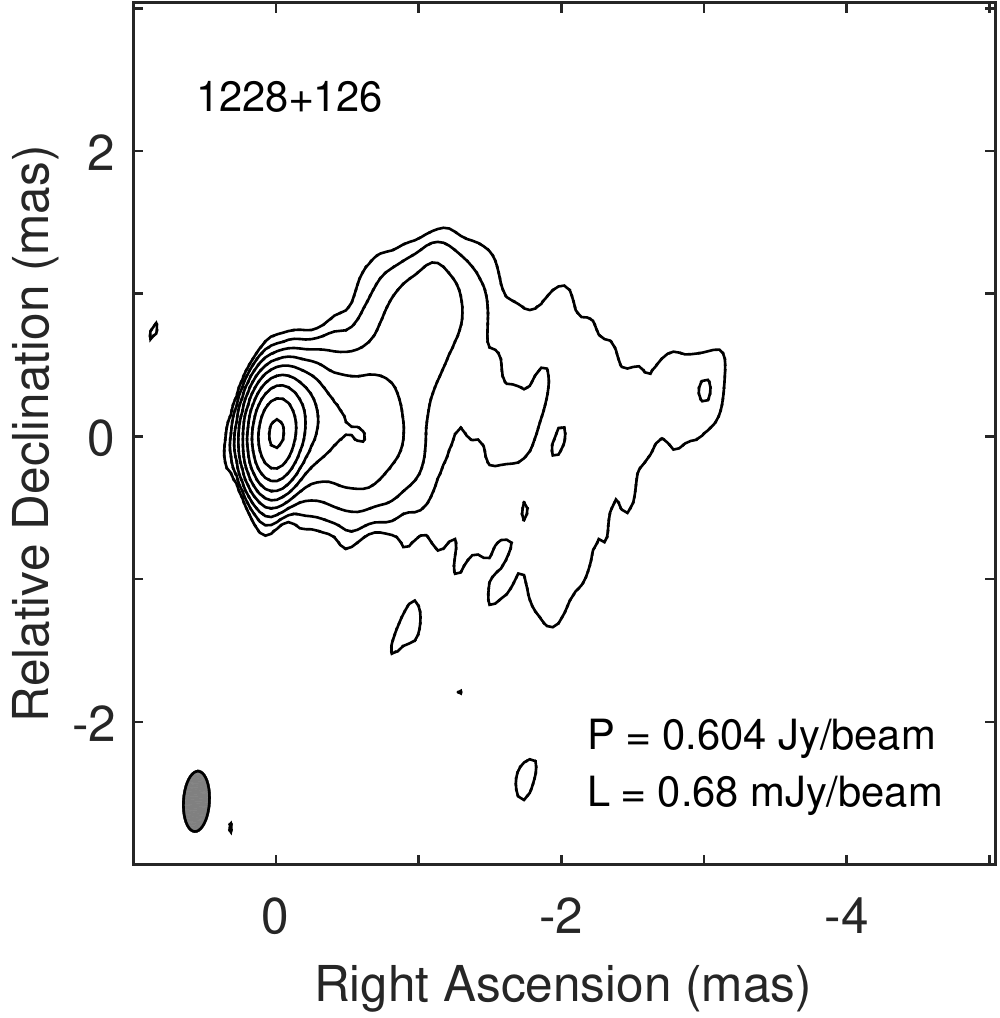}\\
    \end{tabular}
   \caption{VLBA images --- Continued}
  \end{figure}

\begin{figure}
  \centering
  \begin{tabular}{lcr}
  \includegraphics[width=0.3\textwidth]{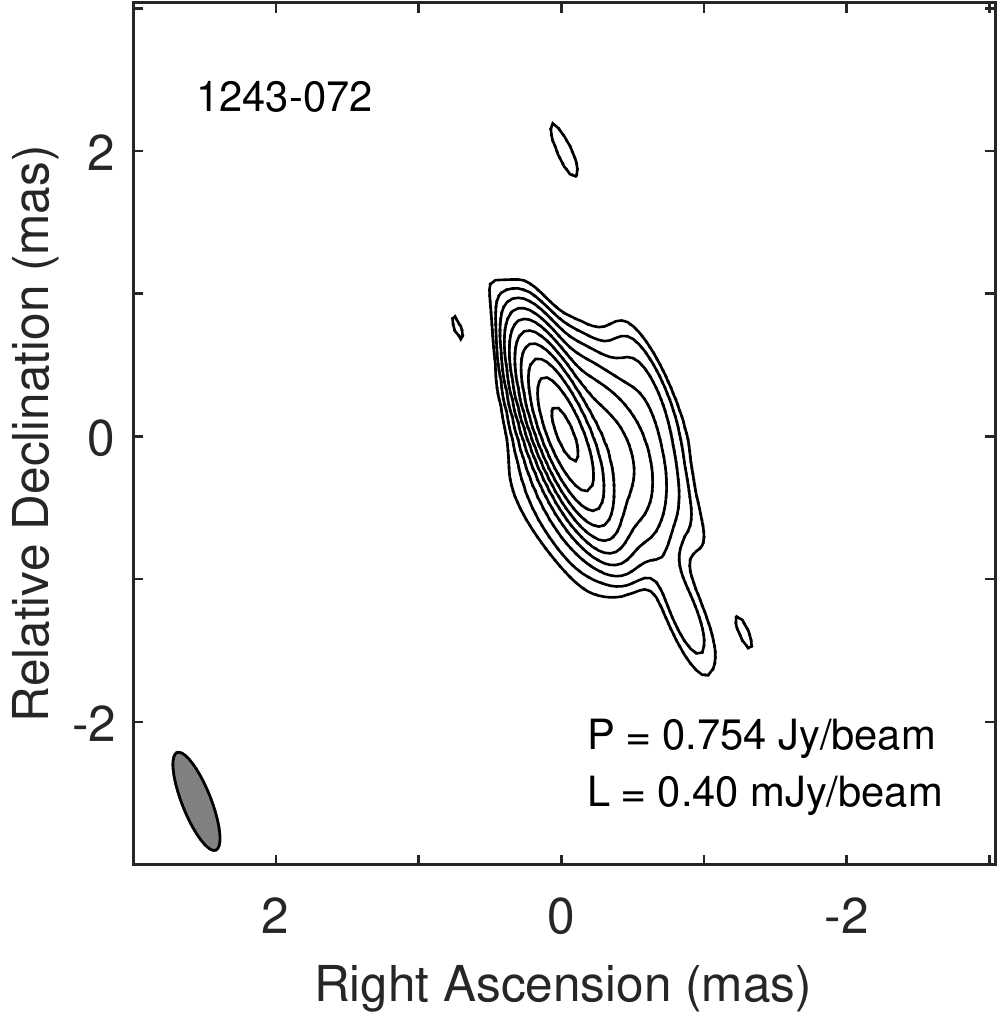} & \includegraphics[width=0.3\textwidth]{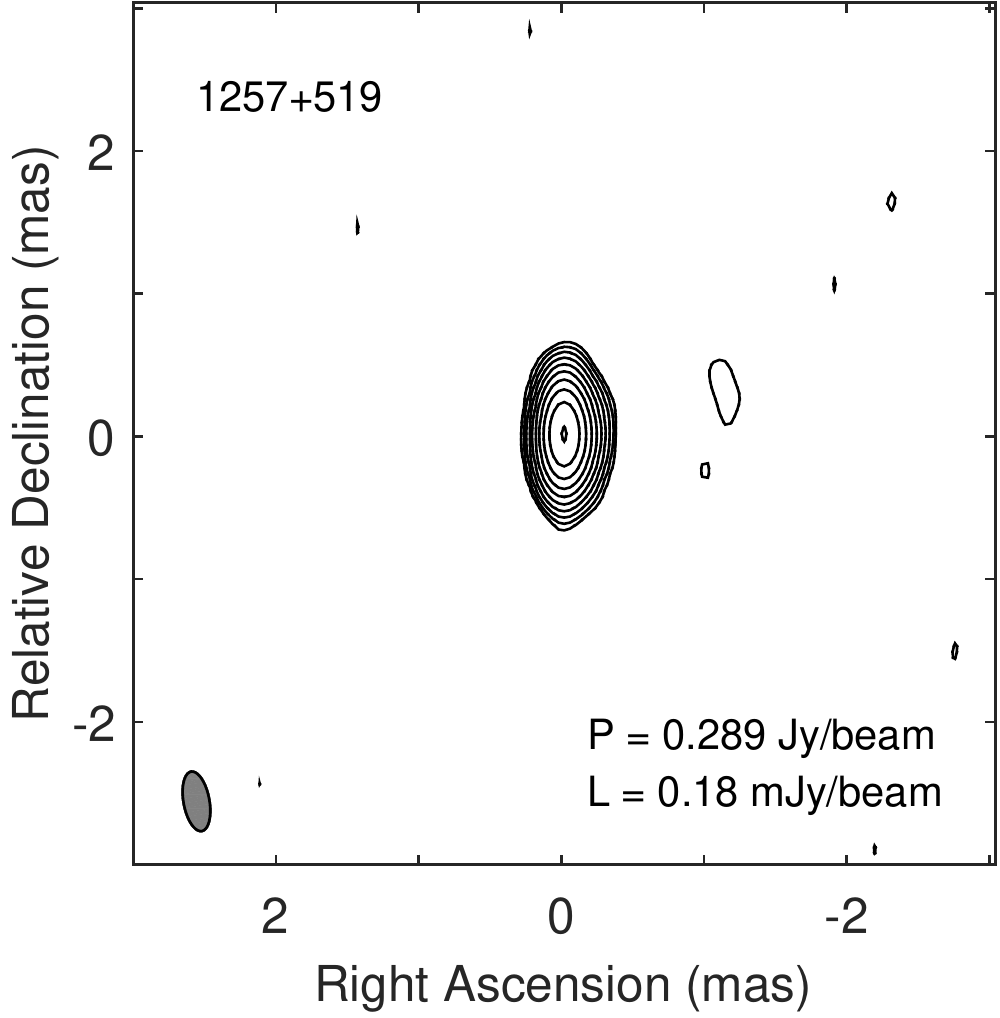}  & \includegraphics[width=0.3\textwidth]{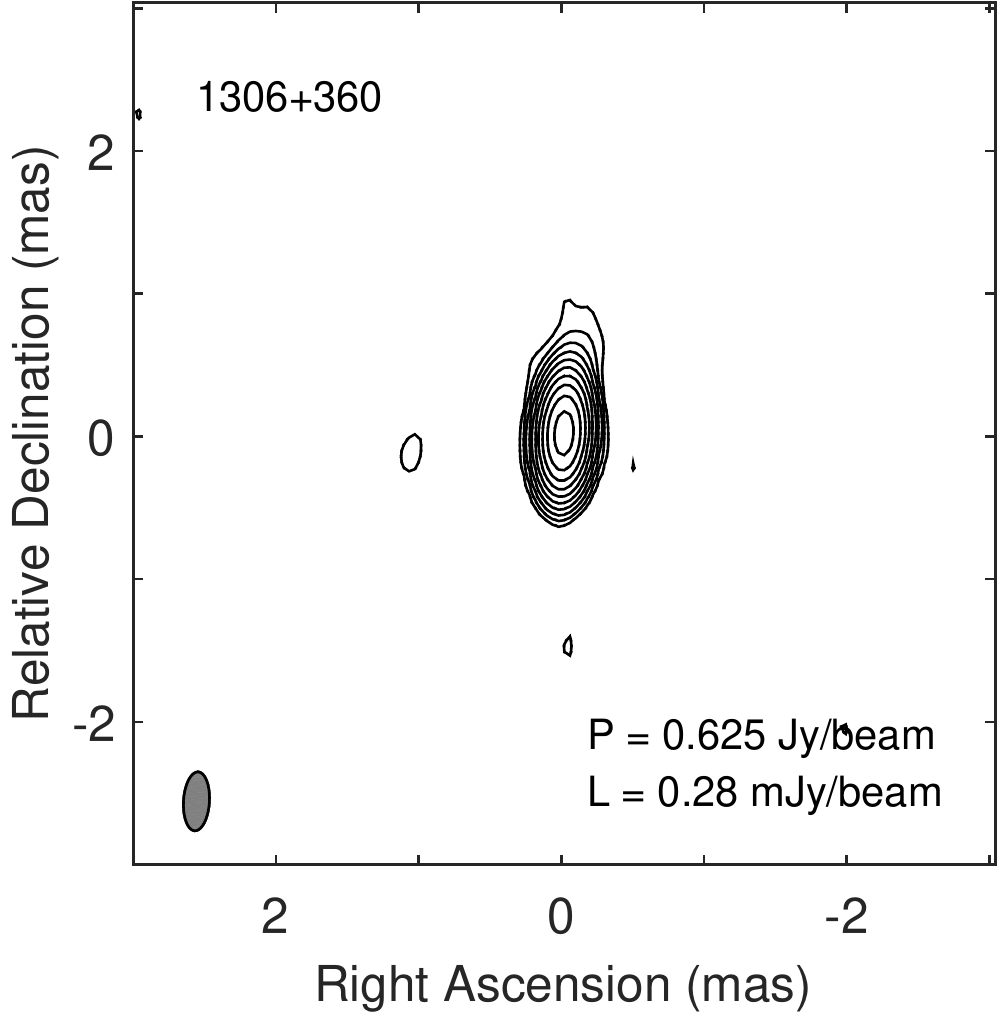}\\
  \includegraphics[width=0.3\textwidth]{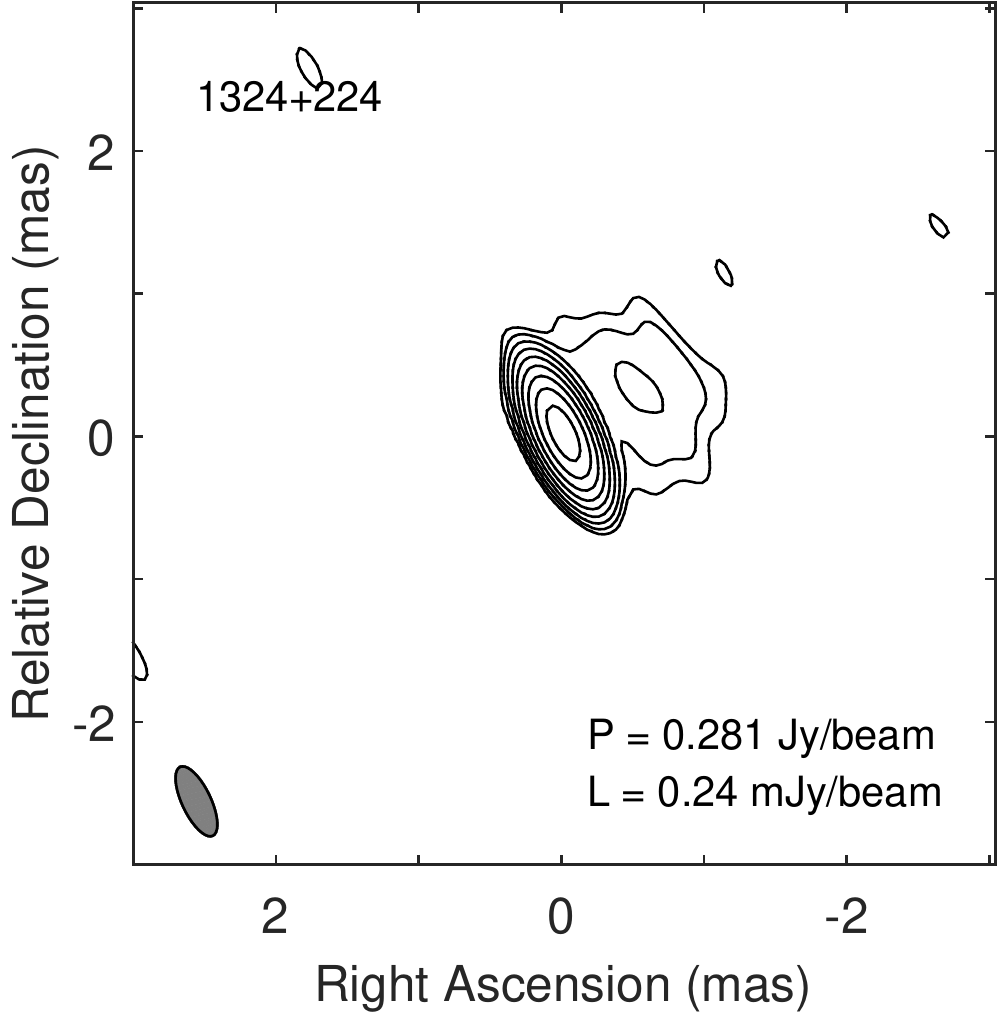} & \includegraphics[width=0.3\textwidth]{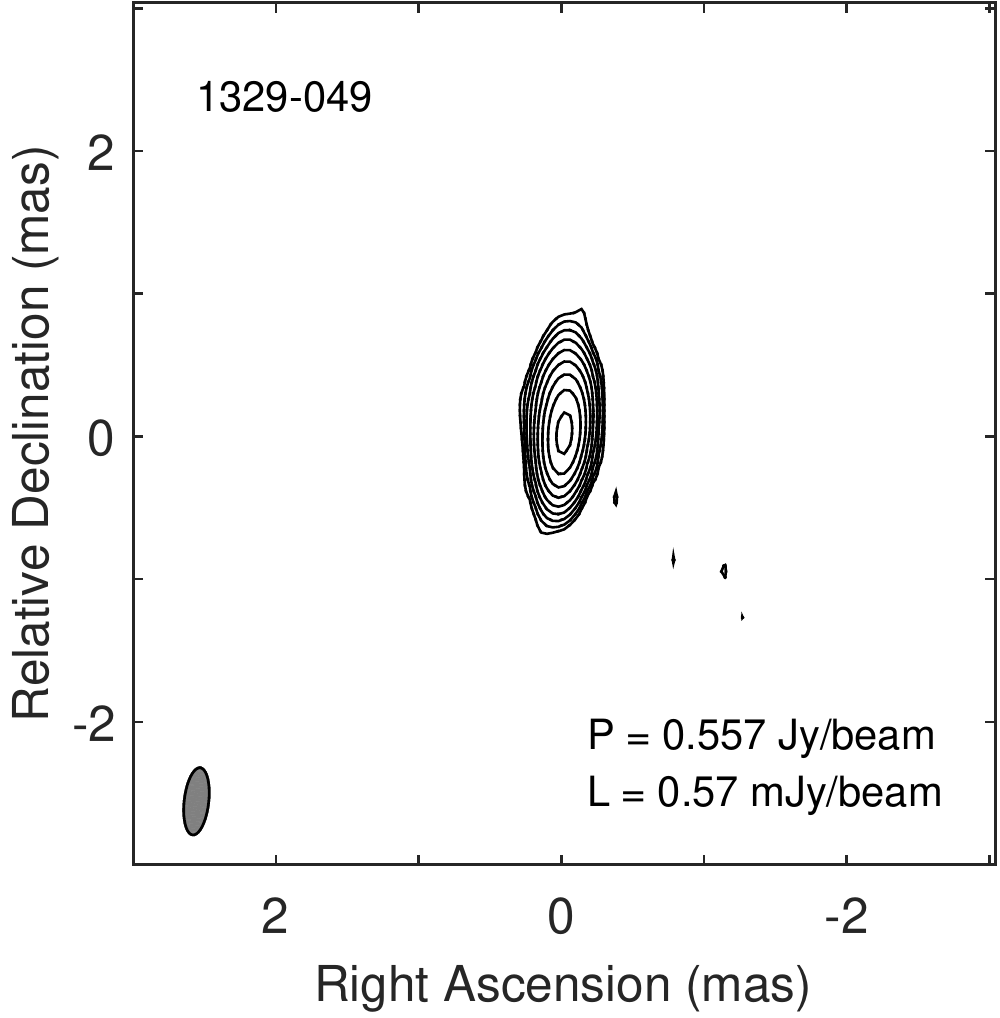}  & \includegraphics[width=0.3\textwidth]{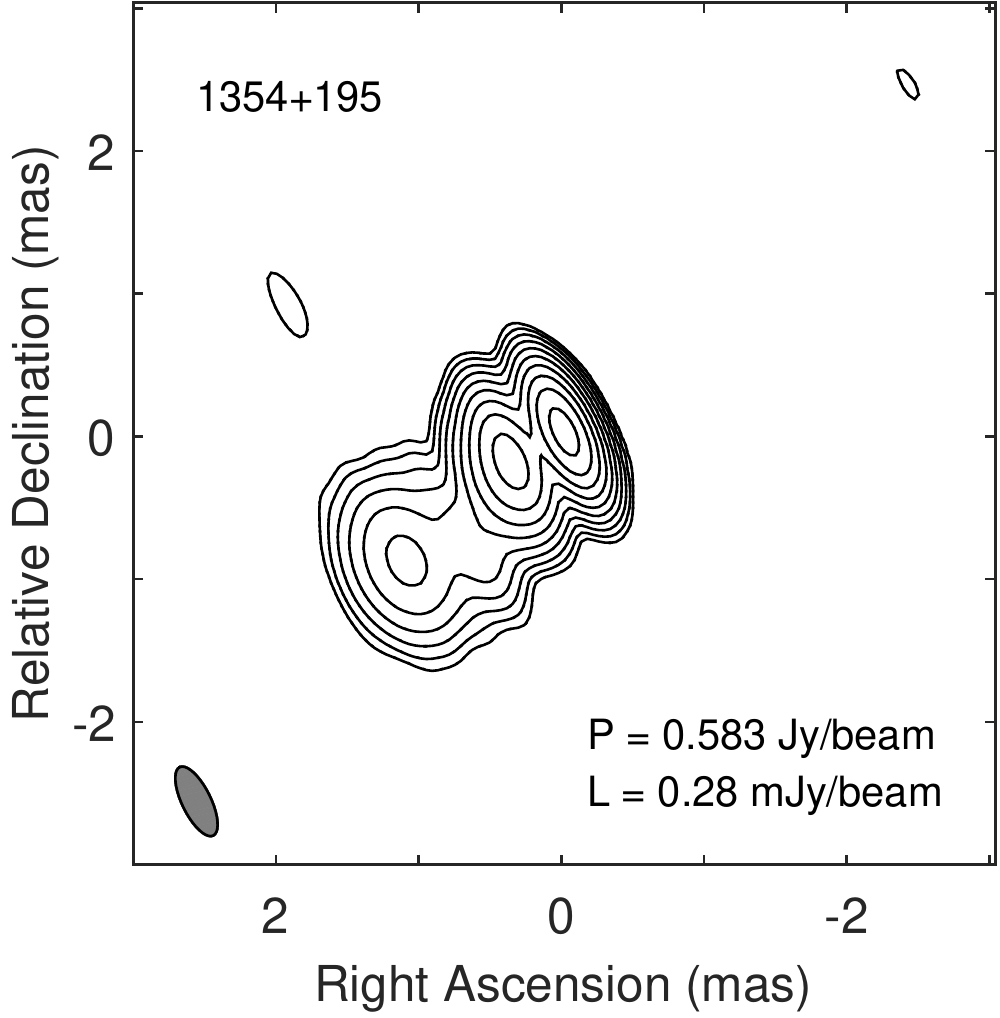}\\ \includegraphics[width=0.3\textwidth]{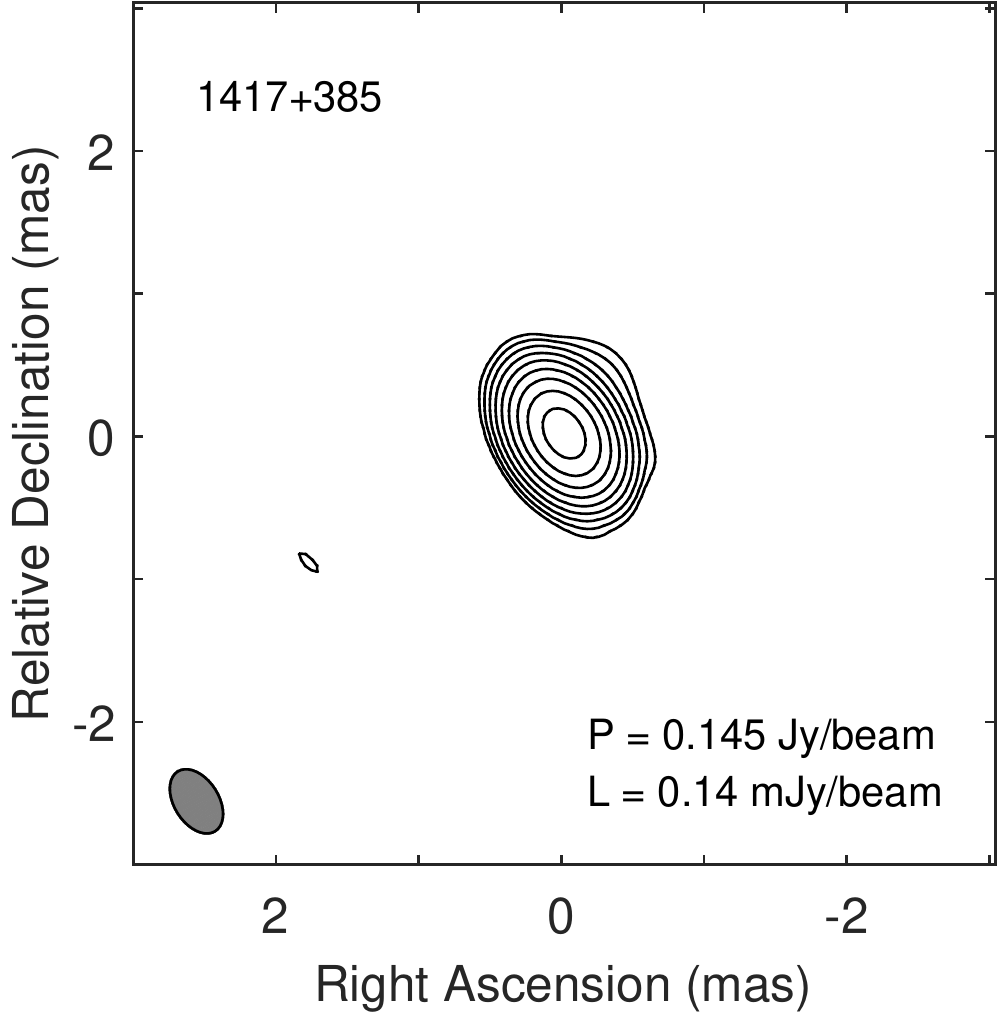} & \includegraphics[width=0.3\textwidth]{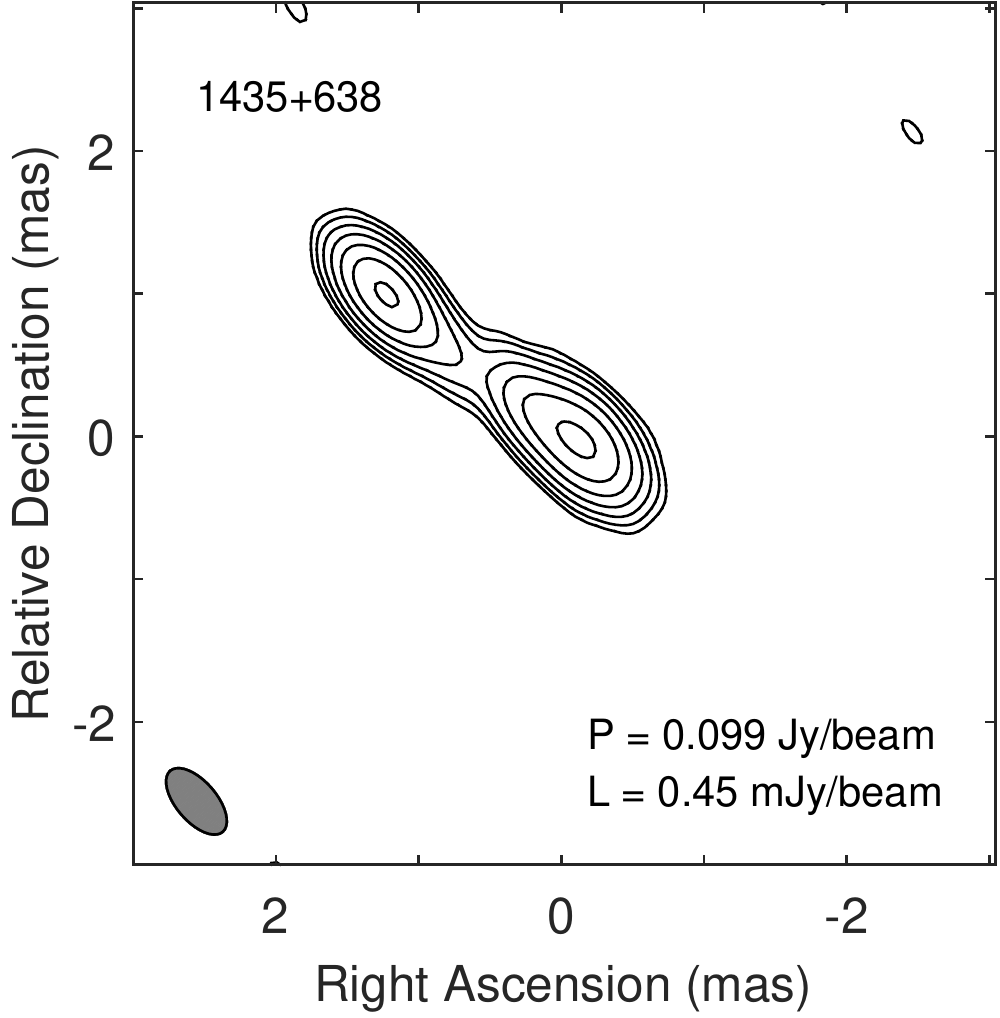}  & \includegraphics[width=0.3\textwidth]{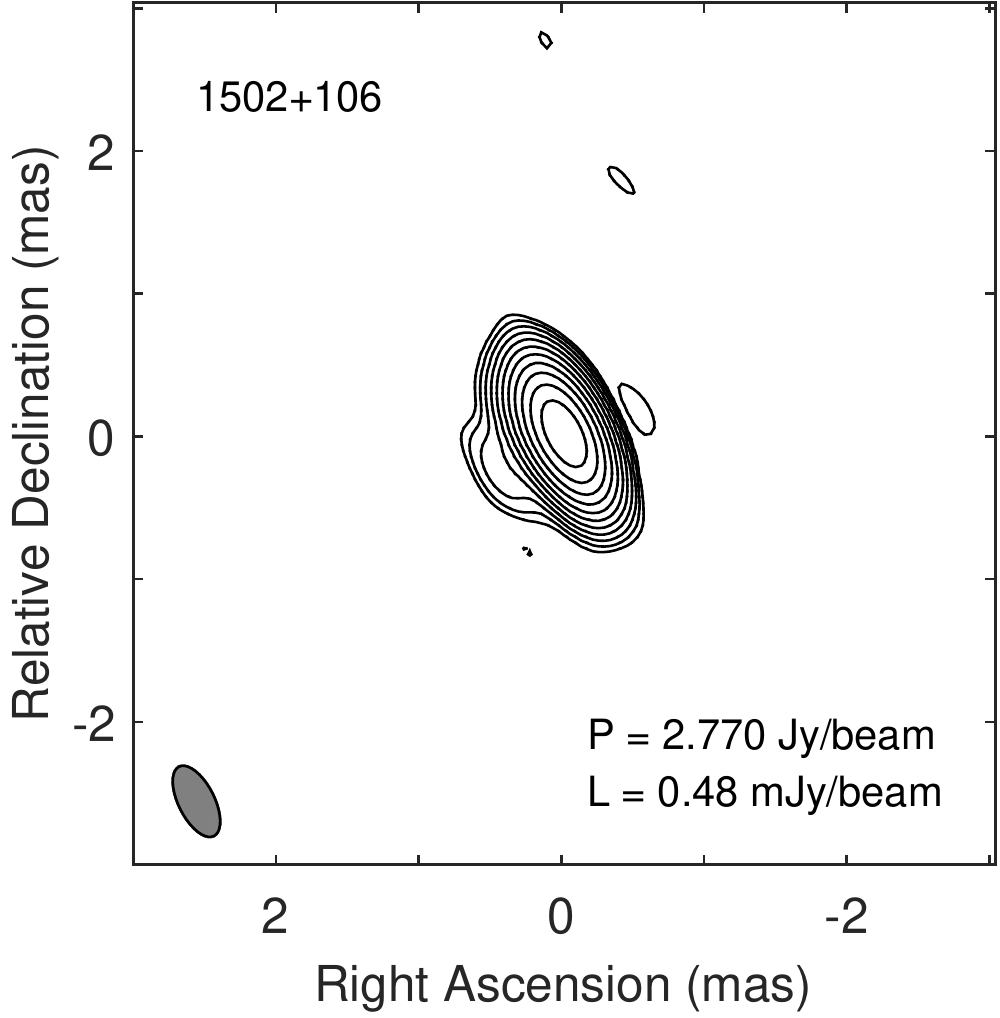}\\
  \includegraphics[width=0.3\textwidth]{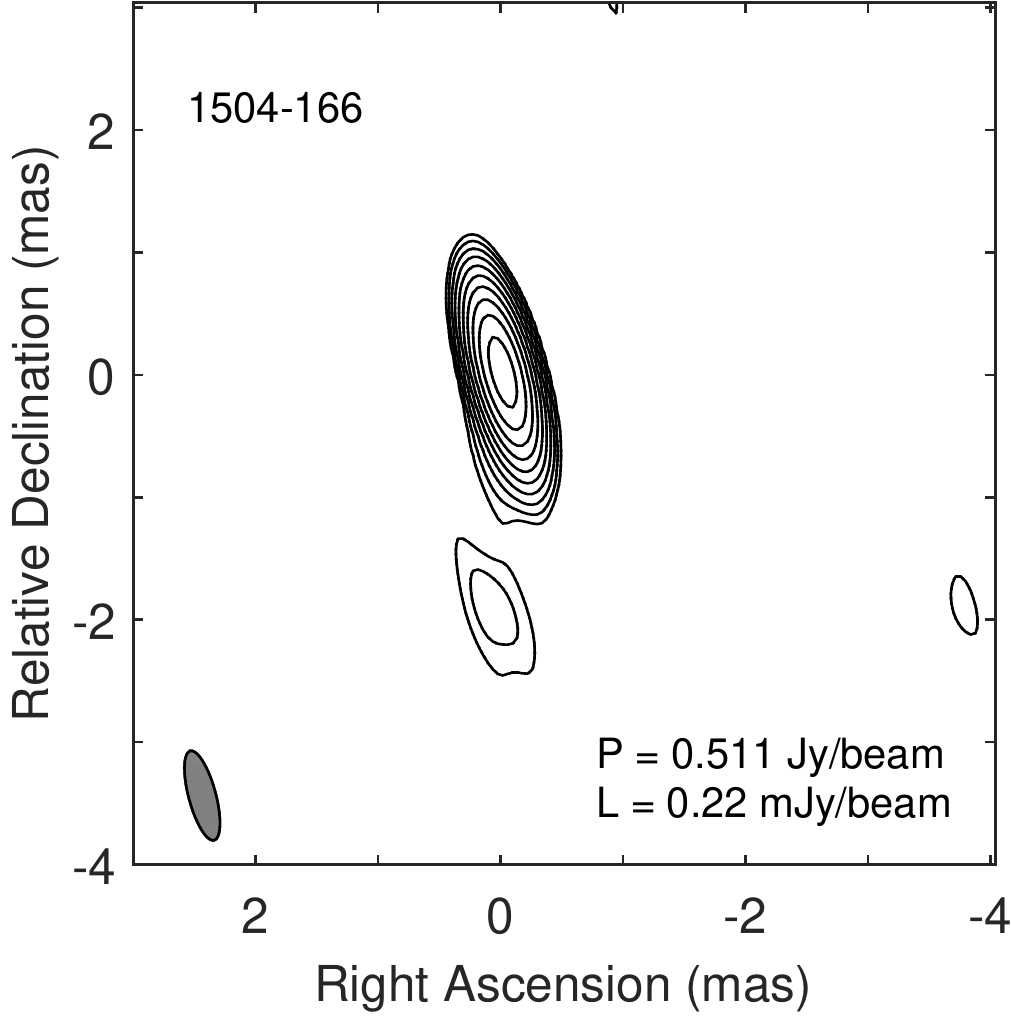} & \includegraphics[width=0.3\textwidth]{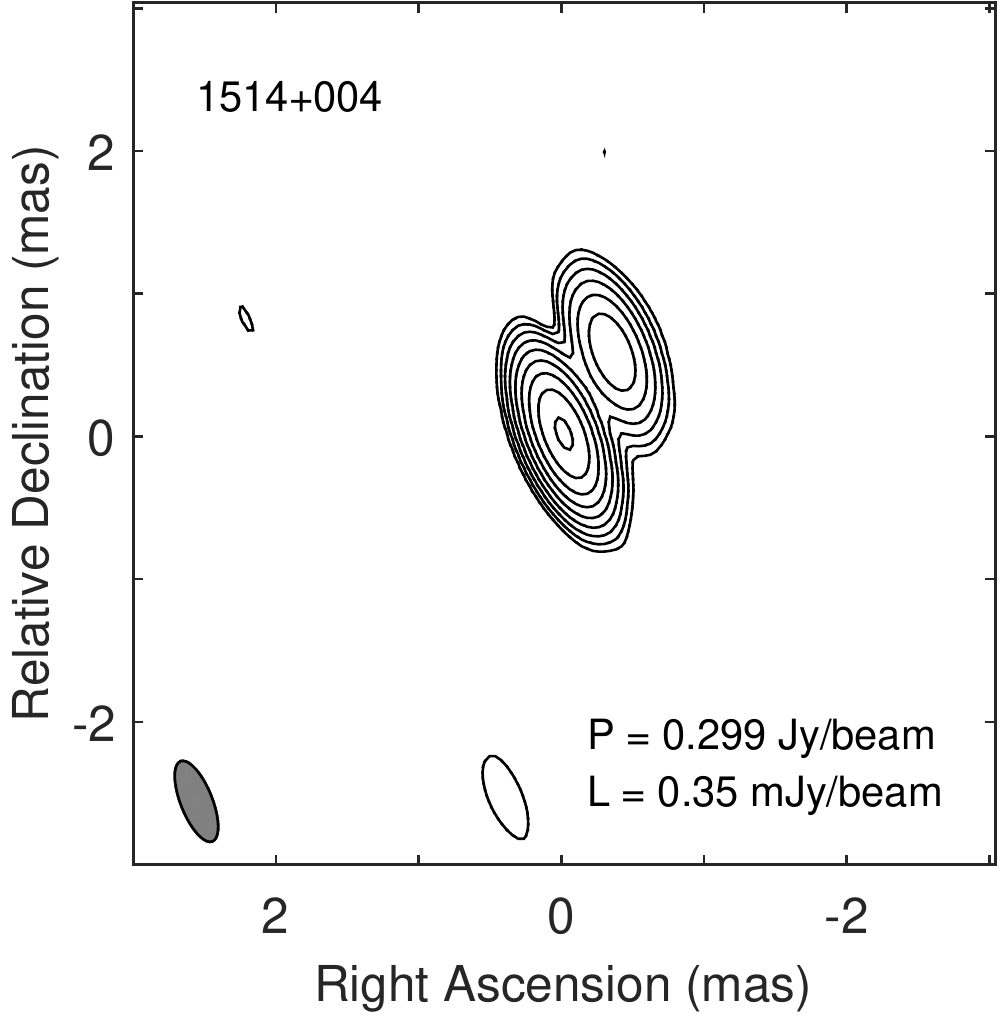}  & \includegraphics[width=0.3\textwidth]{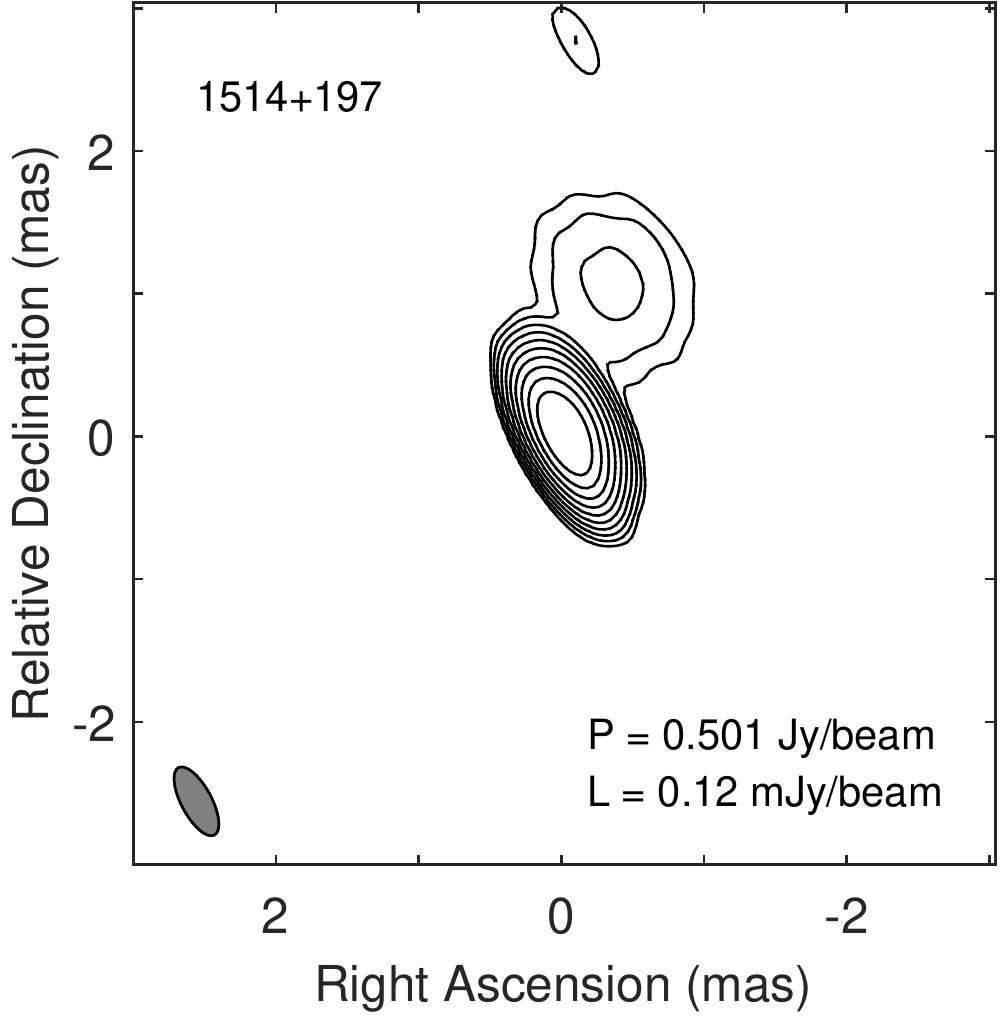}\\
    \end{tabular}
   \caption{VLBA images --- Continued}
  \end{figure}

\begin{figure}
  \centering
  \begin{tabular}{lcr}
   \includegraphics[width=0.3\textwidth]{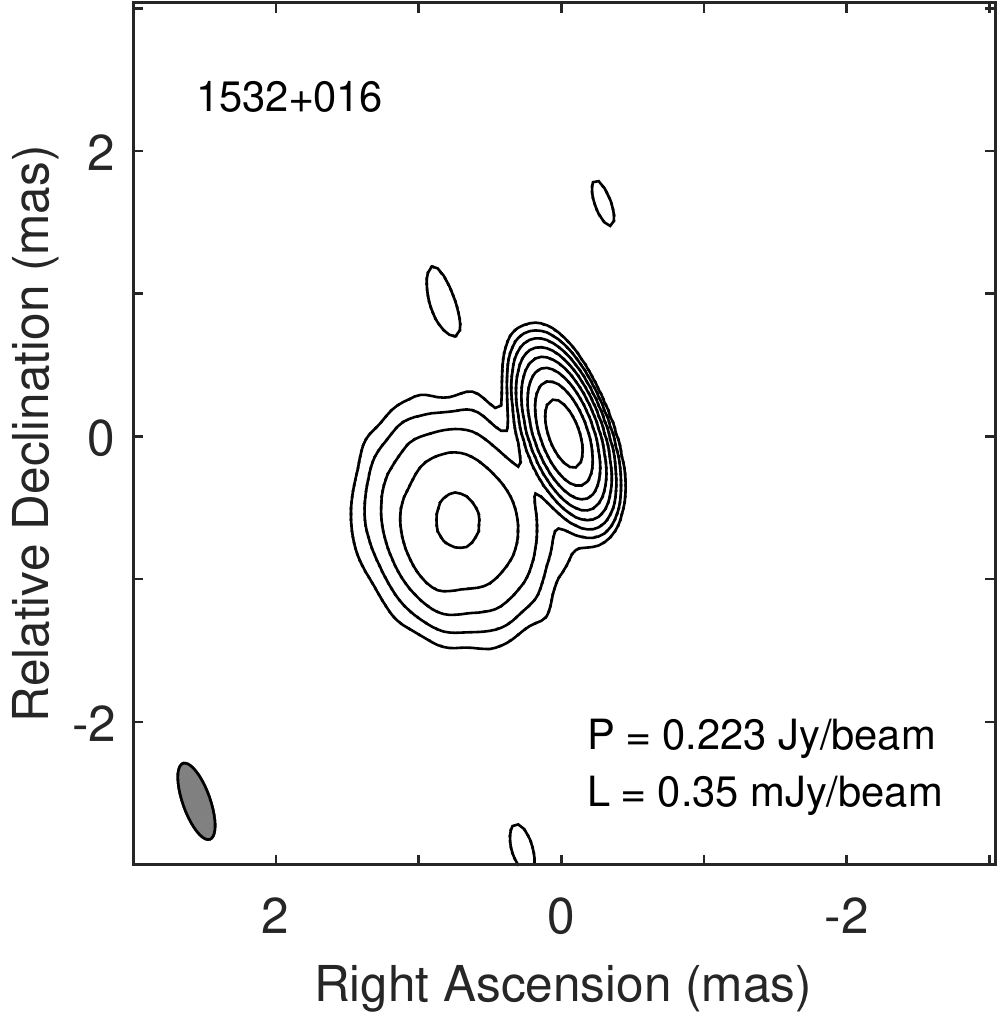} & \includegraphics[width=0.3\textwidth]{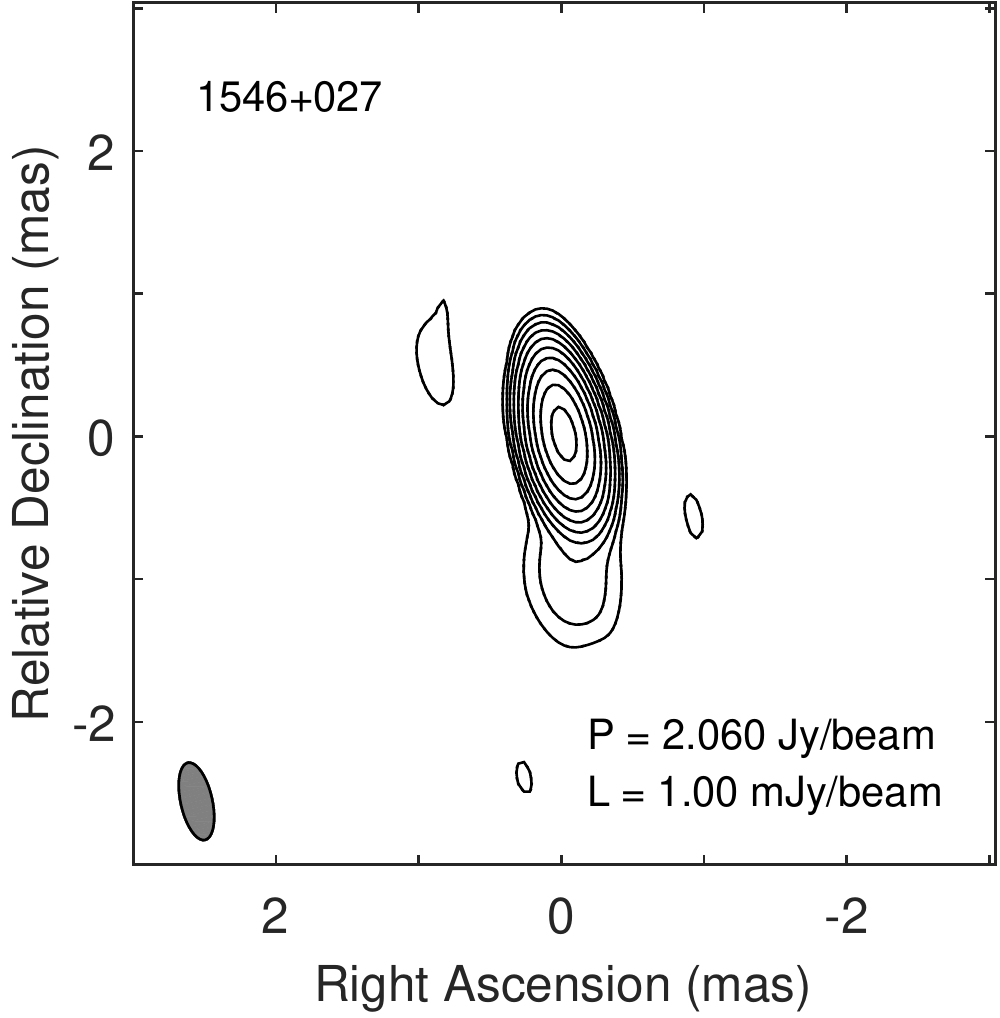}  & \includegraphics[width=0.3\textwidth]{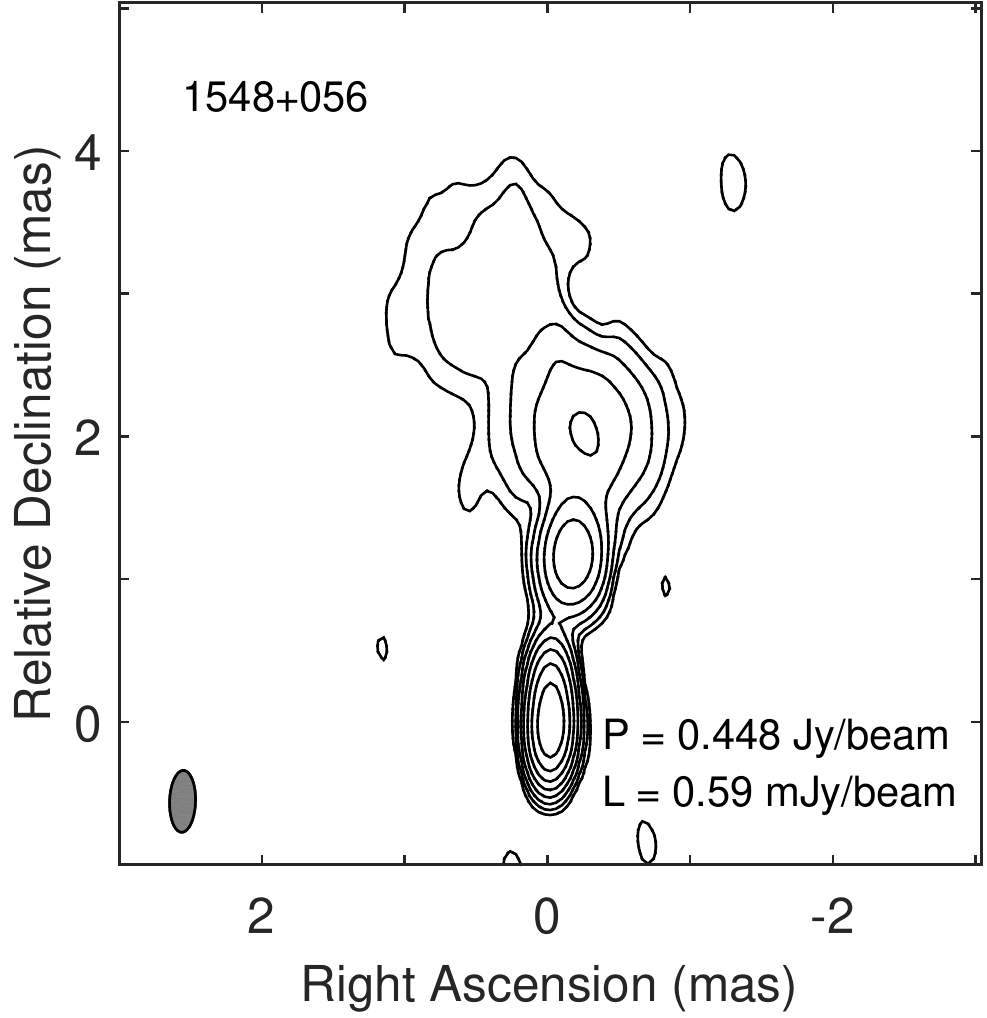}\\
  \includegraphics[width=0.3\textwidth]{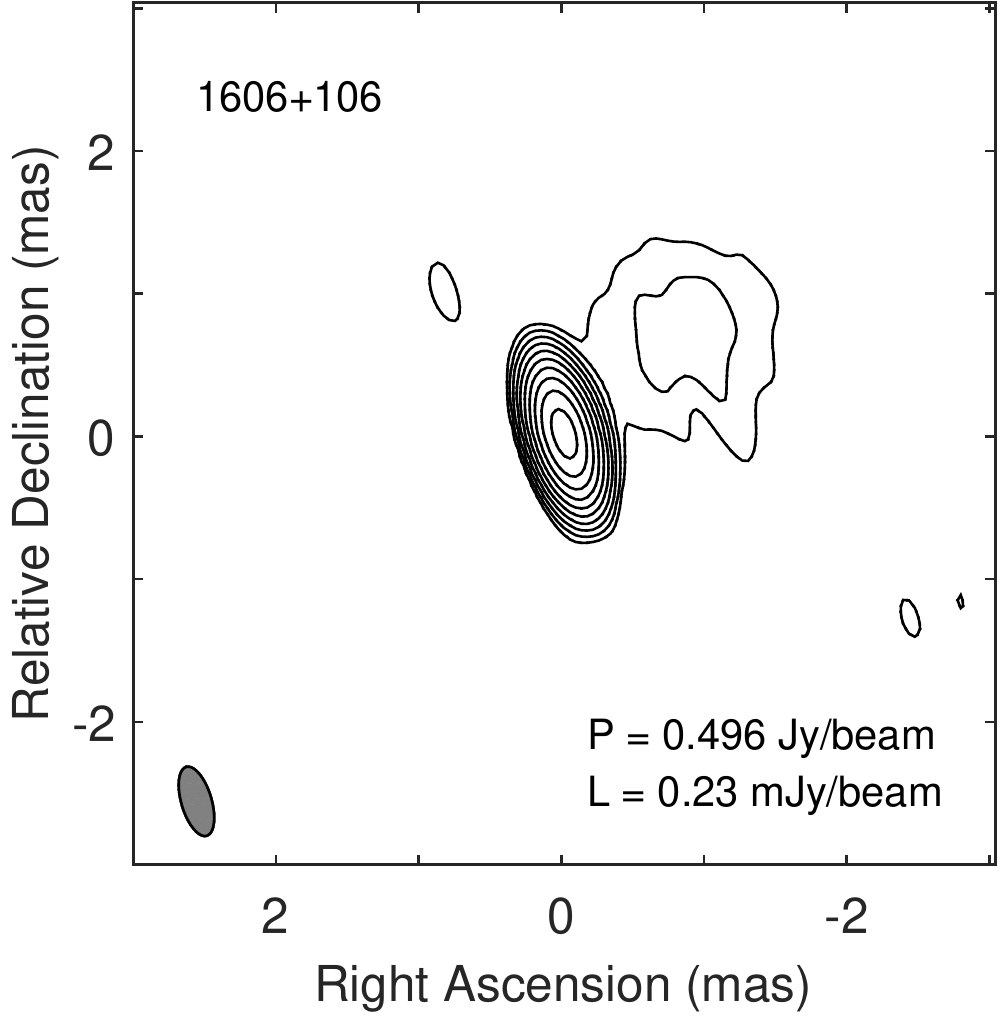} & \includegraphics[width=0.3\textwidth]{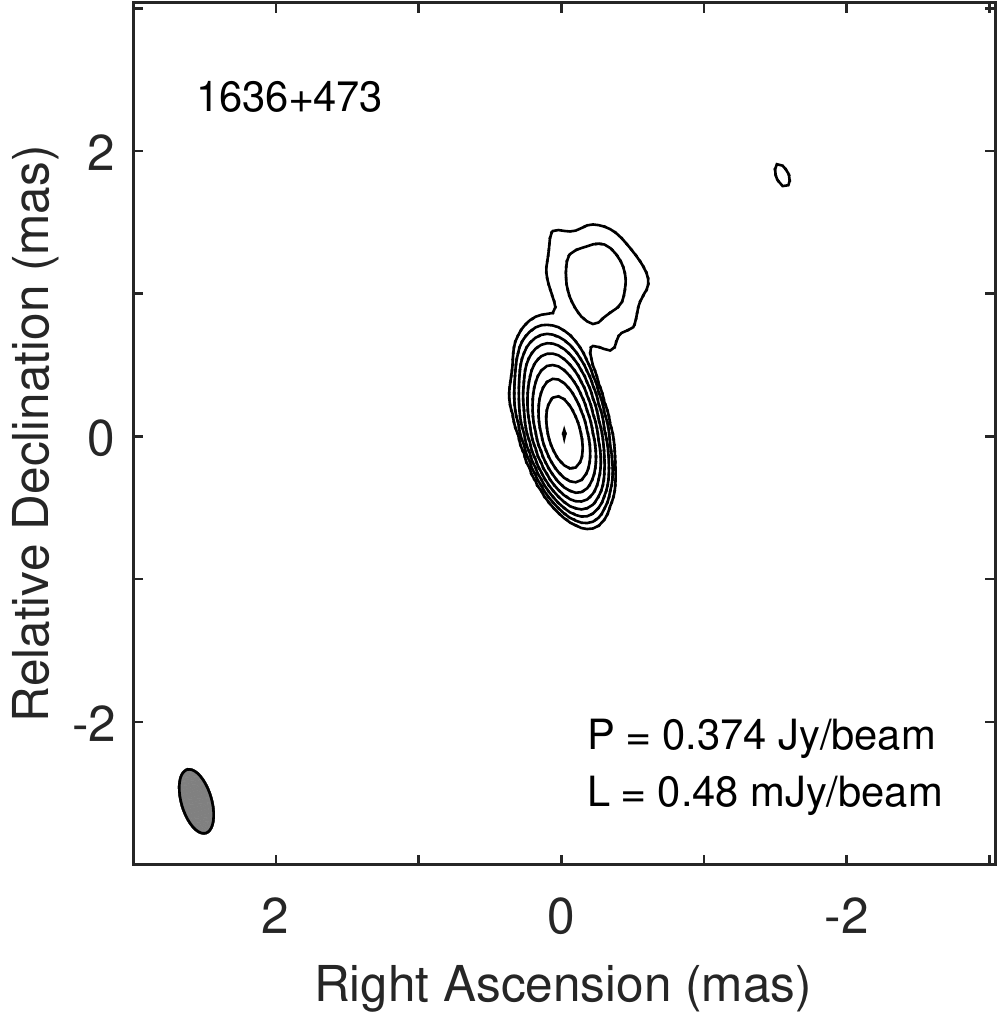}  & \includegraphics[width=0.3\textwidth]{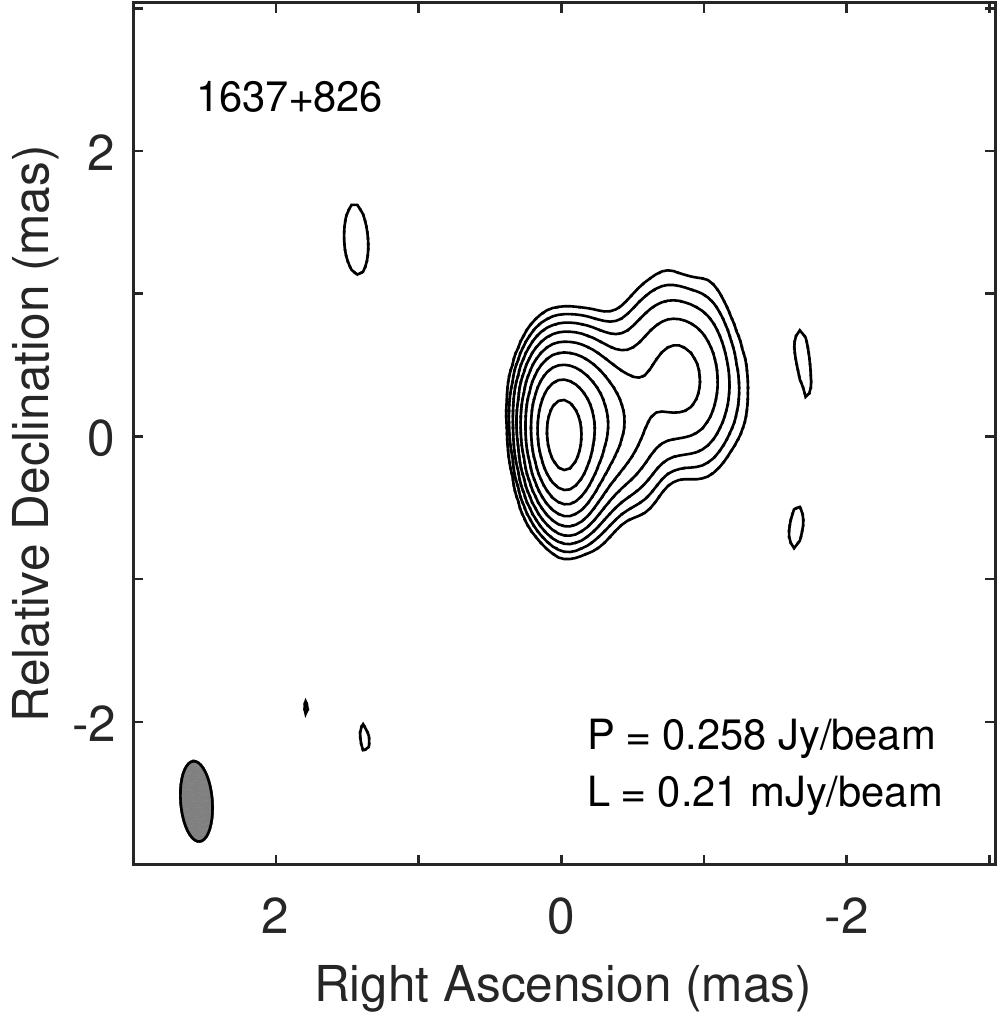}\\ \includegraphics[width=0.3\textwidth]{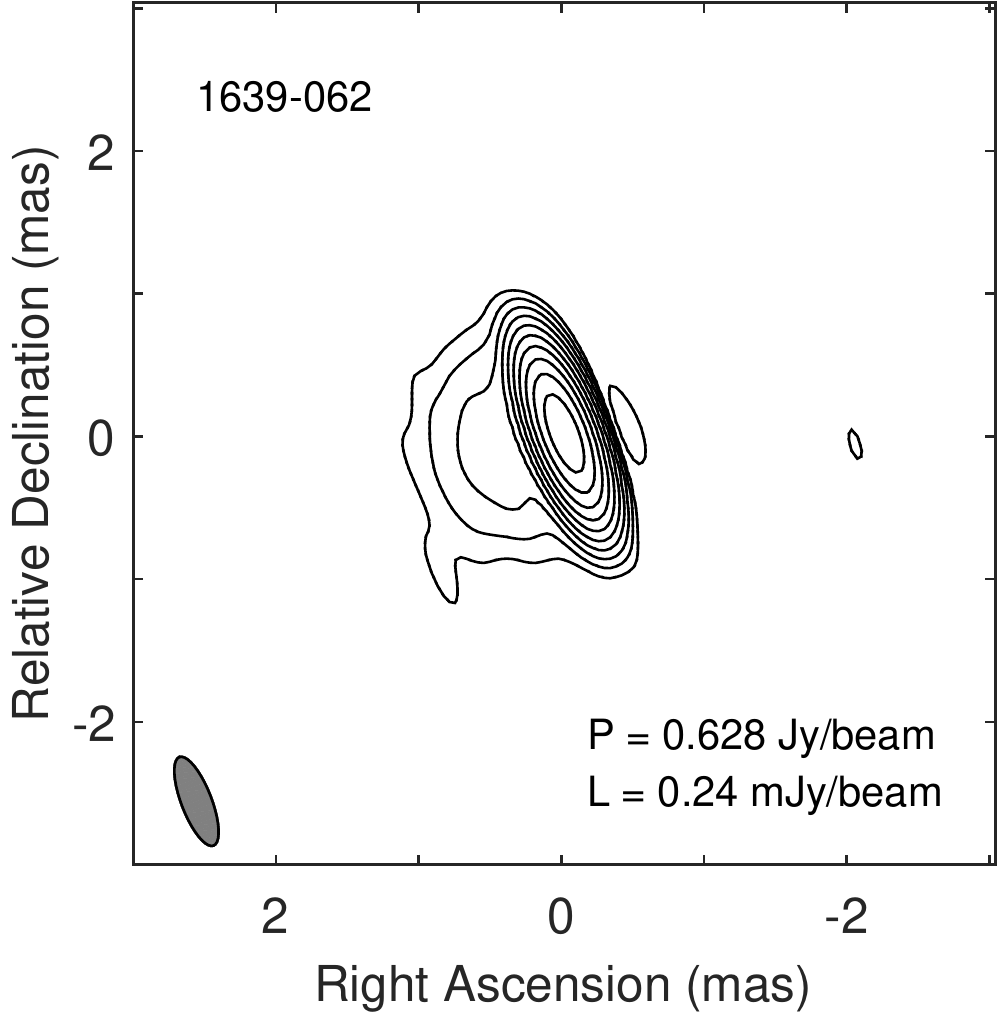} & \includegraphics[width=0.3\textwidth]{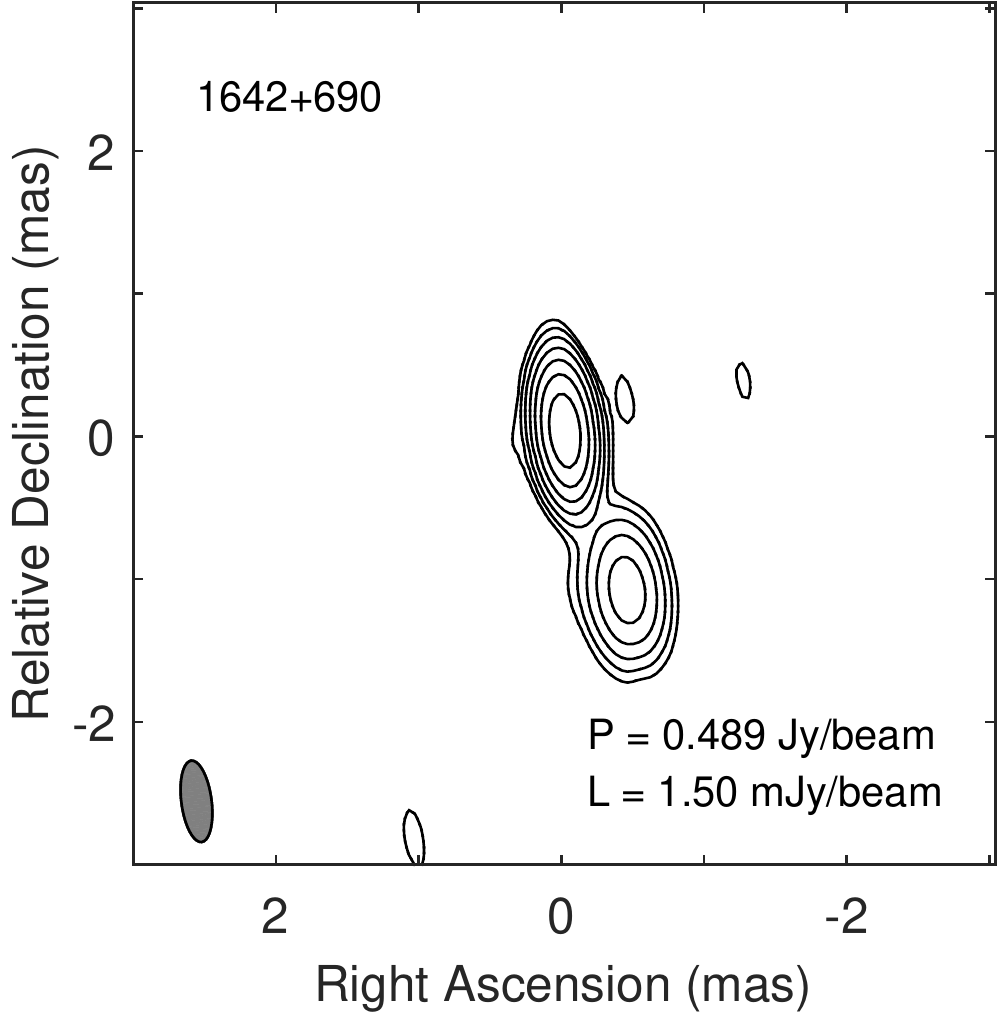}  & \includegraphics[width=0.3\textwidth]{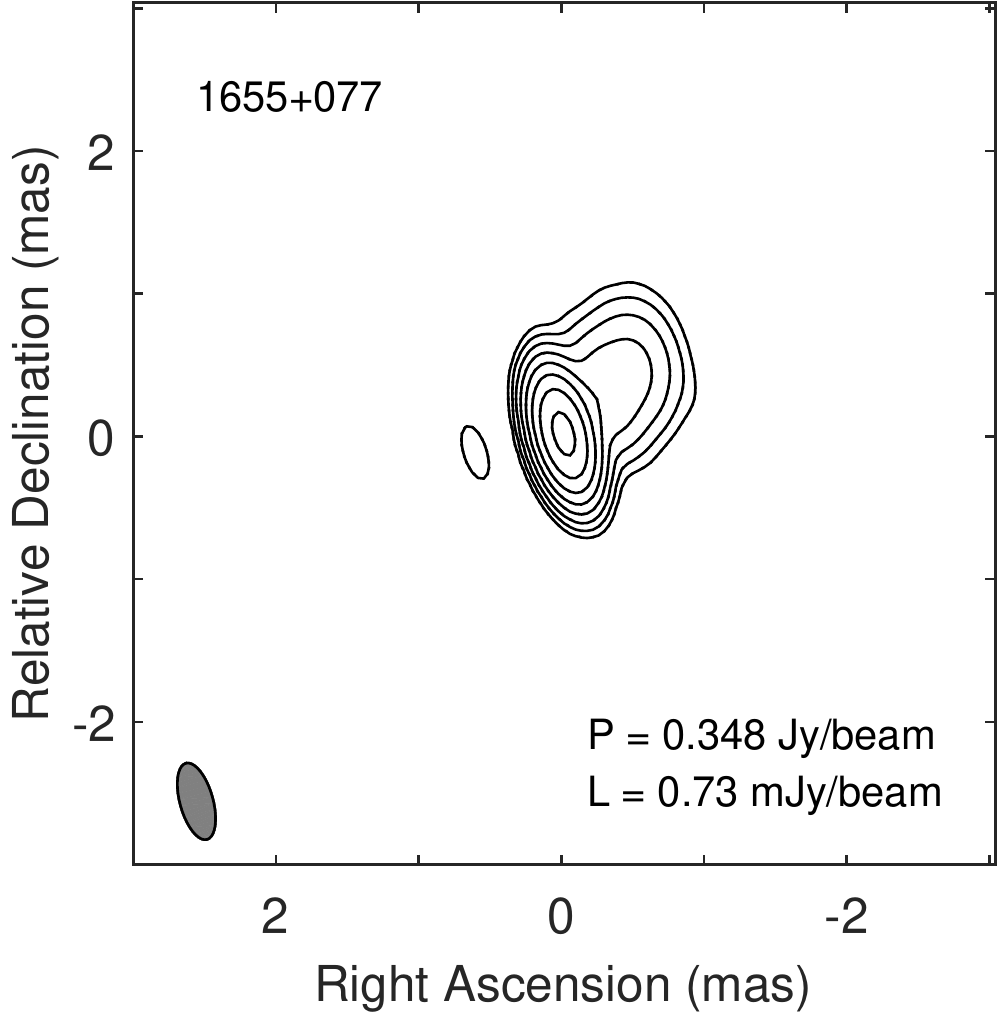}\\
  \includegraphics[width=0.3\textwidth]{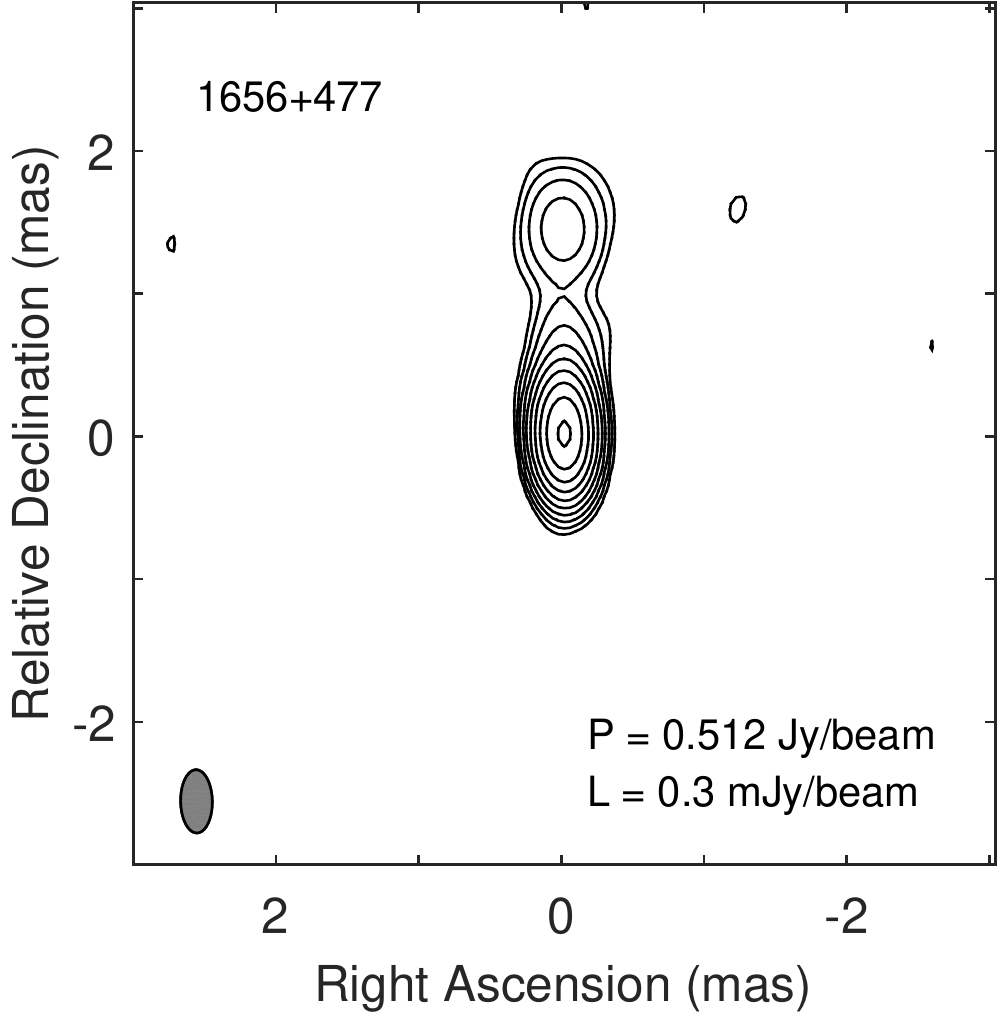} & \includegraphics[width=0.3\textwidth]{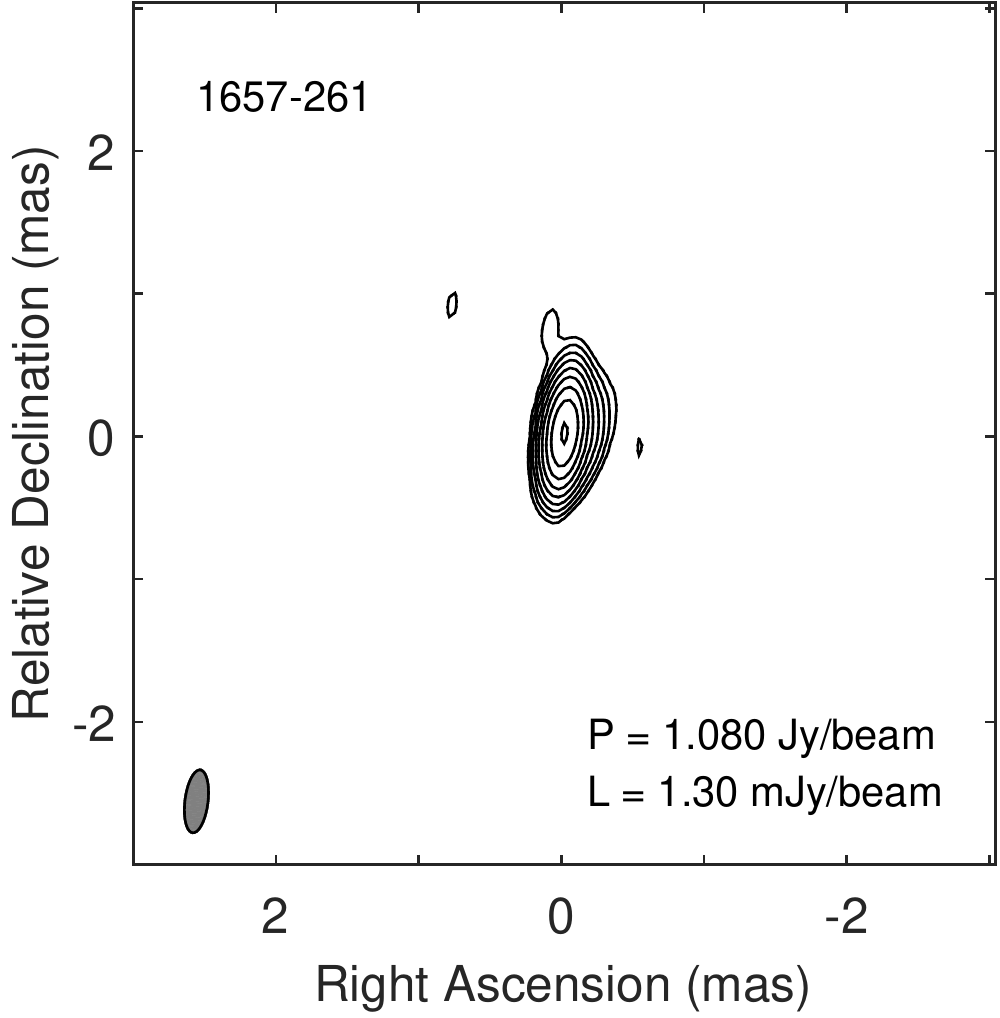}  & \includegraphics[width=0.3\textwidth]{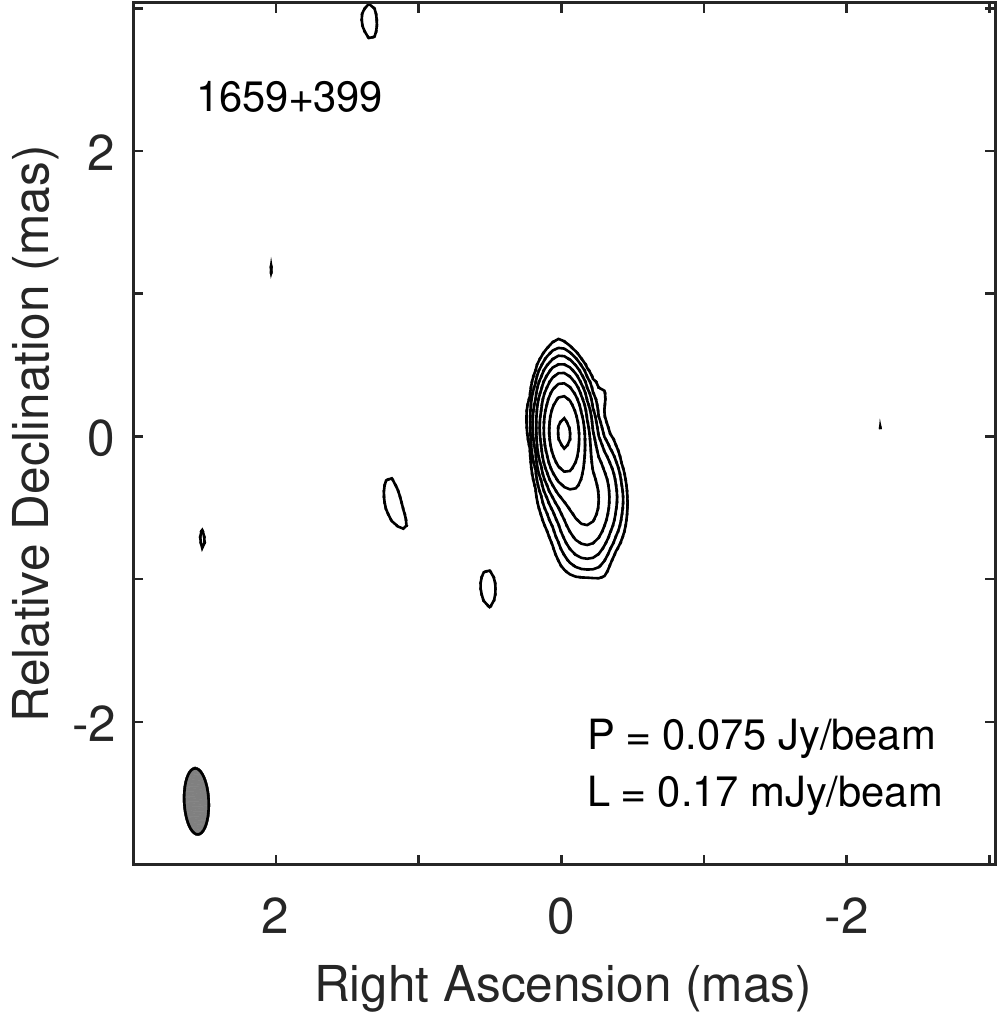}\\
    \end{tabular}
  \caption{VLBA images --- Continued}
  \end{figure}

\begin{figure}
  \centering
  \begin{tabular}{lcr}
  \includegraphics[width=0.3\textwidth]{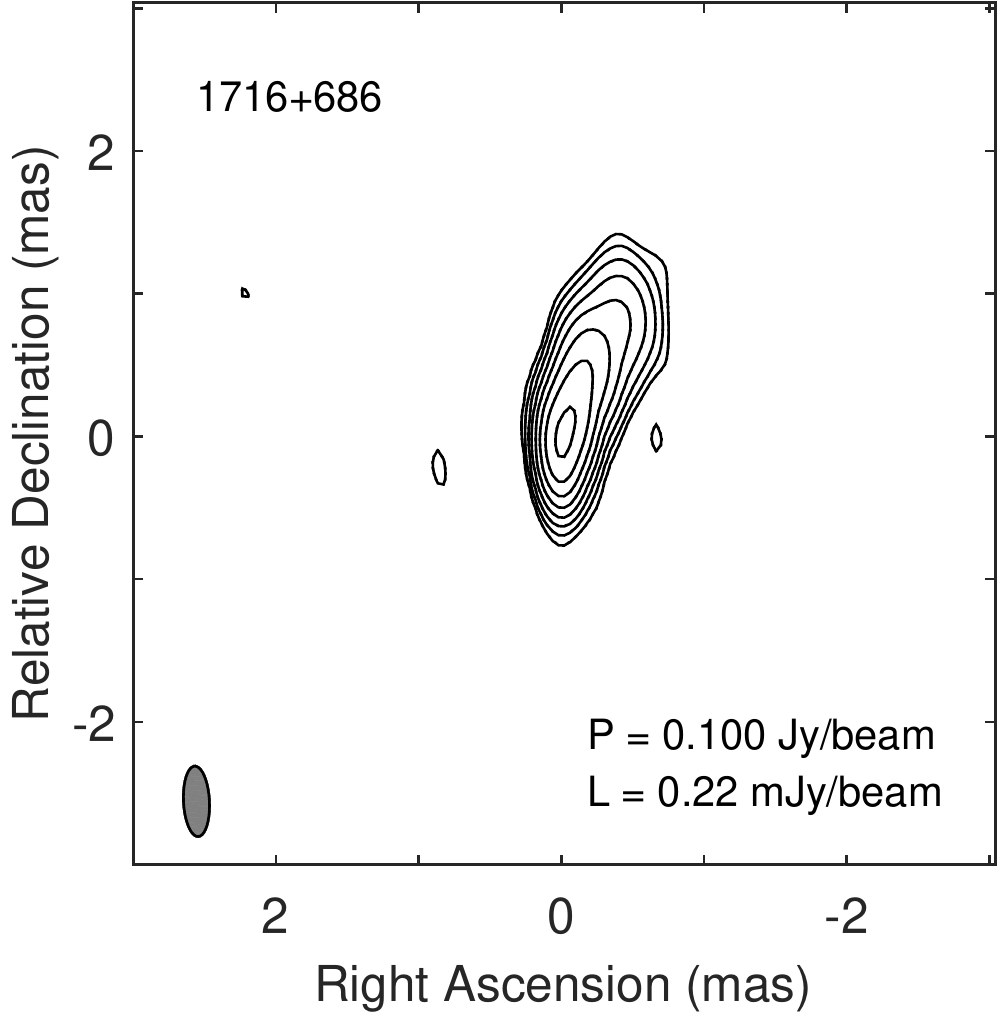} & \includegraphics[width=0.3\textwidth]{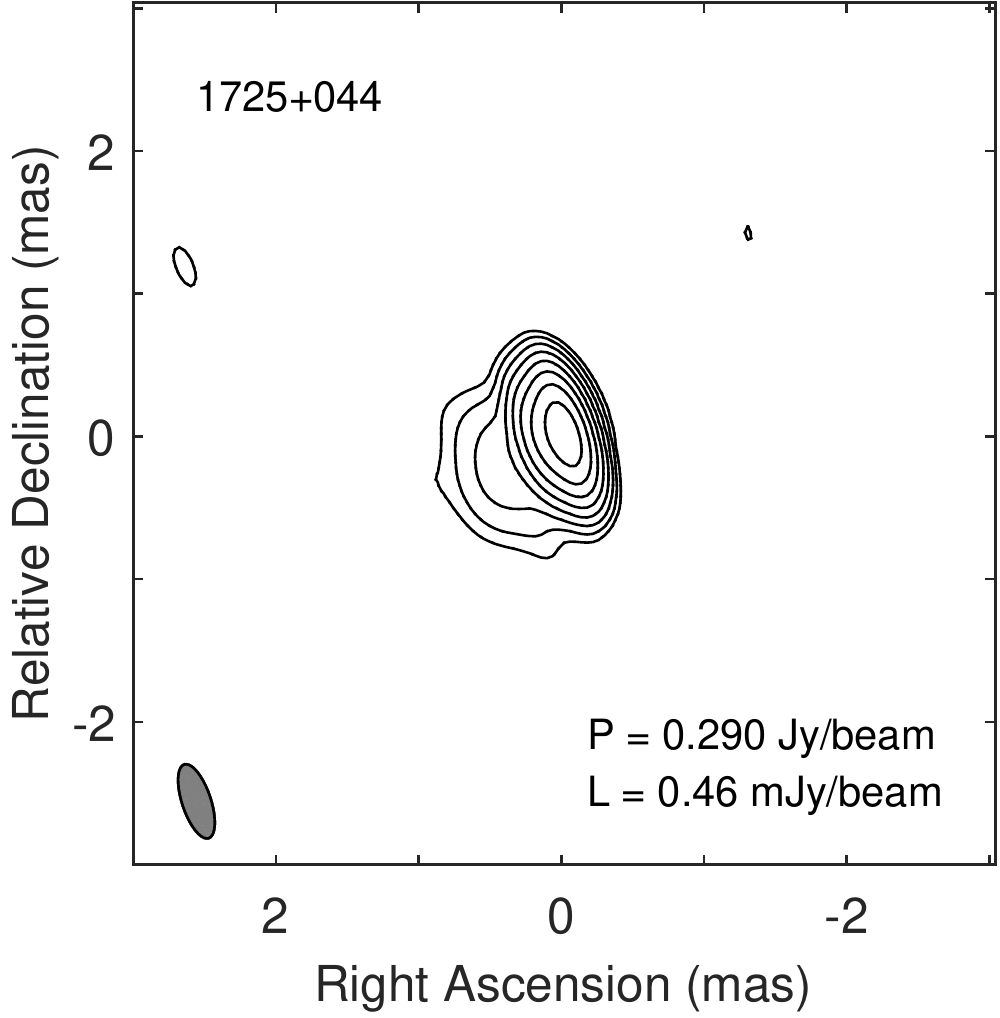}  & \includegraphics[width=0.3\textwidth]{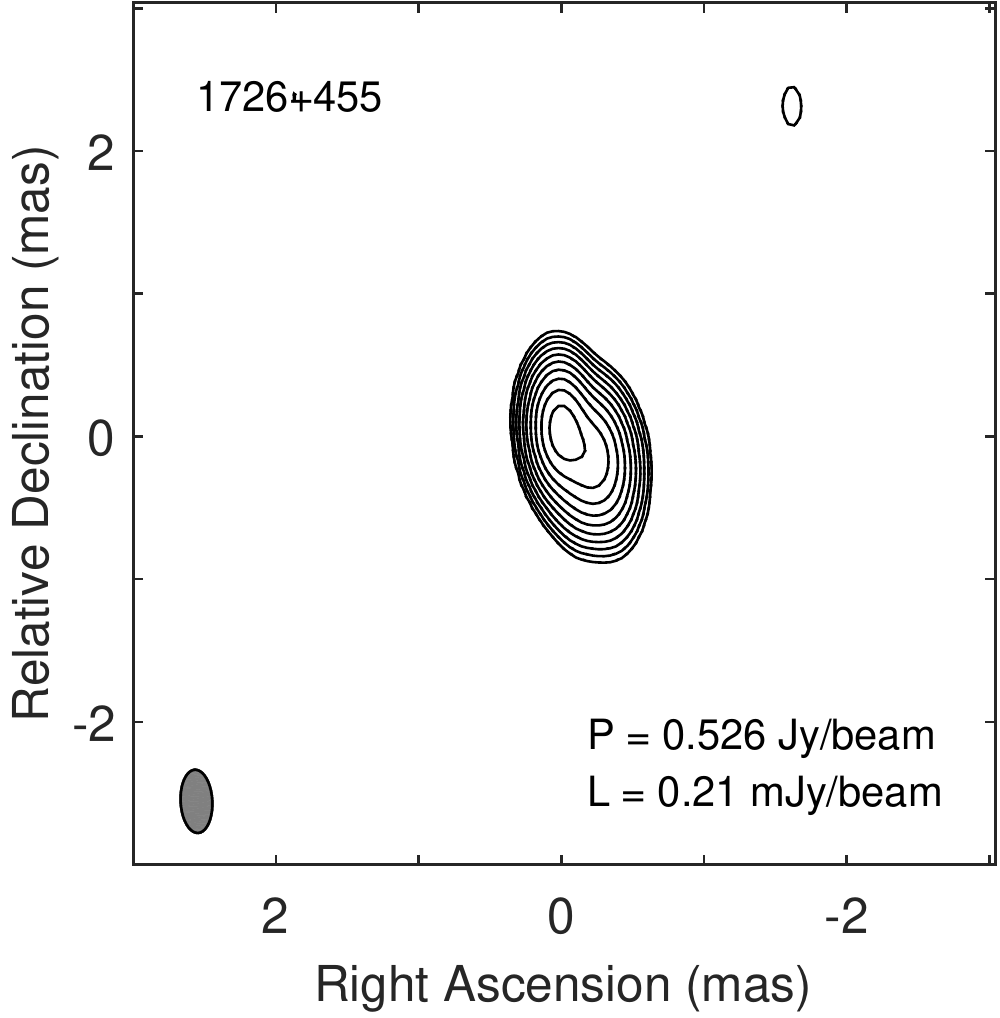}\\
  \includegraphics[width=0.3\textwidth]{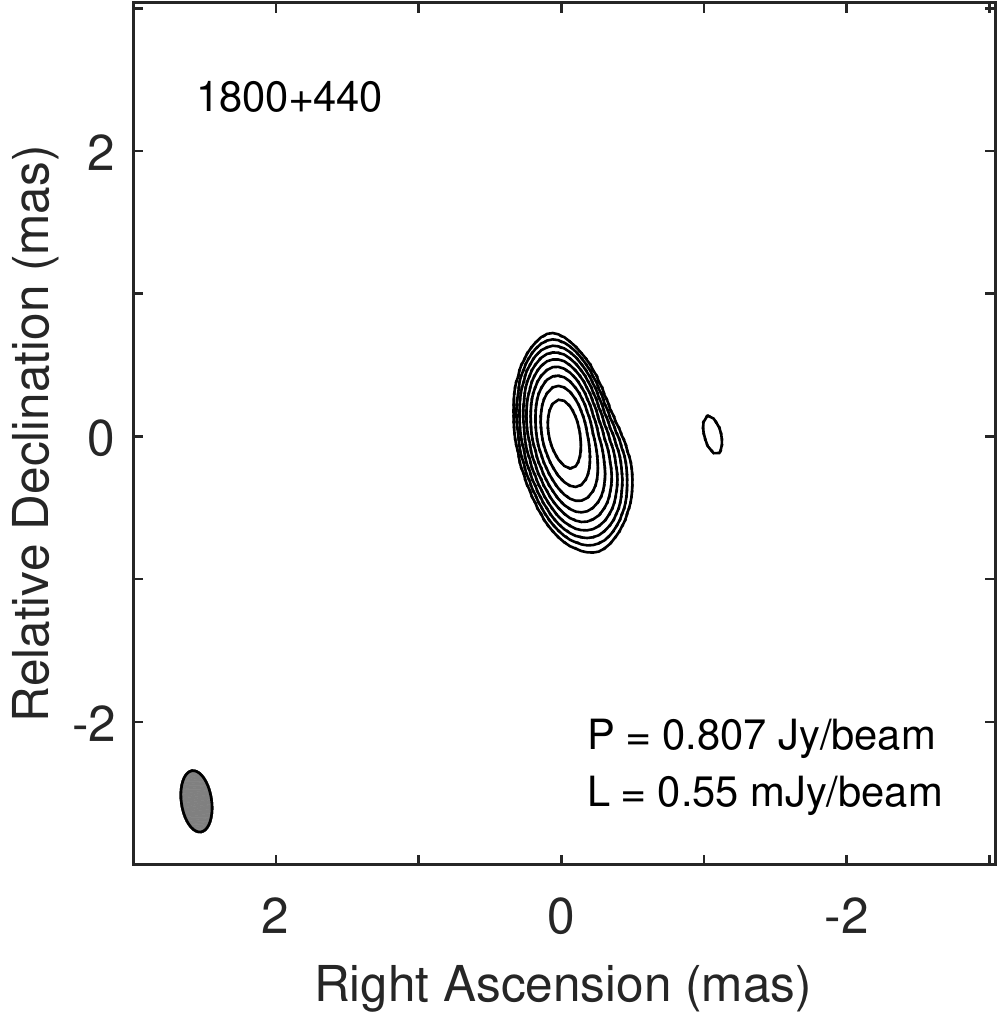} & \includegraphics[width=0.3\textwidth]{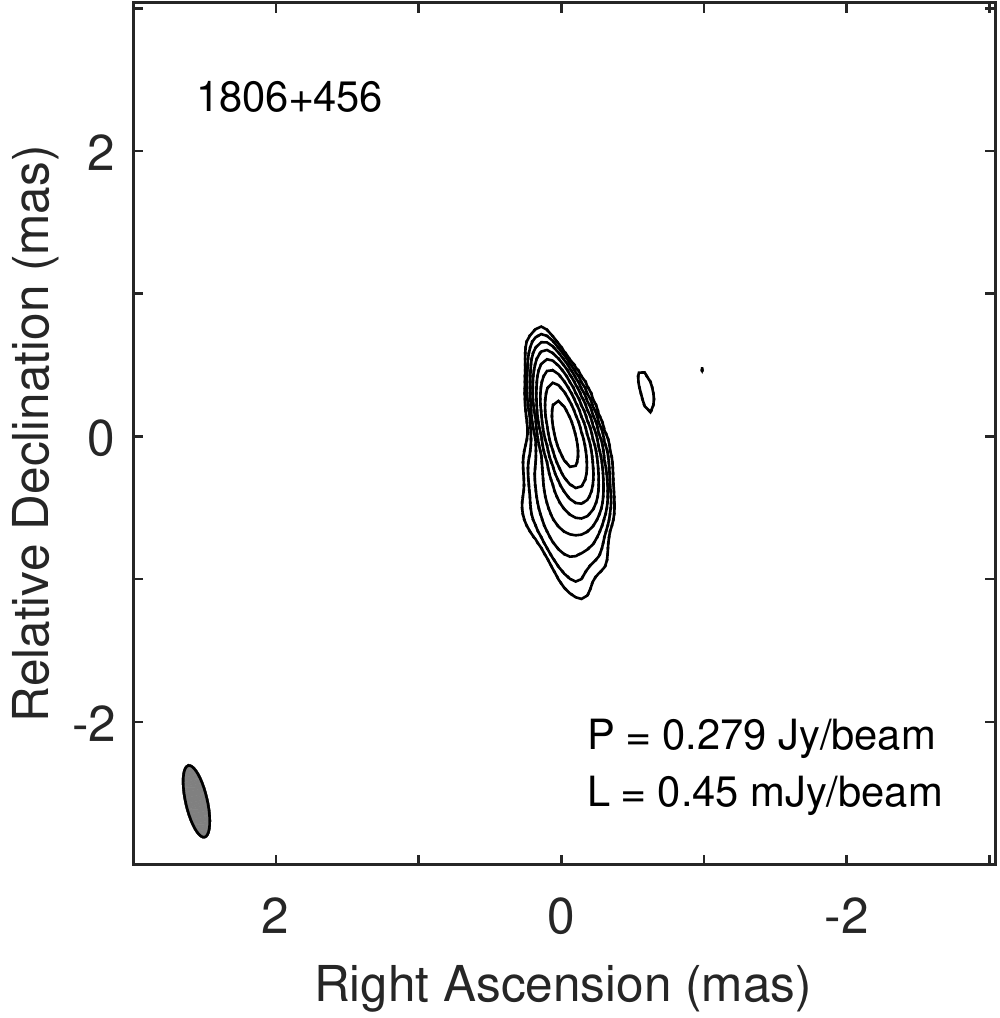}  & \includegraphics[width=0.3\textwidth]{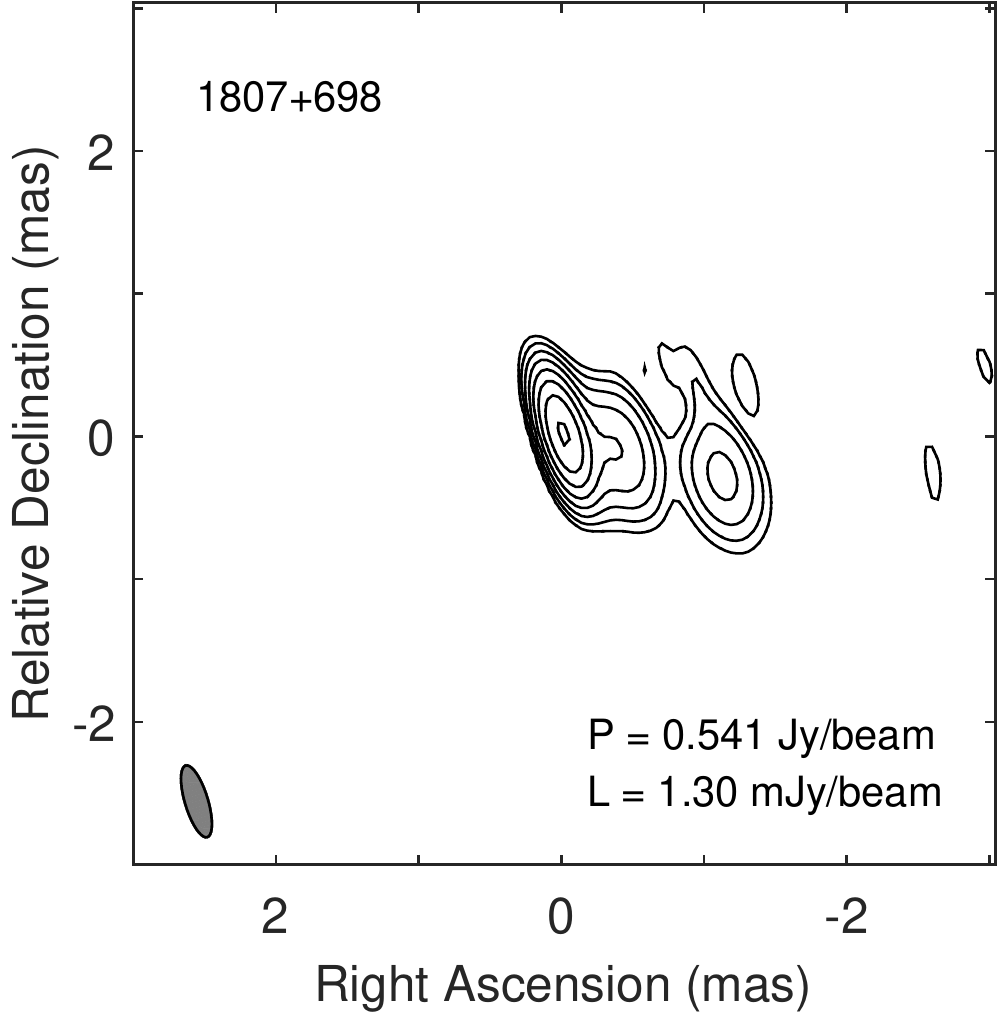}\\ \includegraphics[width=0.3\textwidth]{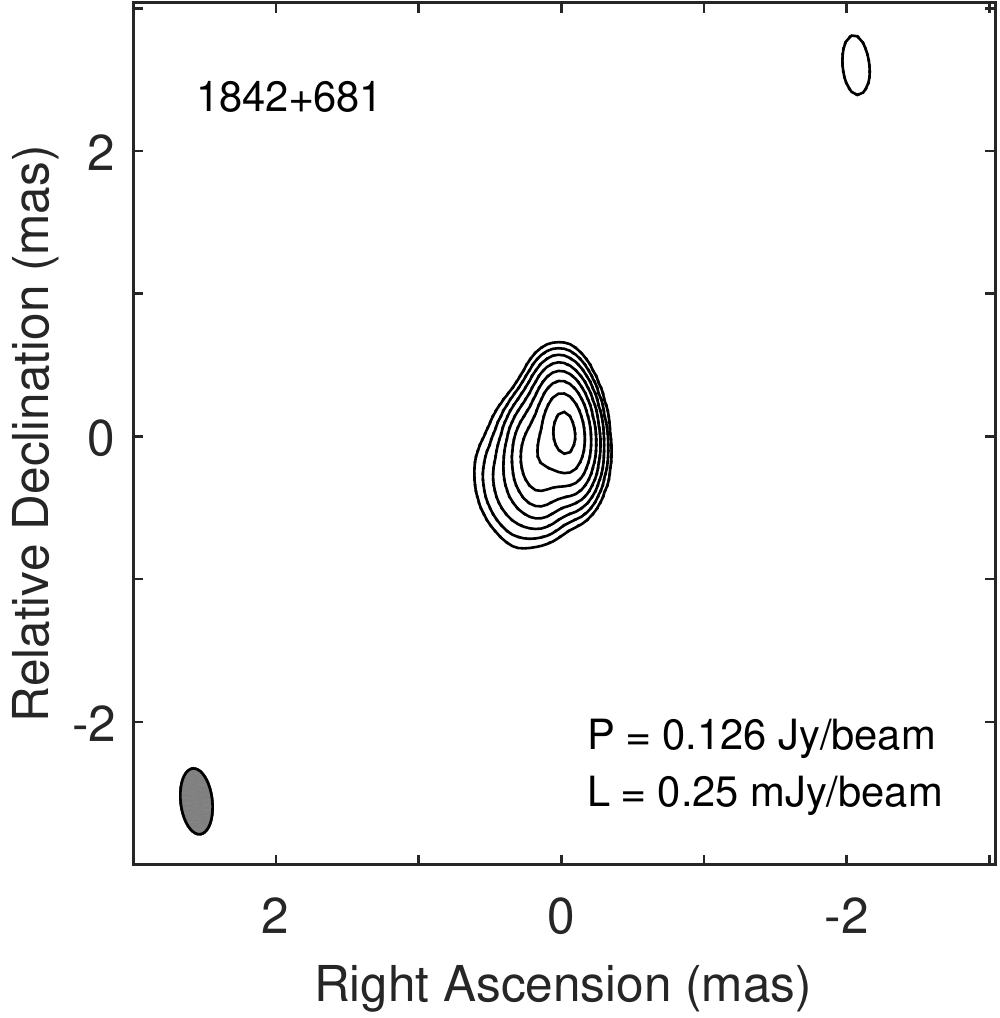} & \includegraphics[width=0.3\textwidth]{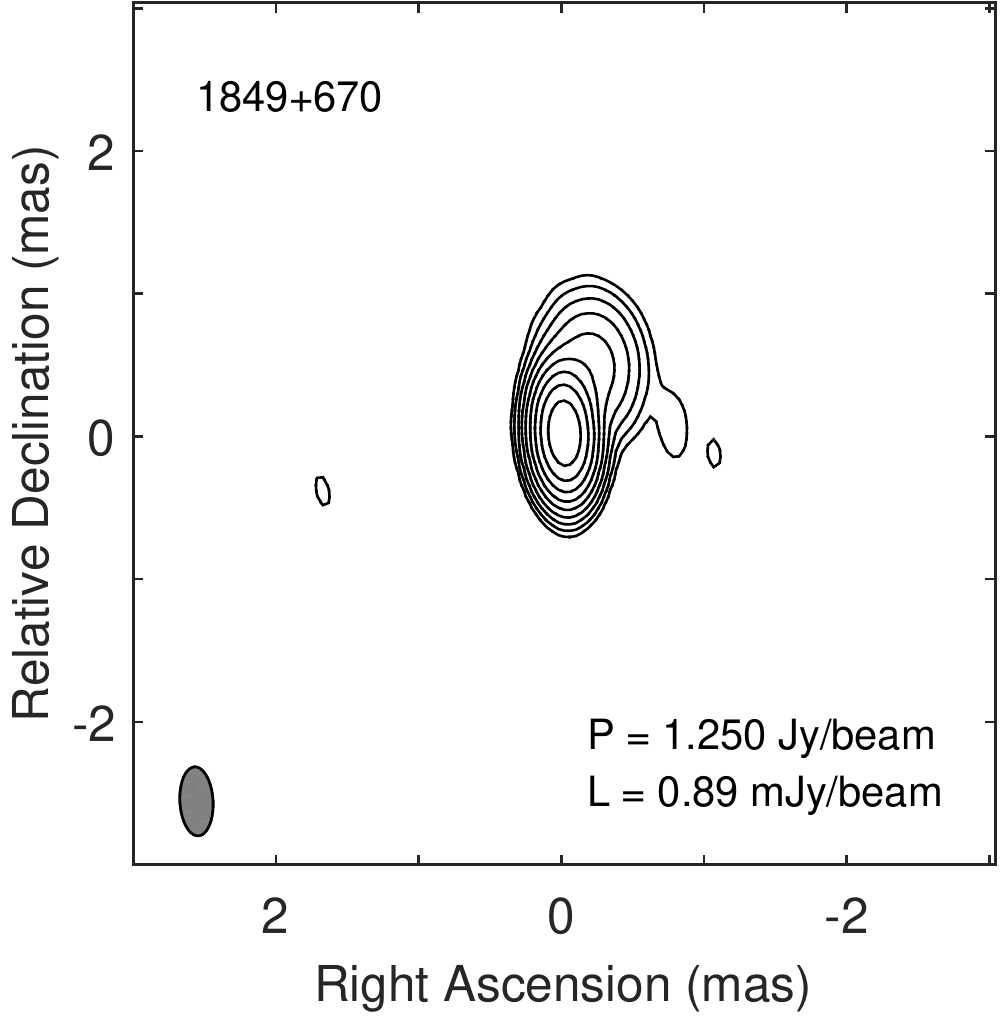}  & \includegraphics[width=0.3\textwidth]{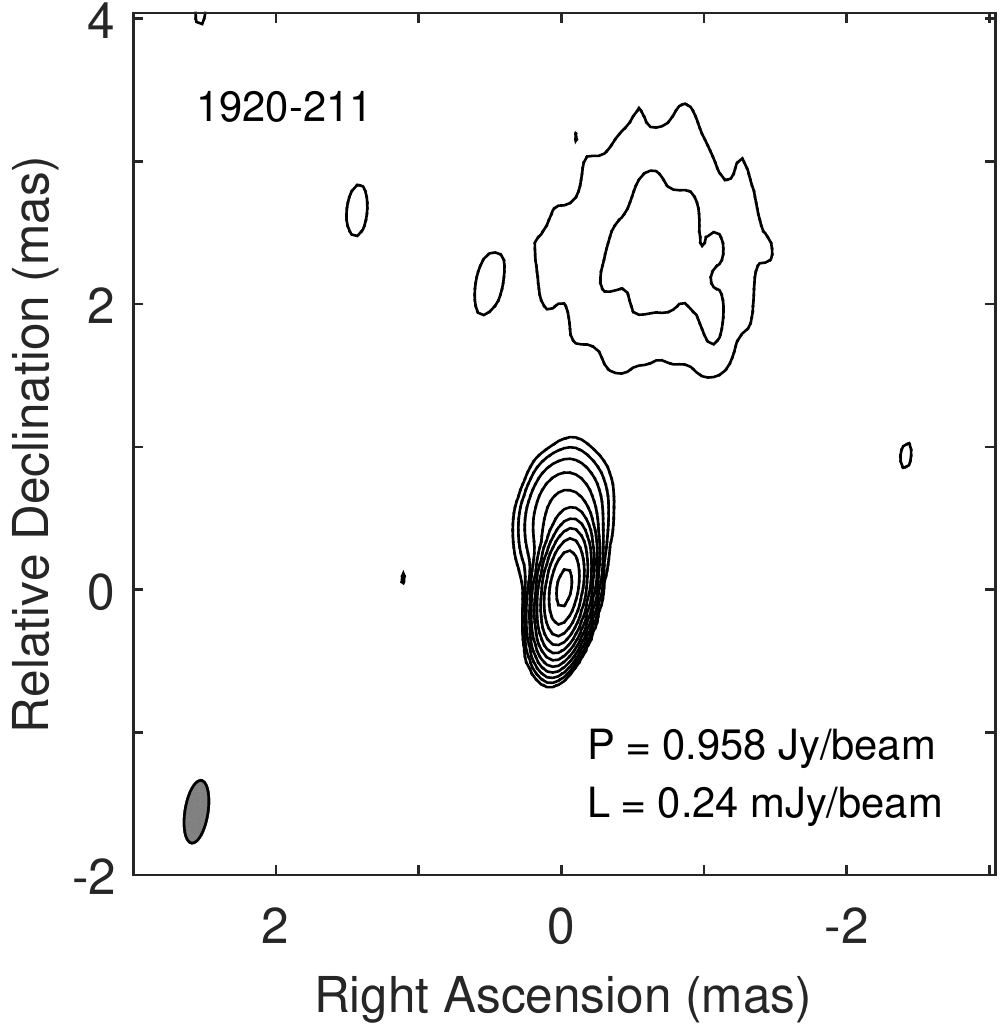}\\
  \includegraphics[width=0.3\textwidth]{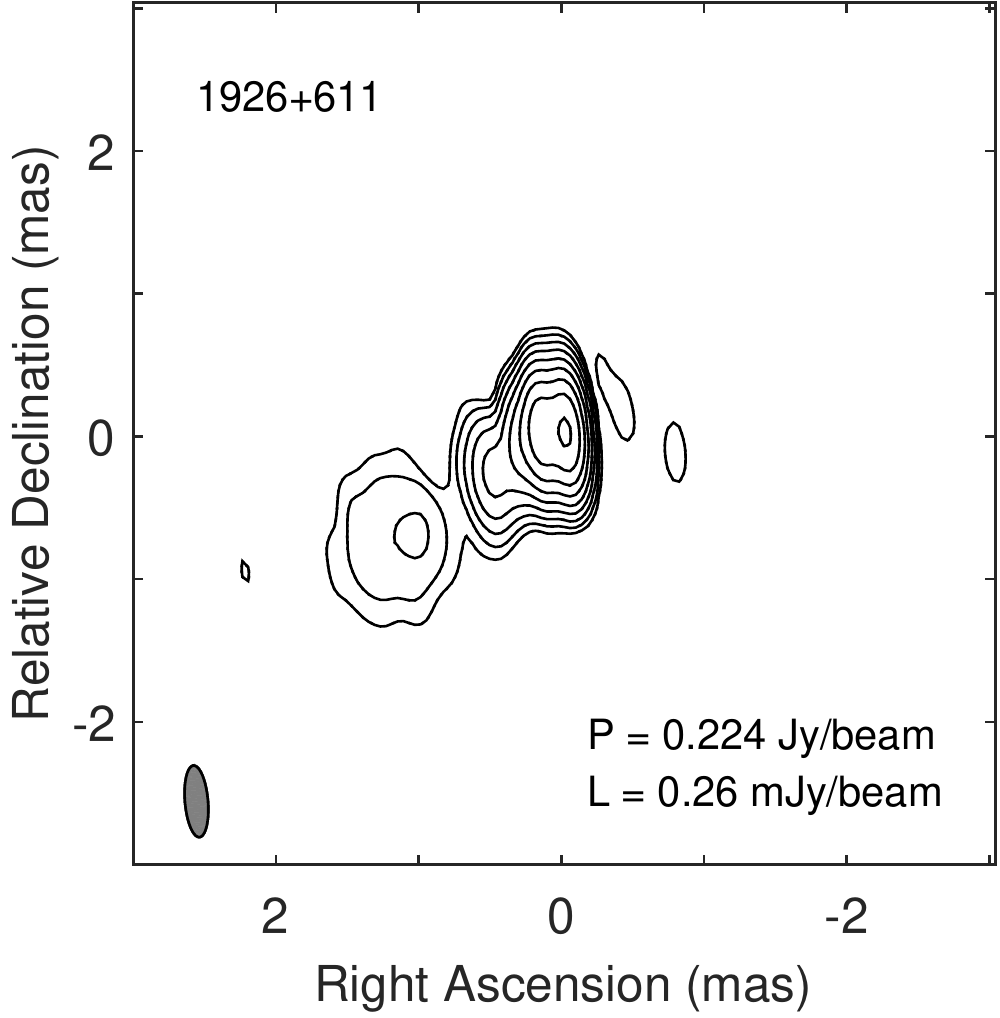} & \includegraphics[width=0.3\textwidth]{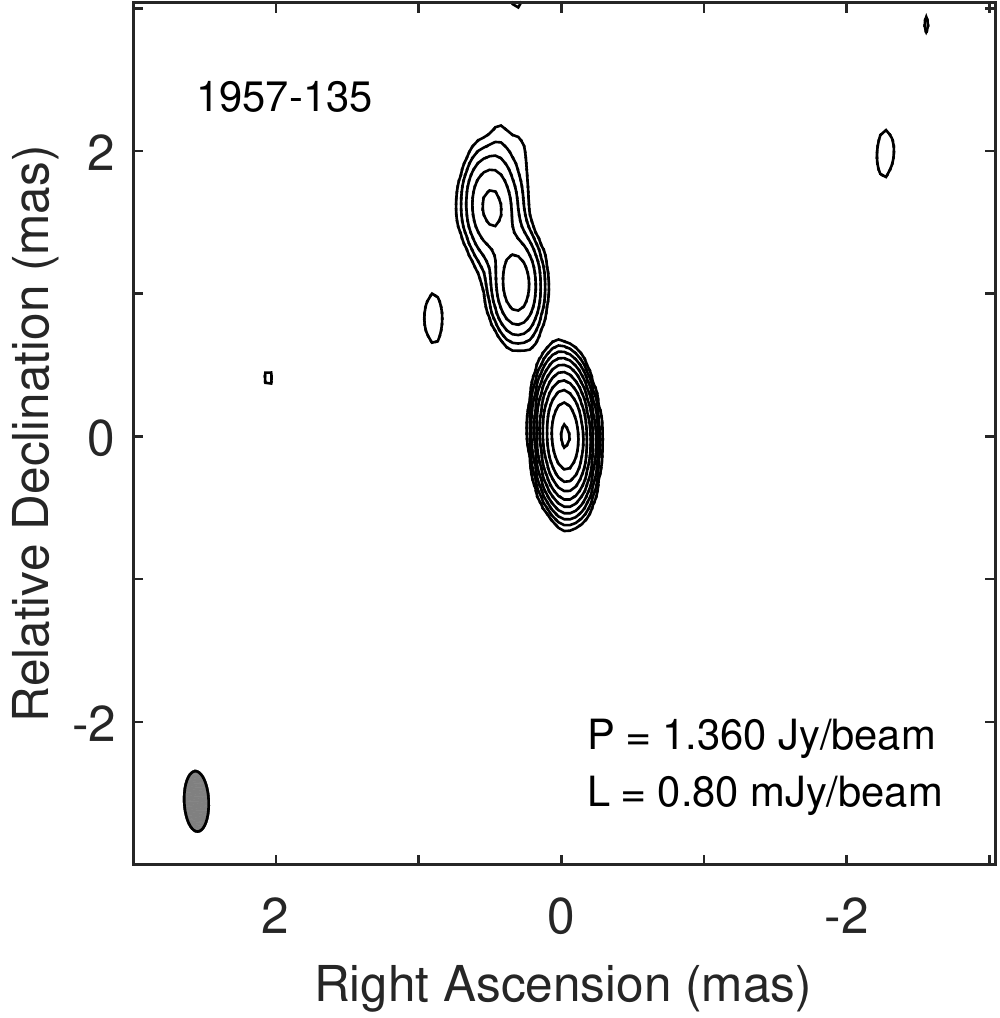}  & \includegraphics[width=0.3\textwidth]{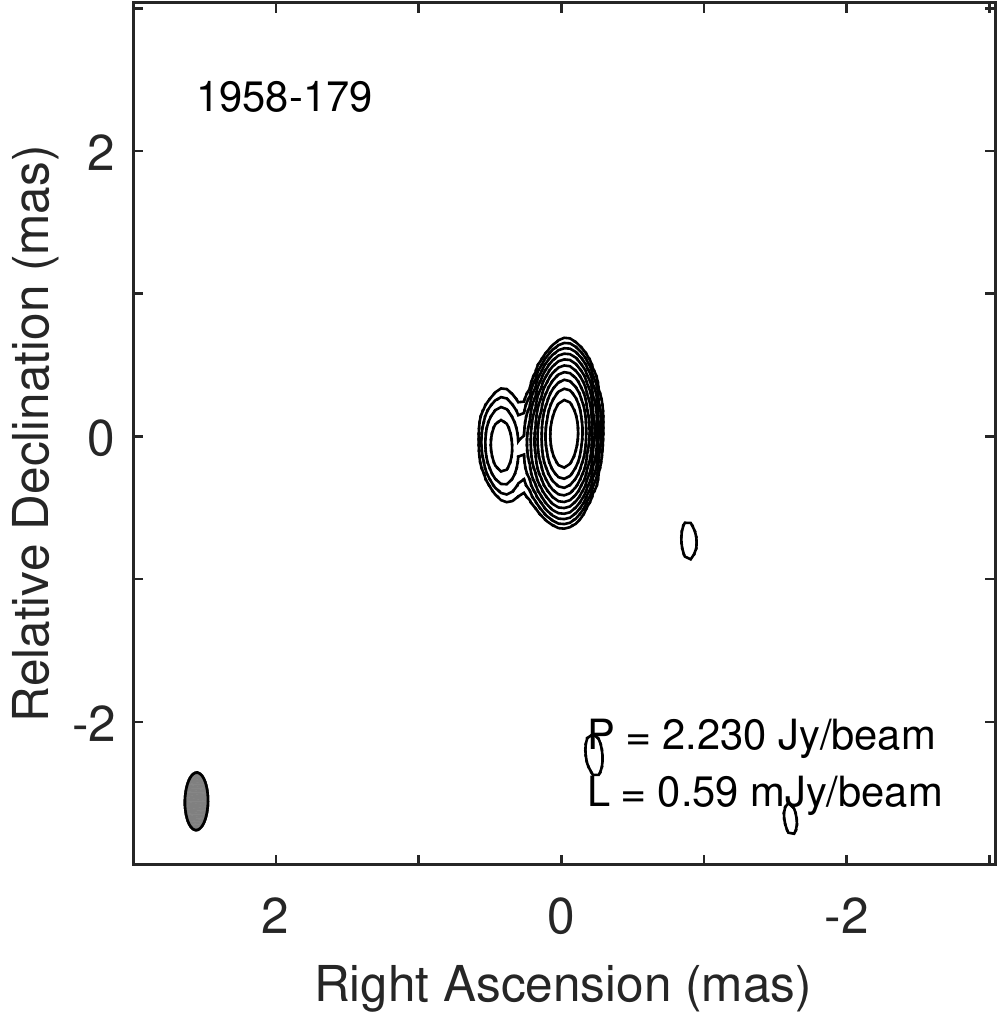}\\
    \end{tabular}
  \caption{VLBA images --- Continued}
  \end{figure}

\begin{figure}
  \centering
  \begin{tabular}{lcr}
  \includegraphics[width=0.3\textwidth]{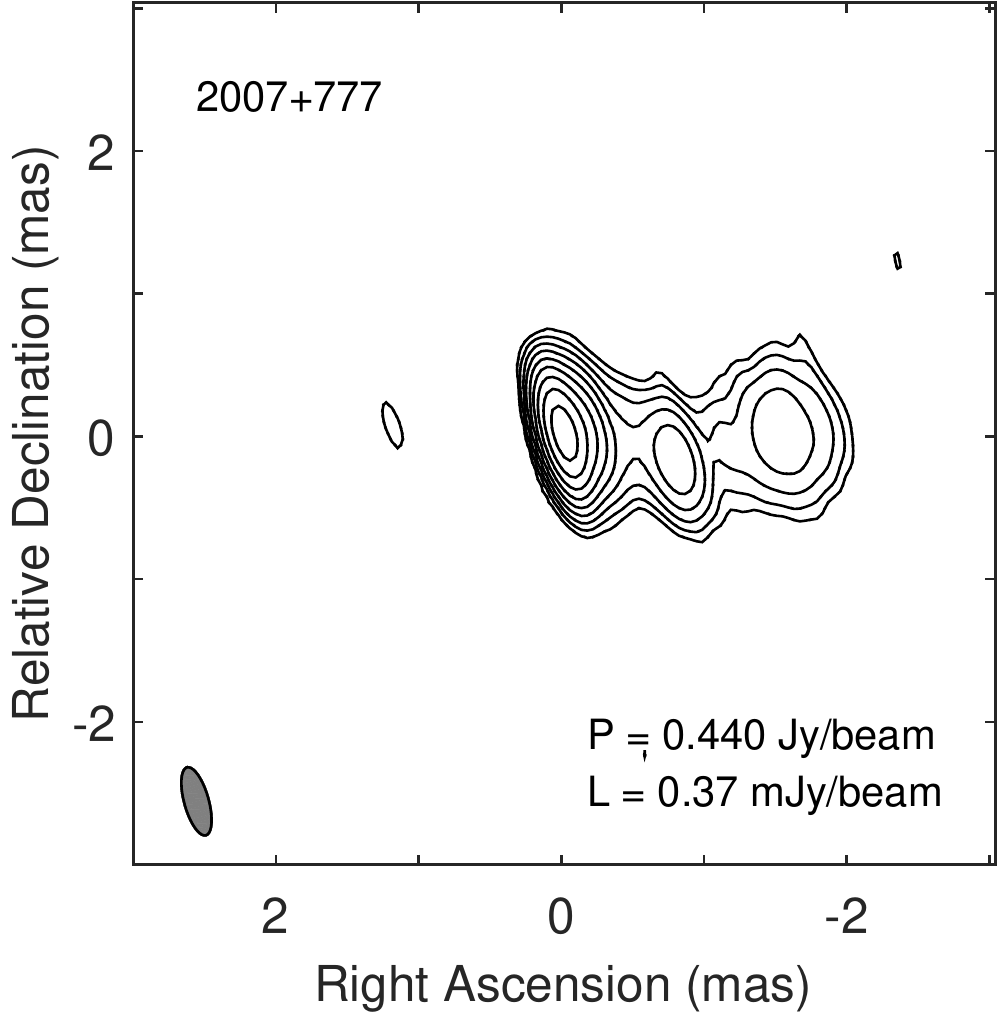} & \includegraphics[width=0.3\textwidth]{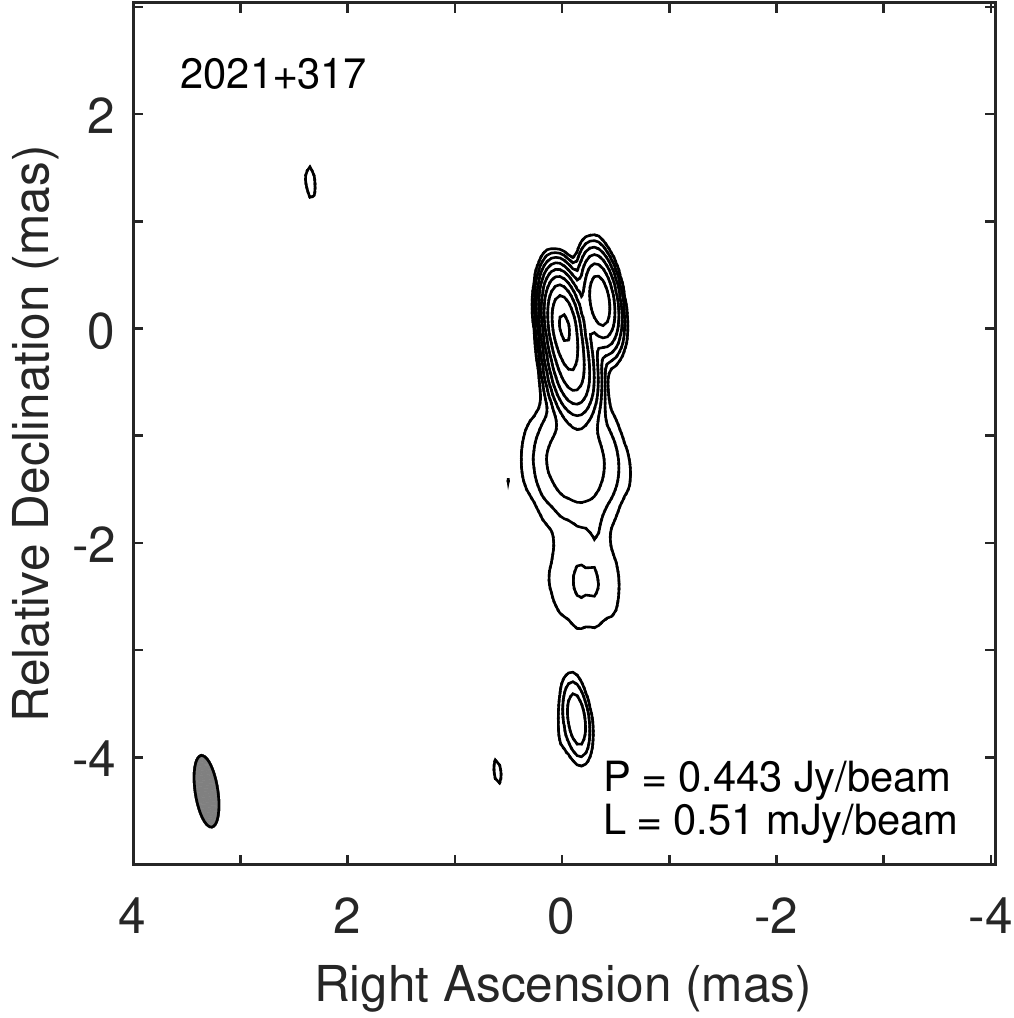}  & \includegraphics[width=0.3\textwidth]{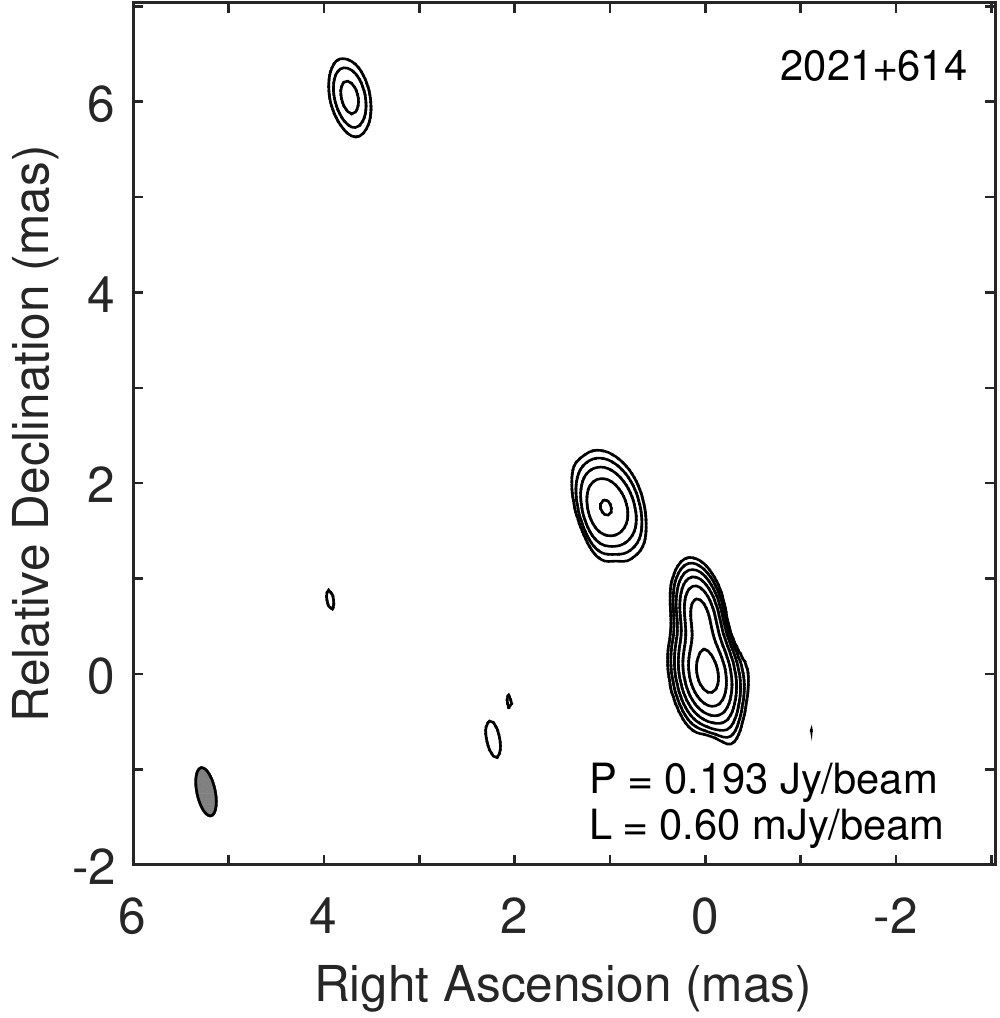}\\
  \includegraphics[width=0.3\textwidth]{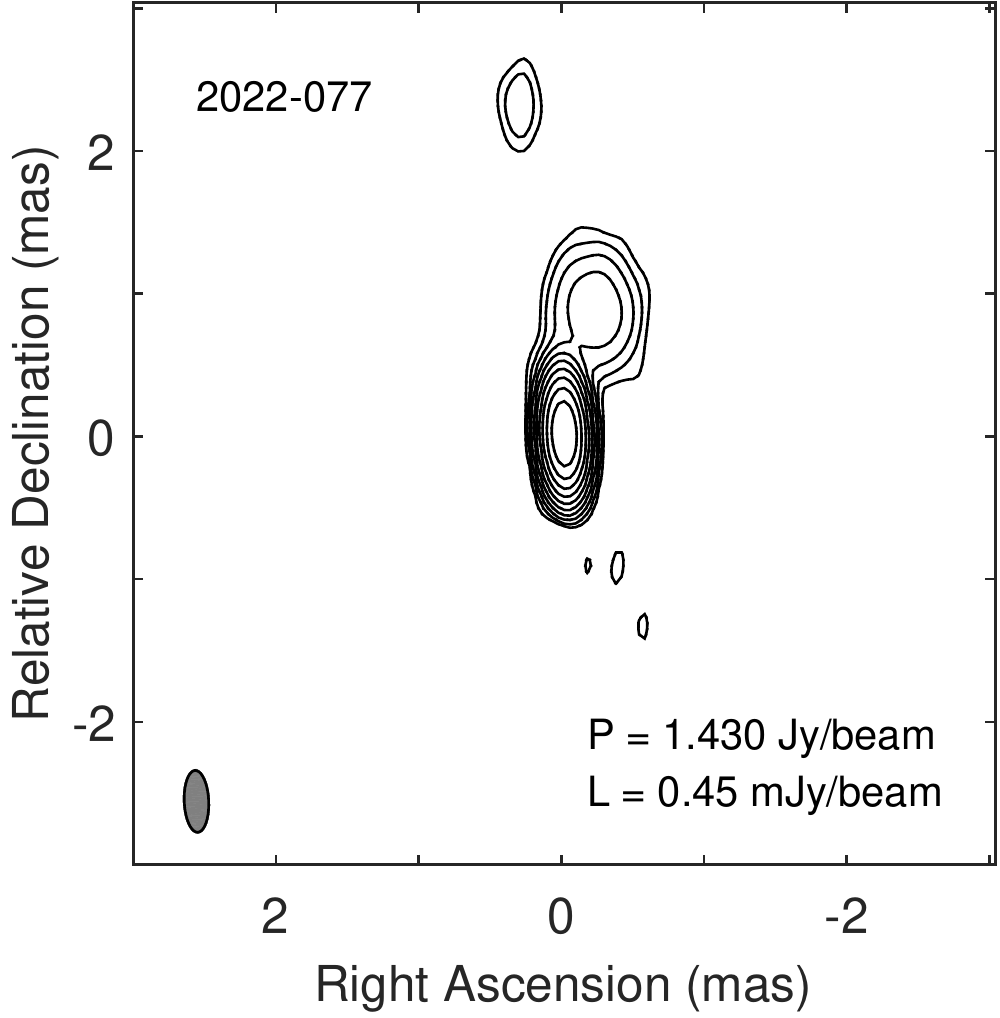} & \includegraphics[width=0.3\textwidth]{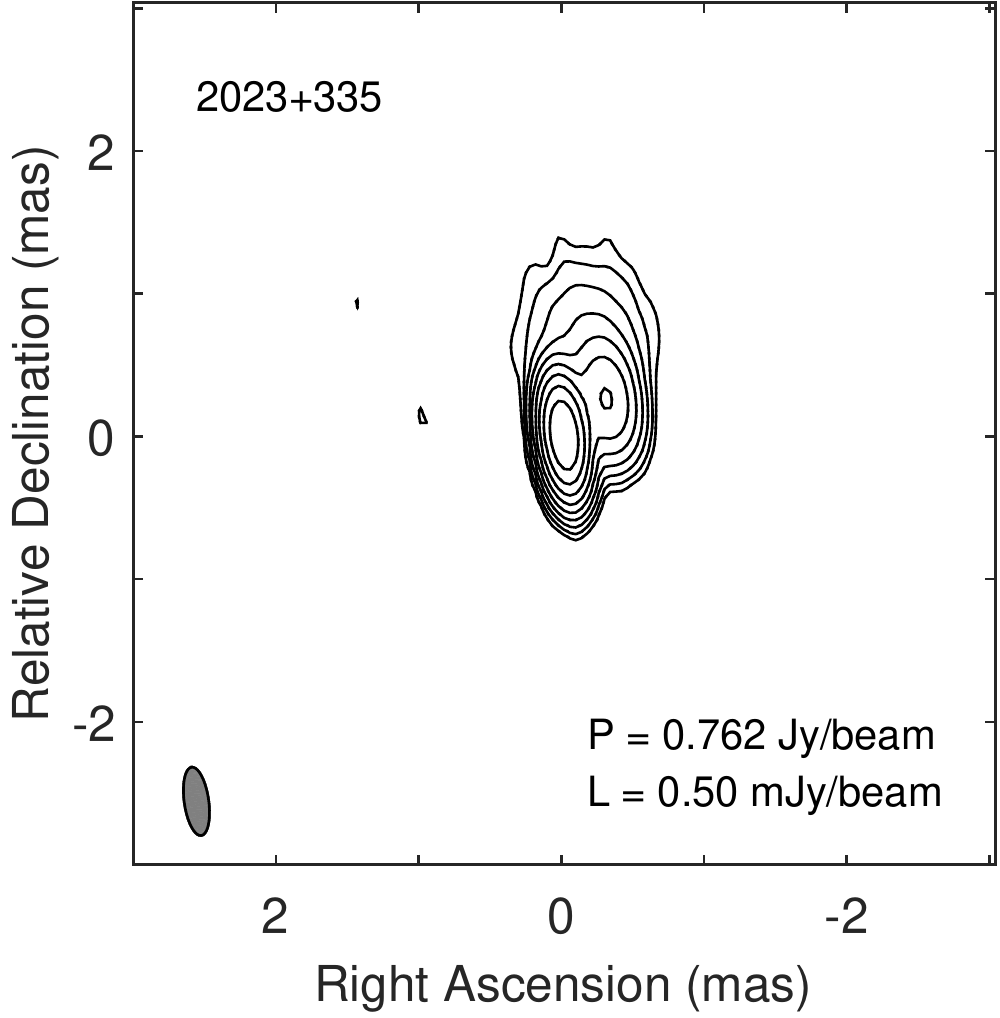}  & \includegraphics[width=0.3\textwidth]{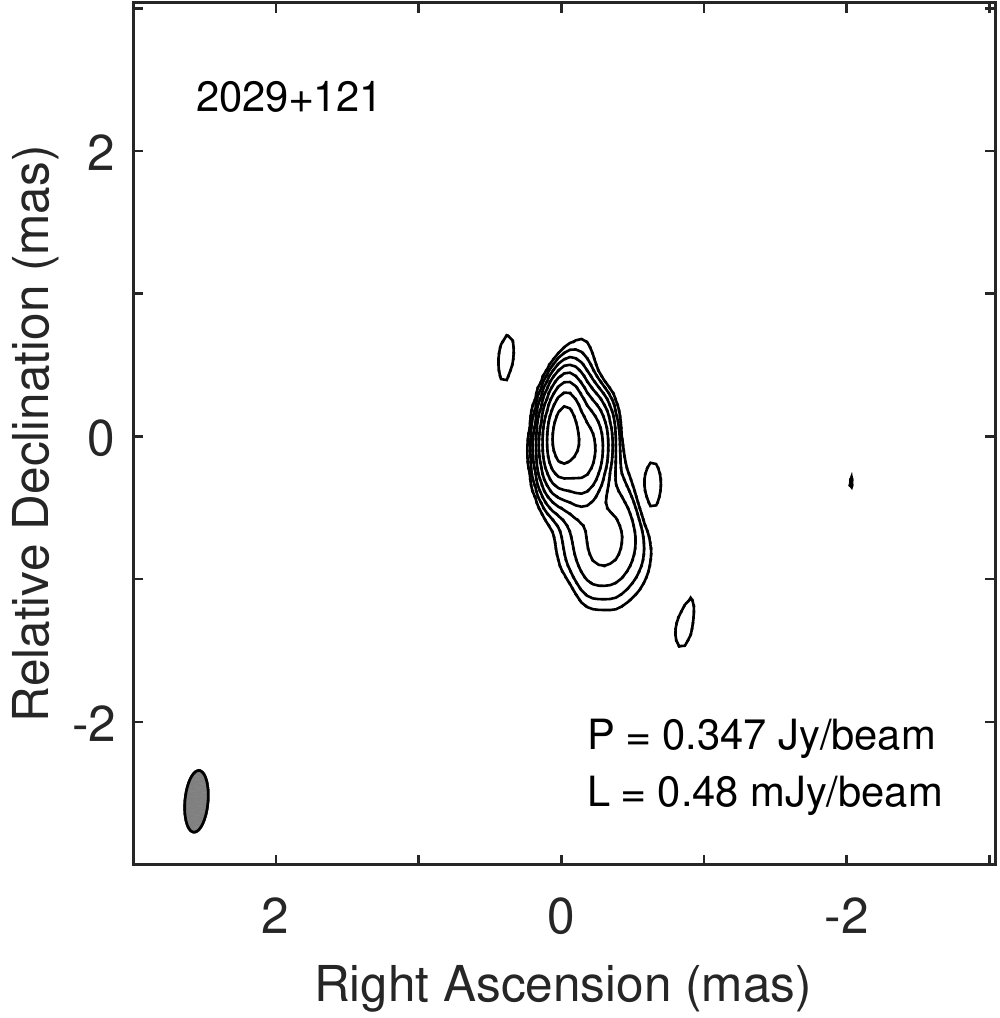}\\ \includegraphics[width=0.3\textwidth]{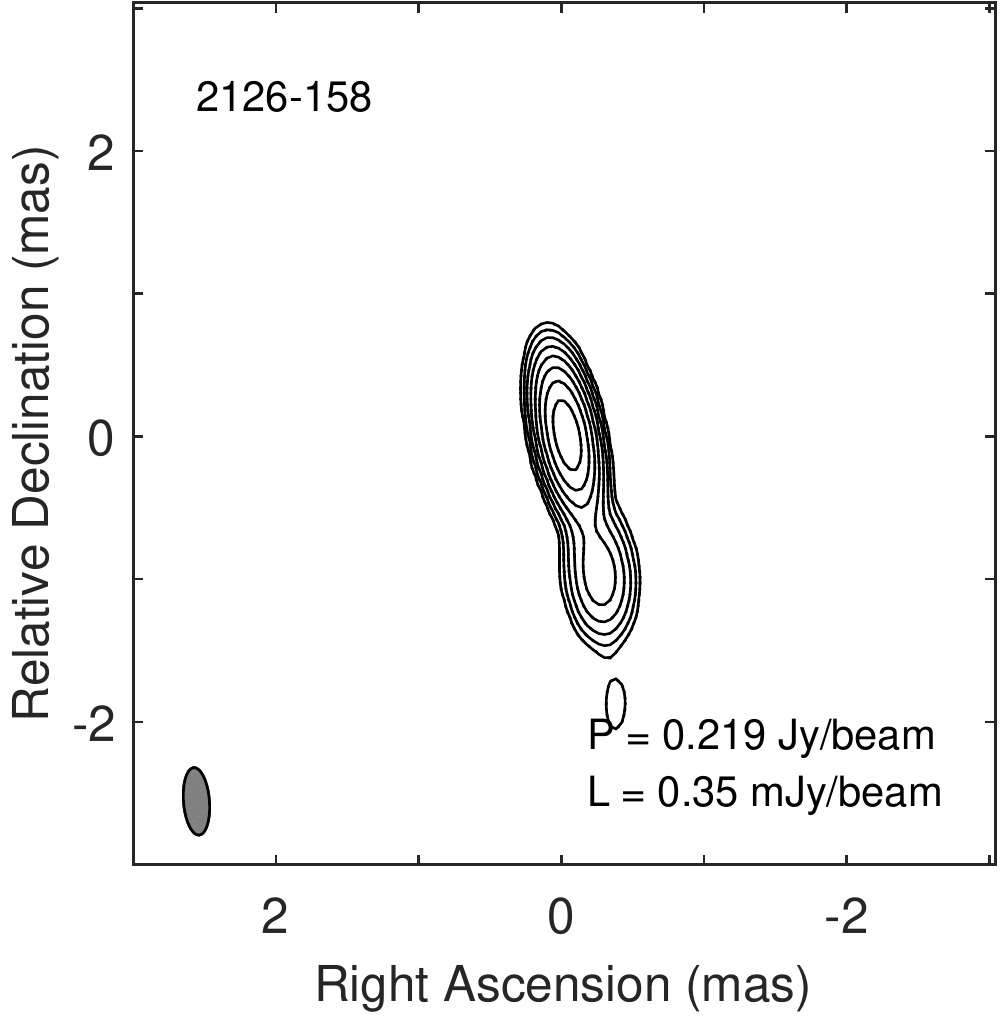} & \includegraphics[width=0.3\textwidth]{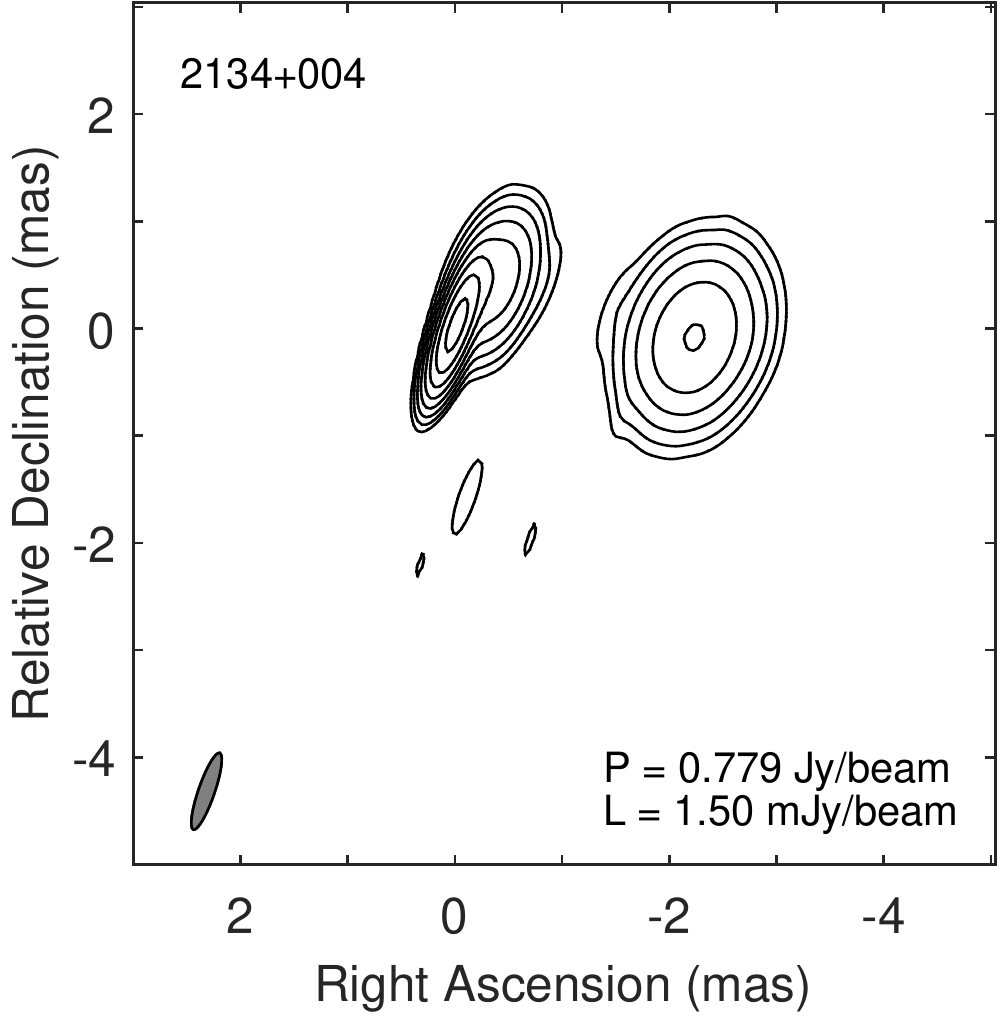}  & \includegraphics[width=0.3\textwidth]{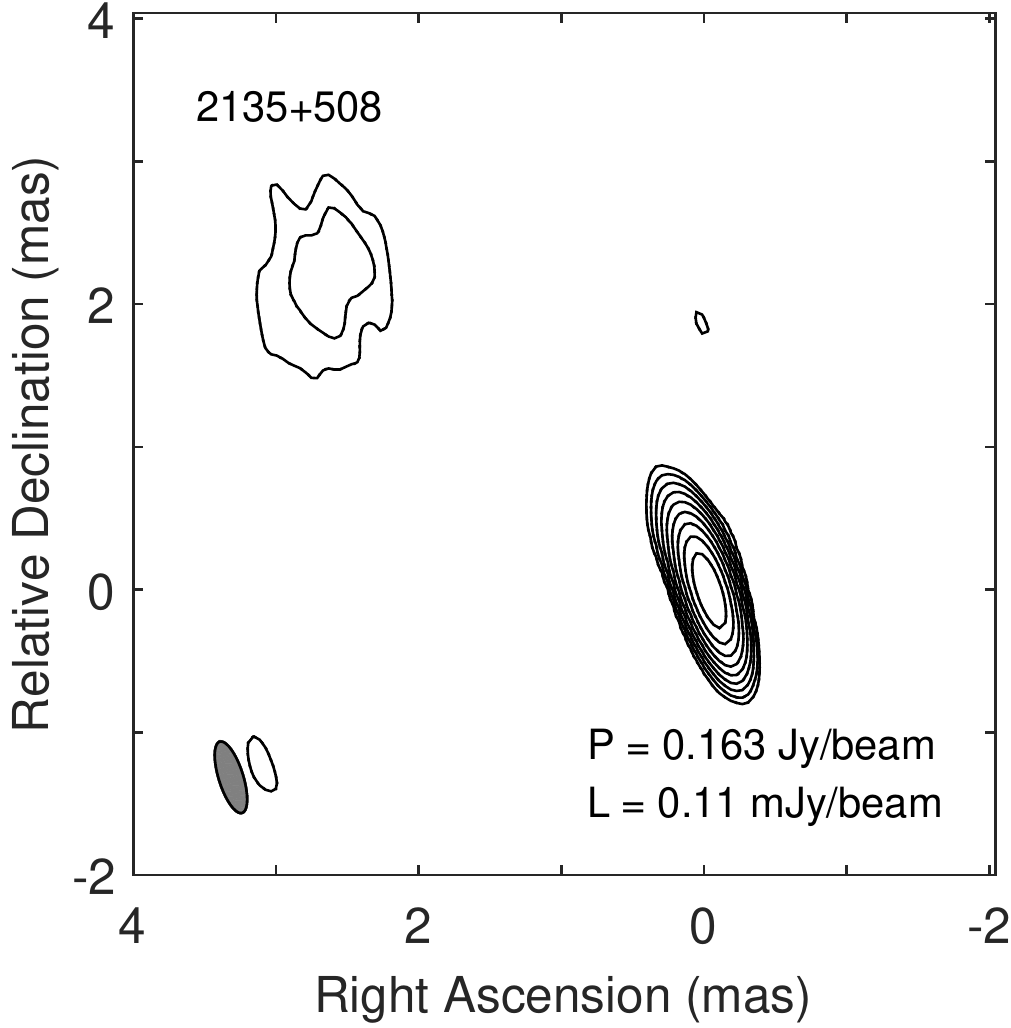}\\
  \includegraphics[width=0.3\textwidth]{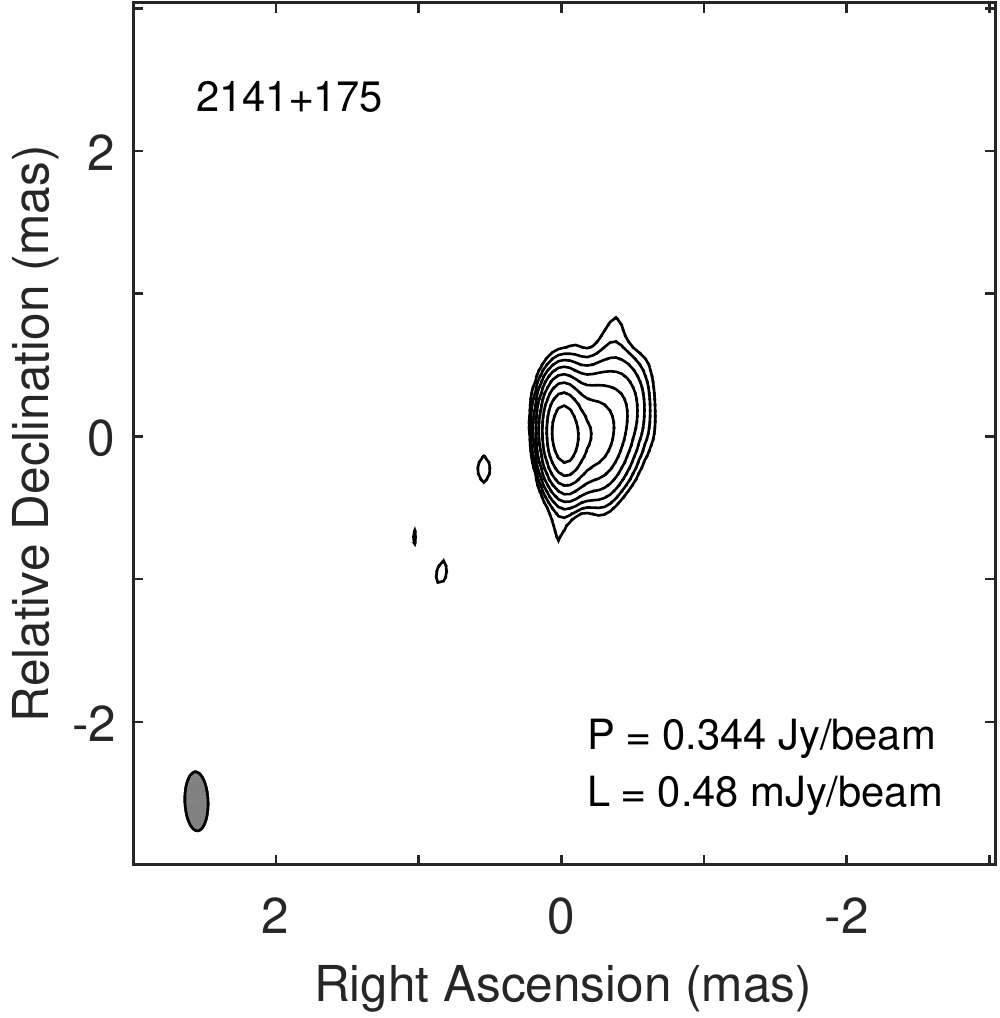} & \includegraphics[width=0.3\textwidth]{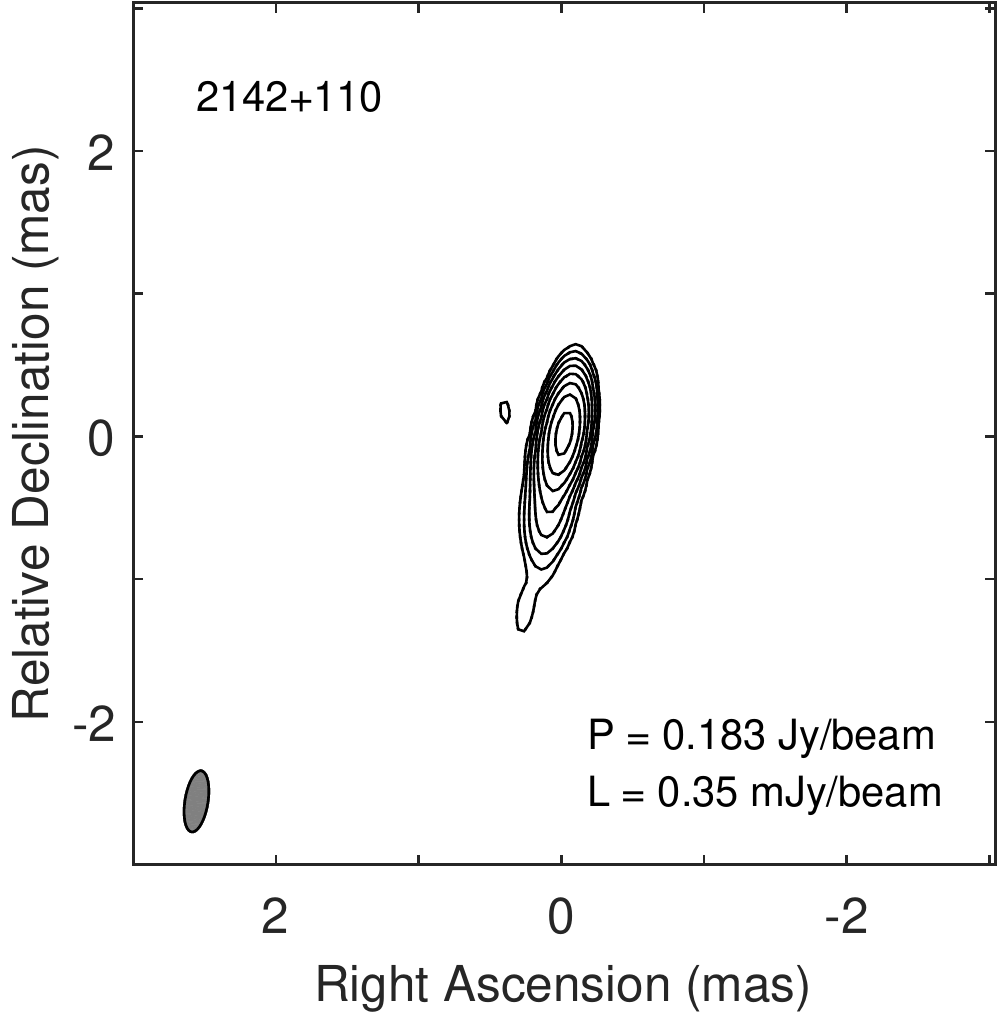}  & \includegraphics[width=0.3\textwidth]{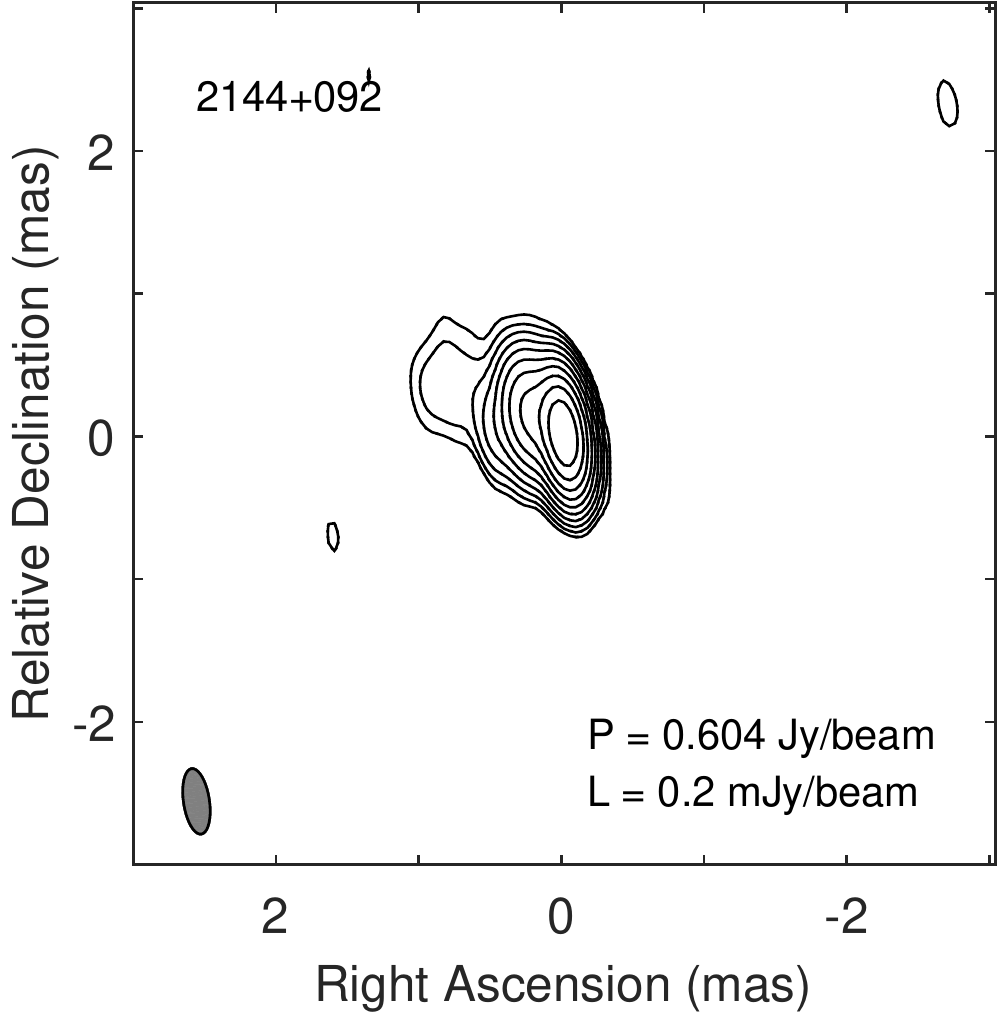}\\
    \end{tabular}
   \caption{VLBA images --- Continued}
  \end{figure}

\begin{figure}
  \centering
  \begin{tabular}{lcr}
  \includegraphics[width=0.3\textwidth]{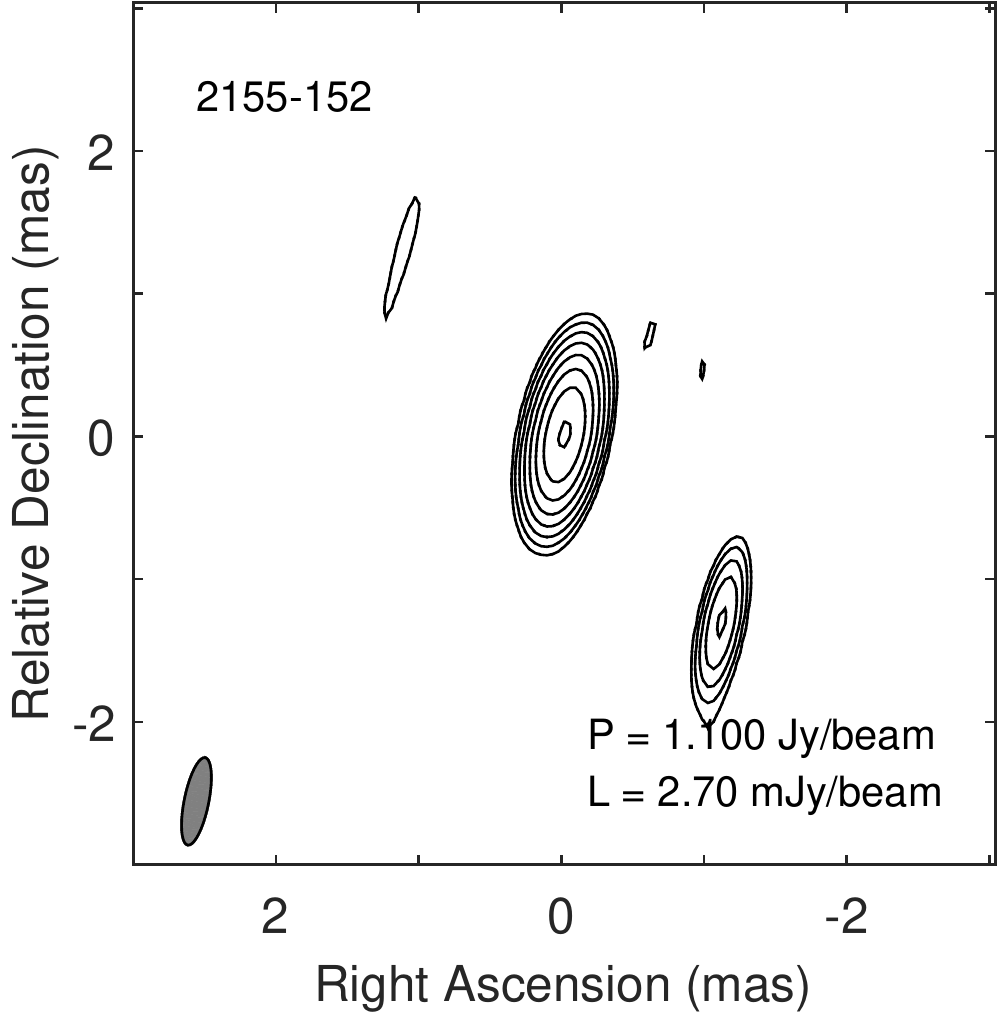} & \includegraphics[width=0.3\textwidth]{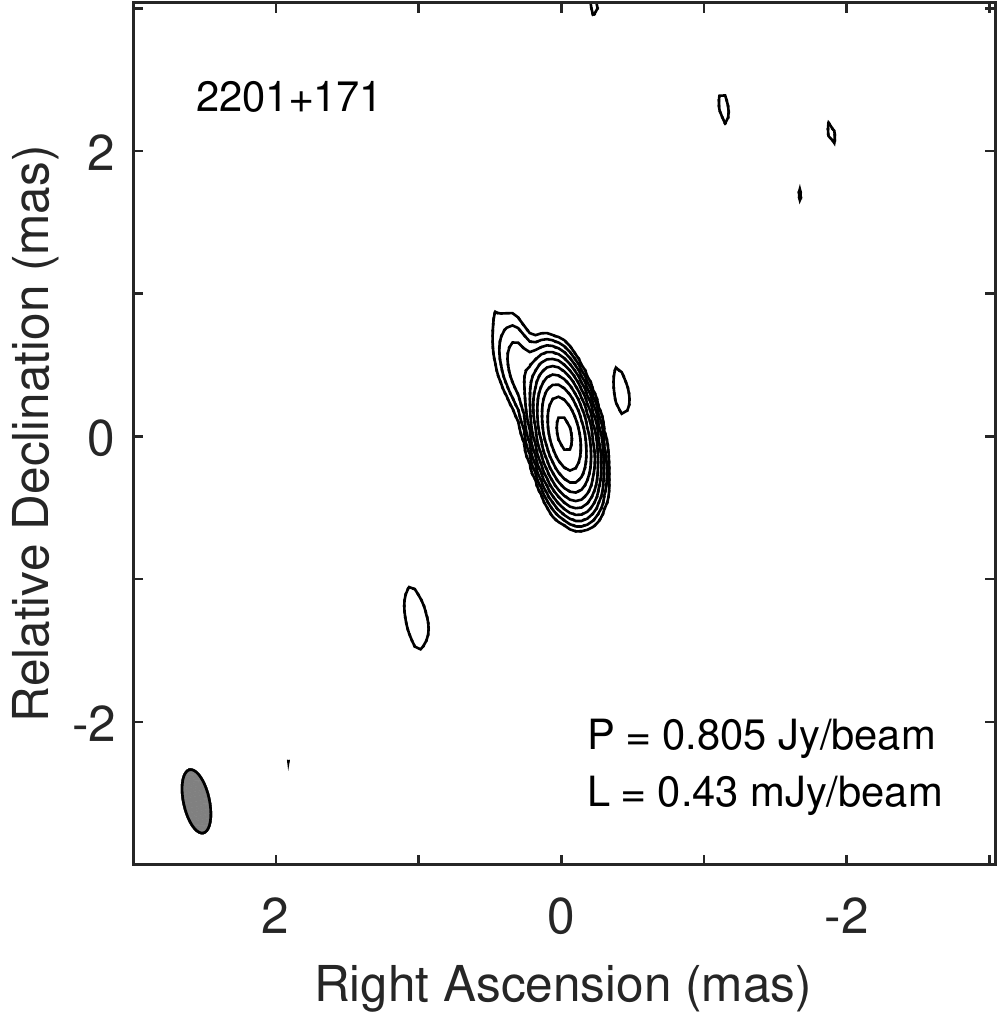}  & \includegraphics[width=0.3\textwidth]{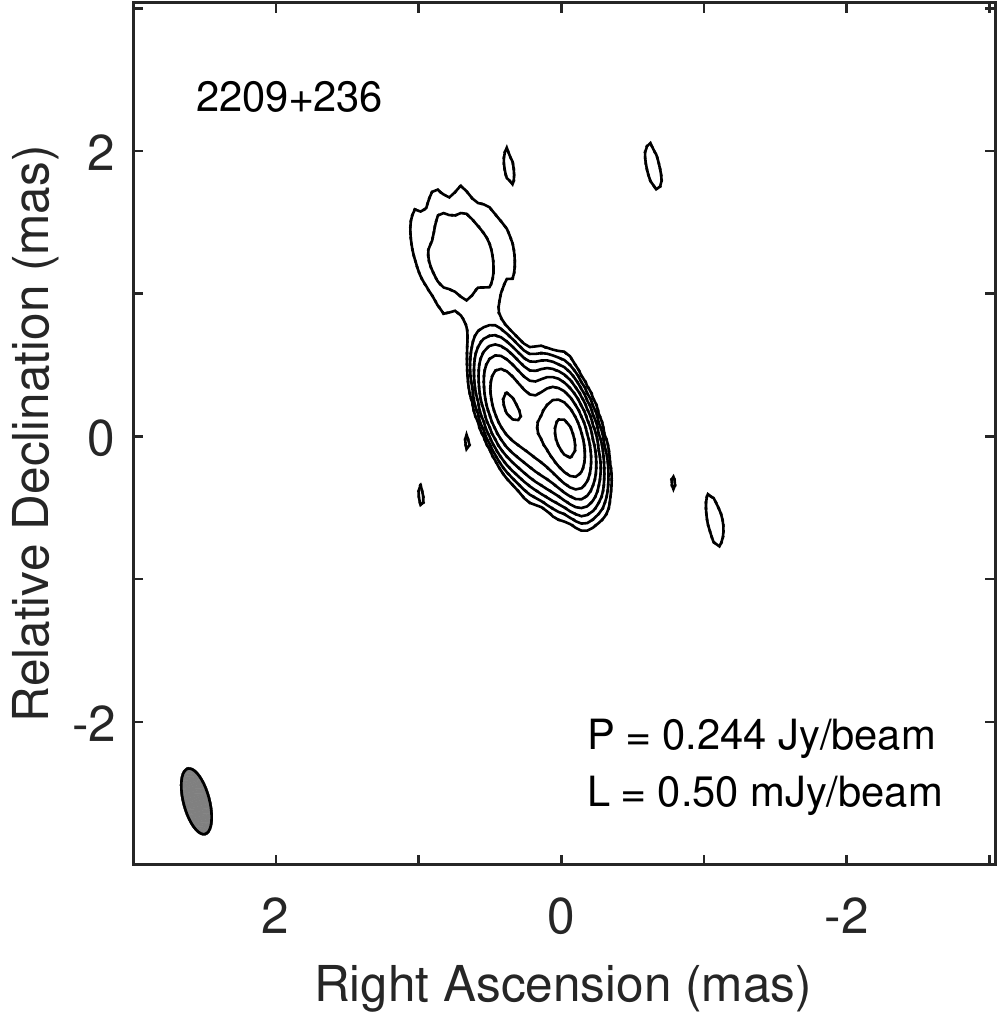}\\
  \includegraphics[width=0.3\textwidth]{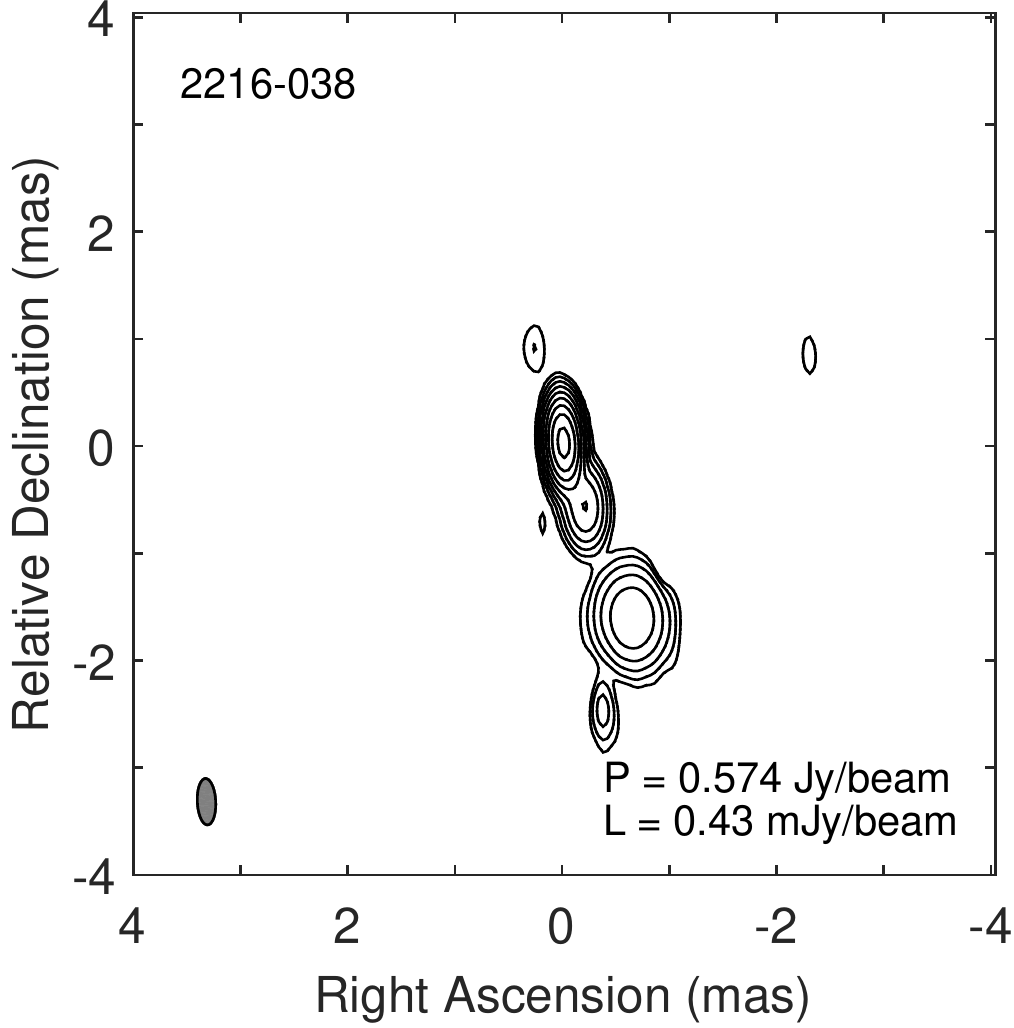} & \includegraphics[width=0.3\textwidth]{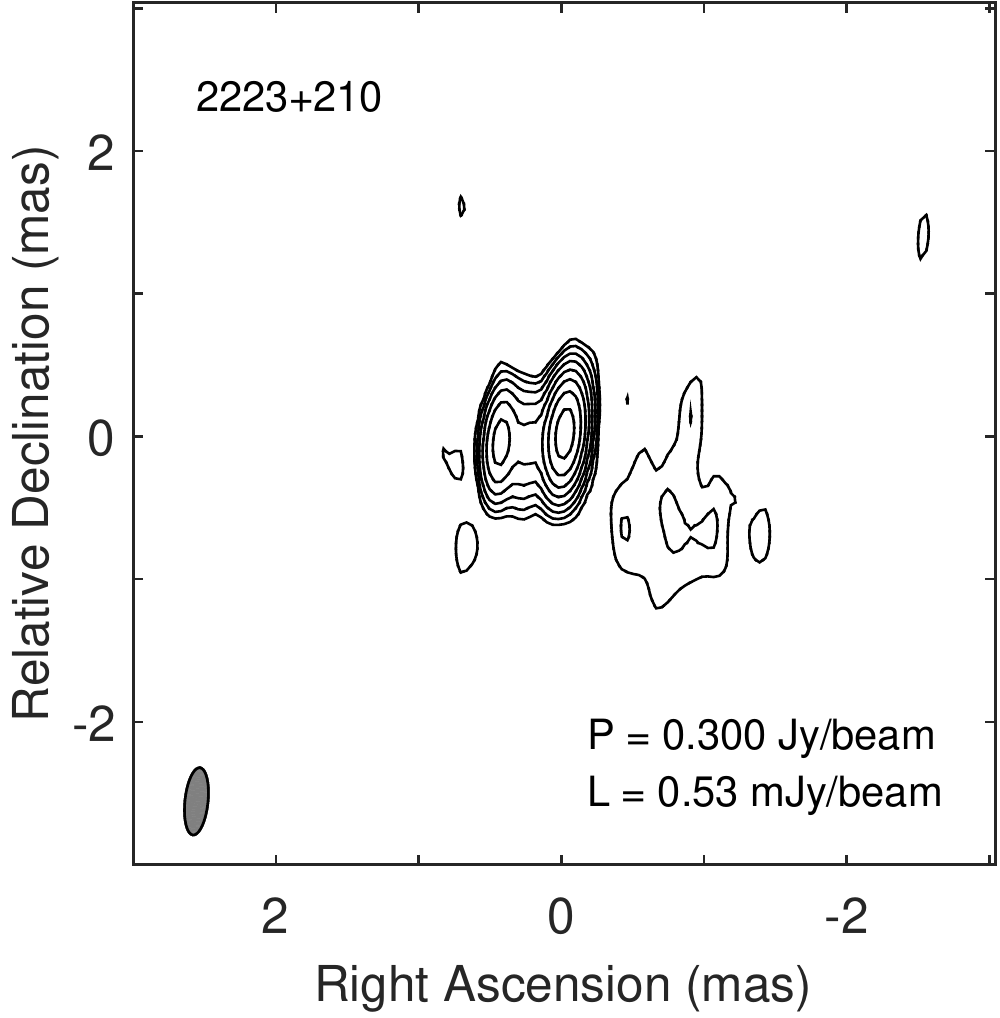}  & \includegraphics[width=0.3\textwidth]{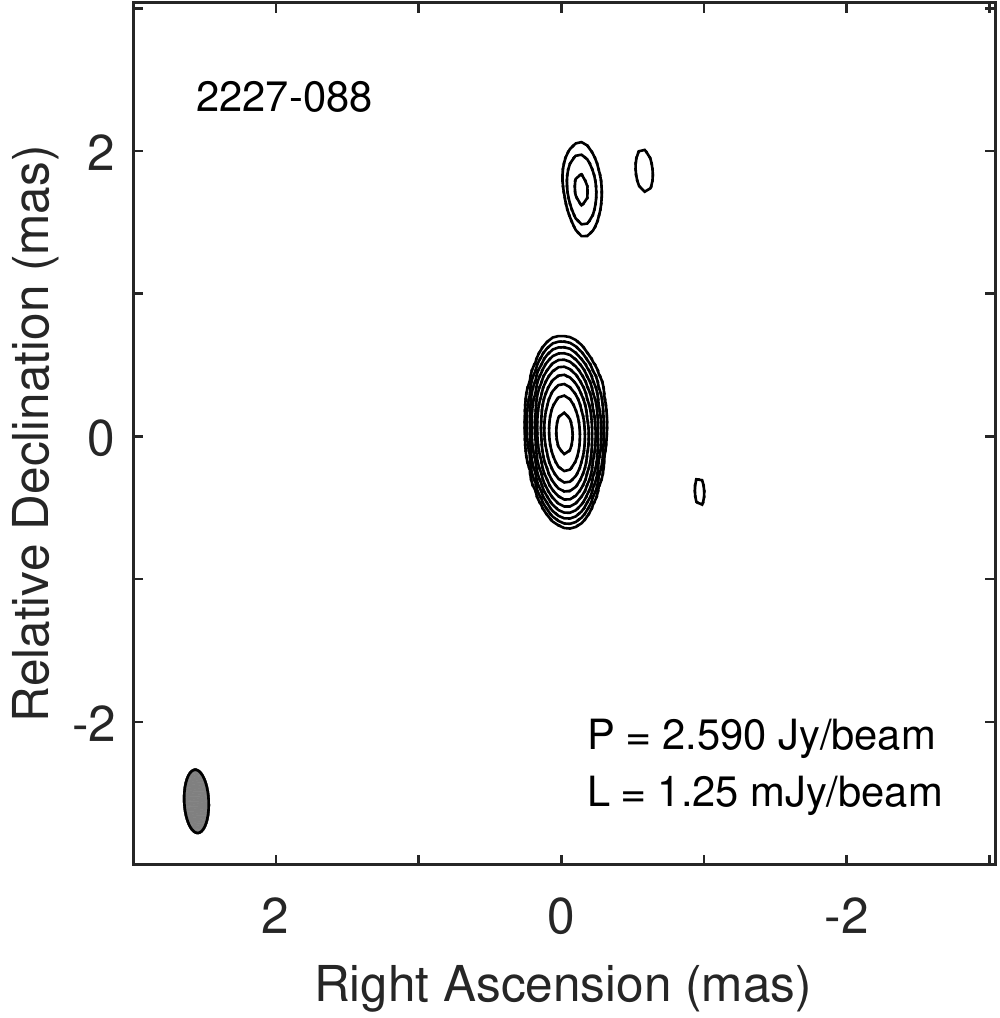}\\ \includegraphics[width=0.3\textwidth]{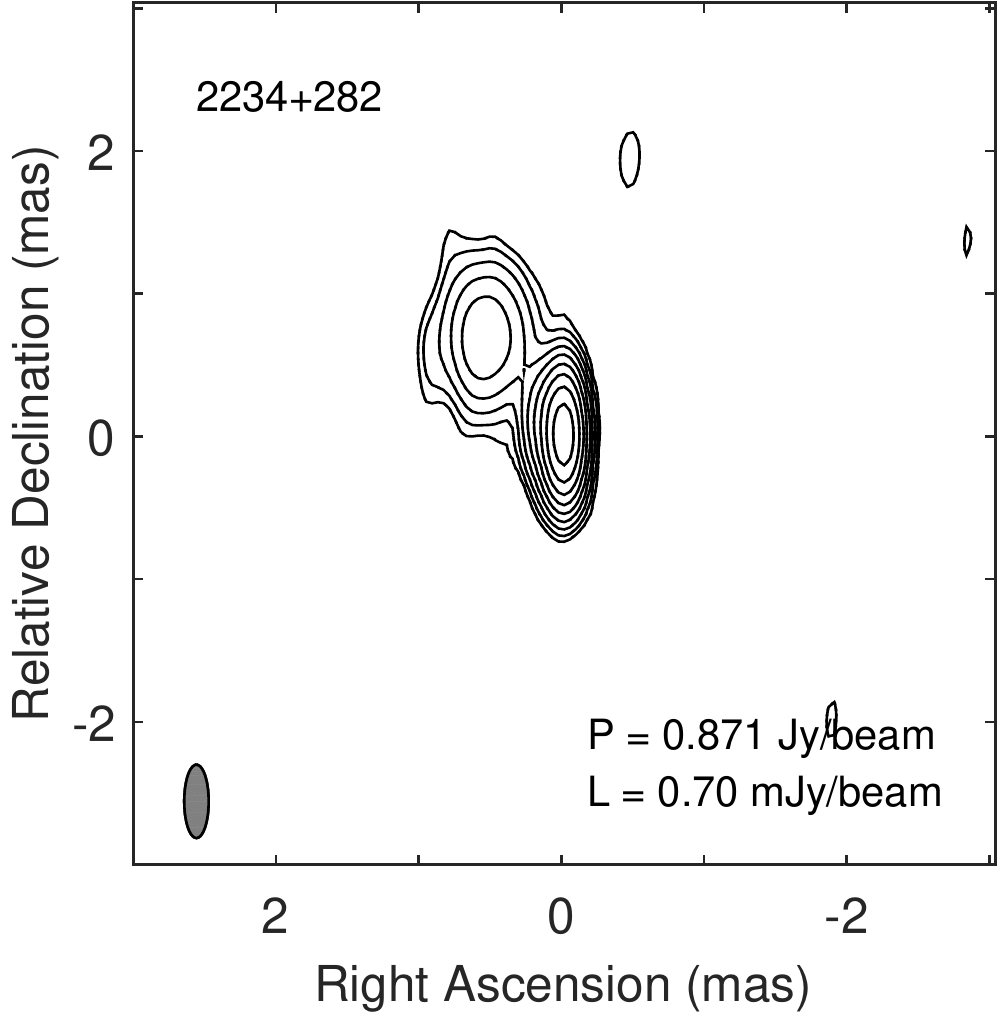} & \includegraphics[width=0.3\textwidth]{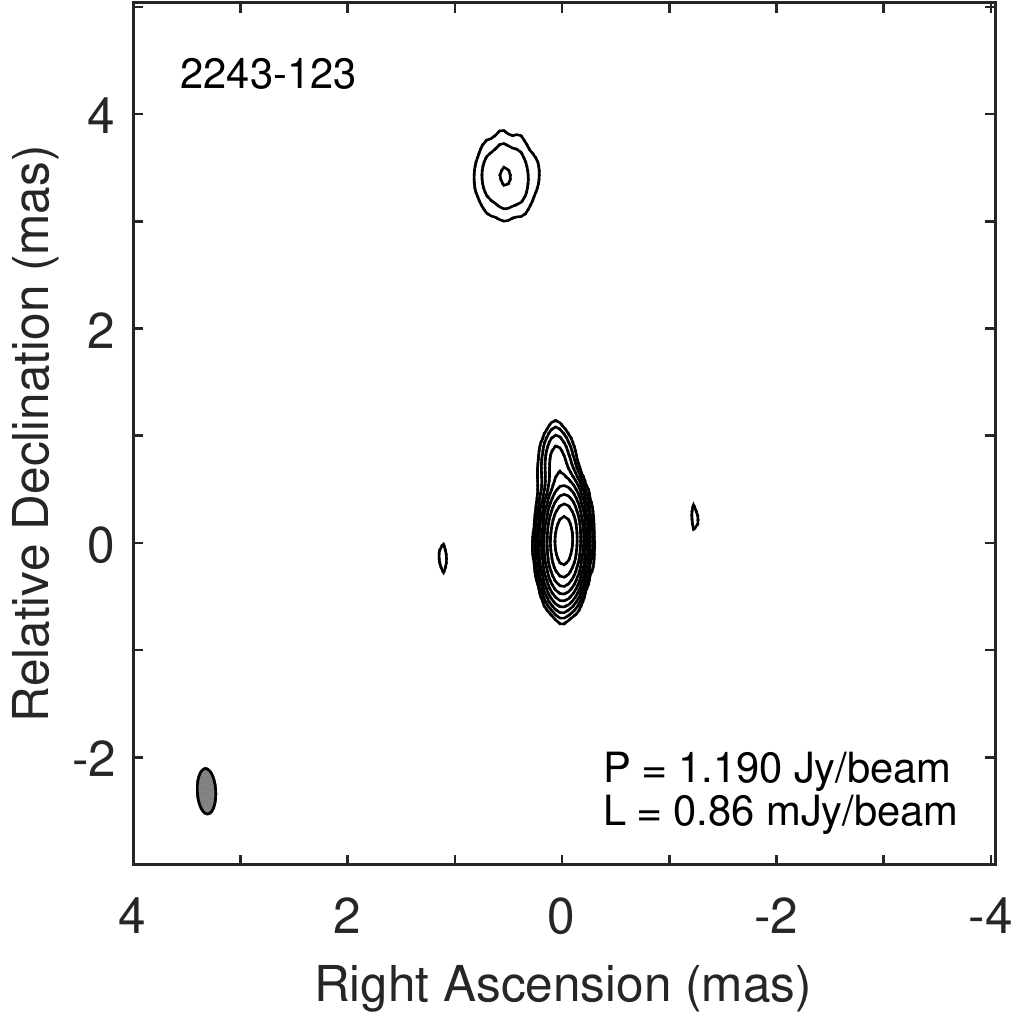}  & \includegraphics[width=0.3\textwidth]{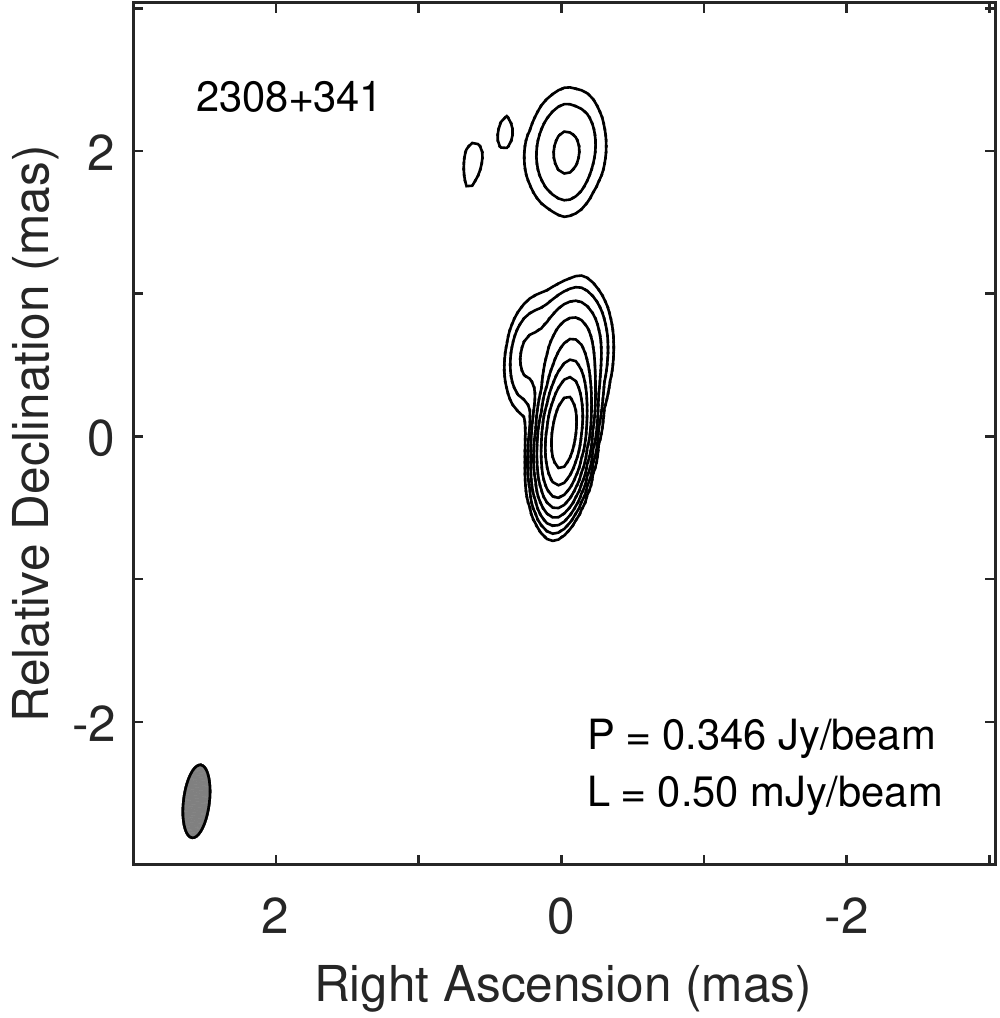}\\
  \includegraphics[width=0.3\textwidth]{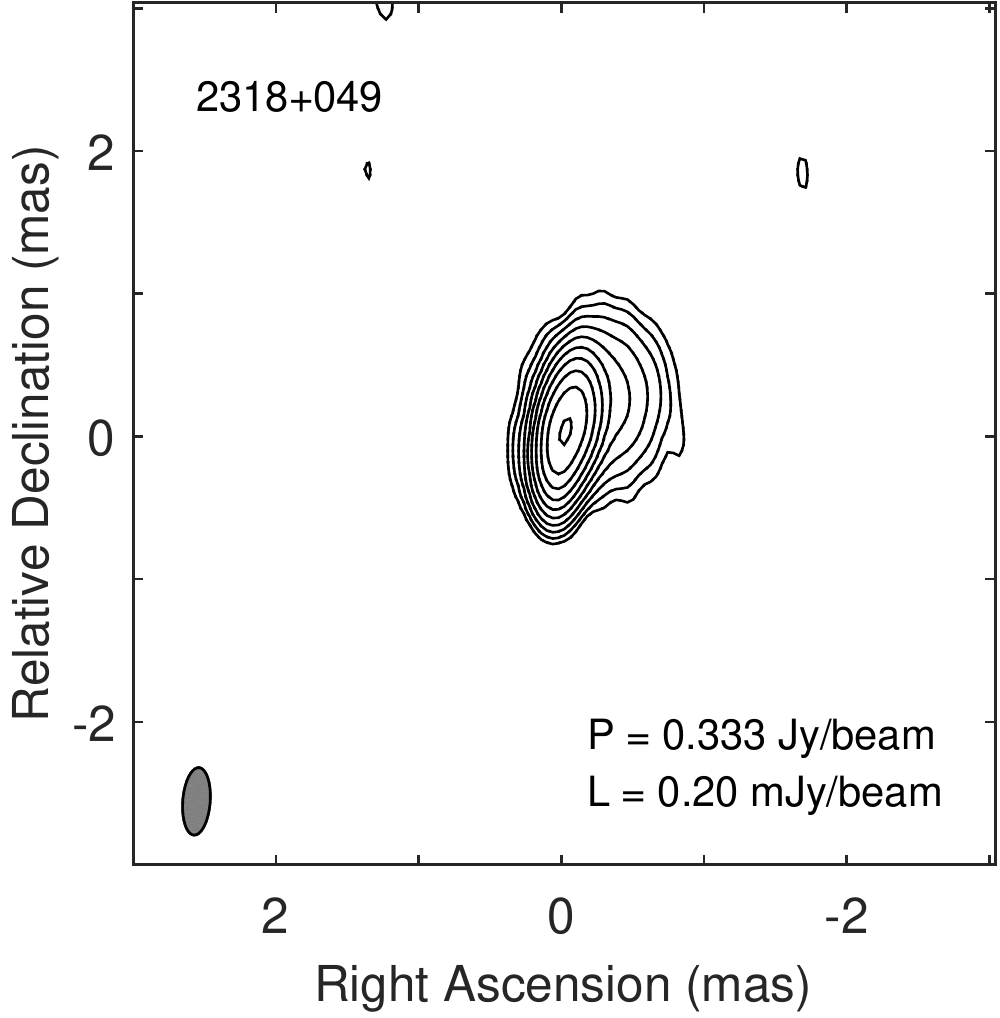} & \includegraphics[width=0.3\textwidth]{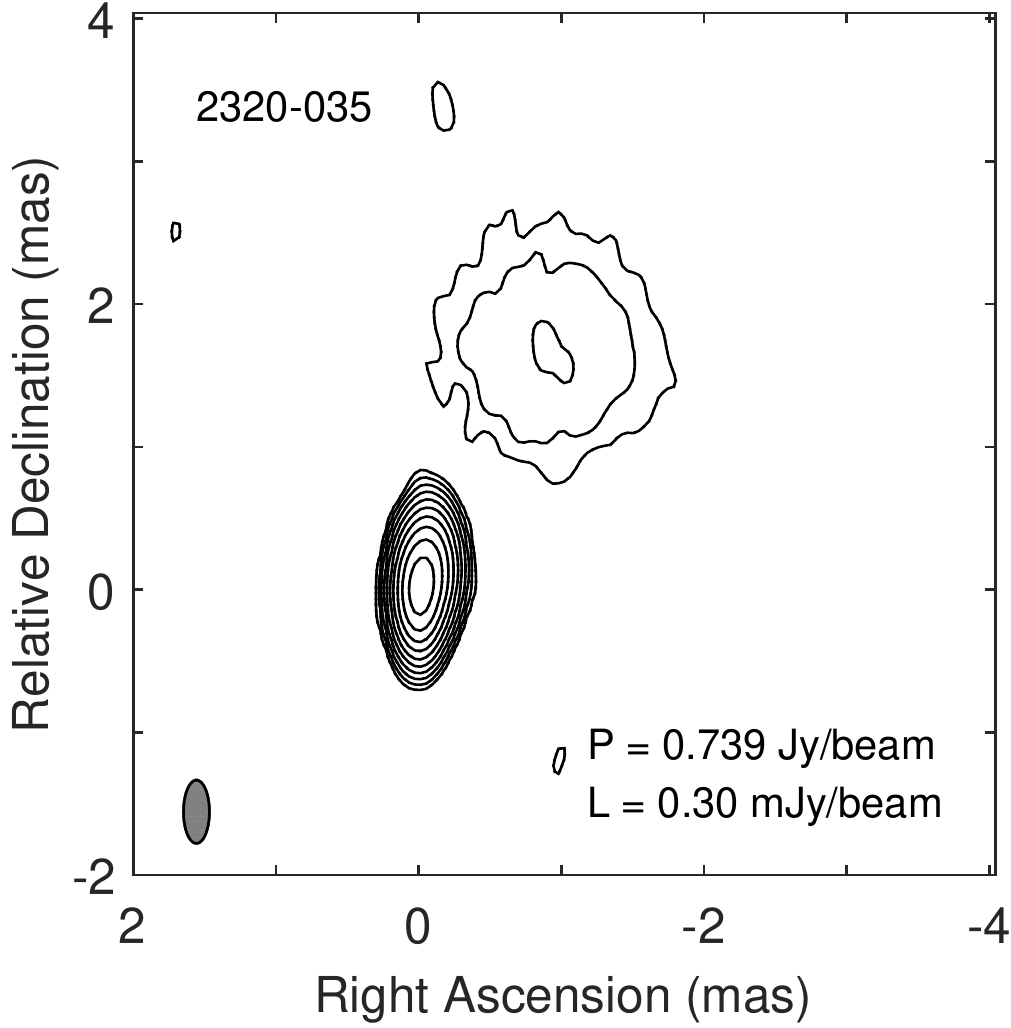}  & \includegraphics[width=0.3\textwidth]{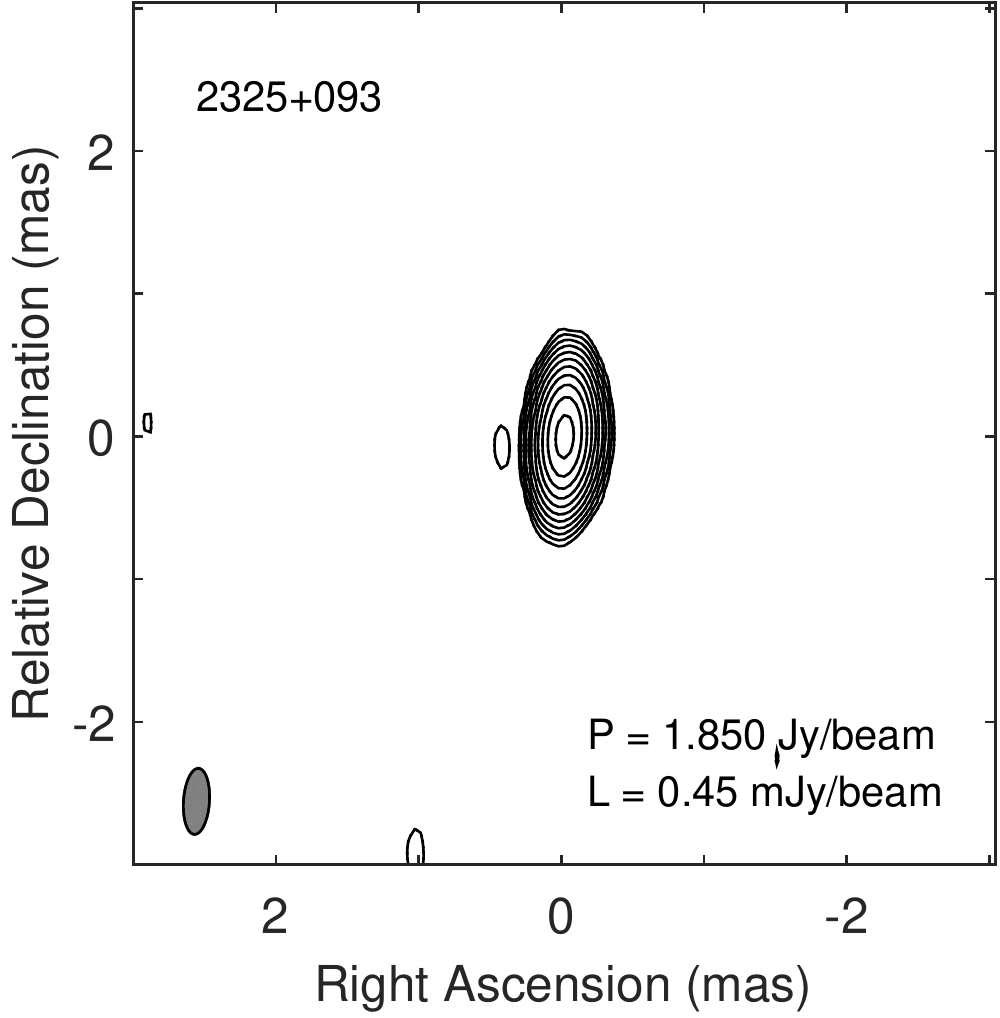}\\
    \end{tabular}
   \caption{VLBA images --- Continued}
  \end{figure}

\begin{figure}
  \centering
  \begin{tabular}{lcr}
  \includegraphics[width=0.3\textwidth]{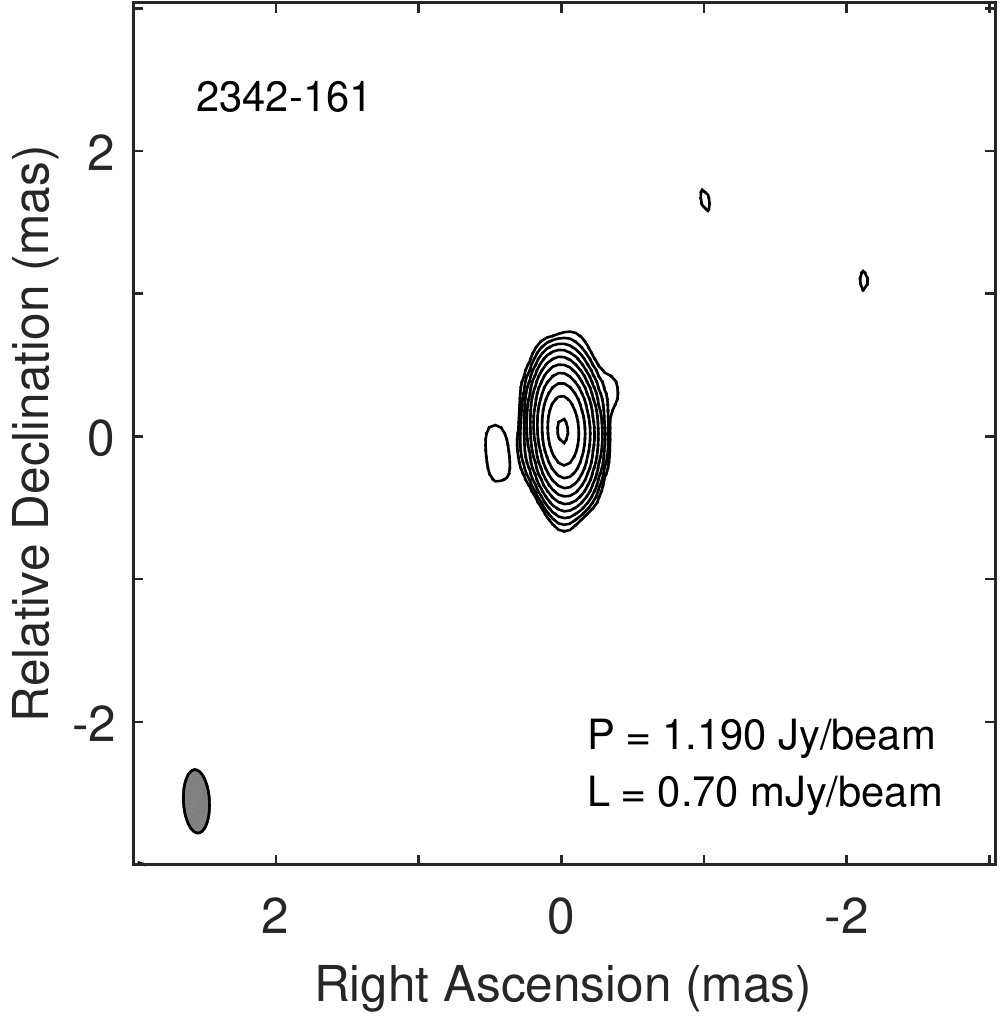} & \includegraphics[width=0.3\textwidth]{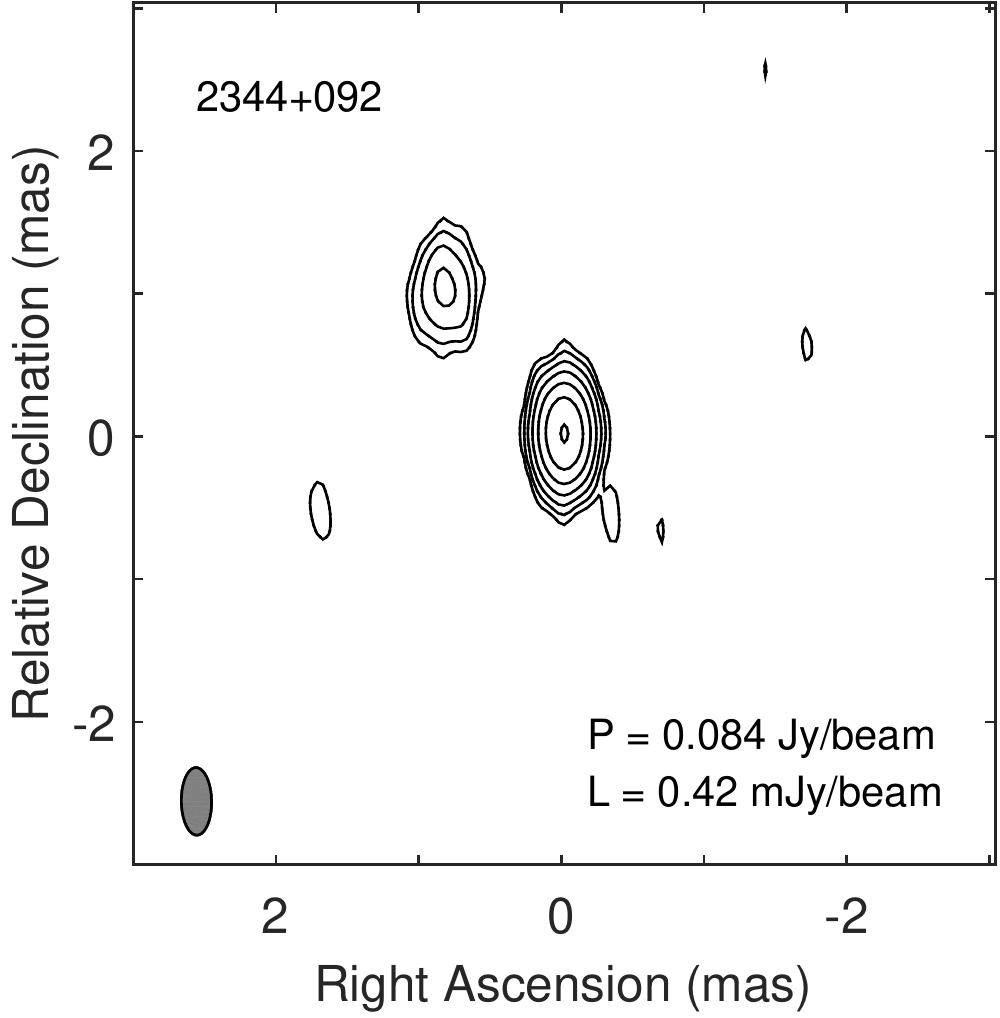}  & \includegraphics[width=0.3\textwidth]{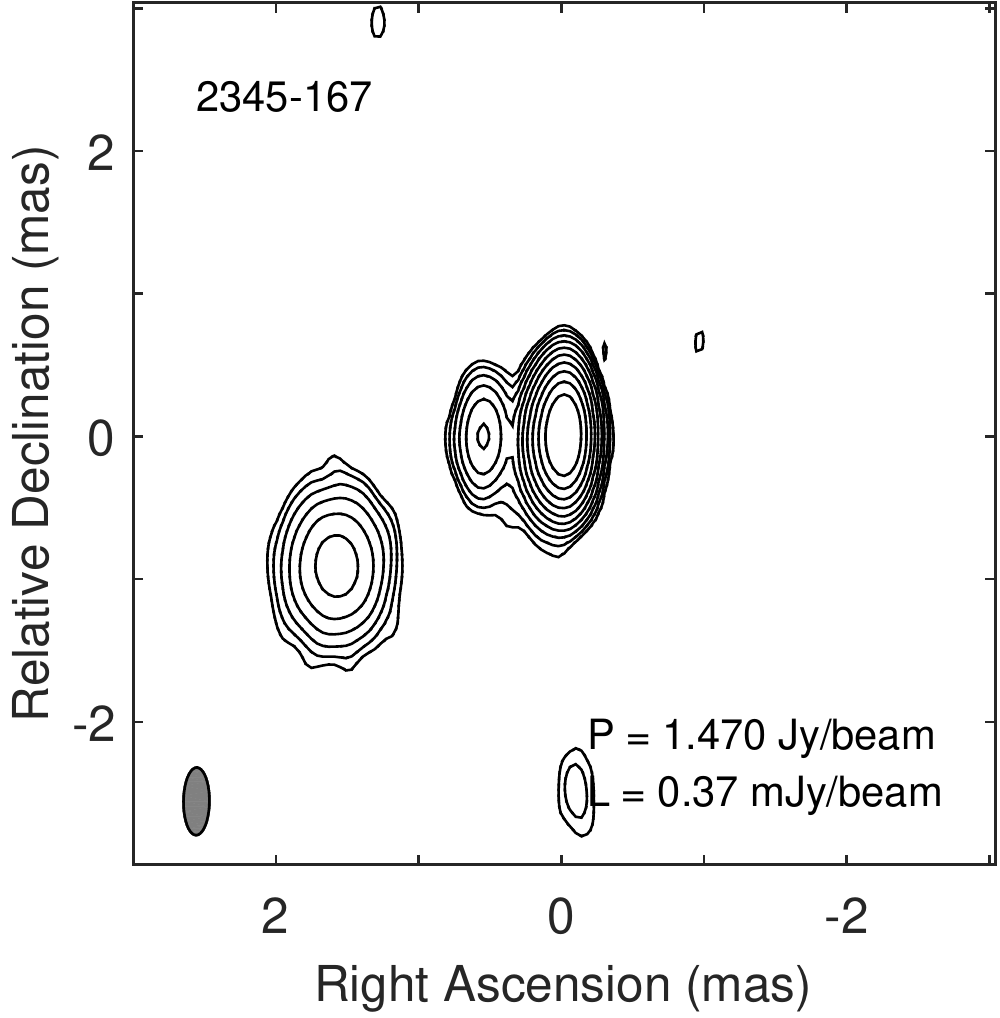}\\
  \includegraphics[width=0.3\textwidth]{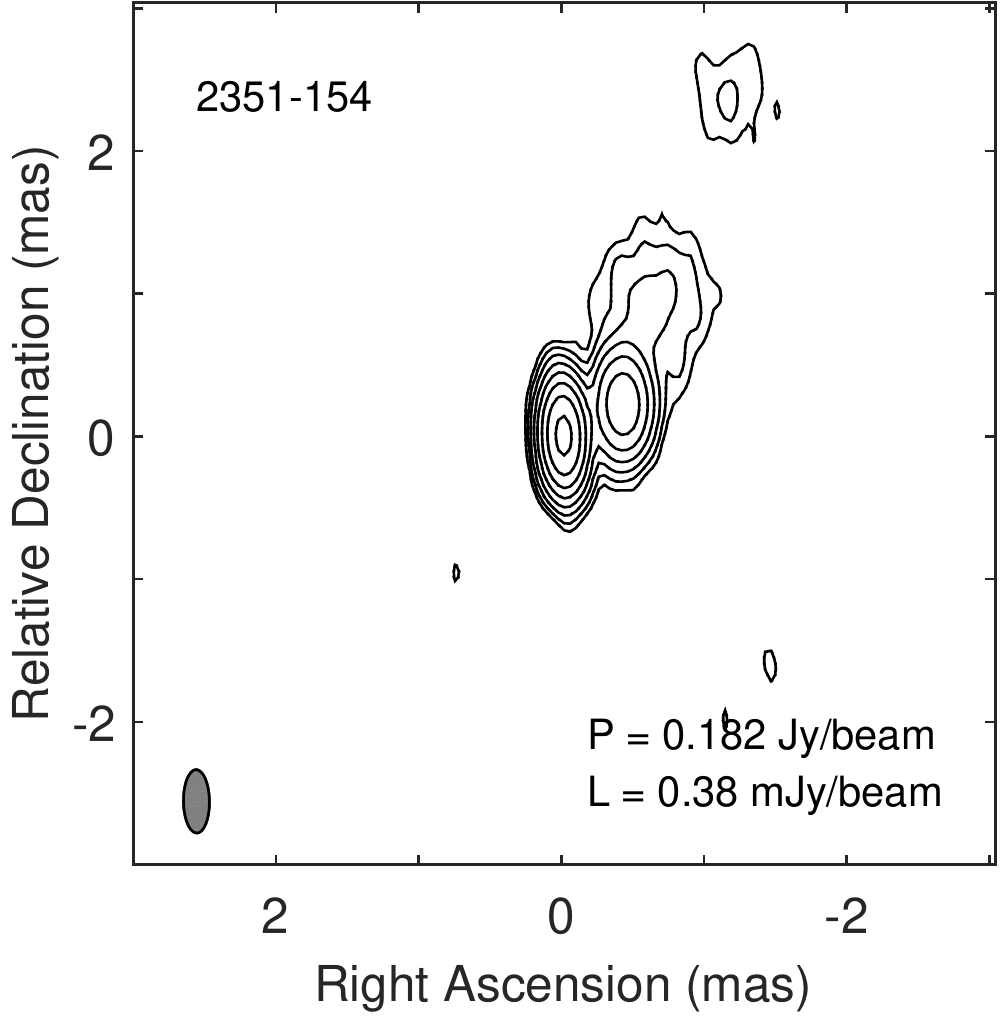} \\
    \end{tabular}
  \caption{VLBA images --- Continued}
  \end{figure}

\end{document}